%% file: 43511corr.tex
\documentclass[longauth]{aa}  
\usepackage{graphicx}
\usepackage{txfonts}
\usepackage{hyperref}
\hypersetup{
    colorlinks=true,
    linkcolor=blue,
    urlcolor=magenta,
}
\usepackage{listings}

\newcommand{\Gaia}{{\it Gaia}}
\newcommand{\gspspec}{{\it GSP-Spec}}
\newcommand{\gspphot}{{\it GSP-Phot}}

\newcommand{\T}{$T_{\rm eff}$}
\newcommand{\g}{log($g$)}

\newcommand{\meta}{[M/H]}
\newcommand{\alphaFe}{[$\alpha$/Fe]}
\newcommand{\Vrad}{$V_{\rm Rad}$}

\newcommand{\XFe}{[X/Fe]}
\newcommand{\NFe}{[N/Fe]}
\newcommand{\MgFe}{[Mg/Fe]}
\newcommand{\SiFe}{[Si/Fe]}
\newcommand{\SFe}{[S/Fe]}
\newcommand{\CaFe}{[Ca/Fe]}
\newcommand{\TiFe}{[Ti/Fe]}
\newcommand{\CrFe}{[Cr/Fe]}
\newcommand{\NiFe}{[Ni/Fe]}
\newcommand{\ZrFe}{[Zr/Fe]}
\newcommand{\CeFe}{[Ce/Fe]}
\newcommand{\NdFe}{[Nd/Fe]}

\newcommand{\zmax}{{\rm Z}$_{max}$}
\newcommand{\vphi}{{\rm V}$_\phi$}

\newcommand{\orcit}[1]{\protect\href{https://orcid.org/#1}{\protect\includegraphics[width=8pt]{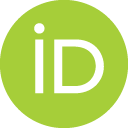}}}

\makeatletter
\renewcommand*\maketitle{%
  \thispagestyle{firstpage}
\begingroup
    \if@wideboxfn
    \setlength\bibindent{1.4\parindent}
    \else
    \setlength\bibindent{\parindent}
    \fi
    \renewcommand*\thefootnote{\@fnsymbol\c@footnote}%
    \renewcommand\@makefntext[1]{%
    \ifaa@longfn\hsize\textwidth\fi
    \noindent
    \hb@xt@\bibindent{\hss\@makefnmark\enspace}##1}
  \ifaa@twocolumn
  \begingroup
    \begin{aa@strip}
          \aa@maketitle
    \end{aa@strip}
    \@thanks            
  \endgroup
  \else
    \begingroup
      \let\thanks\footnote
      \aa@maketitle
    \endgroup
  \fi
\endgroup
  \setcounter{footnote}{0}%
}
\makeatother
\usepackage[normalem]{ulem}

\begin{document}

\definecolor{dkgreen}{rgb}{0,0.6,0}
\definecolor{gray}{rgb}{0.5,0.5,0.5}
\definecolor{mauve}{rgb}{0.58,0,0.82}

   \title{\Gaia\ Data Release 3: Chemical cartography of the Milky Way}

\input{authors_d_nosp}

\date{Received February ??, 2022; accepted ?? ??, 2022}

 
  \abstract
   {The motion of stars  has been used to reveal details of the complex history of the Milky Way, in constant interaction with its environment. Nevertheless, to reconstruct the Galactic history puzzle in its entirety, the chemo-physical characterisation of stars is essential. Previous \Gaia\ data releases were supported  by a smaller, heterogeneous,  and spatially biased mixture of chemical data from ground-based observations. 
   }
   {\Gaia\ Data Release~3 opens a new era of all-sky spectral analysis of  stellar populations thanks to the  nearly 5.6 million stars observed by the Radial Velocity Spectrometer (RVS) and parametrised by the \gspspec\ module. In this work, we aim to demonstrate the scientific quality of \Gaia’s Milky Way chemical cartography through a chemo-dynamical
analysis of disc and halo populations.}
   {Stellar atmospheric parameters and chemical
abundances provided by \Gaia\ DR3 spectroscopy are combined with DR3 radial velocities and EDR3 astrometry to analyse the relationships between chemistry and Milky Way structure, stellar kinematics, and orbital parameters.}
   {The all-sky \Gaia\ chemical cartography allows a powerful and precise chemo-dynamical view of the Milky Way with unprecedented spatial coverage and statistical robustness. First, it reveals  the strong vertical symmetry of the Galaxy and the flared structure of the disc. Second, the observed kinematic disturbances of the disc ---seen as phase space correlations--- and kinematic or orbital substructures are associated with chemical patterns that favour stars with enhanced metallicities and lower \alphaFe\ abundance ratios compared to the median values in the radial distributions. This is detected both for young objects that trace the spiral arms and  older populations.  Several $\alpha$,  iron-peak elements and at least one heavy element trace the thin and thick disc properties in the solar cylinder. Third, young disc stars show a recent chemical impoverishment in several elements. Fourth, the largest chemo-dynamical sample of open clusters analysed so far shows a steepening of the radial metallicity gradient with age, which is also observed in the young field population.
   Finally, the \Gaia\ chemical data have the required coverage and precision to unveil galaxy accretion debris and heated disc stars on halo orbits through their \alphaFe\ ratio, and to allow the study of the chemo-dynamical properties of globular clusters. }
   {\Gaia\ DR3 chemo-dynamical diagnostics open new horizons before the era of ground-based wide-field spectroscopic surveys. They unveil a complex Milky Way that is the outcome of an eventful evolution, shaping it to the present day.}


   \maketitle
   \titlerunning{Chemical cartography of the Milky Way}
   \authorrunning{Gaia Collaboration, Recio-Blanco et al.}   
%

\section{Introduction}

The European Space Agency \Gaia\ mission has transformed our understanding of the Milky Way, thanks to its ability to trace the motion of stars in the sky \citep[][]{EDR3}. The observation of these movements has allowed us to see the Galaxy as an evolving system. Components that were previously thought to be distinct (the thin disc in the Galactic plane with ongoing star formation, the more diffuse and older thick disc, the central bulge, and the extended stellar halo) now appear to be interlinked formation phases of a system in clear interaction with its environment. In particular, studies of stellar orbits and kinematics have uncovered a considerable proportion of merger debris in the halo \citep[e.g.][and references therein]{Helmi2018, Belokurov18, Malhan18, Myeong19, Helmi_20} and the Galactic disc \citep[e.g.][]{Sestito2020, ReFiorentin2021}. Additionally, investigations of stellar motions have revealed the Galactic disc phase mixing process, which is subsequent to a mildly disturbed state \citep[][]{Antoja2018}. A massive disc-crossing perturber  \citep[e.g.][]{ BinneySchoenrich18} ---possibly akin to the Sagittarius dwarf galaxy \citep{Laporte19Sag, BlandHawthorn19}--- or a strong buckling of the stellar bar \citep[e.g.][]{Khoperskov20} are the most likely interpretations of this phenomenon. A recent or ongoing encounter with a satellite galaxy seems also to be responsible for the rapidly precessing disc warp \citep[][]{PoggioWarp,PoggioLaporte}. In summary, the picture of a  `living and breathing Galaxy' has emerged thanks to \Gaia\ data \citep{BelokurovNat,Brown21}.

Despite the above-mentioned transformational results, stellar motions alone do not allow a complete reconstruction of the intricate puzzle of Galactic history. 
The orbit of a star evolves in response to fluctuations in the Galaxy’s gravitational field \citep[e.g.][]{Sellwood2002}. As a consequence, the reconstruction of
the history of the Milky Way based on the interpretation of stellar motions in terms of evolutionary processes  is hampered by degenerate explanations and the complex interplay of different physical mechanisms.

Indeed, understanding how galaxies like the Milky Way form and evolve remains a challenge. In the cold dark matter Universe, galaxies grow through a sequence of merger and accretion events. However, the impact of these events on the  evolution of a galaxy is extremely difficult to predict because of the complex physics of baryons. As a consequence, studying the chemo-physical properties of matter is essential to comprehend the Galaxy in which we live. Fortunately, we have a powerful tool at our disposal:  stellar spectroscopy. 

The study of stellar spectra gave rise to the science of astrophysics in the 19th century \citep[e.g.][]{Huggins1864, Huggins1899} and, since then, the varying characteristics of spectral absorption lines have been used by researchers to decipher the physical properties of stars \citep[][]{Maury1897,Cannon1928}. Stellar parametrisation became a powerful decryption tool, allowing us to unveil the chemical composition of stellar atmospheres \citep[][]{PaynePhD, Payne1925}, and to provide observational evidence of the stellar nucleosynthesis theory \citep{Burbidge57}. 
Stars form during the collapse of molecular clouds of gas and dust. 
Like alchemists, stars of different masses synthesise all chemical elements except hydrogen\footnote{More particularly, the Big Bang nucleosynthesis produced H, He and Li, cosmic rays contribute to Li, Be and B production and stellar nucleosynthesis concerns all chemical elements except H.}; they partially return them in the later stages of their life into the interstellar medium, from which new stars are born. As a consequence, the stellar chemical composition evolves from one generation to the next, and reflects the gas conditions at the time and place of the formation of a star. Moreover, contrary to stellar motions, the chemical abundances of a star's atmosphere are conserved\footnote{With the exception of some chemical species whose surface abundance can be modified in certain stellar evolution phases.} from its birth, and can therefore be used to trace its origin. Chemical abundances break degenerated dynamical scenarios with a variety of conserved parameters \citep[e.g. they play a key role in merger debris studies; e.g.][]{Helmi_20}. 
Therefore, stellar atmospheres record the past in their chemical abundances, allowing a look-back time that varies between a few hundred million years and the age of the first stars in the Universe.

In this framework, previous intermediate \Gaia\ data releases had to be complemented with chemical data from ground-based observations. However, ground-based spectroscopic surveys like  GALAH \citep[e.g.][]{Buder21}, APOGEE \citep{APOGEE17}, \Gaia-ESO Survey \citep{GilmoreGES,RandichGES}, and RAVE \citep{RAVE}, despite the recent improvement of multiplex capabilities, are still hampered by spatially biased samples. In addition, the inhomogeneity induced by different analysis procedures, targeted stellar types, and spectral configurations blur the collected chemical information. Moreover, ground-based spectroscopy suffers from time-dependent effects such as the  Earth's atmospheric absorption and instrumental systematic effects, which are difficult to model with discontinuous data collections. 

Fortunately, the context is now evolving favourably.  \Gaia\ Data Release 3 \citep[DR3;][]{GDR3} opens a new era of all-sky spectroscopy and chemo-physical characterisation of Galactic stellar populations, and includes a new transformational data set that confirms \Gaia's leading role in the golden age of Galactic archaeology: the largest homogeneous spectral analysis performed so far with a total of 5\,594\,205 stars observed by the Radial Velocity Spectrometer \citep[RVS;][]{Cropper2018,Katz22} and parametrised by the General Stellar Parametriser - spectroscopy \citep[\gspspec;][]{GSPspecDR3}. With continuous data collection for 34 months outside the Earth's atmosphere, and a large 
volume coverage reaching distances of about 8~kpc from the Sun (thanks to the population of giant stars), the \Gaia\ DR3 spectroscopic survey provides stellar parameters and chemical abundances in all major Galactic populations, sharpening our global view of the Milky Way. In addition to the sky coverage advantage, it is worth comparing this \Gaia\ DR3 \gspspec\ catalogue with  high-resolution ground-based surveys in other crucial characteristics for Milky Way studies, as the number of analysed stars, the limiting magnitude, and the explored chemical diagnostics. For magnitudes brighter than G$=$13.6\footnote{In DR3, 99\% of the \gspspec\ catalogue has Gmag$<$13.6. This will strongly evolve in future releases, reaching much fainter magnitudes thanks to the continuous and on-going \Gaia\ observations}, there are more stars in the \Gaia\ DR3 \gspspec\ catalogue than in any other ground-based survey (with both GALAH and APOGEE representing only $\sim$8\% of \Gaia\ \gspspec). For magnitudes fainter than 13.6, \Gaia\ \gspspec\ has about 61\,000 stars (reaching G$=$16.1~mag), GALAH has about 130\,000 stars (20\% of the survey, reaching G$=$18~mag) and APOGEE about 314\,000 stars (43\% of the survey, reaching G$\simeq$20~mag). Concerning the nucleosynthesis diagnostics\footnote{The following general nucleosynthetic channels are considered:  (i) cosmic rays spallation, (ii) nuclear burning in low- and intermediate-mass stars, (iii) $\alpha$-process in core collapse supernovae (massive stars), (iv) neutrino process in core collapse supernovae, (v) explosions of Type Ia supernovae (C+O white dwarfs in binary systems), (vi) slow neutron capture (s-process), and (vii) rapid neutron capture (r-process).} through individual abundance estimates, \Gaia\ DR3 \gspspec\ explores five different nucleosynthetic channels with 13 chemical elements, while GALAH covers seven nucleosynthetic channels with 21 elements, and APOGEE six channels with 24 elements.

The goal of this paper is to demonstrate the scientific quality of \Gaia's chemical cartography through a chemo-dynamical analysis of disc and halo populations. To this purpose, Sect.~\ref{Sec:Data} presents the data that are used, including (i) DR3 atmospheric parameters and chemical abundances (Sect.~\ref{Sec:APs}), (ii) DR3 radial velocities (Sect.~\ref{sec:RVs}), (iii) EDR3 astrometric data and distances (Sect.~\ref{sec:Astrometry}), (iv) a  set of stellar velocities and orbits specifically derived for this work (Sect.~\ref{Sec:Orbital_params}), and (v) the definition of working subsamples (Sect.~\ref{Sec:samples}) allowing us to optimise the scientific analysis, and illustrating the use of quality flags defined in \citet[][]{GSPspecDR3}. 

 In Sect.~\ref{Sec:GlobalView} we present the global chemical properties of the Milky Way through sky and Galactic maps (Sect.~\ref{Sec:globalMaps}) and explore selection function effects (Sect.~\ref{Sec:selFunc}). Section~\ref{Sec:Gradients} presents the radial and vertical chemical gradients of field stellar populations. In Sect.~\ref{sec:5} we present our analysis of large-scale chemo-kinematical correlations, 
while in Sect.~\ref{Sec:Dynamics} we explore the relation between the orbital parameters and  stellar chemistry. 
Subsequently, Sect.~\ref{Sec:SolarCylinder} is dedicated to chemo-dynamical relations in solar cylinder populations using individual element abundances, and in Section~\ref{Sec:clusters} we use the open clusters population to study chemo-kinematical correlations and the temporal evolution of disc radial chemical gradients.

Finally, Sect.~\ref{Sec:Summary} summarises the results of our Galactic chemo-dynamical analysis using the \Gaia\ RVS \gspspec\ database. In particular,  we discuss the observed chemical markers of Milky Way structure (Sect.~\ref{Sec:MWstruct}), disc kinematic disturbances (Sect.~\ref{Sec:ChemSeis}), and satellite accretion (Sect.~\ref{Sec:ChemMergers}). This is completed with the examination of the detected  chemo-dynamical trends of the last billion years (Sect.~\ref{Sec:1Gyr}) and, finally, with a discussion of the Sun's chemo-dynamical properties
in the context of its Galactic environment (Sect.~\ref{sec:Sun}). 
Our overall conclusions are presented in Sect.~\ref{Sec:Conclusions}.


\section{Data}
\label{Sec:Data}

\subsection{Stellar atmospheric parameters and chemical data }
\label{Sec:APs}

This work makes use of the stellar physical parameters and chemical abundances derived from the \Gaia\ RVS spectra by the  \gspspec\ module and available through the {\it astrophysical\_parameters} table of \Gaia\ DR3.  It is worth mentioning that the present work does not use the global metallicity \meta\ derived from BPRP spectra by the General Stellar Parametrizer from Photometry (\gspphot) and published in the \texttt{GaiaSource} and \texttt{AstrophysicalParameters} tables (\verb|mh_gspphot| field). Although \gspphot\ metallicities are suitable for different scientific purposes, their application to large-scale Galactic chemo-dynamical studies requires a calibration that at the time of writing this paper was not available.

\gspspec\ estimates the main atmospheric parameters (effective temperature \T , stellar surface gravity \g\footnote{$g$ being expressed in cm.s$^{-2}$}, global metallicity \meta \footnote{We adopt the standard abundance notation for a given element $X$: [X/H] $= \log{({\rm X/H})_\star} - \log{({\rm X/H})_\odot}$, where (X/H) is the abundance by number, and $\log \epsilon(X) \equiv \log{({\rm X/H})} + 12$.}, and the global abundance of $\alpha$-elements\footnote{$\alpha$-elements are O, Ne, Mg, Si, S, Ar, Ca, and Ti.} with respect to iron  \alphaFe), together with the individual abundances of 12 different chemical elements
from RVS spectra of single stars. The RVS wavelength range is $[845 - 872]$~nm, and its medium resolving power is $R=\lambda / \Delta \lambda \sim 11\,500$ \citep{Cropper2018}. 
This spectral parametrisation is based on the MatisseGauguin  \gspspec\  workflow and is described in detail in the  \gspspec\  processing paper \citep{GSPspecDR3}. It is worth recalling that the \gspspec\ \meta\ estimation considers all the non-$\alpha$ metallicity indicators in the observed spectra  thanks to a four-dimensional synthetic spectra grid including not only the \meta\ dimension but also the \alphaFe\ one. Non-$\alpha$ indicators are dominated by Fe lines in the RVS domain. As a consequence, the estimated \meta\ value follows the [Fe/H] abundance with a tight correlation.

In the following sections, \T\ is taken from the \verb|teff_gspspec| field; \g\ comes from the \verb|logg_gspspec| field; \meta\ is taken from  \verb|mh_gspspec|;  and \alphaFe\ corresponds to  \verb|alphafe_gspspec| with a calibration\footnote{It is important to note that \Gaia\ DR3 catalogue values are not calibrated.} imposed that requires a zero average value for \alphaFe\ in the solar neighbourhood for any gravity \citep[see][Table~16]{GSPspecDR3}. 

In a similar way, all the stellar individual chemical abundances come from the  \gspspec\ estimates. In particular, this paper makes use of [N/Fe] (\verb|nfe_gspspec|), [Mg/Fe] (\verb|mgfe_gspspec|), [Si/Fe] (\verb|sife_gspspec|), [S/Fe] (\verb|sfe_gspspec|), [Ca/Fe] (\verb|cafe_gspspec|), [Ti/Fe] (\verb|tife_gspspec|), [Cr/Fe] (\verb|crfe_gspspec|), [Fe/M] (\verb|fem_gspspec|), [Ni/Fe] (\verb|nife_gspspec|), and [Ce/Fe] (\verb|cefe_gspspec|).  As for the \alphaFe\ estimates, these individual abundances were calibrated following the prescriptions  indicated in Table~16 of \citet[][]{GSPspecDR3}. It is important to note here that {\it GSP-spec} assumes the reference solar abundances of \cite{Grevesse2007}.

In addition, we make use of the \gspspec\  quality flags reported in the \verb|flags_gspspec| string chain (which consists of 41 characters) defined in \cite{GSPspecDR3}. For example, we make use of the first three characters in this chain (that is, \verb|flags_gspspec[0]|,  \verb|flags_gspspec[1]|, and \verb|flags_gspspec[2]|, reporting on the degree of parameter biases from line broadening) and called \verb|vbroadT|, \verb|vbroadG,| and \verb|vbroadM,|  respectively \citep[see naming convention in][] {GSPspecDR3}. 

Finally, the uncertainty on any derived parameter (\verb|Param_unc|) or abundance (\verb|[X/Fe]_unc|) is defined as half of the difference between its upper and lower confidence levels (e.g. \verb|Teff_unc| $=$ [\verb|teff_gspspec_upper| $-$ \verb|teff_gspspec_lower|]/2).

\subsection{Radial velocities}\label{sec:RVs}

The complete \gspspec\ sample contains 5\,594\,205  stars in total (based on the query \verb|flags_gspspec| is not null). DR3 provides radial velocities (\verb|radial_velocity|, \Vrad) for 33\,834\,834 stars \citep[][]{Katz22} through the {\it gaia\_source} table.  The \gspspec\ sample is a subset of the \Vrad\ sample because the 
\gspspec\ sample was selected based on the signal-to-noise ratio (S/N) of the mean RVS spectra.  An unpublished \gspspec\ S/N $> 20$ was used \citep{GSPspecDR3}.  This was found to be very similar to \verb|rv_expected_sig_to_noise|\footnote{\texttt{rv\_expected\_sig\_to\_noise} propagates median noise properties, whereas \texttt{rv\_spec\_sig\_to\_noise} calculates the noise from the scatter on the signal in each wavelength bin and provides the median of these.} in the {\it gaia\_source} table. \Vrad\ is used to Doppler shift RVS CCD spectra to rest before combining them into a mean RVS spectrum \citep[][]{Seabroke22}.
The sensitivity of \gspspec\ parametrisation to spectra that are not perfectly corrected for their radial velocity shift is flagged in the \gspspec\ \verb|flags_gspspec| string. In particular, characters 3 to 5 (\verb|flags_gspspec[3]|,  \verb|flags_gspspec[4]| and \verb|flags_gspspec[5]|), called \verb|vradT|, \verb|vradG,| and \verb|vradM| respectively, report on the degree of parameter biases from \Vrad\ uncertainties \citep[see][for more details on these flags]{GSPspecDR3}.

\begin{figure}[t]
\includegraphics[width=0.49\textwidth]{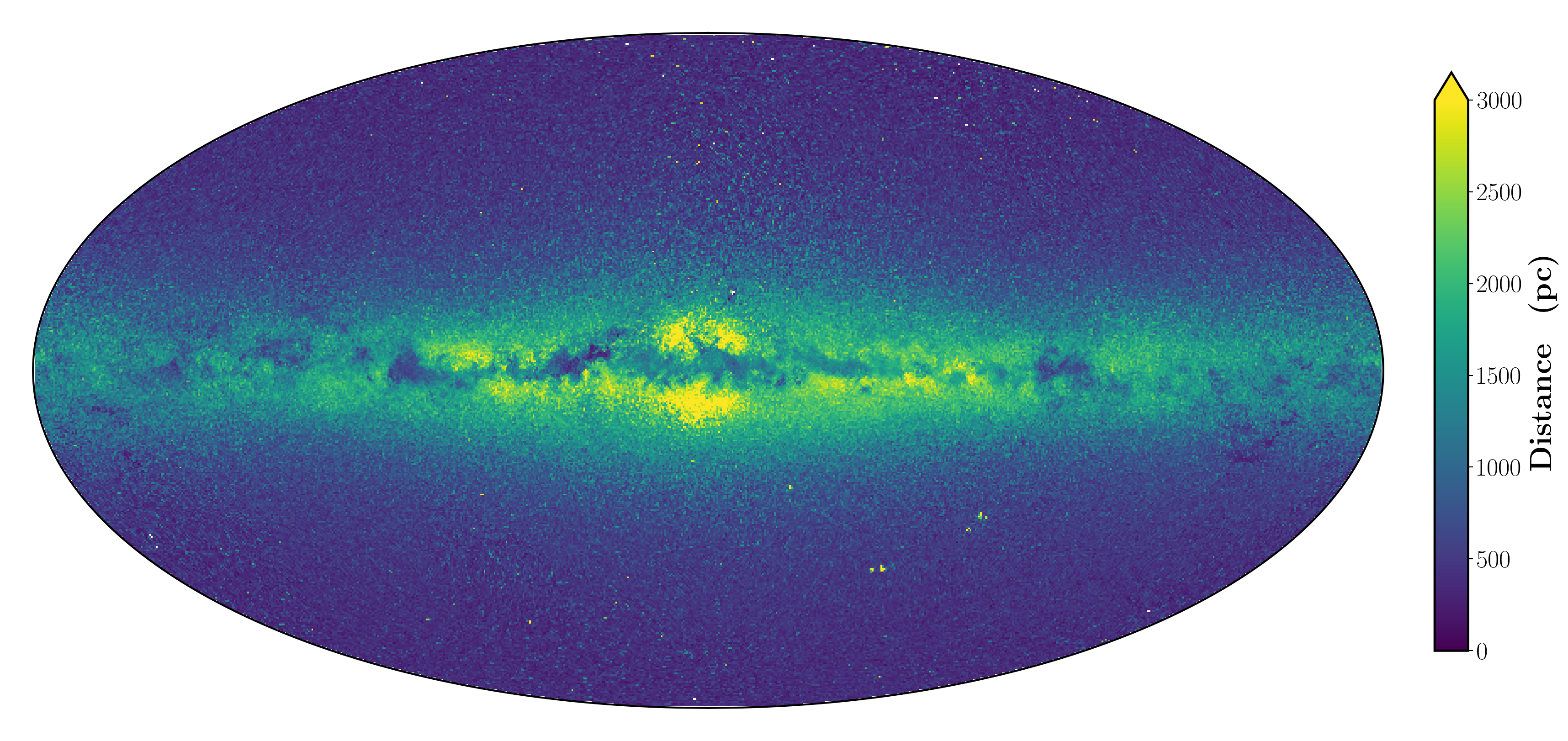}
\caption{Milky Way stars in the   \gspspec\ database (General sample). This HEALPix map in Galactic coordinates  has a resolution of 0.46$^\circ$. The colour code corresponds to the median of the stellar distance from the Sun. This colour code enhances thin and thick disc populations in the Galactic plane, far from the solar neighbourhood. The Galactic bulge stars are also visible in the central region. Finally, the higher distance of the Magellanic Clouds and of several globular clusters reveal their presence in the regions far from the Galactic plane otherwise dominated by foreground stars of the solar neighbourhood. }
\label{Fig.lb}
\end{figure}

\begin{figure*}[ht]
\sidecaption
\includegraphics[width=0.7\textwidth, trim=0 0 0.2cm 0, clip]{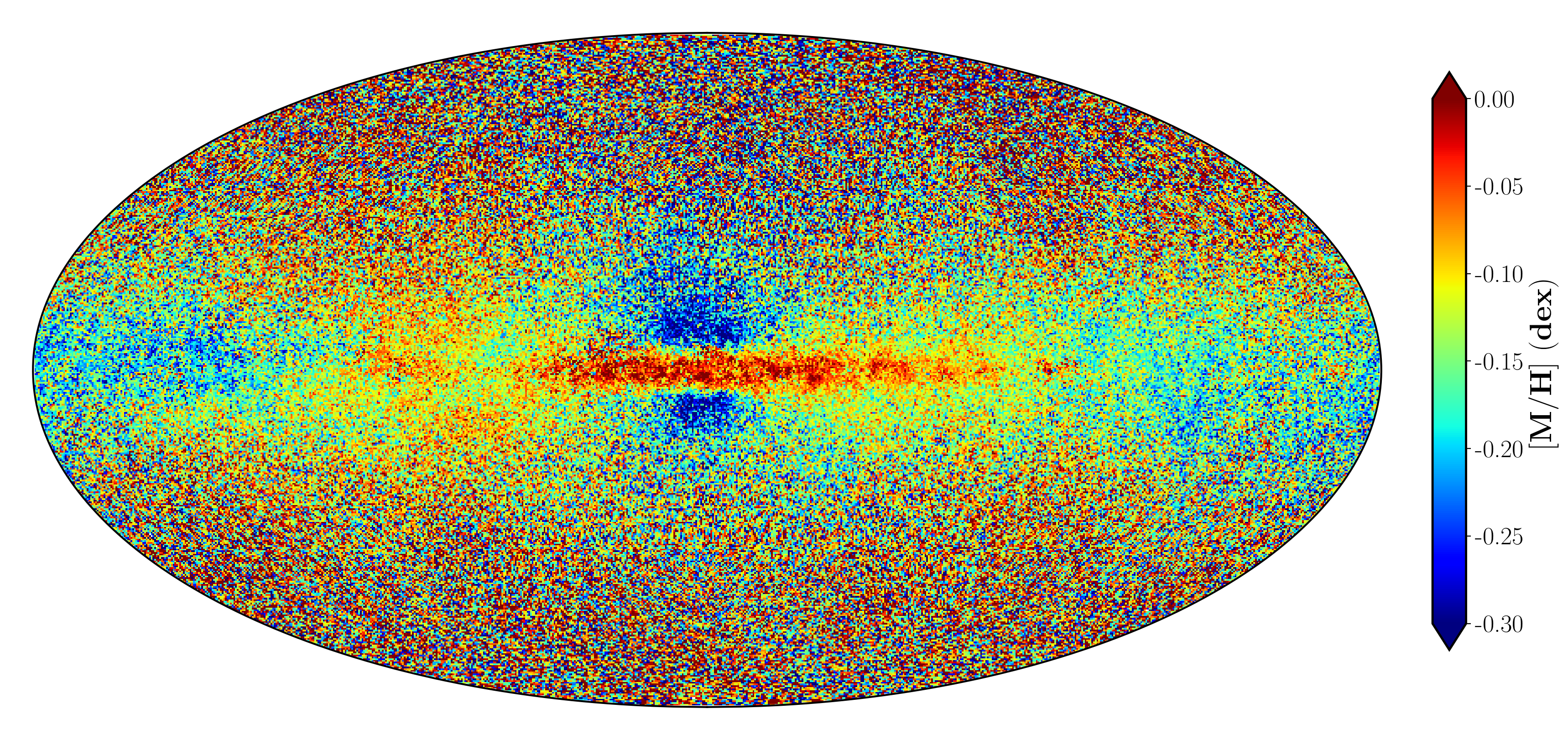}
\caption{Same as Fig.\ref{Fig.lb} but colour coded with the median of the
stellar metallicity \meta\ in each pixel. In the Galactic plane, the higher metallicity values of the thin disc are  visible. In the central Galactic regions, a more metal-poor mix of bulge and thick disc populations is present. Far from the disc plane, high-metallicity thin disc stars at low distances from the Sun and more distant metal-poor halo stars are present.}
\label{Fig.lbBIS}
\end{figure*}

\begin{figure*}[ht]
\sidecaption
\includegraphics[width=0.7\textwidth, trim=0 0 0.2cm 0, clip]{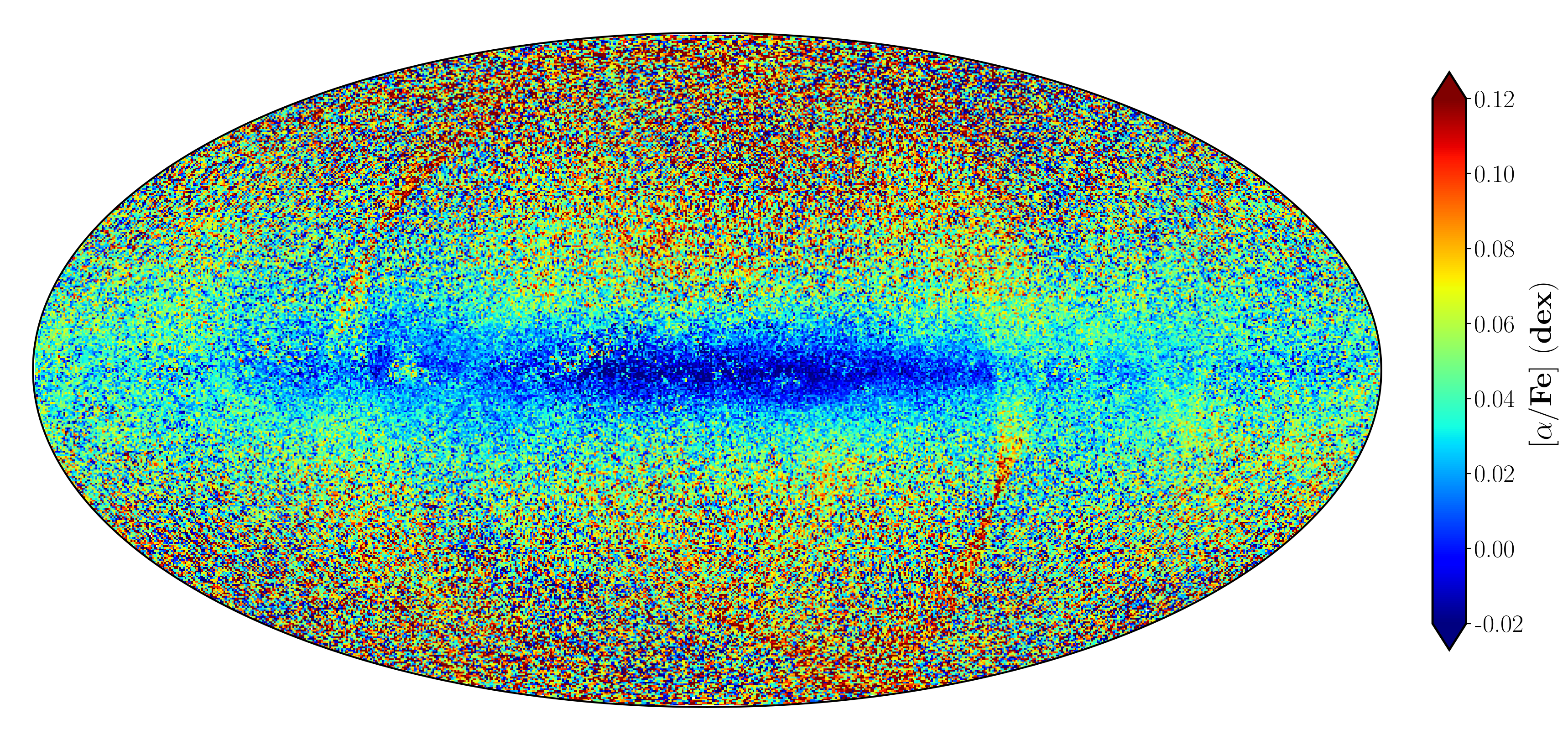}
\caption{Same as Fig.\ref{Fig.lbBIS} but colour-coded with the median of \alphaFe\ in each pixel. The large structures close to the ecliptic poles are artifacts caused by the \Gaia\ scanning law (see text).  Thin disc stars are visible in the plane thanks to their low \alphaFe\ values.}
\label{Fig.lbTER}
\end{figure*}

\begin{figure*}[h]
\includegraphics[width=0.49\textwidth]{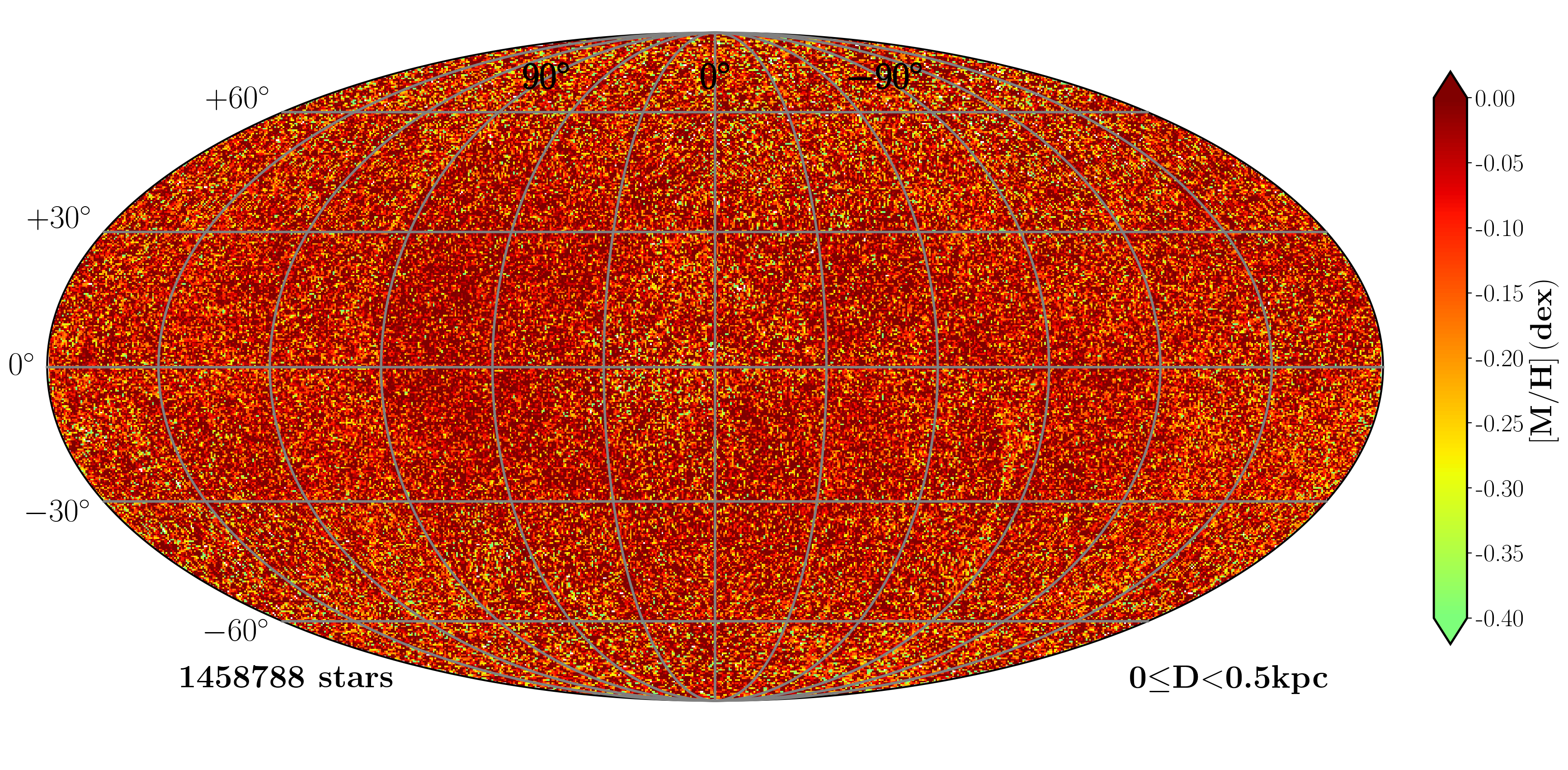}
\includegraphics[width=0.49\textwidth]{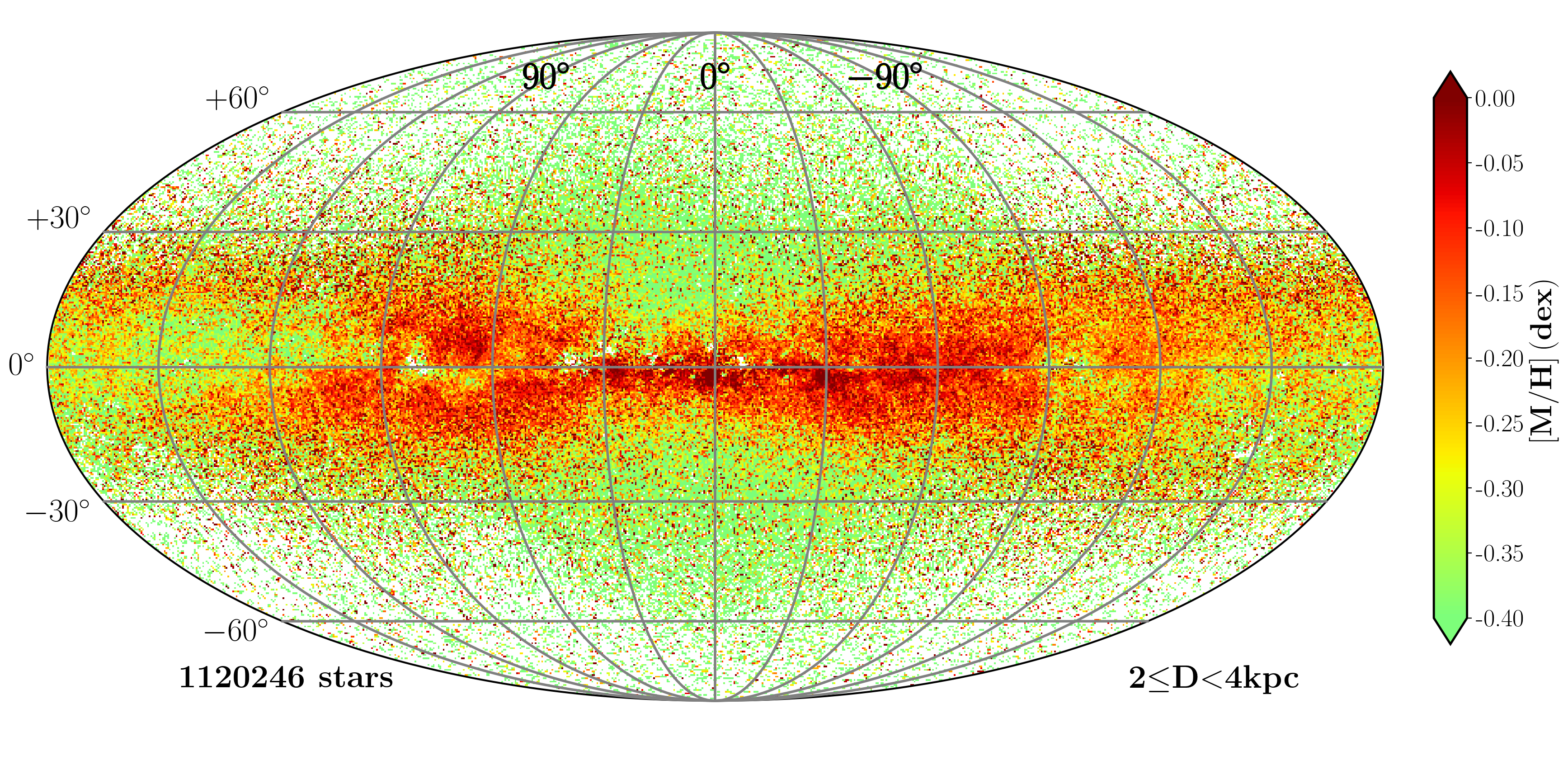}
\includegraphics[width=0.49\textwidth]{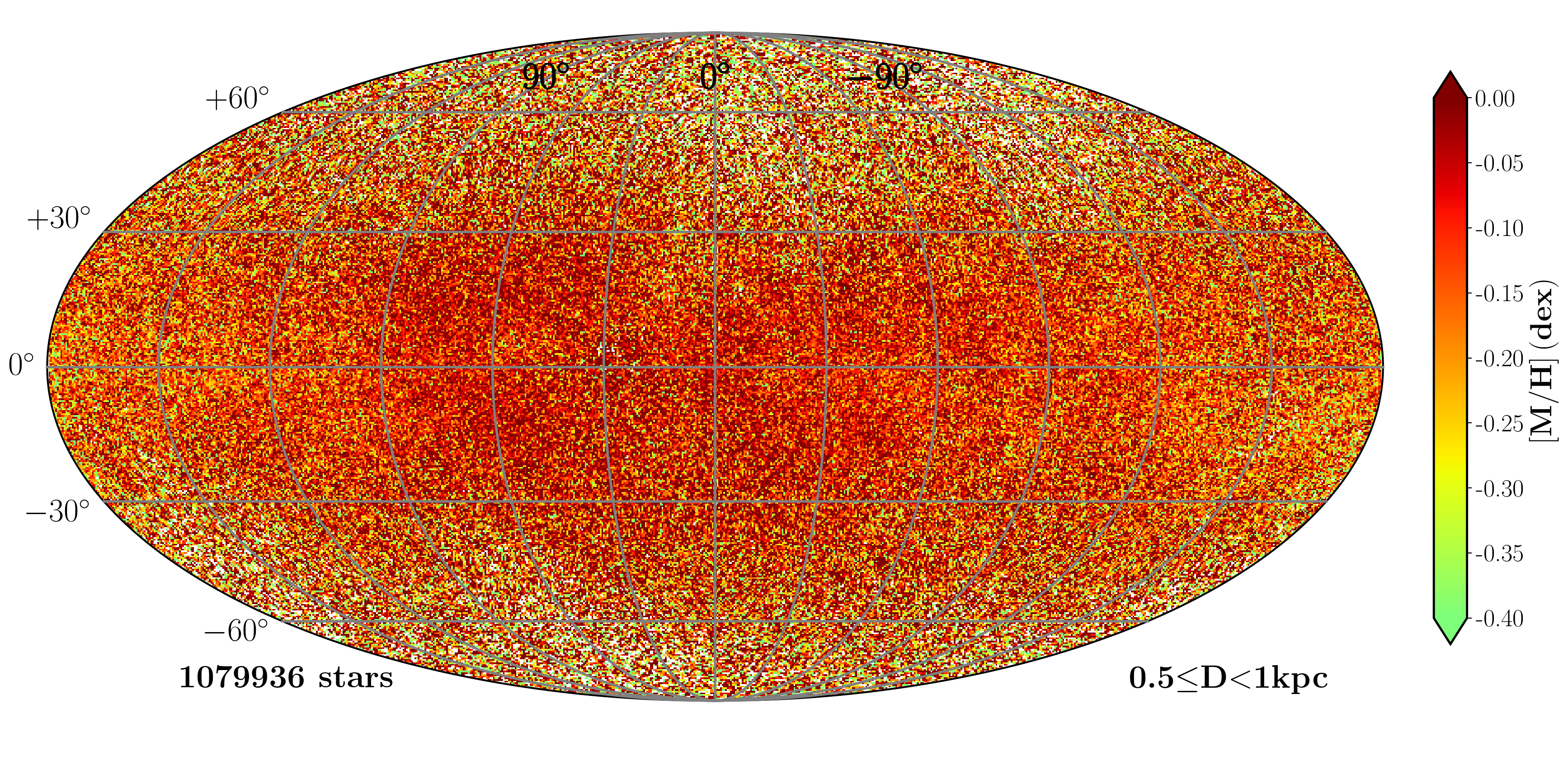}
\includegraphics[width=0.49\textwidth]{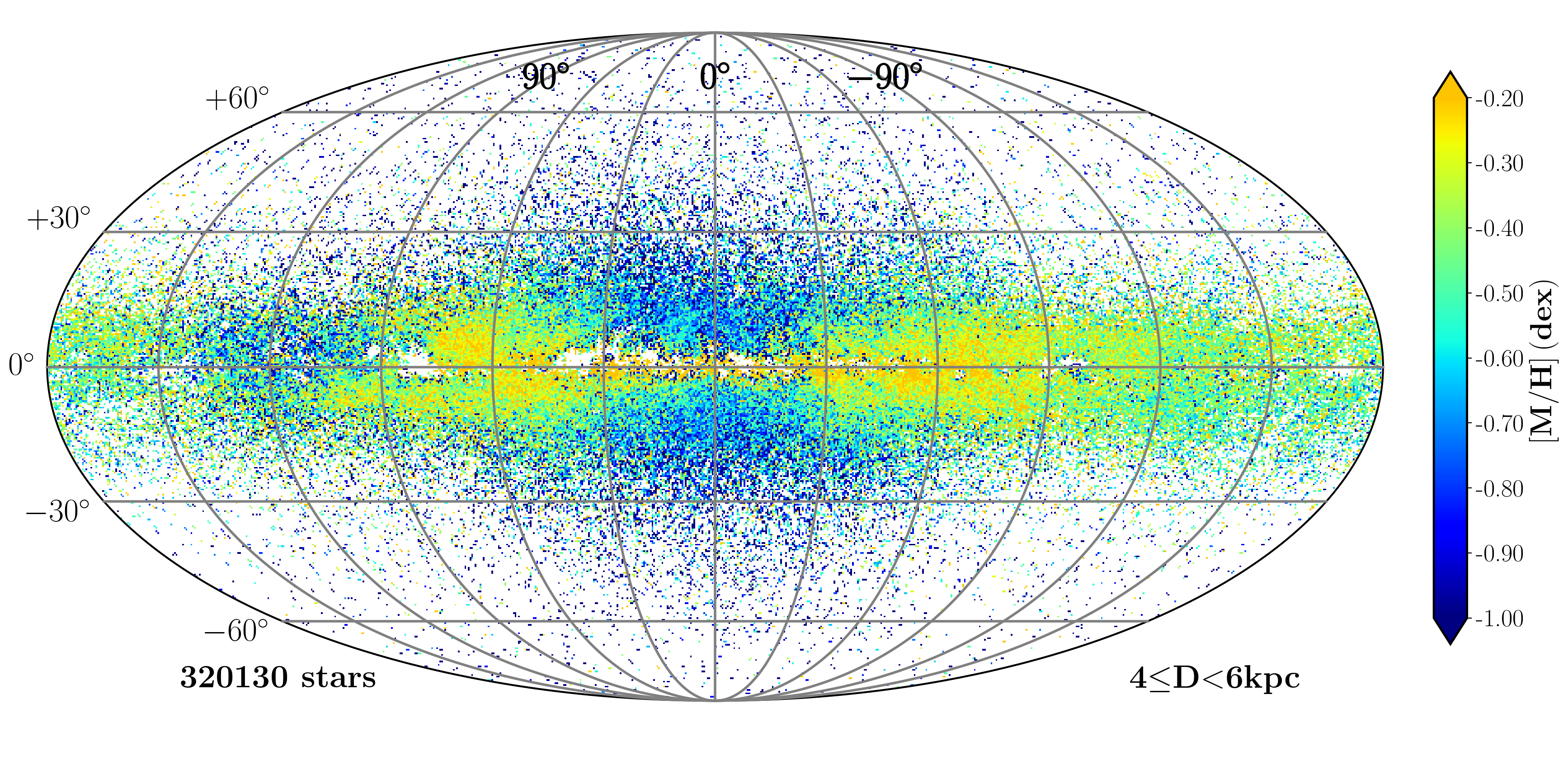}
\includegraphics[width=0.49\textwidth]{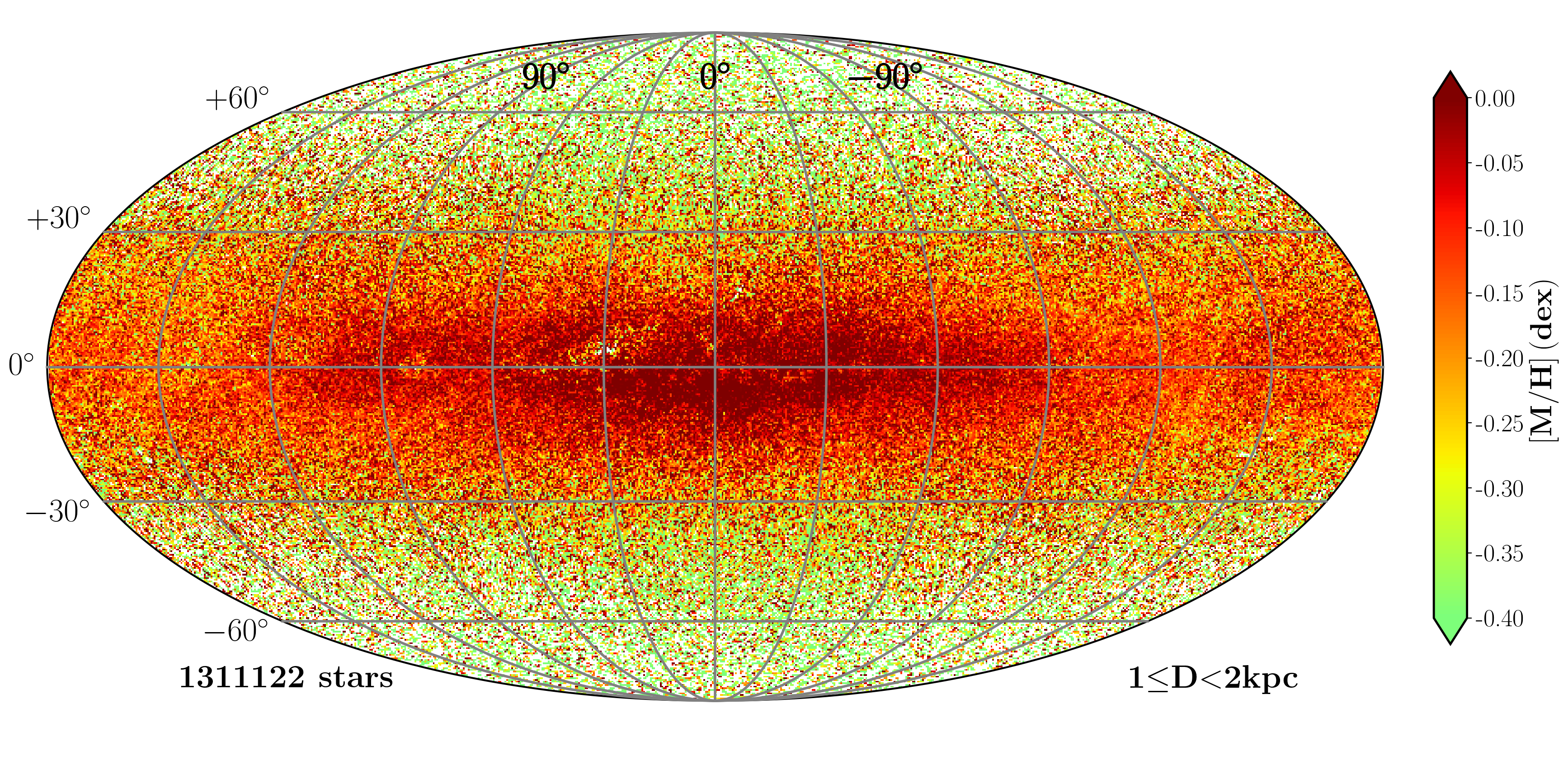} 
\hspace{0.2cm}
\includegraphics[width=0.49\textwidth]{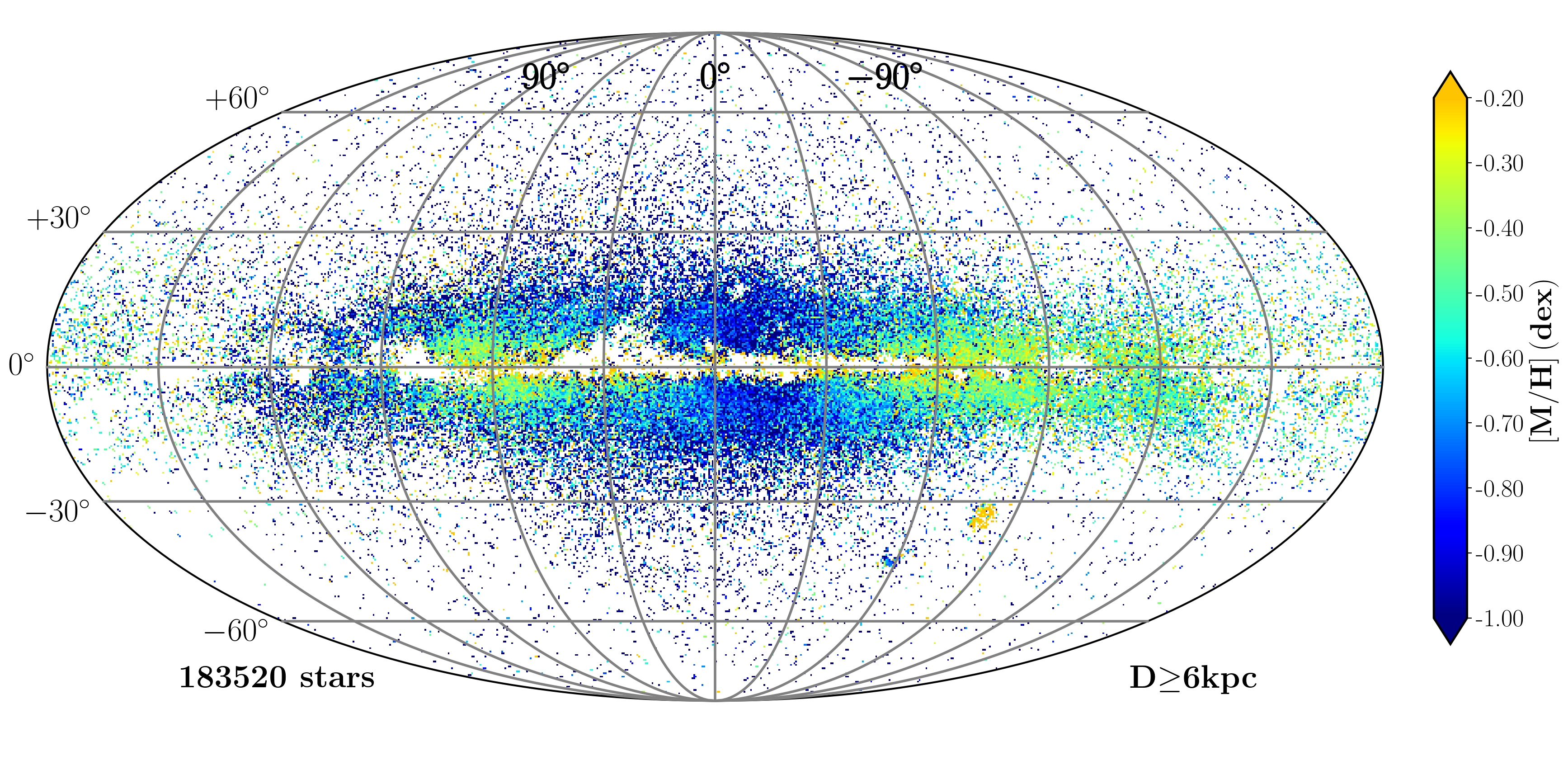}
\caption{Same as Fig.~\ref{Fig.lbBIS} but for different distance intervals shown from the closest (top panel of left column) to the most distant (bottom panel of right column). The distance ranges adopted in each panel are indicated in their lower right corner together with the number of stars in their bottom left corner. The colour code corresponds to the median of the 
metallicity and is continuous between the first four subpanels showing rather metal-rich stars, and the two last panels dominated by metal-poor populations. }
\label{Fig.lb2}
\end{figure*}

\begin{figure*}[ht]
\includegraphics[width=0.32\textwidth]{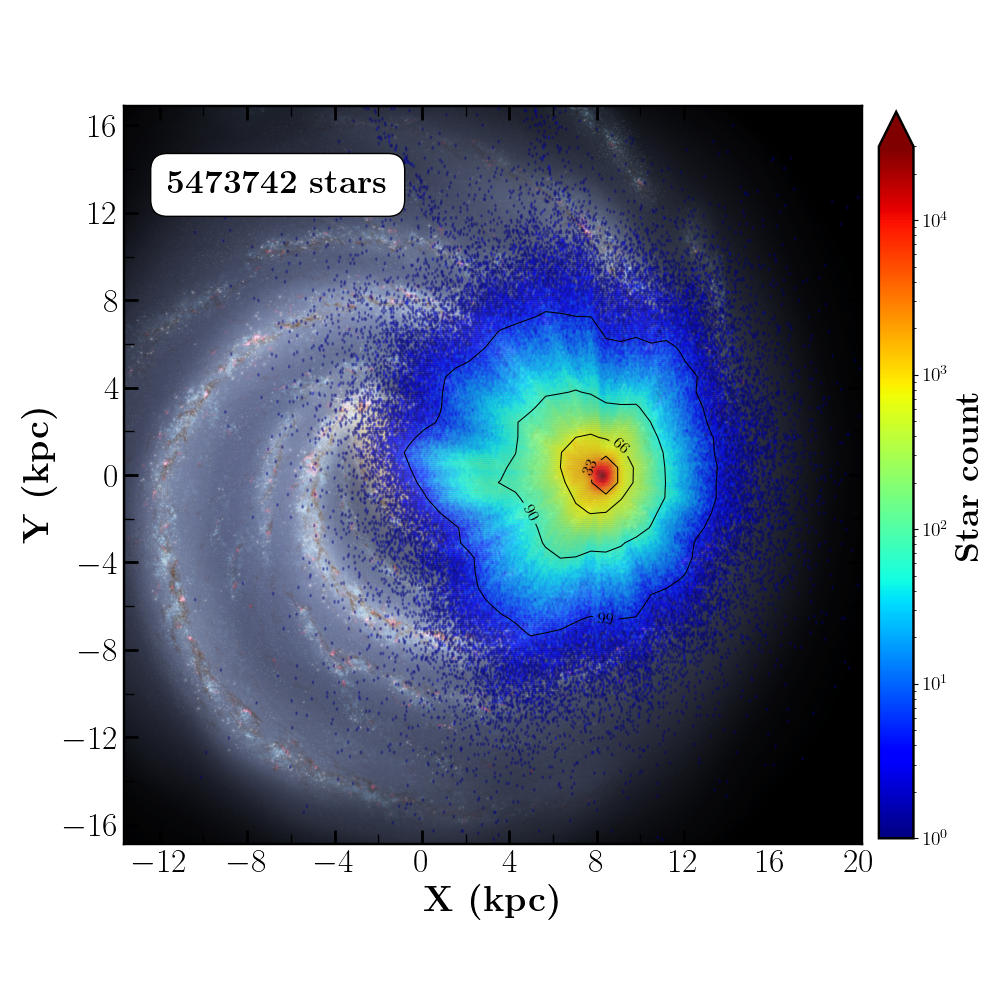}
\includegraphics[width=0.32\textwidth]{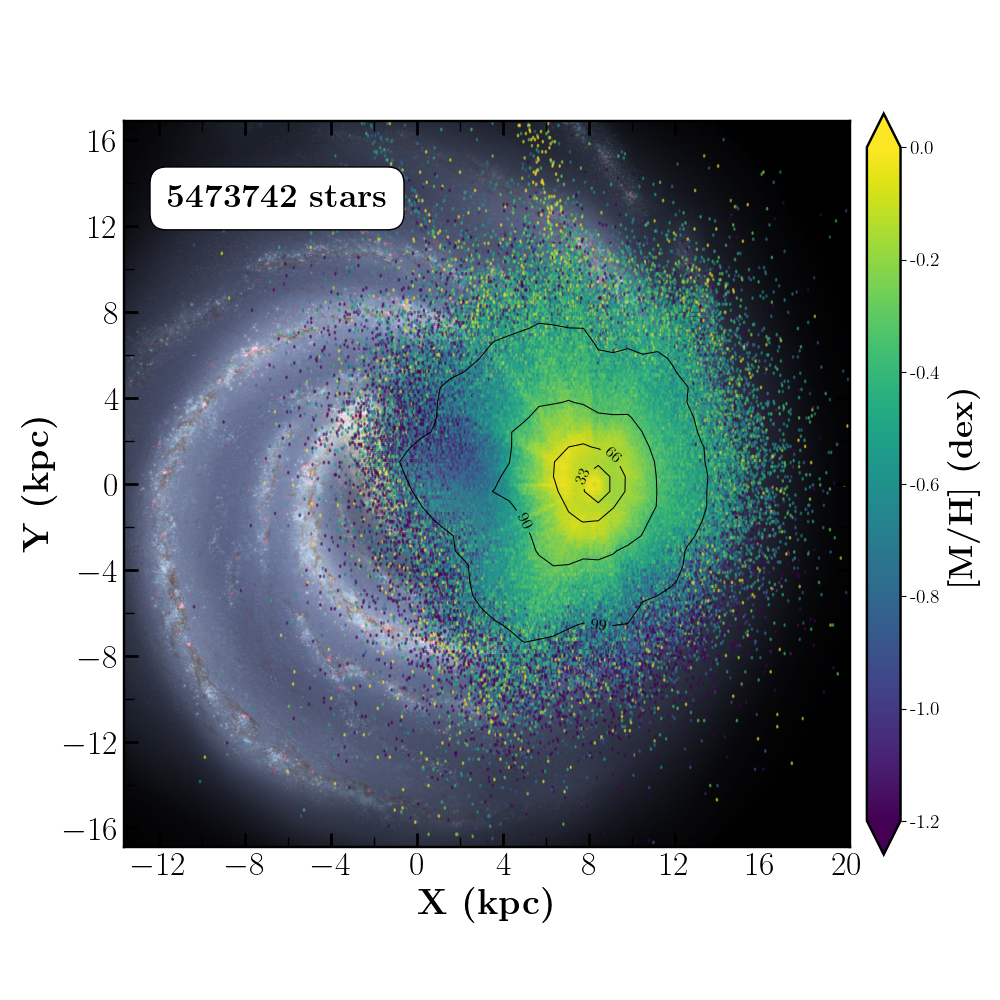}
\includegraphics[width=0.32\textwidth]{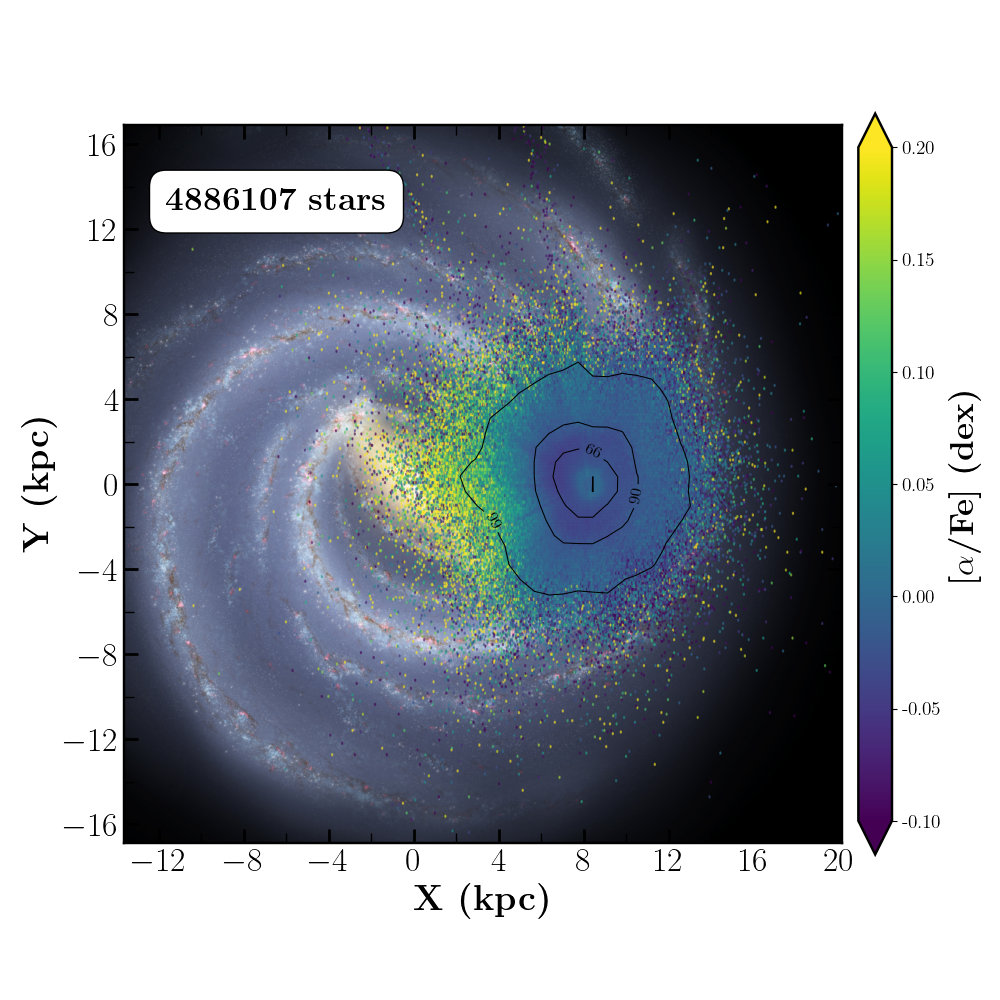}
\includegraphics[width=0.32\textwidth, trim=0 -3.5cm 0 0]{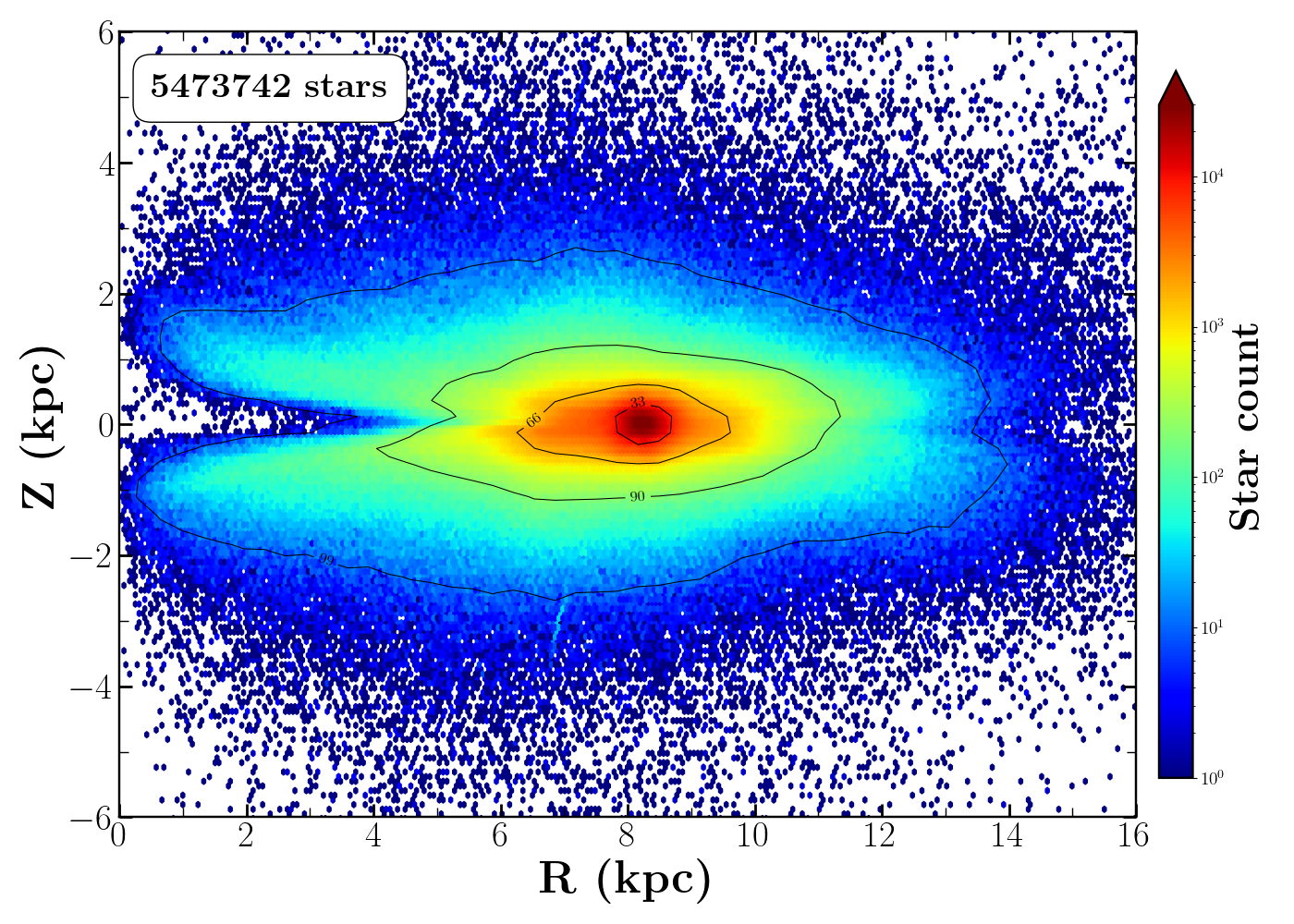}
\hspace{0.4cm}
\includegraphics[width=0.32\textwidth, trim=0 -3.2cm 0 0]{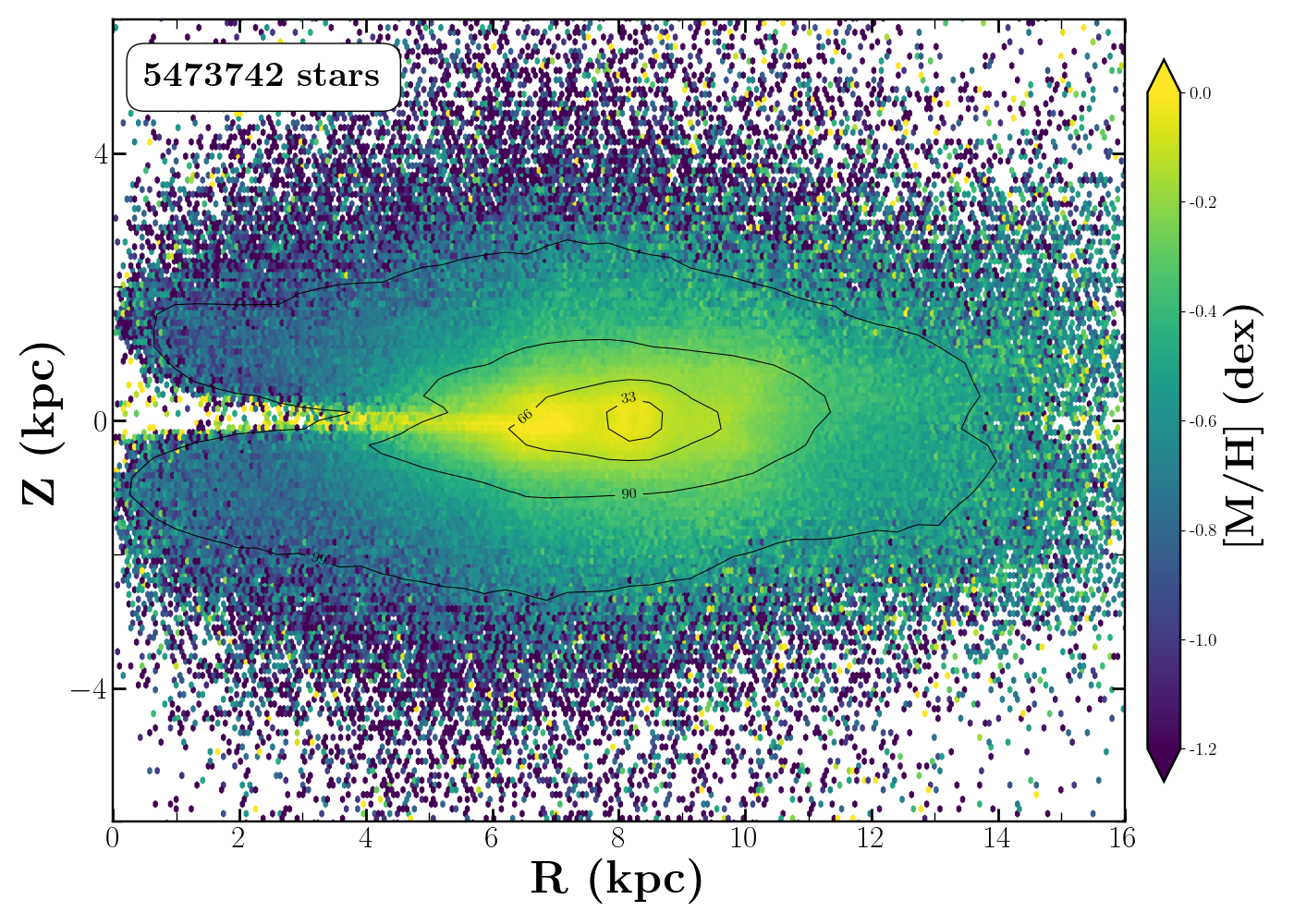}
\hspace{0.3cm}
\includegraphics[width=0.32\textwidth, trim=0 -3.2cm 0 0]{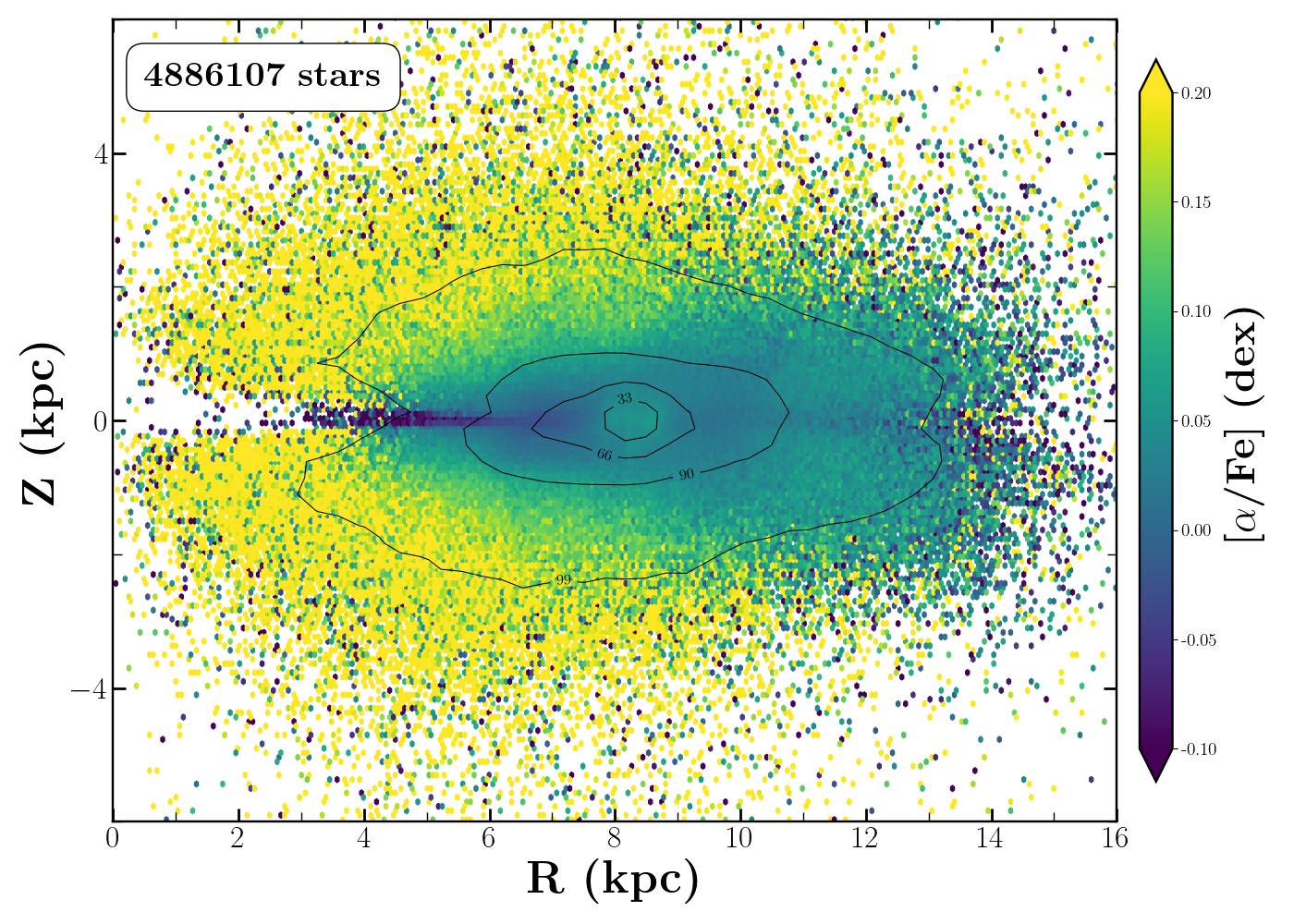}
\caption{Galactic maps for the General sample stars 
colour coded according to stellar count, the median of \meta,\ and the median of \alphaFe\ (from left to right, respectively).
The top row shows the maps in Cartesian coordinates (X,Y), with the background image being the Milky Way sketch from \cite{Churchwell2009}. In the bottom row, $R$ and $Z$ are the distances from the Galactic centre and plane, respectively. }
\label{Fig.XY}
\end{figure*}

\subsection{Astrometric data and distances}\label{sec:Astrometry}
High-precision astrometric parameters $(\alpha, \delta, \mu_{\alpha*}, \mu_\delta, \varpi)$ from \Gaia\ EDR3 \citep[][]{EDR3} complement the above-described radial velocities, allowing three-dimensional study of stellar velocity. Thanks to the bright magnitude limit of the spectroscopic sample, the  median parallax uncertainty is better than 20~$\mu$as for most of our targets and the median uncertainty increases up to a maximum of $\sim40$~$\mu$as for the faintest stars with magnitude $G\simeq 14$. 
 
Based on these \Gaia\ EDR3 astrometric data, \citet{Bailer-Jones21} geometric distances, $r_{\rm Geo}$, are adopted in this study.
To test the implications on our analysis of the Galaxy prior (a 3D model of the stellar distribution and extinction) adopted by  \citet{Bailer-Jones21}, 
we defined an astrometric control sample (ACS, Appendix \ref{Sect:ACS}). This sample is composed of stellar distances, $r_\varpi$, exclusively based on the observed \Gaia\ trigonometric parallaxes after removing the individual parallax zero points, $\Delta_\varpi$, following  \citet[][]{Lindegren2021}. In addition, the ACS selection fulfills the following criteria:
(i) a parallax uncertainty of less than 20\% derived from (ii) five-parameter astrometric solutions with renormalised unit weight error (RUWE) $<1.4$, (iii) sources not flagged as {\it duplicated} by the \Gaia\ cross-matching algorithm \citep[][]{Torra21}, and (iv) with full photometry ($G$, $G_{\rm BP}$, $G_{\rm RP}$) and radial velocity available. 
To compare the two distance estimates, we consider the fractional residual $(r_{\rm Geo}-r_\varpi)/r_\varpi$ for objects in common. The median value of the residuals is  $-0.04$\% and its median absolute deviation is 0.2\%, showing very good overall agreement between both estimates. 
As shown in Appendix \ref{Sect:ACS}, the main discrepancy occurs in the Galactic plane, where Bayesian distances are greater than the pure trigonometric parallax-based distances by $\lesssim$~4\%. 
These very small discrepancies have a negligible impact on the results of this work and, as a consequence, $r_{\rm Geo}$ distances are adopted in the following.

\subsection{Galactocentric positions, velocities, and orbits}
\label{Sec:Orbital_params}

For the computation of the stellar positions (Galactocentric Cartesian $X$, $Y$,  and $Z$ positions, and cylindrical radius $R$) and the  Galactocentric cylindrical velocities ($V_R$, $V_\phi$, and $V_Z$ in a right-handed frame with positive $V_\phi$  for most of the disc stars), we adopted the Sun's Galactocentric position $(R, Z)_{\odot}=(8.249, 0.0208)$\,kpc \citep[][]{Gravity20, Bennett19} and  Galactocentric cylindrical velocities 
$(V_R, V_\phi, V_Z)_\odot=(-9.5, 250.7, 8.56)$~km\,s$^{-1}$ \citep[][]{Reid20, Gravity20}, as well as \Gaia's $\alpha$, $\delta$, line-of-sight velocity, and \citet{Bailer-Jones21} geometric distances for each target. 
The orbital parameters (actions $J_R, J_\phi, J_Z$, apocenter and pericenter distance $R_a$, $R_p$, maximum distance from the plane reached during the star's orbit $Z_{\rm max}$) are computed using the Stäckel fudge method \citet[][]{Binney12} implemented in the Galpy code \citep[][]{Bovy14}, combined with a rescaled version of the \citet[][]{McMillan17} axisymmetric Galactic potential. In this model, the Galactocentric distance of the Sun is $R_{\odot}=8.249$\,kpc while the circular velocity of the Local Standard of Rest is set to 238.5~km\,s$^{-1}$ \citep{Gravity20, SchBinDeh2010}.
The lower and upper confidence limits  for each of these parameters are obtained by drawing 
Monte Carlo realisations of $\mu_{\alpha*}, \mu_\delta$, \Vrad, and distance, and propagating them through the positions, velocities, and orbits. We assumed Gaussian symmetric distributions for the proper motions and line-of-sight velocities, and a broken Gaussian distribution for distance (based on the lower and upper confidence limits of \citealt{Bailer-Jones21}). 

\subsection{Working samples} \label{Sec:samples}
To optimise the analysis of this work, different data selections and quality cuts are required. To this purpose, we defined several data samples with the following characteristics. The precise conditions used for these sample selections are presented in Appendix~\ref{Appendix:queries}, including the corresponding catalogue queries.

\medskip

\noindent  General sample (5\,527\,090 stars; Sects.~\ref{Sec:GlobalView} and \ref{Sec:Dynamics}): From the complete \gspspec\ sample (containing 5\,594\,205  stars in total),
we excluded about 56\,000 stars located near the cool end  of the \gspspec\ parameter space and suffering from biases in different parameters \citep[][] {GSPspecDR3}. 
The remaining objects without $r_{\rm Geo}$ distances were removed. For this sample, the median uncertainty in \meta\ is 0.075~dex and the median uncertainty in \alphaFe\ is 0.05~dex.

\medskip
\noindent Medium quality sample (4\,140\,759 stars; Sect.~\ref{Sec:GlobalView}, Sect.~\ref{sec:5}, Sect.~\ref{Sec:Dynamics}):  This results from additional quality criteria rejecting stars with grid border effects and large parameter uncertainties. For this sample, the median uncertainty in \meta\ is 0.06~dex and the median uncertainty in \alphaFe\ is 0.04~dex.

\medskip
\noindent Gradient analysis sample (2\,762\,809 stars, Sect.~\ref{Sec:Gradients}): This optimises the quality of the \meta\ and \alphaFe\ parameters and the astrometric quality to ensure reliable distance estimates, with the goal of estimating radial and vertical gradients. 
For this sample, the median uncertainty in \meta\ is 0.05~dex and the median uncertainty in \alphaFe\ is 0.035~dex.

\medskip
\noindent High-quality sample (2\,218\,573 stars, Sect.~\ref{sec:5}): This selects the most precise stellar parameters and chemical abundances. We note, in particular, that the choice of the $fluxNoise=0$ filter flag imposes a selection of stars with very low parameter uncertainties. For this sample, the median uncertainty in \meta\ is 0.03~dex and the median uncertainty in \alphaFe\ is 0.015~dex.

\medskip
\noindent Individual abundances samples (Sect.~\ref{Sec:GlobalView} and \ref{Sec:SolarCylinder}): these have been defined to optimise the individual abundance distributions of different chemical elements. The number of selected stars depends on the chemical element.
Specifically, as expected, the precision of abundance measurements is a function of different conditions, such as the number of analysed lines, their equivalent widths, and possible blends, which vary with effective temperature, surface gravity, mean metallicity, and other individual abundances, and  is different from one element to another.


\section{Global chemical properties of the Milky Way}
\label{Sec:GlobalView}
\begin{figure*}[h]
\sidecaption
\includegraphics[width=1.0\textwidth, trim=0 1cm 6.2cm 1.2cm, clip]{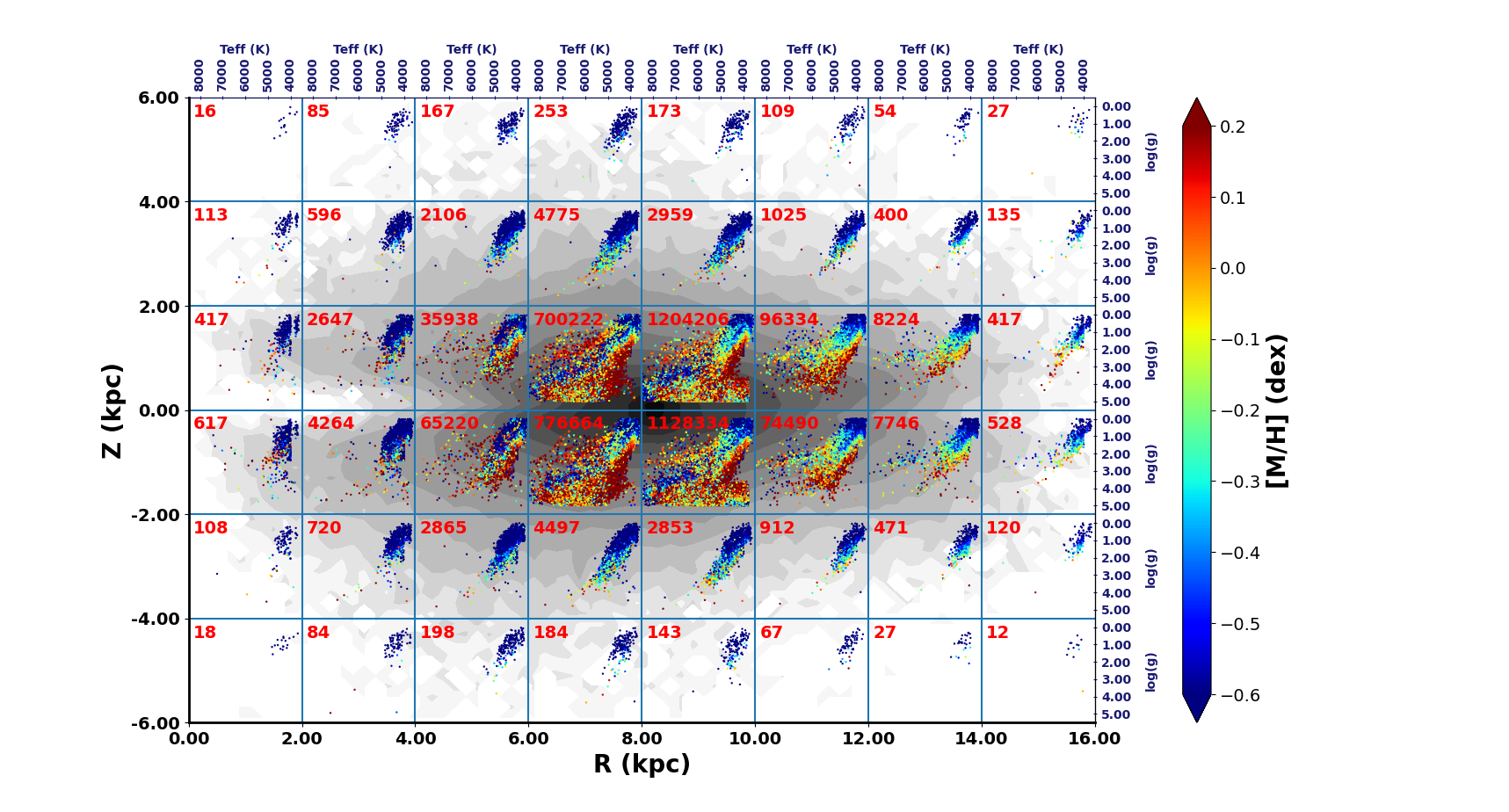}
\caption{Kiel diagrams across the Milky Way for the Medium Quality sample, colour coded according to the median of the mean metallicity. The number of stars in each subpanel is indicated in its upper left corner. The global
($R$, $Z$) distribution of Fig.~\ref{Fig.XY} (bottom row) is shown in the background with grey levels. Moving away from the Sun, the sample becomes dominated by intrinsically bright giants.}
\label{Fig.MW_Kiel}
\end{figure*}

\begin{figure*}
\sidecaption
\includegraphics[width=1.0\textwidth, trim=2.5cm 1.0cm 5.5cm 1.2cm, clip]{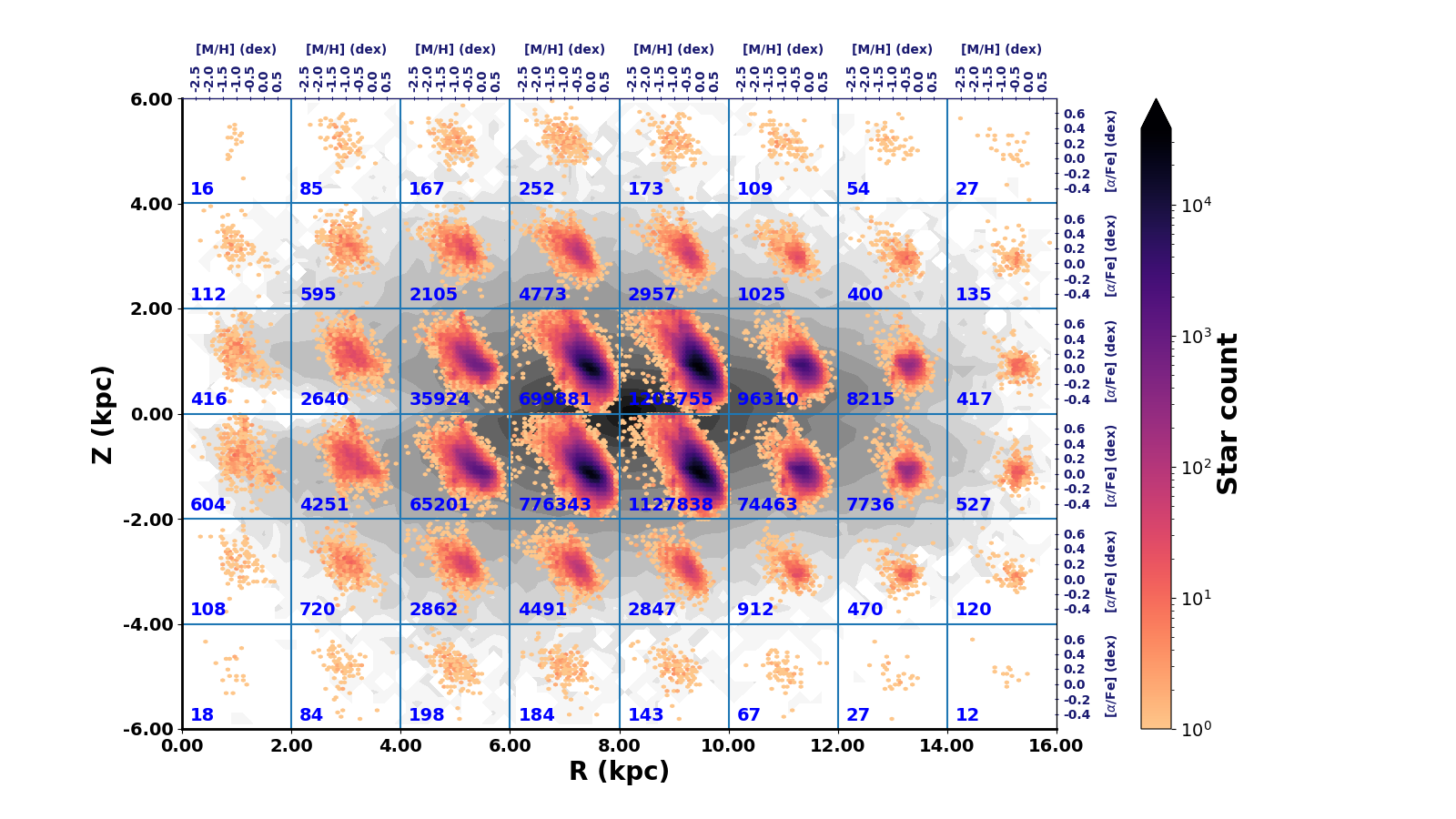}
\caption{Same as Fig.~\ref{Fig.MW_Kiel} but for \alphaFe\ versus \meta\ (star counts).}
\label{Fig.MW_MetaAlpha}
\end{figure*}

The $\sim$5.5~million star database of spectroscopic stellar parameters and chemical abundances (see Sect~\ref{Sec:APs}), as part of the \Gaia\ all-sky data collection, provides significant radial and vertical coverage of our Galaxy, but also probes a considerable range in azimuth. 
In this section, we first analyse the global spatial distribution of the stars in the Milky Way and their chemical properties, and then we illustrate the selection function in radial and vertical bins. Here, we show the chemical properties and spatial coverage of three representative stellar populations selected from the Kiel diagram of the near-plane objects: massive young stars, bright red giant branch (RGB) objects and hot turn-off (TO) stars.

\subsection{Spatial and chemical trends}\label{Sec:globalMaps}

{\it Sky maps in Galactic coordinates}

\smallskip
\noindent The distribution of all the stars in the {\it General} sample is presented in Galactic coordinate maps
colour coded according to distance (Fig.~\ref{Fig.lb}),  metallicity (Fig.~\ref{Fig.lbBIS}), and \alphaFe\ abundance (Fig.~\ref{Fig.lbTER}).

We first note that the ecliptic pole  scanning law \citep[EPSL;  ][]{Boubert20}  pattern is visible in Fig.~\ref{Fig.lbTER} with slightly higher \alphaFe\ values.  Between 25 July and 21 August 2014 (the first weeks of {\it Gaia}’s nominal mission), \Gaia\ operated in EPSL mode \citep{Prusti16}.  During this period, stars close to the ecliptic poles and along great circles connecting them were observed a greater number of times than during Gaia’s nominal scanning law.  This leads to\ spectra with  higher S\textbackslash N towards the EPSL footprint.  As Gaia continues to observe in nominal scanning law, the S/N will homogenise over the sky for future data releases.
The higher S/N of the spectra along the EPSL pattern allows the parameterisation of fainter (and hence more distant) stars in these regions, leading to a slightly lower mean metallicity and especially larger mean \alphaFe\ values per pixel. 

 The metal-rich and relatively $\alpha$-low thin disc is easily  identified in Figs. 1, 2, and 3. In the  direction of the Galactic bulge, outside the Galactic plane, one sees a dominant metal-poor population. Indeed, due to the fact that the bulge is more radially concentrated, many thick disc stars are present in these regions, decreasing the median metallicity. The regions located far from the plane are filled with a mixture of nearby metal-rich $\alpha$-low disc stars and more distant metal-poor $\alpha$-rich halo stars (see also Sect.~\ref{Sec:Gradients}). 

To complement this first global view, we cut Fig.~\ref{Fig.lbBIS} into  different distance intervals from the Sun (see Fig.~\ref{Fig.lb2}). 
Similar figures for the \alphaFe\ distribution are provided in the Appendix (Fig.~\ref{Fig.lb3}). 
The scanning law features described above are visible in  the  closest distance bin (D$<$0.5~kpc, upper left panel). 
Figures~\ref{Fig.lb2} and~\ref{Fig.lb3} in the Appendix show that, whereas the sky is filled by metal-rich and  $\alpha$-low thin disc stars within 500~pc of the Sun, the Galactic discs start to progressively emerge with increasing distance. The more metal-poor and $\alpha$-rich halo dominates beyond $\sim$4~kpc. In the more distant panel, beyond 6~kpc, the thick disc is well seen on both sides of the Galactic centre and the thin disc has almost disappeared because of interstellar extinction. 
An asymmetry can also be seen in the metallicity map close to the Galactic centre (with the central region being more metal-rich at  positive longitudes than at negative ones), which may result from the presence of the central bar and an inhomogeneous distance distribution around the Galactic centre in this more distant bin due to extinction.
Finally, the Small and Large Magellanic Clouds (SMC and LMC, respectively) are also clearly visible in this panel, both being rather $\alpha$ poor. We note that the LMC appears rather metal rich. Our LMC sample is indeed dominated by bright massive and young stars \citep[referred to as blue loop stars by][]{Luri21}, which likely sample the metal-rich LMC populations.

\medskip 
\noindent {\it Galactic maps in Cartesian coordinates}

\smallskip
\noindent Figure~\ref{Fig.XY} (top panels) shows three maps in (X-Y) Cartesian coordinates of all the General sample stars superimposed on the Milky Way sketch of Hurt \& Benjamin \citep{Churchwell2009}. From left to right, the maps are colour coded according to stellar counts, mean metallicity, and enrichment in $\alpha$-elements with respect to iron. We note that, in the panels colour coded according to \meta\ and \alphaFe,\, the apparent bulge asymmetry described in Figs.~\ref{Fig.lb2}  and~\ref{Fig.lb3} is again seen between the Sun and the Bar.

The bottom row of Fig.~\ref{Fig.XY} shows the distribution of the same stars in the ($R$, $Z$) plane, where $R$ is the Galactic centre distance and $Z$ the distance from the Galactic plane. It can be seen that the spatial coverage of this sample with measured \meta\ and \alphaFe\ is very large, spanning several kiloparsec. Close to the Galactic plane and towards the inner regions, a thin stripe of metal-rich  low-$\alpha$ stars is clearly seen. Its vertical extension is
about 300~pc below and above the plane, which is the typical scale height of the thin disc. This stripe is therefore the thin disc seen edge-on. It is not detected for $R\la$~3.5~kpc because of interstellar extinction and the resultant lack
of sufficiently bright stars for analysis by \gspspec\ at smaller Galactic radii.
Towards the external regions near the plane, this thin disc is also seen up to $\sim$14~kpc from the Galactic centre. Its flare in the outer regions is traced by the progressive increase of high-metallicity low-\alphaFe\ stars farther away from the plane, as we move towards higher $R$ values (see Sect. \ref{sec:5} for a chemo-kinematical perspective of the disc flare). Moreover, the stars located close
to the Galactic plane but in the external parts of the  disc appear more metal poor and slightly more $\alpha$ rich: a clear gradient in \meta\ and \alphaFe\ versus $R$ is thus revealed (see also Sect.\ref{Sec:Gradients}). 
As a consequence, at large heights from the Galactic plane, internal regions are dominated by low-metallicity thick disc and halo stars, while for R$>$12 kpc, flared disc stars of higher metallicity and lower \alphaFe\ values dominate. 

\begin{figure}[t]
\includegraphics[height=7.0cm, width=0.44\textwidth]{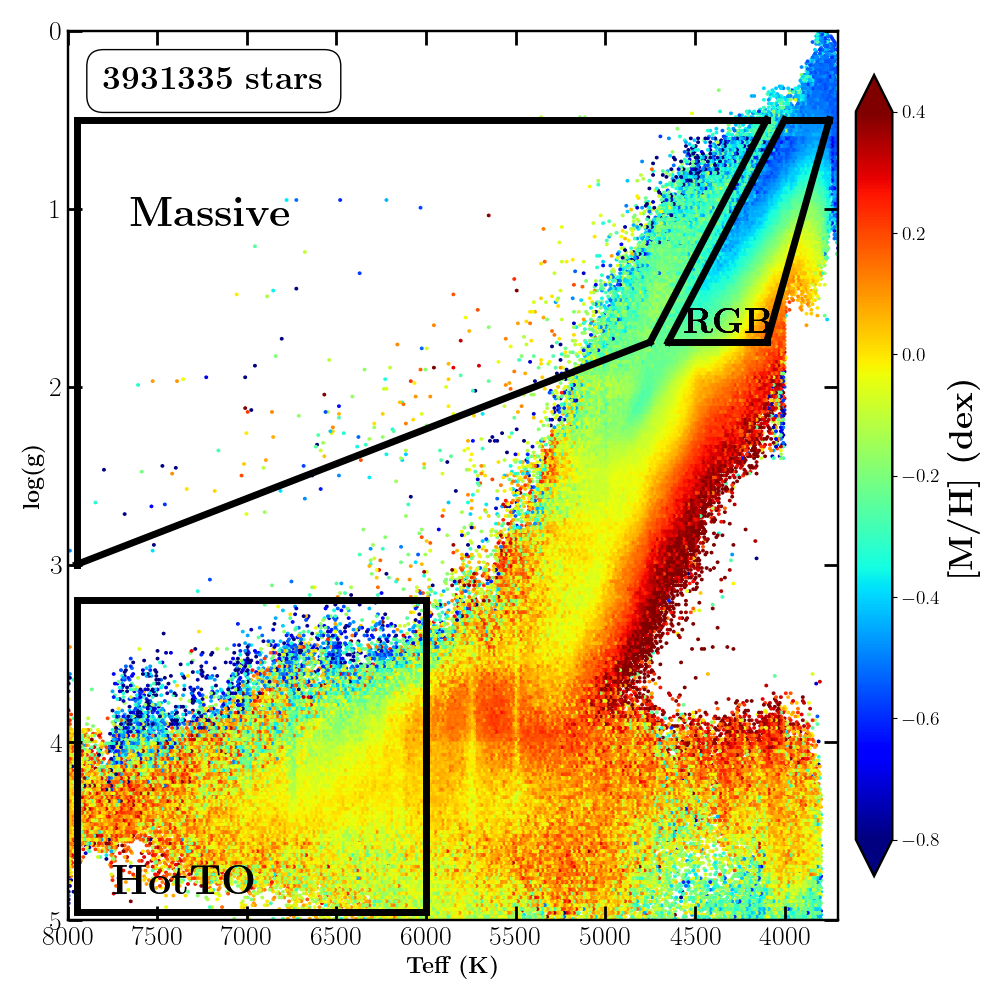}
\caption{Kiel diagram of all the Medium Quality stars located within 1~kpc of the Galactic plane. The location of the HotTO,  RGB, and Massive subsamples is illustrated. 
Their spatial distribution and chemical properties are detailed in Figs.~\ref{Fig.SelFuncXY1}, \ref{Fig.SelFuncXY2}, and \ref{Fig.SelFuncXY3}, respectively.}
\label{Fig.SelFuncKiel}
\end{figure}

\begin{figure*}[ht]
\includegraphics[width=0.3\textwidth]{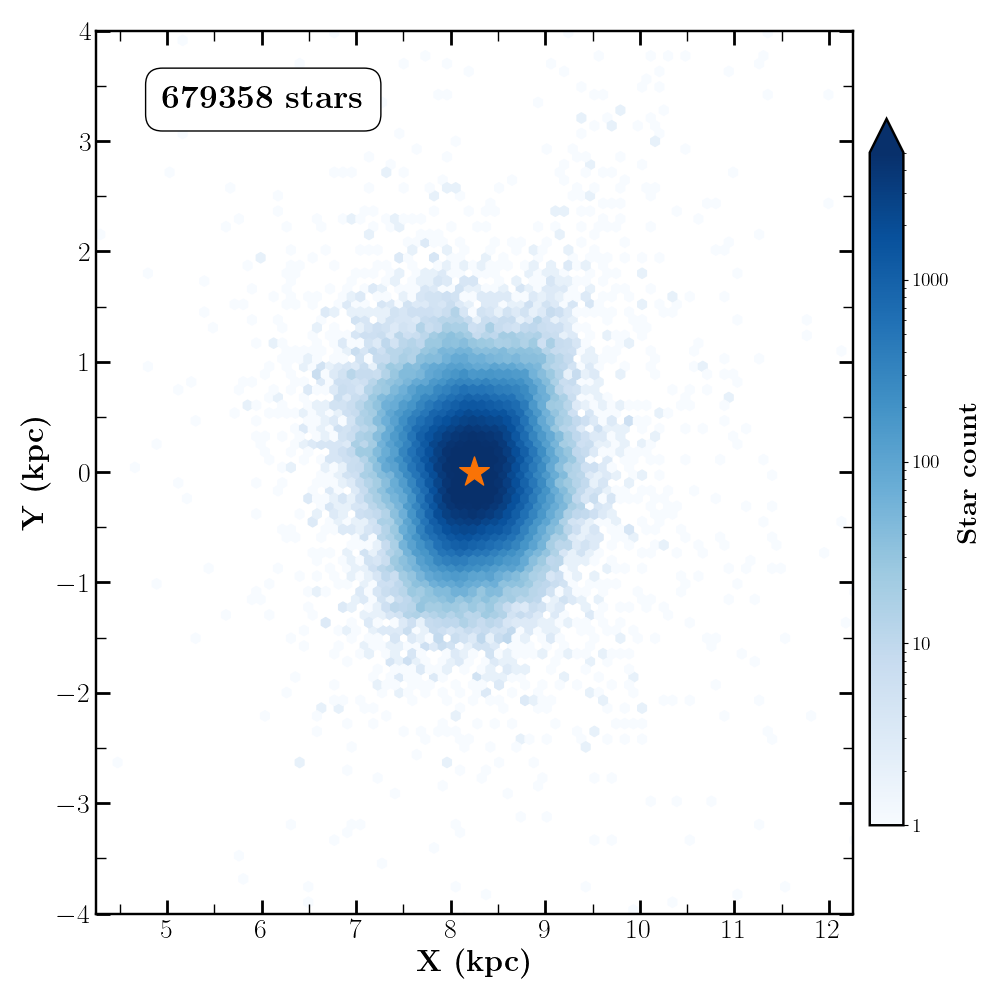}
\includegraphics[width=0.3\textwidth]{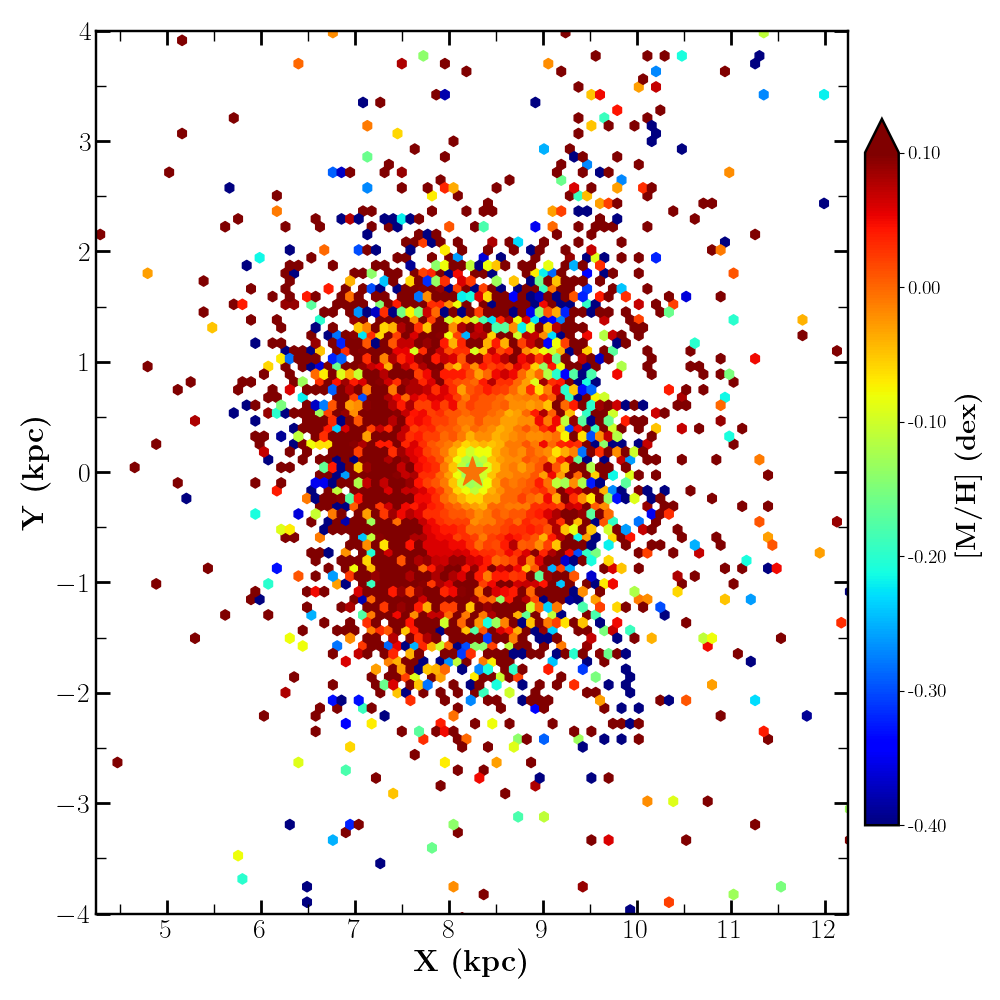}
\includegraphics[width=0.3\textwidth]{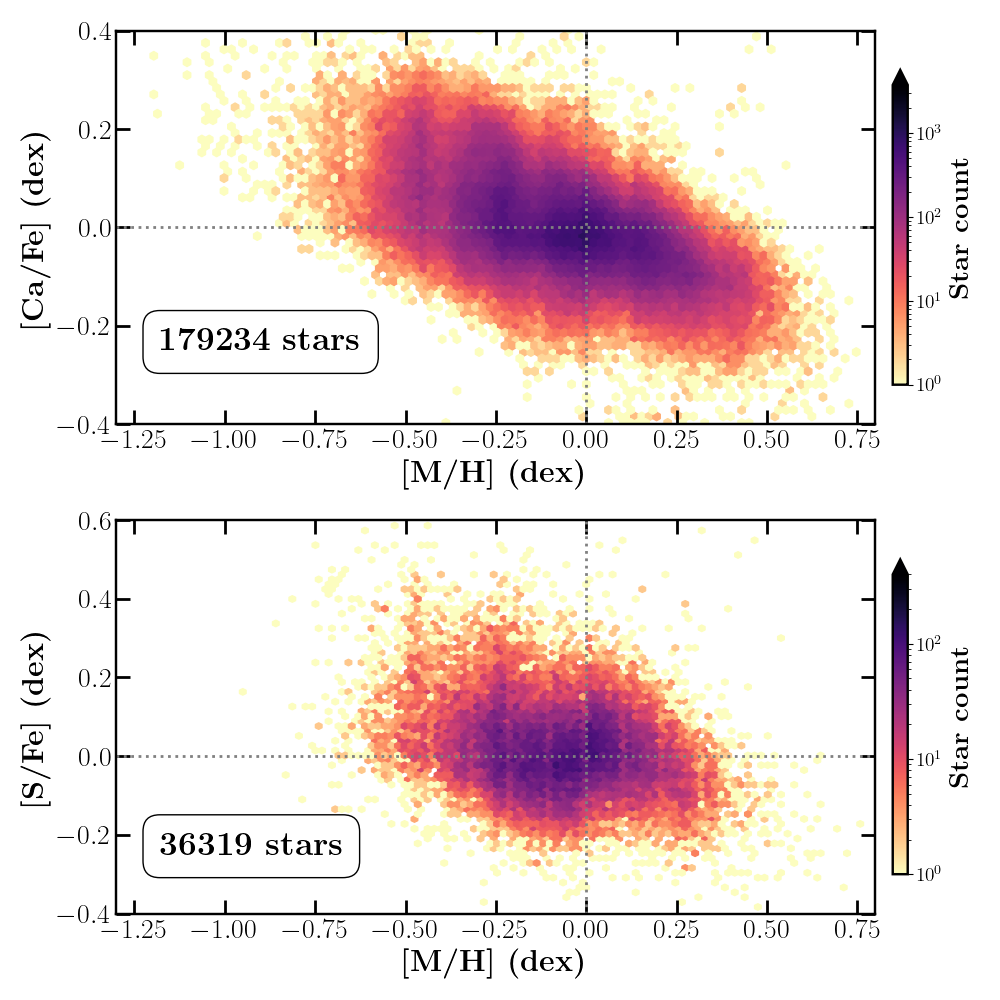}
\caption{HotTO stars of the Medium Quality sample located within 1~kpc of the Galactic plane. Their XY spatial distribution is shown within $\pm$4~kpc of the Sun, colour coded according to stellar count and the median of the mean metallicity (left and central panels, respectively). The solar position is indicated by a filled star coloured according to solar metallicity (\meta=0.0~dex). The right panel shows the  evolution of calcium (top) and sulfur (bottom) abundances with respect to iron abundances versus the mean metallicity.}
\label{Fig.SelFuncXY1}
\end{figure*}

\begin{figure*}[h]
\includegraphics[width=0.3\textwidth]{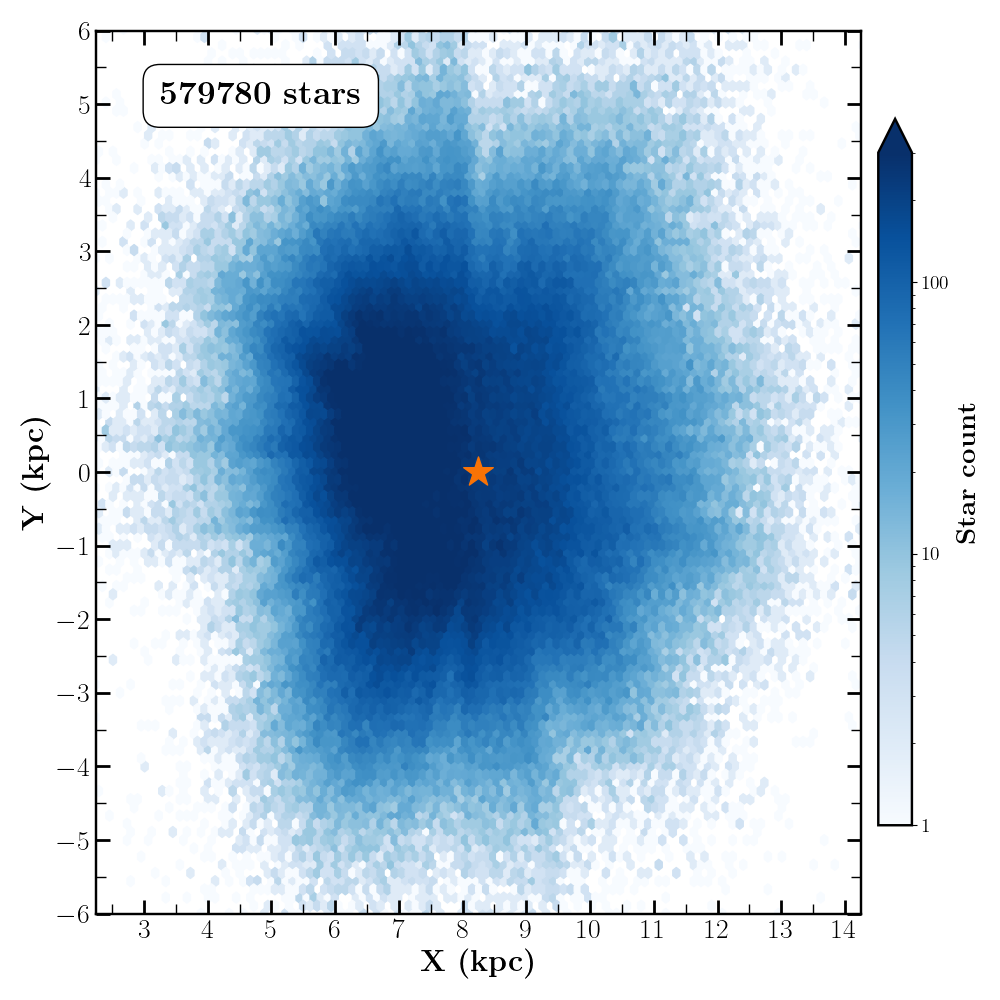}
\includegraphics[width=0.3\textwidth]{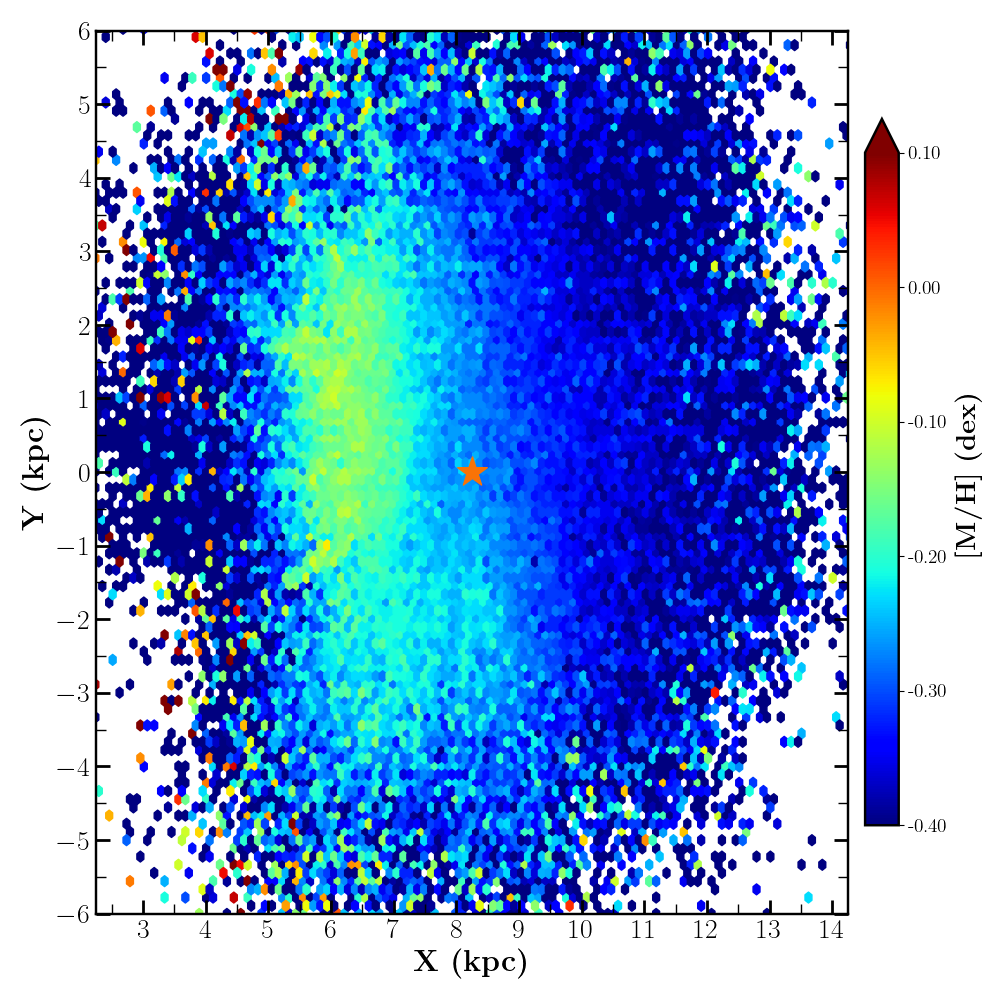}
\includegraphics[width=0.3\textwidth]{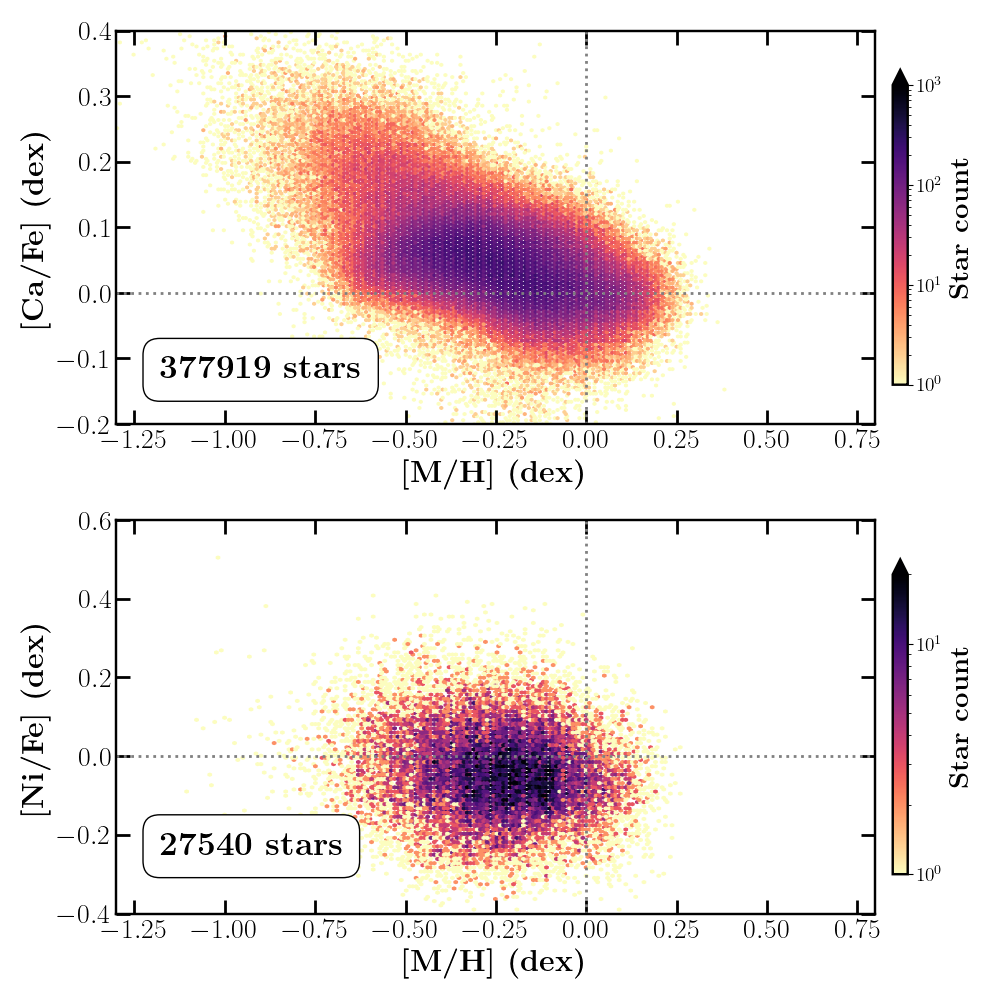}
\caption{Same as Fig.~\ref{Fig.SelFuncXY1} but for RGB stars within 1~kpc of the Galactic plane} and over a more extended XY domain ($\pm$6~kpc from the Sun).  Nickel abundances are shown in the bottom-right panel.
\label{Fig.SelFuncXY2}
\end{figure*}

\begin{figure*}[h]
\includegraphics[width=0.3\textwidth]{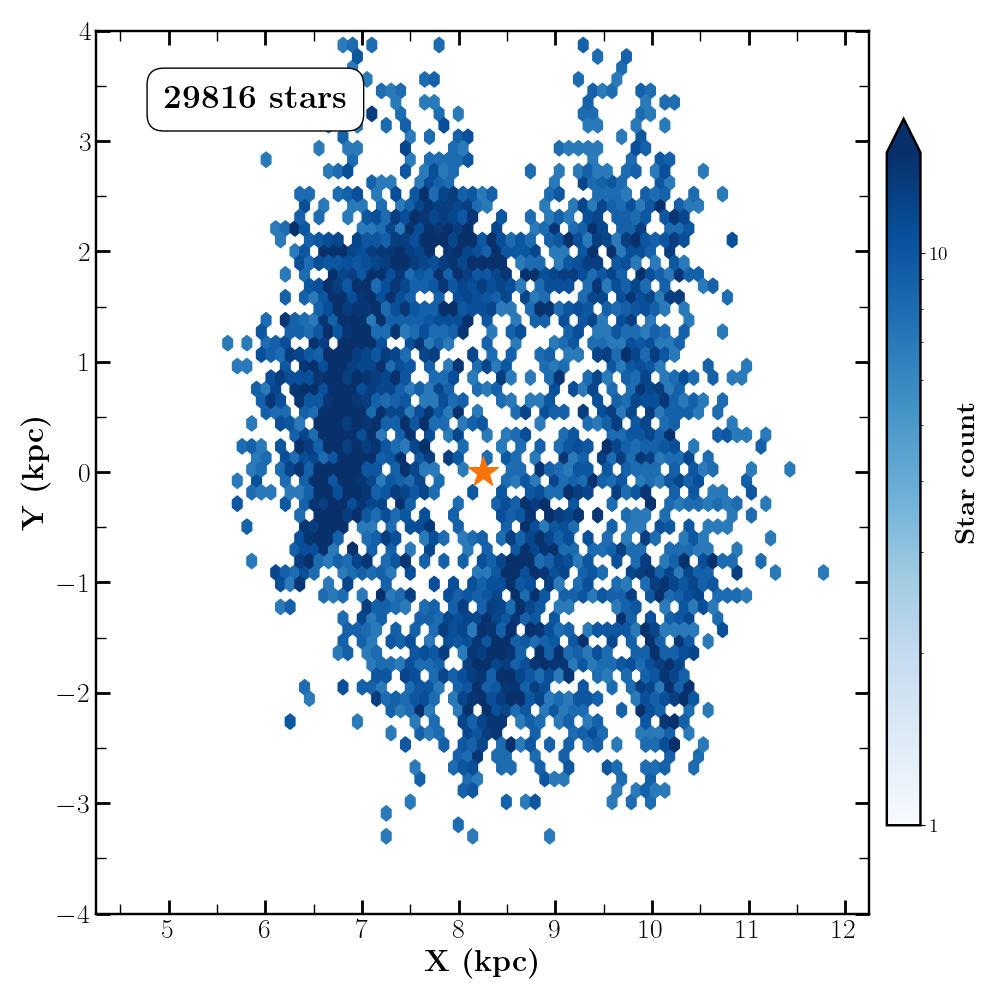}
\includegraphics[width=0.3\textwidth]{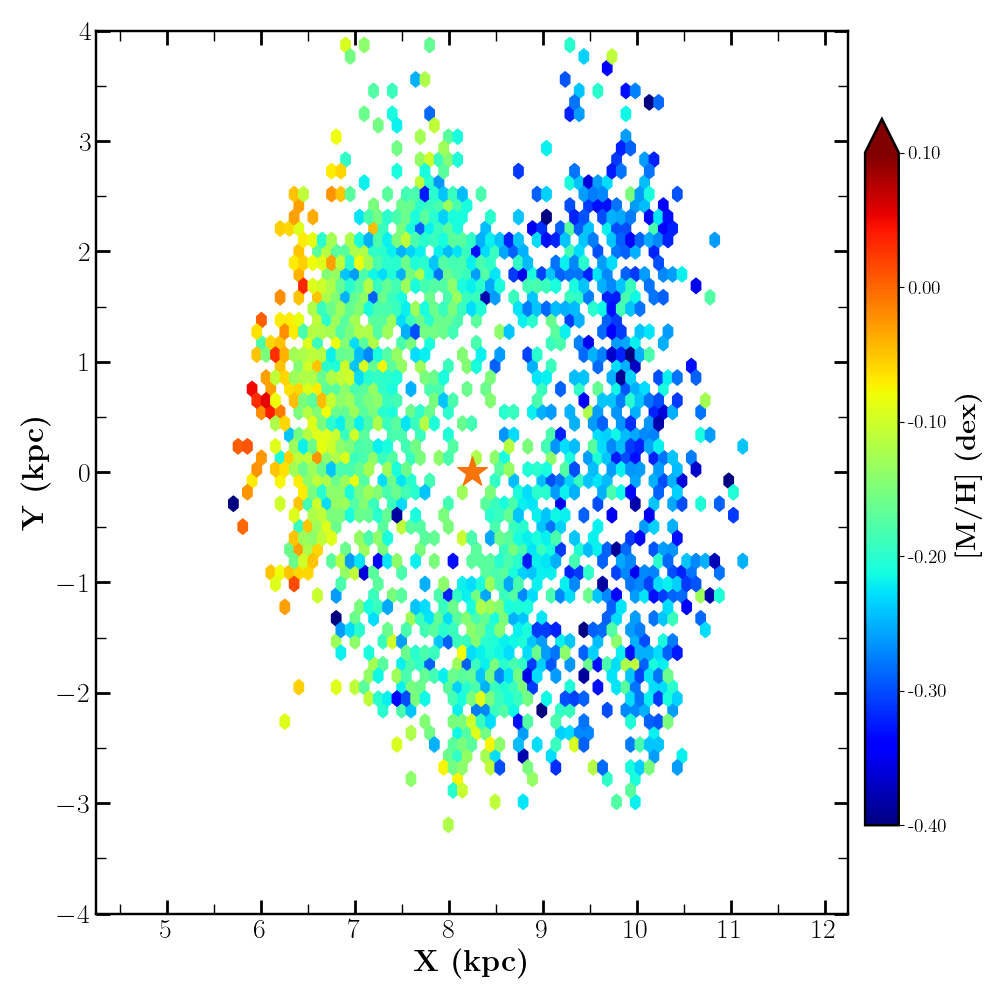}
\includegraphics[width=0.3\textwidth]{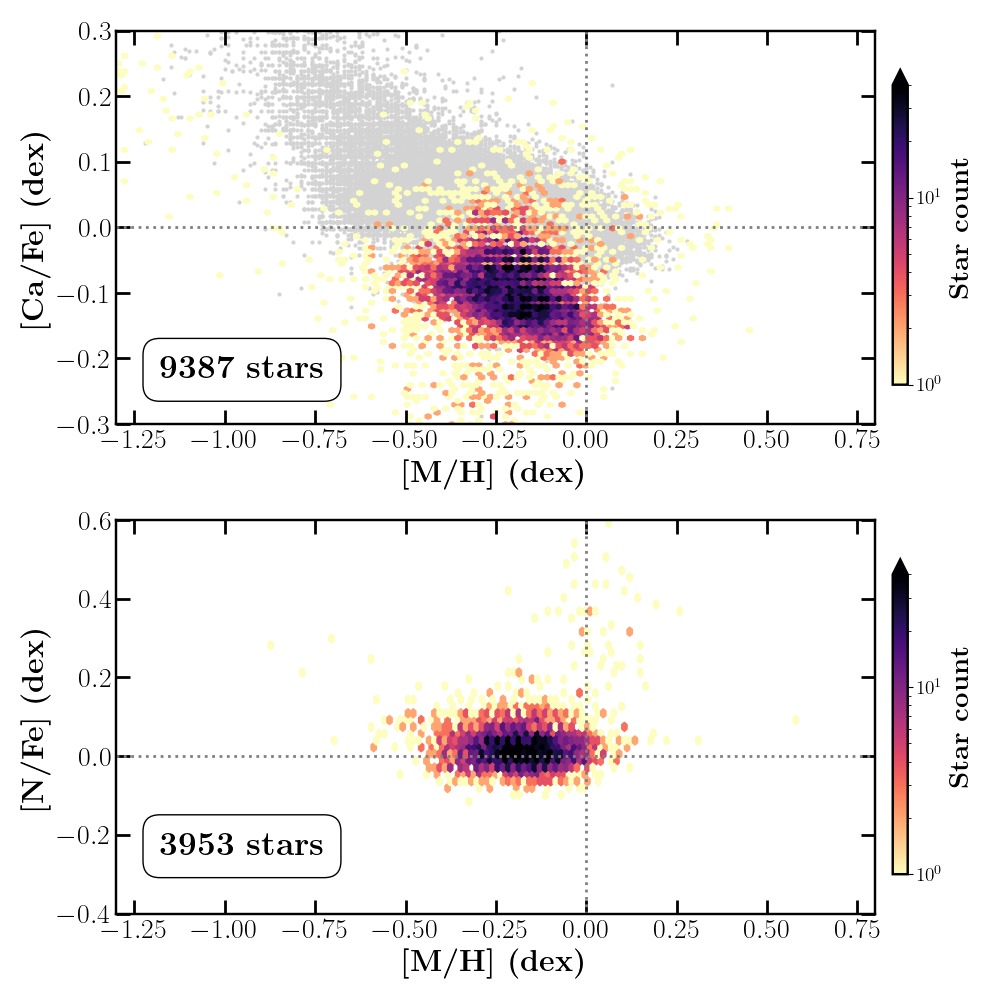}
\caption{Same as Fig.~\ref{Fig.SelFuncXY1} but for Massive
stars located within 400~pc of the Galactic plane, thus focusing attention on the thin disc. The spiral arm structure is visible in the left and central panels. Calcium and nitrogen abundance distributions are shown in the right top and bottom panels, respectively. For comparison, the distribution of the RGB stars in terms of calcium abundance is shown in grey in the top panel}.
\label{Fig.SelFuncXY3}
\end{figure*}

\begin{figure}[h]
\includegraphics[width=0.48\textwidth]{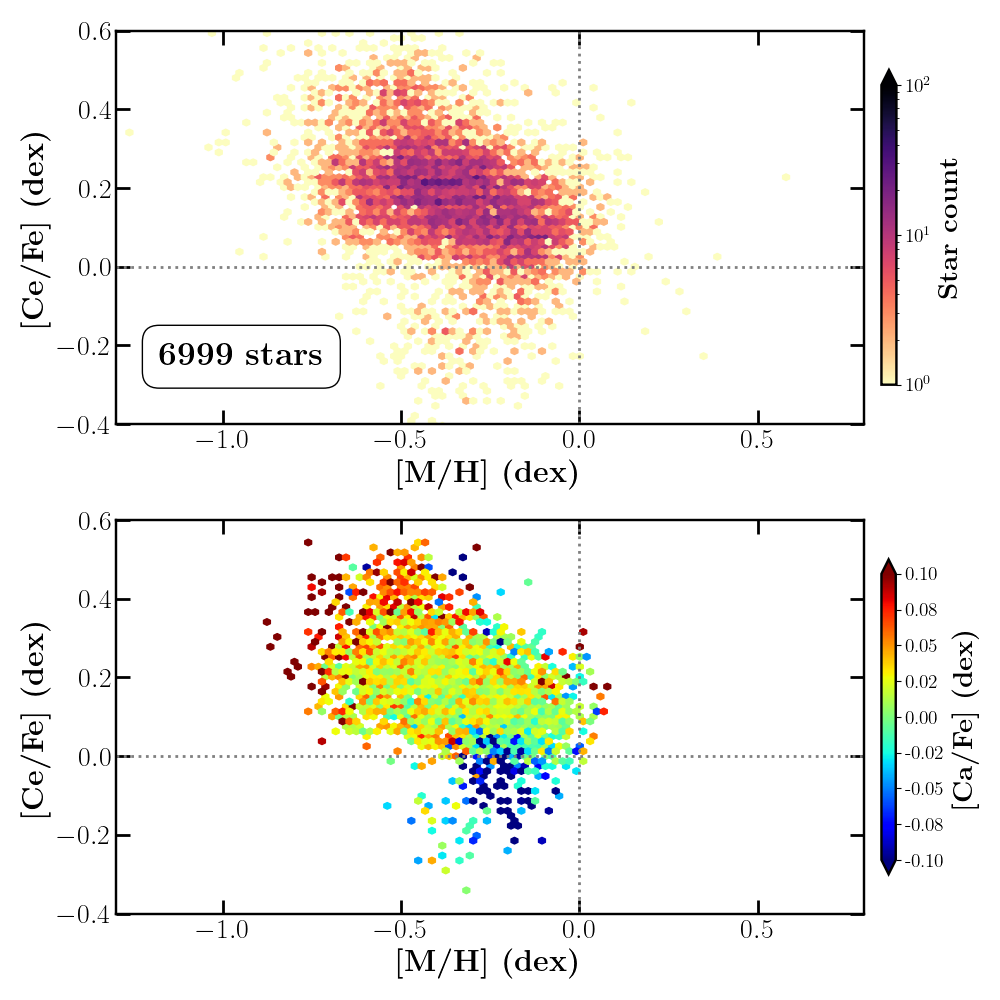}
\caption{Cerium abundances for the coolest Massive and RGB stars of the Medium Quality sample located close to the Galactic plane and colour coded according to stellar density and the median of \CaFe\ (top and bottom panel, respectively).}
\label{Fig.SelFuncCe}
\end{figure}

\subsection{Exploration of the selection function}\label{Sec:selFunc}
The large spatial coverage of the Gaia chemical database requires a global presentation of the underlying parameter distribution across the Galaxy. In addition, the large variety of stellar types included in the catalogue demands that we understand the underlying spatial distribution across the Kiel diagram. Without entering into a detailed characterisation of the selection function, in this section we explore its impact so as to help DR3 users and to support the chemo-dynamical analysis of this paper.

\subsubsection{Stellar parameter distribution across the Galaxy}
To dissect the covered spatial volume, we divided the ($R$, $Z$) plane into bins of 2x2~kpc$^2$. The resulting [$R$, $Z$] grid of Kiel diagrams colour coded according to \meta\ is shown in  Fig.~\ref{Fig.MW_Kiel},  and that of \alphaFe \ versus \meta\ distributions is presented in Fig.~\ref{Fig.MW_MetaAlpha}. In both figures, the global ($R$, $Z$) density contour plot is shown in the background. For the purpose of clarity, only the stars in the  Medium Quality sample (4\,140\,759 objects) are included.
As anticipated from the previous section, about 90\% of this sample is found between 6 and 10~kpc in $R$ and  $|Z|<2$~kpc. 

The majority of the stars are  metal-rich and alpha-poor dwarfs on the main sequence and belonging to the thin disc. Close to the Sun, all evolutionary stellar stages from main sequence stars up to the tip of the asymptotic giant branch (AGB) are encountered with a large mixture of \meta\ and \alphaFe\ combinations. Moving away from the solar location in both directions, the selection function progressively favours intrinsically brighter stars with lower \g\ values. The radial and vertical metallicity gradients can also be qualitatively appreciated (see Sect.~\ref{Sec:Gradients} for quantification of the gradients). 
For $|Z|>4$~kpc, only metal-poor RGB and AGB stars are observed and most of them have metallicities of between $\sim$-2.5 and $\sim$-0.5~dex with high $\alpha$-enrichment. Nevertheless, a few more metal-rich stars from the flared disc are visible in the outer regions. We also note a high symmetry  with respect to the Galactic plane for the Kiel 
and \alphaFe\ versus \meta\ diagrams. As illustrated in Fig.~\ref{Fig.MW_MetaAlpha}, the global standard decreasing trend of \alphaFe\ with metallicity can be seen across the Galaxy.  

\subsubsection{Kiel diagram selections near the Galactic plane }
To further illustrate the selection function close to the Galactic plane, where stellar type variations are more important, we first show in Fig.~\ref{Fig.SelFuncKiel} the Kiel diagram of the Medium Quality stars with $|Z|<1$~kpc. We then define three subsamples by selecting: (i) the hottest dwarfs of the main sequence around the turn-off, (ii) the bright part of the RGB above the clump, and (iii) the massive bright stars hotter than the RGB, including blue loop stars and variables such as  Cepheids. This last selection, which is biased towards young stars, is indeed dominated by stars located very close to the Galactic plane and an additional filter keeping stars at $|Z|<$~400~pc is applied to avoid contaminants (mainly from the RGB population that is nearby in the Kiel diagram).
Hereafter, these subsamples are respectively labelled hot turn-off (HotTO), RGB, and Massive as indicated in Fig.~\ref{Fig.SelFuncKiel}. 

The (X,Y) spatial distribution and chemical properties of the three populations are shown in the left and central panels of Figures \ref{Fig.SelFuncXY1}, \ref{Fig.SelFuncXY2}, and \ref{Fig.SelFuncXY3}, respectively. The metallicity colour scale is the same in the three figures for comparison purposes. In addition, the right top panels in the three figures show the evolution of the calcium abundance ratio with respect to iron, \CaFe, as a function of metallicity for the three populations selected in the Kiel diagram of Fig.~\ref{Fig.SelFuncKiel}. The abundance of this chemical element is the easiest to derive from RVS spectra thanks to the presence of the strong absorption Ca~{\sc ii} IR triplet  lines.
Finally, the abundance trends of additional elements are shown in the right bottom panels as a function of stellar type and therefore the presence or absence of the lines of an element in the related spectra.  Sulfur lines are only detected for hot dwarfs (because of its rather high excitation energy) and its abundance is illustrated in Fig.~\ref{Fig.SelFuncXY1}. The only nickel line selected in the RVS range  provides reliable estimates for bright cool giants and its corresponding abundance is shown in Fig.~\ref{Fig.SelFuncXY2}. Finally, nitrogen abundances are nicely illustrated using the hot massive population in Fig.~\ref{Fig.SelFuncXY3}.

To optimise the individual chemical abundance plots, the {\it Individual abundance sample} defined in Sect.~\ref{Sec:samples} is used. We note that, even after applying rather strict selection criteria (c.f. Sect.~\ref{Sec:samples}), the number of reported abundances is huge (from $\sim 10^4$ to a couple of $\sim 10^6$, depending on the chemical species) and generally much larger than any other published abundance catalogue. As shown below, the behaviour of these chemical species is in perfect agreement with previously published catalogues of stellar abundances and with Galactic chemical evolution  models \citep[see e.g.][and references therein for observational datasets]{Prantzos18, Koba20, Matteucci21}, confirming their high quality.  Finally, we note here that the individual abundance calibrations proposed by \cite{GSPspecDR3} have been applied.

\bigskip
\noindent {\it Spatial distributions in the disc plane}

\smallskip
\noindent First,  the left panels of Figs.~\ref{Fig.SelFuncXY1}-\ref{Fig.SelFuncXY3} reveal that almost 680\,000 HotTO stars are found within about 1~kpc of the Sun, because they are not bright enough to sample more distant regions. In contrast, the RGB sample contains somewhat fewer stars (about 580\,000) but allows us to explore much more extended regions up to 5-6~kpc from the Sun. Similarly, Massive stars are distributed up to 
about 3~kpc from the Sun. Their spatial distribution is not homogeneous and nicely reveals the closest spiral arms, that is, the Sagittarius/Carina, Local, and Persus arms, 
in agreement with the spatial maps derived by \cite{Eloisa21} from Gaia EDR3 astrometry and photometry. These massive stars also show the enhanced equivalent width of the diffuse interstellar band at 862\,nm, confirming their nature as tracers of the spiral arms (\citealt{PVPDIB}).

\bigskip
\noindent {\it Metallicity distributions in the Galactic plane}

\smallskip
\noindent
The metallicity distributions 
 in the (X,Y) plane of the three samples are shown in the central panels of Figs.~\ref{Fig.SelFuncXY1}-\ref{Fig.SelFuncXY3}. Most strikingly, we see that a large fraction of the HotTO stars are metal rich  (up to \meta$\sim$+0.5~dex). Indeed, the selection of the HotTO sample  favours the younger and therefore more chemically enriched TO stars because of the temperature cut at \T$>$6000~K. Finally, the slightly more metal-poor yellowish feature centred on the solar position comes from an inhomogeneous $Z$ distribution of these stars. In fact, near the Sun, the sample reaches slightly greater distances from the plane, with median values of around 0.2~kpc, while outside the solar vicinity, and probably driven by the younger objects in the sample, most of the HotTO stars are slightly hotter (in order to be brighter) and therefore closer to the plane  at $|Z|\la$0.1~kpc. Therefore, the vertical metallicity gradient (see also Sect.\ref{Sec:Gradients}) seems to explain the central metallicity feature. Nevertheless, the solar metallicity (represented by the colour of the central starred marker) is in agreement with the global \meta\ distribution of HotTO stars, as expected.

Regarding the RGB sample (cf. Fig~\ref{Fig.SelFuncXY2}), stars are on average more metal-poor 
than the HotTO ones, with values around \meta$<-0.4$~dex in most parts of the (X,Y) plane. However, there is a metallicity gradient that is visible for $X\gtrsim$5~kpc, with \meta\ values decreasing with distance from the Galactic centre, as expected (see also Sect.~\ref{Sec:Gradients}). In this framework, the Sun's metallicity is clearly higher that the average observed one in the solar surroundings. This is probably due to the underlying age distribution of RGB stars. In fact, the selected locus of the Kiel diagram is populated by a large range of stellar masses, corresponding to a wider age distribution of these  objects. In addition, the $Z$ distribution of this RGB sample includes stars farther away from the plane that reduce the median metallicity.

Interestingly, we clearly see that the distribution is not azimuthally symmetric. Stars with $X$ in the range between 5 and 7~kpc and $Y$ within [-2.0,+4.0]~kpc seem to define a metal-rich sustructure, which probably coincides with the Sagittarius arm position.
Moreover, stars with $X$ in the range between 7 and 10~kpc show an azimuthal metallicity trend, with objects in azimuthal angles closer to the major axis of the Galactic bar  (lower $Y$ values in the figure) having higher metallicities. This could also be the consequence of the presence of the  Local Arm and generally of the pitch angle of the  spiral arms. For the innermost regions of the Galaxy, it could also partially result from a spatially inhomogeneous extinction distribution. In any case, these azimuthal dependencies reveal an inhomogeneous metallicity distribution, even for this upper RGB population for $R\gtrsim$5-6~kpc.
In this sense, it is worth noting that the innermost regions ($X$<5~kpc) in this figure appear metal poor and therefore do not follow the radial metallicity gradient. However, as is visible in the bottom panels of Fig.~\ref{Fig.XY}, the inner Galaxy sample is dominated by metal-poor thick-disc stars, particularly at  $|Z|$>0.5 kpc. 

Finally, the central panel of  Fig.~\ref{Fig.SelFuncXY3} shows that the arms internal to the position of the Sun and mapped by the Massive stars are much more metal-rich than the external ones, following again the expected radial gradient. 
It can also be noted that, as already discussed for the RGB sample, Massive stars  show azimuthal dependencies in their metallicity distribution. 
In addition, it can be seen that the Sun lies between two distinct spiral arms and that its mean metallicity is higher than that of the surrounding  younger Massive stars. This seems to indicate that the younger populations are chemically impoverished, which is in contradiction with a continuous chemical evolution where  metallicity increases with time. 
It is also in contrast with the older but more metal-rich turn-off stars discussed above. This point is discussed in more detail below, when looking at other chemical diagnostics. It is worth noting that although the parametrisation of young stars  from high-resolution  spectra (R$\ga$100,000) can be difficult because of a combination of intrinsic factors \citep[activity, fast rotation, magnetic fields; see for instance][]{Zhang21,Spina20}, this is mainly affecting particularly young stars (age $\la$200~Myr) and does not alter the RVS parametrisation, which is performed at medium resolution (R$\sim$11\,500). Visual inspection of the \gspspec\ parameter and individual abundance solution and synthetic spectral fitting for one typical star of the Massive population can be found in \cite{GSPspecDR3}. No particular parametrisation problems are reported for these stars; they have generally very high S/N values. Additional support to the derived metallicities for this population comes from the validation by \cite{ClementiniDR3} of RR Lyrae metallicities. The authors show that spectroscopic \gspspec\ metallicities are in good agreement with those derived from the period--luminosity relation. RR Lyrae stars are indeed a subsample of our Massive population.

\bigskip
\noindent {\it [Ca/Fe] distributions}

\smallskip
\noindent
We now look at the \CaFe\  abundances presented in the upper right panels of Figures~\ref{Fig.SelFuncXY1}, \ref{Fig.SelFuncXY2}, and \ref{Fig.SelFuncXY3} (individual abundances samples). Concerning HotTO stars, Figure~\ref{Fig.SelFuncXY1} presents the \CaFe\ trend with metallicity. 
As expected for this faint nearby population, the distribution is dominated by a low-$\alpha$ thin-disc sequence, with \alphaFe\ decreasing as metallicity increases and reaching \meta\ values as high as +0.5~dex. The relatively large metallicity range of this sample is therefore probably due to stars of different origins visiting the solar neighbourhood as already observed for ground-based surveys of dwarf stars.
We emphasise that the decrease in \CaFe\ with \meta\ is continuous even for super-solar metallicity stars, in perfect agreement with Galactic evolution chemical models. This is seen for the first time with Gaia DR3 data  thanks to an optimised spectral normalisation of metal-rich stars \citep{Pablo20}. 

\CaFe\ abundances are presented for the RGB population in the right upper panel of Fig.~\ref{Fig.SelFuncXY2}. Two sequences are easily identifiable, with the thick disc population being more numerous for the RGB selection than for the HotTO one. As already mentioned, the metallicity distribution of the RGB is shifted to more metal-poor values than that of HotTO stars.
We also note that the continuous decline in \CaFe\ abundance with metallicity seen for the HotTO sample is also visible here; however it is less clearly observable due to  the presence of \CaFe-\-poor RGBs at 
metallicities larger than $\sim$-0.3~dex coming from the Massive sample, which is contiguous in our Kiel diagram selection. The 
\CaFe\ decline with \meta\  in the thin-disc sequence can also be seen by examining the  upper envelope of the \CaFe\ distribution, which reaches 
\CaFe$\sim$+0.0dex at \meta$\sim$+0.25dex.

Finally, Fig.~\ref{Fig.SelFuncXY3} illustrates the \CaFe\ abundance for about 7\,400 Massive stars. The corresponding distribution for the stars in the RGB population is shown in grey for reference. Surprisingly, most stars in the Massive sample are Ca poor, presenting  \CaFe\ values down to $\sim$-0.3~dex with \meta\ varying between -0.5 and +0.0~dex. This confirms the apparent chemical impoverishment of this young population discussed above. We investigated whether these peculiarly low abundances could be due to a zero-point offset for the Massive stars or an artifact of the abundance calibration. The different tests that we performed, which are presented in Sect.\ref{Sec:calibsTests}, allow us to conclude that, for all the combinations
of calibration flavours including no calibration, the Massive sample is impoverished in \alphaFe\ with respect to the thin disc sequence. In addition, the maximum metallicity reached by the Massive sample is also lower than that of the RGB and HotTO samples, independently of the chosen calibration flavour. It is also interesting to consider the fact that this type of young cool massive star in the spiral arms is rare (because of its shorter lifetime). Therefore, only a high number statistics survey, as in the Gaia/RVS one, can observe enough stars of this type to be statistically relevant in the different abundance distributions.  

As noted above, the observed chemical impoverishment of the Massive population is in contradiction with a continuous chemical evolution of the thin disc. Moreover, because of the decreasing stellar density with Galactic radius, the mechanism whereby inward migration favours lower metallicities (as required to explain the chemical pattern of the Massive population) should not dominate. Chemical evolution models and/or simulations, which are beyond the scope of this paper, will be necessary to develop the interpretation of the chemical pattern seen in Massive stars.

\bigskip
\noindent {\it Other individual abundance distributions}

\smallskip
\noindent

We now discuss the additional individual chemical abundance distributions presented in the right lower panels of Figures~\ref{Fig.SelFuncXY1}, \ref{Fig.SelFuncXY2}, and \ref{Fig.SelFuncXY3}. For HotTO stars, Figure~\ref{Fig.SelFuncXY1} shows
the variation of sulfur abundance with metallicity for more than 36\,000 stars. The behaviour of sulfur with metallicity, although exhibiting a larger spread than the \CaFe , is in agreement with the observed global \alphaFe\ trend, again showing a continuous decrease with metallicity. Indeed, the ratio [S/Ca] (and the [S/$\alpha$] one) is almost equal to zero (with a median value of 0.02~dex) over the whole metallicity range, with a rather small median absolute deviation (MAD=0.06~dex). This confirms the $\alpha$-like nature of sulfur in agreement with previous studies \cite[see][for a recent review and consistent results derived from high-resolution spectra]{Jeremy21}. 

Regarding the RGB sample, about 28\,000 nickel abundances are reported for the RGB sample in Fig.~\ref{Fig.SelFuncXY2} (we adopt here [Ni/Fe]$_{\rm unc}$<0.15~dex to increase the size of the sample). [Ni/Fe] stays close to zero for all metallicities, in agreement with the iron-peak nature for this chemical element. Regarding the Massive population in Fig.~\ref{Fig.SelFuncXY3}, the approximately 4\,000 calcium-poor stars located in the spiral arms have nitrogen abundances close to zero (with a very small dispersion) for metallicities varying between -0.5 and 0.0~dex. 

Finally, we show in Fig.~\ref{Fig.SelFuncCe} the behaviour of cerium with metallicity for the coolest part of the Massive stars plus the whole RGB sample. This allows us to select a disc stellar sample with a wide age distribution, from the younger populations in the spiral arms to the older stars in the RGB sample. The two populations show a large, and very similar spatial coverage.
We remind the reader that cerium is a neutron-capture element that is
predominantly produced by slow-neutron captures and belongs to the second peak of the $s$-elements. We find about 7\,000 stars with Ce estimates in these two subsamples, relaxing the Ce upper limit flag to allow values equal to 2. The upper panel of Fig.~\ref{Fig.SelFuncCe}, which is colour coded according to stellar density, shows that the  [Ce/Fe] abundance decreases with mean metallicity in agreement with previous studies (see also Sect.~\ref{Sec:SolarCylinder}) and Galactic chemical evolution models. Moreover, there is a group of stars with very negative [Ce/Fe] for \meta\ between about -0.5 and -0.25~dex. As shown in the lower panel of Fig.~\ref{Fig.SelFuncCe}, in which the [Ce/Fe] versus \meta\ distribution is colour coded according to [Ca/Fe] abundance,  these low-[Ce/Fe] stars correspond to the Massive  Ca-poor population located in the Galactic arms. Therefore,   Ce abundances confirm the chemical impoverishment seen in metallicity and other chemical elements for this Massive star sample. 


\begin{figure}
\includegraphics[width=0.49\textwidth]{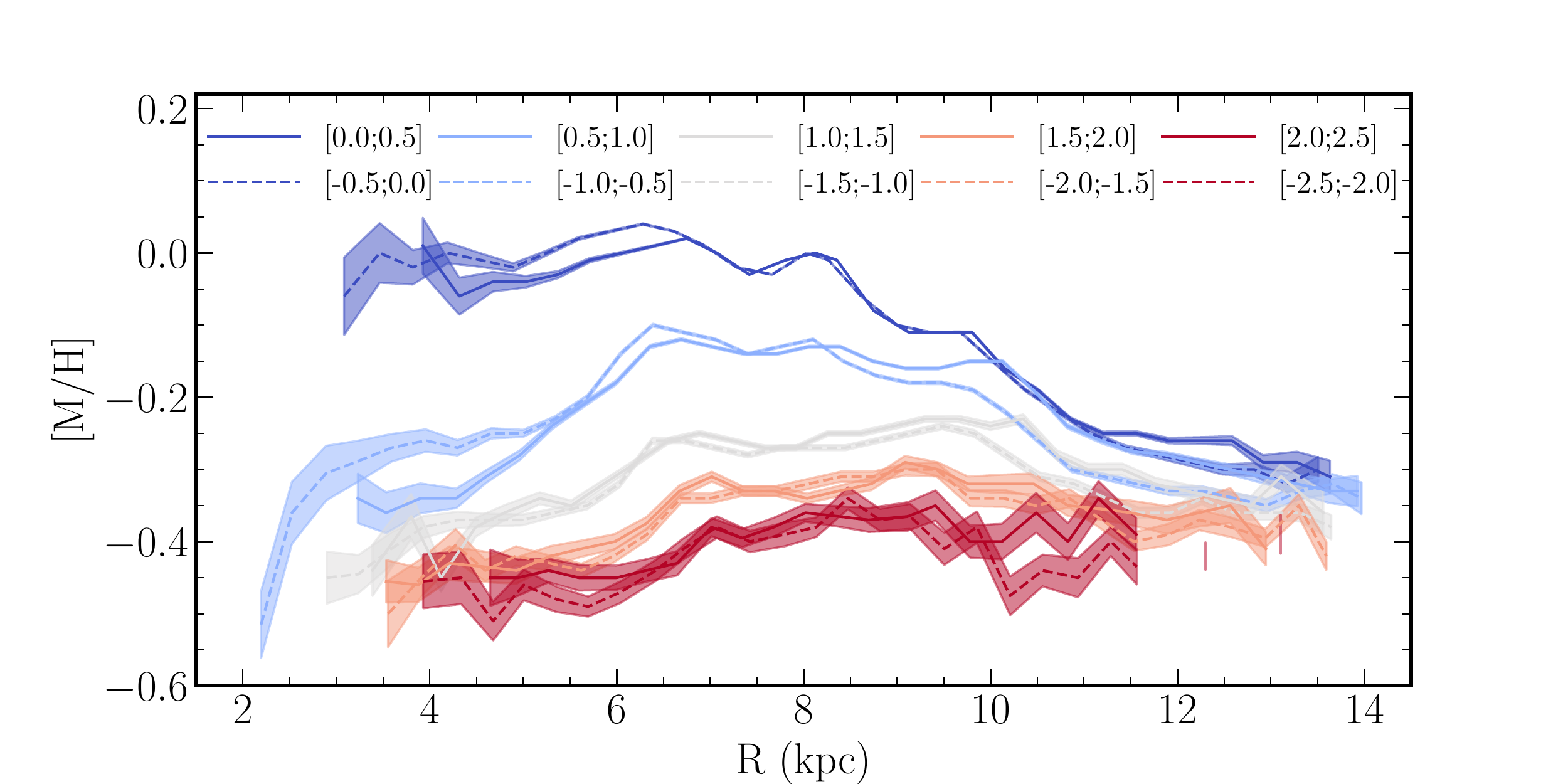}

\includegraphics[width=0.49\textwidth]{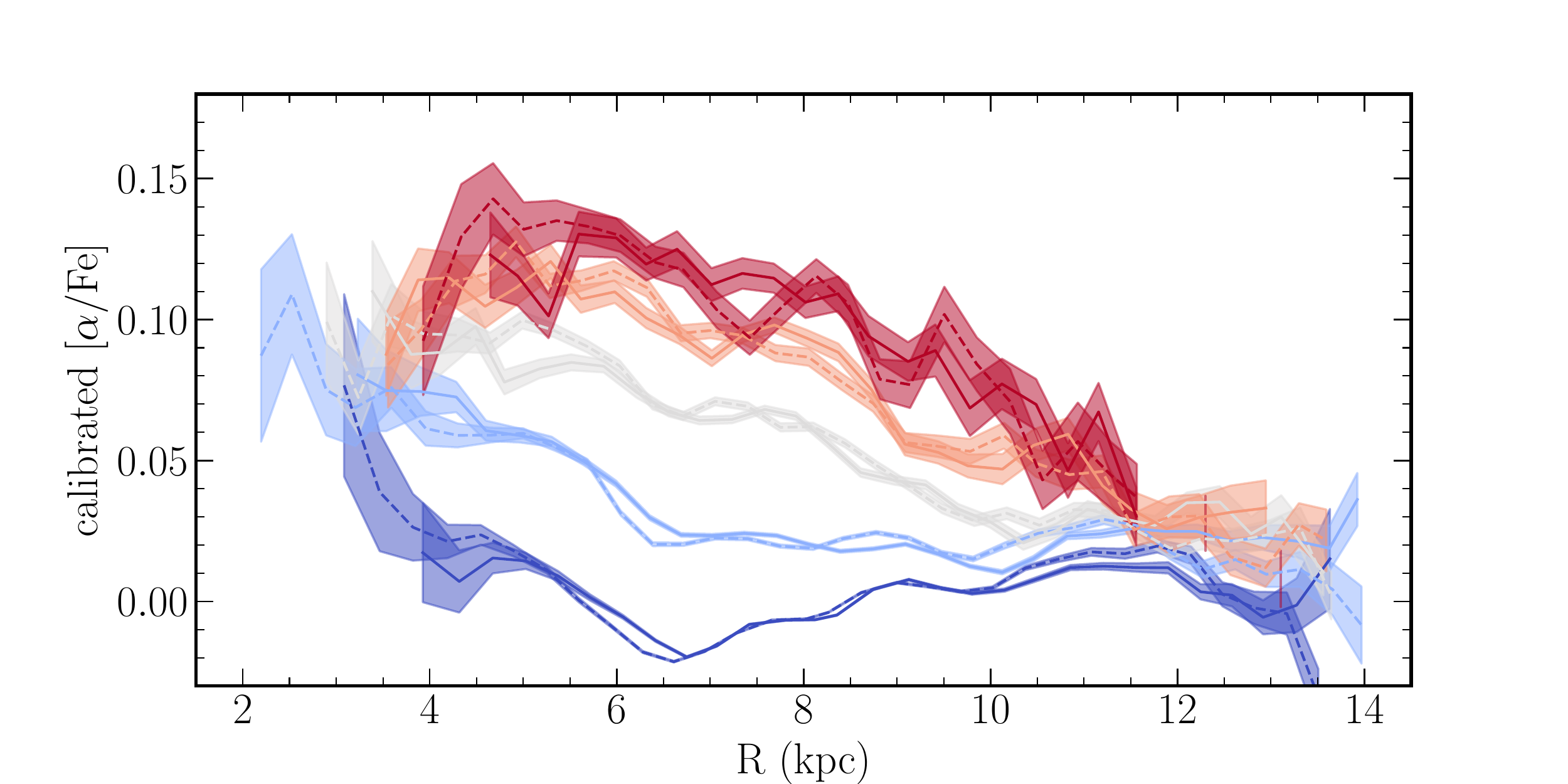}
\caption{Radial gradients for metallicity (top) and \alphaFe\ (bottom) for different distances from the Galactic midplane (in kpc; see the legend). The trends are computed as running medians in bins of 0.5\,kpc, with a 40 percent overlap, provided that at least 50 stars are available to compute the median. The shaded areas represent the Poisson uncertainty on the trends. The colours, which are associated to $Z$ distances from the plane, are similar if the $|Z|$ range is the same.   }
\label{fig:radial_gradients}
\end{figure}

\begin{figure}
\includegraphics[width=0.49\textwidth]{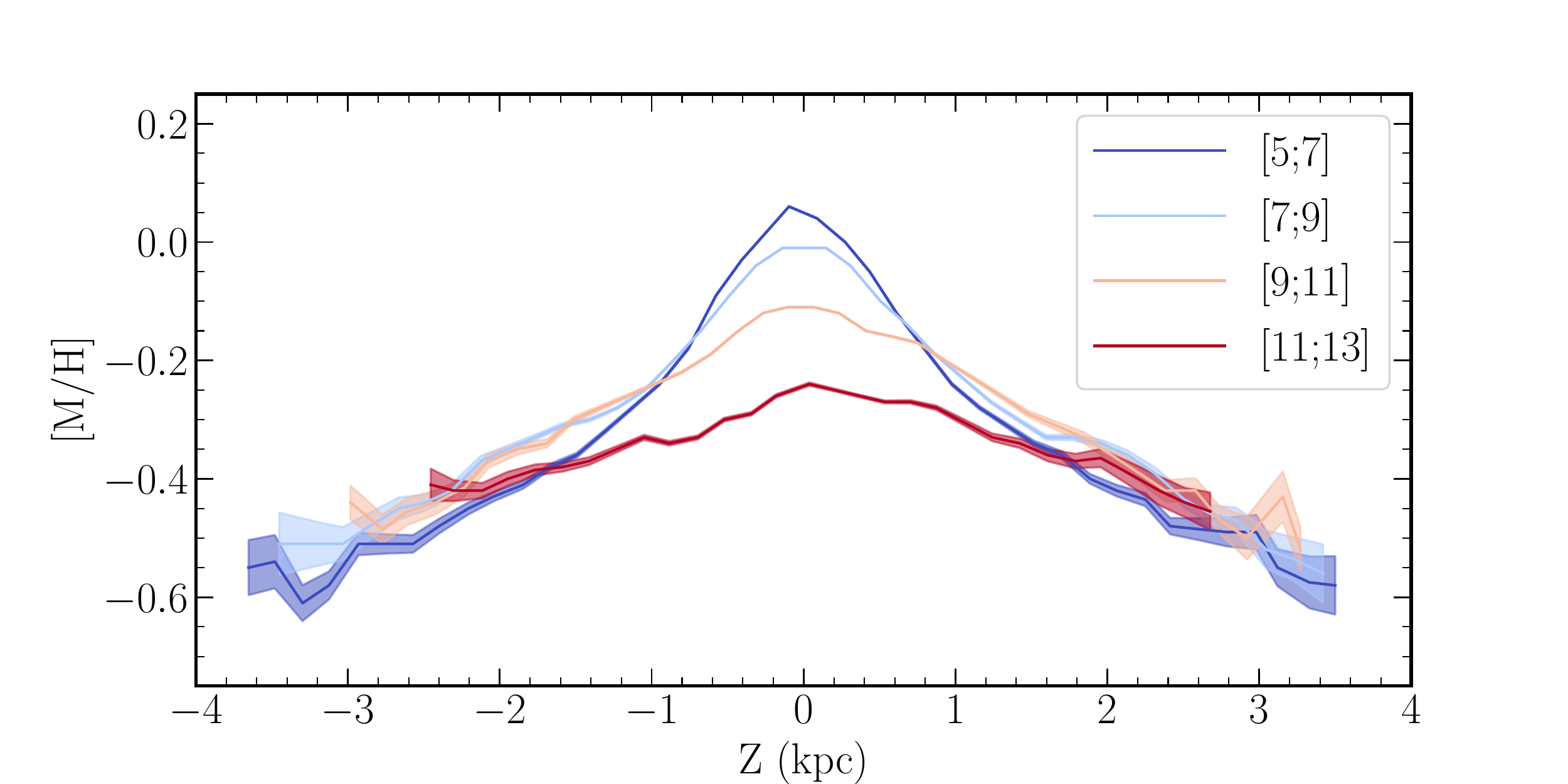}

\includegraphics[width=0.49\textwidth]{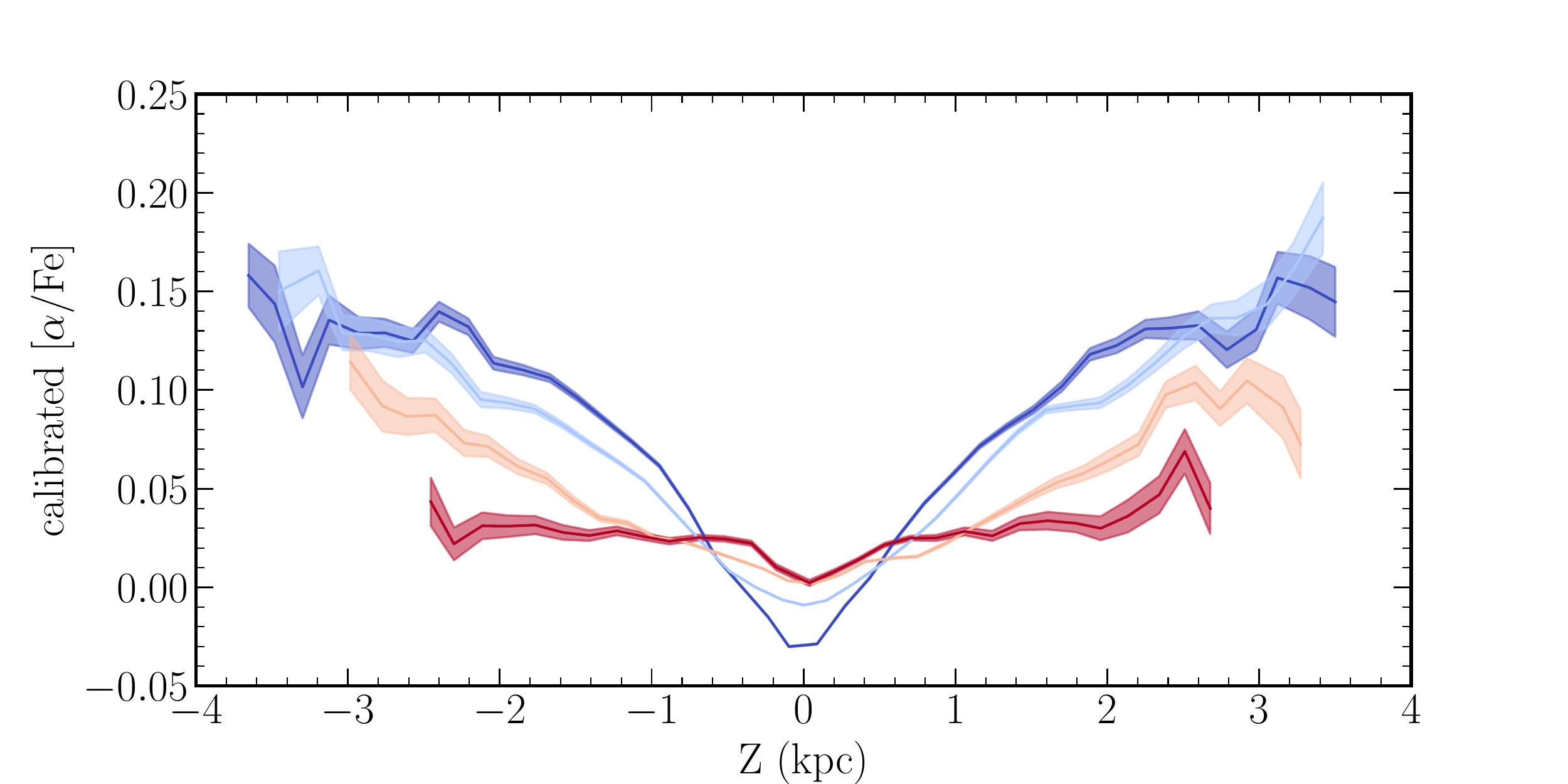}
\caption{Vertical gradients for metallicity (top) and \alphaFe\ (bottom), for different Galactocentric radial ranges. The trends are computed as running medians in bins of 0.5\,kpc, with a 40 percent overlap, provided that at least 50 stars are available to compute the median. Shaded areas represent the Poisson uncertainties on the plotted values. }
\label{fig:vertical_gradients}
\end{figure}

\section{Radial and vertical gradients}
\label{Sec:Gradients}
In this section, we investigate the overall radial and vertical gradients for metallicity and calibrated \alphaFe\ \citep[cf. Sect.~\ref{Sec:APs} and][]{GSPspecDR3}, linking these results to the ones previously found  in the literature. For this purpose, we use the Gradient analysis sample defined in Sect.~\ref{Sec:samples}.

\begin{table*}
\caption{Radial metallicity and \alphaFe\ gradients for $R>7$ kpc and all of the stellar types, at different distances from the plane. The ($a,b$) columns refer to the linear coefficients of the fit and the ($e_a$, $e_b$) ones to their associated errors.}
\centering
\label{Tab:radial_gradients}
\begin{tabular}{c|cccc|cccc}
& \multicolumn{4}{c}{[M/H] } & \multicolumn{4}{c}{\alphaFe } \\
 $<Z>$ &  $a$ &     $b$ &  $e_a$ &  $e_b$ &  $a$ &     $b$ &  $e_a$ &  $e_b$ \\
 (kpc) & (dex\,kpc$^{-1}$) & (dex) & (dex\,kpc$^{-1}$) & (dex) & (dex\,kpc$^{-1}$) & (dex) & (dex\,kpc$^{-1}$) & (dex) \\
 \hline
 0.17 &  -0.055 &  0.418 &       0.007 &   0.058 &   0.007 & -0.062 &       0.001 &   0.007 \\
-0.17 &  -0.057 &  0.436 &       0.007 &   0.061 &   0.007 & -0.063 &       0.001 &   0.006 \\
 0.66 &  -0.024 &  0.055 &       0.004 &   0.034 &  -0.002 &  0.034 &       0.001 &   0.006 \\
-0.65 &  -0.039 &  0.174 &       0.003 &   0.029 &  -0.000 &  0.024 &       0.001 &   0.005 \\
 1.17 &  -0.006 & -0.204 &       0.004 &   0.038 &  -0.010 &  0.140 &       0.001 &   0.011 \\
-1.17 &  -0.014 & -0.151 &       0.004 &   0.033 &  -0.011 &  0.144 &       0.001 &   0.008 \\
 1.68 &  -0.005 & -0.283 &       0.003 &   0.027 &  -0.014 &  0.195 &       0.002 &   0.014 \\
-1.67 &  -0.010 & -0.242 &       0.003 &   0.025 &  -0.014 &  0.192 &       0.001 &   0.007 \\
 2.18 &   0.002 & -0.388 &       0.004 &   0.031 &  -0.017 &  0.238 &       0.002 &   0.014 \\
-2.18 &  -0.009 & -0.310 &       0.004 &   0.040 &  -0.016 &  0.228 &       0.002 &   0.019 \\
\hline
\end{tabular}
\end{table*}

\begin{table*}
\caption{Radial metallicity and \alphaFe\ gradients for $R>7$ kpc, at different distances from the plane when only giants are considered.}
\centering
\label{Tab:radial_gradients_giants}
\begin{tabular}{c|cccc|cccc}
& \multicolumn{4}{c}{[M/H] } & \multicolumn{4}{c}{\alphaFe } \\
 $<Z>$ &  $a$ &     $b$ &  $e_a$ &  $e_b$ &  $a$ &     $b$ &  $e_a$ &  $e_b$ \\
 (kpc) & (dex\,kpc$^{-1}$) & (dex) & (dex\,kpc$^{-1}$) & (dex) & (dex\,kpc$^{-1}$) & (dex) & (dex\,kpc$^{-1}$) & (dex) \\
 \hline
 0.19 &  -0.044 &  0.203 &       0.002 &   0.018 &   0.005 & -0.044 &       0.001 &   0.008 \\
-0.20 &  -0.043 &  0.200 &       0.001 &   0.012 &   0.006 & -0.048 &       0.001 &   0.006 \\
 0.67 &  -0.020 & -0.077 &       0.001 &   0.008 &  -0.001 &  0.031 &       0.000 &   0.002 \\
-0.66 &  -0.028 & -0.017 &       0.002 &   0.014 &  -0.000 &  0.025 &       0.001 &   0.005 \\
 1.18 &   0.000 & -0.354 &       0.001 &   0.010 &  -0.010 &  0.144 &       0.001 &   0.007 \\
-1.18 &   0.000 & -0.369 &       0.001 &   0.010 &  -0.011 &  0.151 &       0.001 &   0.006 \\
 1.68 &   0.012 & -0.538 &       0.003 &   0.024 &  -0.017 &  0.230 &       0.001 &   0.010 \\
-1.68 &   0.010 & -0.518 &       0.002 &   0.019 &  -0.016 &  0.221 &       0.001 &   0.005 \\
 2.19 &   0.014 & -0.602 &       0.005 &   0.041 &  -0.023 &  0.308 &       0.002 &   0.019 \\
-2.20 &   0.014 & -0.607 &       0.004 &   0.037 &  -0.018 &  0.258 &       0.004 &   0.032 \\
\end{tabular}
\end{table*}

Figure~\ref{fig:radial_gradients}  shows the median metallicity (upper panel) and \alphaFe\ (lower panel) radial values for five different distances below (dashed lines) and above (full lines) the Galactic midplane (i.e. in a total of ten different bins in $Z$).  This is indeed achievable with good statistics,  given the large number of stars for which we have parameters; however, we note that we do not separate our sample into chemically thin or thick disc (i.e. based on their \alphaFe-\meta\ sequence). It is interesting to note the limited range in \alphaFe\ enhancement in this figure, peaking at about $\sim$0.2~dex. This is explained by the fact that \gspspec\ \alphaFe\ abundances are governed by calcium indicators dominating the RVS $\alpha$-element information in many pixels. The calcium abundance range in the thin- and thick-disc distributions is more limited than in other elements such  as magnesium, as already observed in the data from ground-based surveys such as the Gaia-ESO Survey \citep[][]{Mikolaitis14} and APOGEE \citep{APOGEE17}.
Figure~\ref{fig:vertical_gradients} is similar to Fig.~\ref{fig:radial_gradients}, showing this time the vertical gradients in metallicity (upper panel) and \alphaFe\ (lower panel) for four different $R$ bins. 

\begin{figure}
\includegraphics[width=0.49\textwidth]{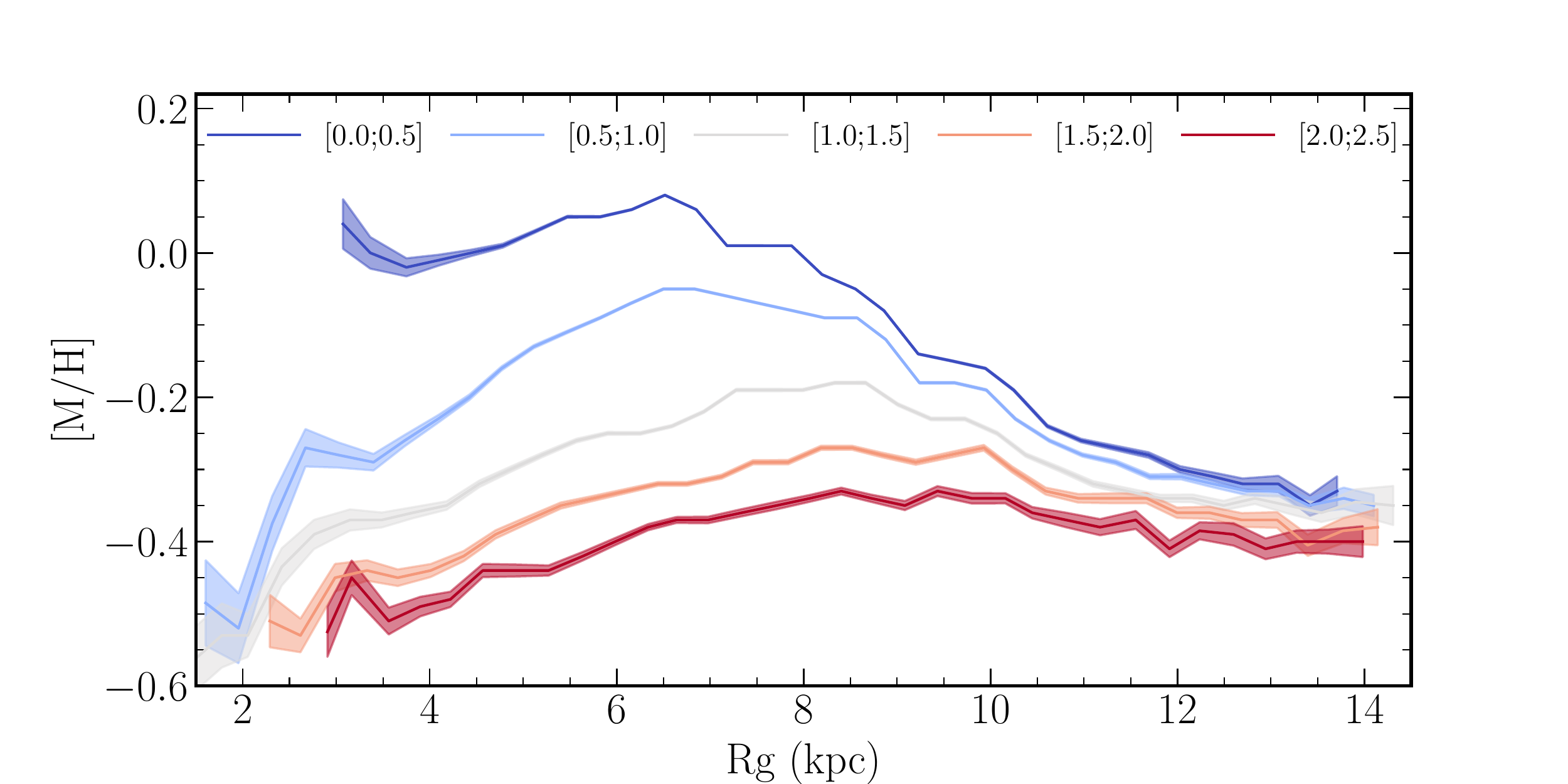}

\includegraphics[width=0.49\textwidth]{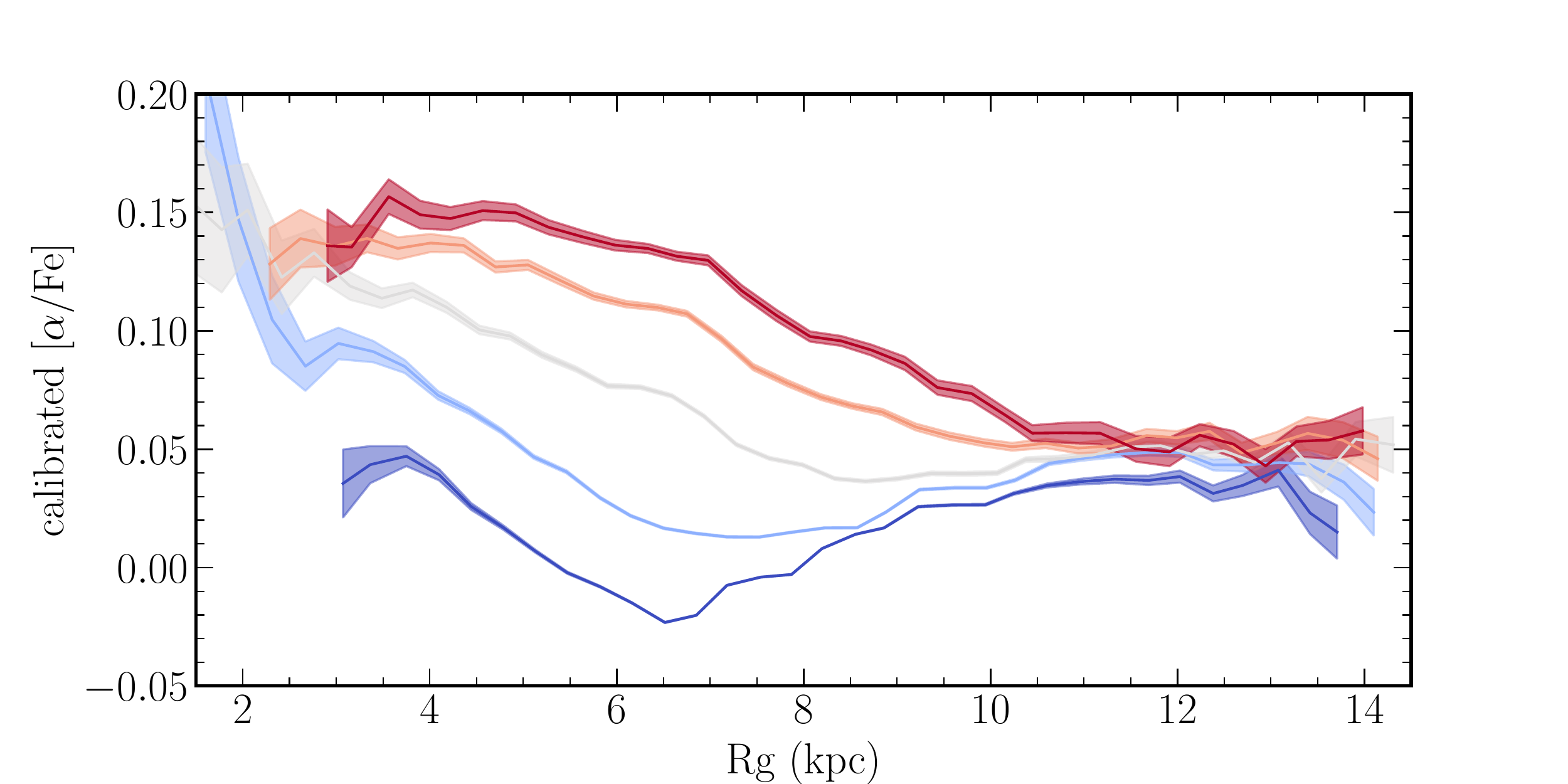}
\caption{Same as Fig.~\ref{fig:radial_gradients}, selecting this time the stars as a function of $Z_{\rm max}$ and plotting the guiding radius 
(defined as ($R_{\rm apo}+R_{\rm peri}$)/2) instead of the observed radius.}
\label{fig:radial_gradients_Zmax_Rguide}
\end{figure}

It is remarkable to find that both metallicity and \alphaFe\ trends show similar behaviours above and below the plane, when the same $|Z|$ range is considered.
More specifically, we find that both  metallicity and \alphaFe\ gradients have a break in their slope  at galactocentric radii $R\sim7$\,kpc, which is particularly visible for $|Z|\lesssim 1$\, kpc. This behaviour is in agreement with results from \citet{Haywood19}  using $\alpha$-low stars from the APOGEE-DR14 catalogue,  \citet{Kordopatis20} from a compilation of spectroscopic catalogues of field stars (LAMOST, RAVE, GALAH-DR2, and APOGEE-DR14), and \citet{ Katz21} using both $\alpha-$high and $\alpha-$low stars from the APOGEE-DR16 catalogue. More specifically, we find that the median metallicity either flattens or even decreases towards the inner Galaxy, while, at the same $R-$position, the median \alphaFe\ increases (see lower plot of Fig.~\ref{fig:vertical_gradients}). 
    We note that this change of regime stays roughly similar when one selects stars based on $Z_{\rm max}$ and guiding radius (see Fig.~\ref{fig:radial_gradients_Zmax_Rguide}). 

    Table~\ref{Tab:radial_gradients} reports the linear \meta\ and \alphaFe\ trends as a function of $R$, in the form of $ax+b$ (and their associated uncertainties $e_a$ and $e_b$, respectively), only considering the galactocentric radii larger than 7\,kpc.  We find that the metallicity slope for the closest bin to the plane 
    is $\approx -0.056 \pm 0.007$ dex\,kpc$^{-1}$.
    This is slightly different from but still consistent with the metallicity gradients found using Cepheids \citep[e.g. $-0.060\pm0.002$ dex\,kpc$^{-1}$][ and $-0.045\pm0.07$ dex\,kpc$^{-1}$, \citealt{Lemasle18} ]{Genovali14}, open clusters \citep[e.g. $-0.076\pm0.09$ dex\,kpc$^{-1}$][ and $-0.068 \pm 0.001$ dex\,kpc$^{-1}$, \citealt{Donor20}]{Spina21}, or thin disc (i.e. low $\alpha-$sequence) field stars \citep[e.g. $\sim -0.053$ to $\sim -0.068$  dex\,kpc$^{-1}$][]{Bergemann14, Recio-Blanco14, Hayden15, Anders17}.   The trends become flatter as one moves to larger $|Z|$, eventually becoming null or slightly positive for bins farther than 1.5\,kpc from the plane ($-0.005\pm0.003$ at $1.5<Z<2$\,kpc and $+0.002\pm0.004$ at $1.5<Z<2$\,kpc). This flattening is again qualitatively consistent with previous results using field stars and interpreted as a smooth transition from a thin disc to more radially concentrated thick disc \citep[see e.g.][]{Boeche14, Kordopatis20, Nandakumar20}. We note that the metallicity gradient when considering the Massive stars sample  (see the selection in Fig.~\ref{Fig.SelFuncKiel}; the gradient  close to the plane is flatter than the one found using the RGB sample ($-0.036 \pm 0.002$\, dex\,kpc$^{-1}$). This difference might be due to the difference in age between these two samples, and is  consistent with the gradient found using the youngest sample of open clusters (see Sect.~\ref{Sec:clusters}). 

    As far as the $\partial [\alpha/Fe]/\partial R$ gradients are concerned (see Table~\ref{Tab:radial_gradients}), for $R\gtrsim7$kpc, we find a value of $+0.007\pm0.001$ dex/kpc close to the plane, in agreement with the values published by \citet{Carrera11, Yong12,Mikolaitis14,  Genovali15, Reddy16, Casamiquela19, Donor20}. The trend becomes steeper as one moves farther away from the Galactic plane, reaching eventually $-0.016\pm0.002$ dex/kpc for $2<|Z|<2.5$\,kpc. Despite not reaching typical thick disc \alphaFe\ enhancements at $|Z|$ where the thick disc dominates\footnote{Possibly due to the fact that Ca abundance dominates the \alphaFe\ estimate.} ($>1.5$\,kpc; see e.g. \citealt{Hayden15, Kordopatis15}), the qualitative trends that we find show indeed that one moves from a thin-disc-dominated population to a thick-disc-dominated population as |Z| increases. This is also seen from the bottom plot of Fig.~\ref{fig:vertical_gradients}, for the curves with $R<11$\,kpc.  
    

The  points above, especially the breaks in $\partial [M/H]/\partial R$ and $\partial [\alpha/Fe]/\partial R$, are more clearly illustrated by Fig.~\ref{fig:vertical_gradients}, which shows the vertical gradients as a function of galactocentric radius. Only four  $R$ bins are investigated, each of 2 kpc in width, for which a sufficient number of stars are available at all $Z$ (see Fig.~\ref{Fig.XY}).
As anticipated from the shift in the slopes of Fig.~\ref{fig:radial_gradients}, we find non-zero vertical gradients at all $R$. Considering stars at $|Z|<1$\,kpc, we find that the inner disc shows steeper vertical metallicity and \alphaFe\ gradients. Eventually, at the outermost radii ($R=[11-13]$\,kpc), we see only a mild vertical metallicity gradient and a null  \alphaFe\ gradient. This is in agreement with \citet[][]{Bensby11, Hayden15, Kordopatis15,  Mackereth19, Katz21}, for example, who found that the thick disc is more radially concentrated than the thin disc. 

\begin{figure}
\includegraphics[width=0.49\textwidth]{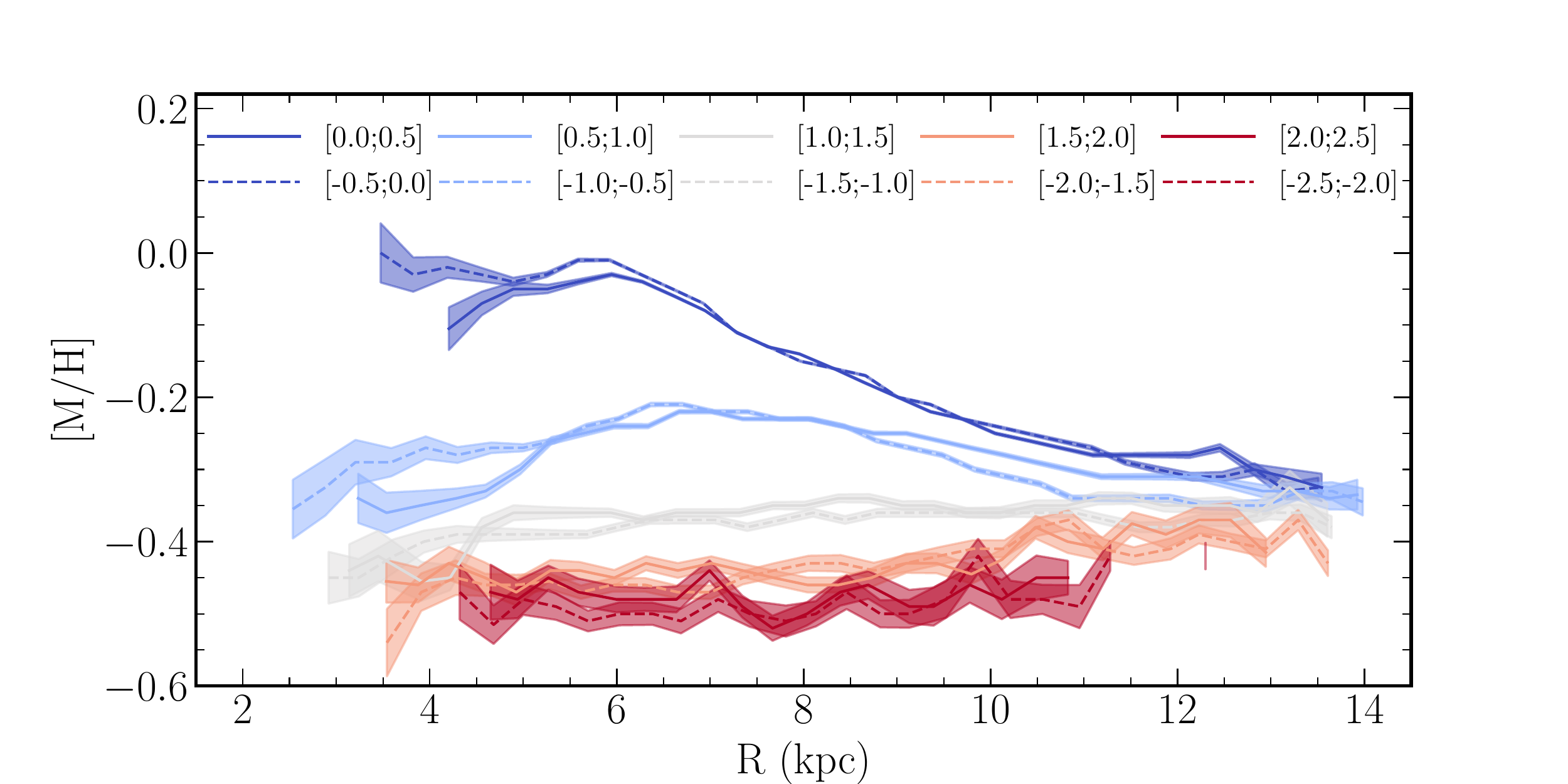}

\includegraphics[width=0.49\textwidth]{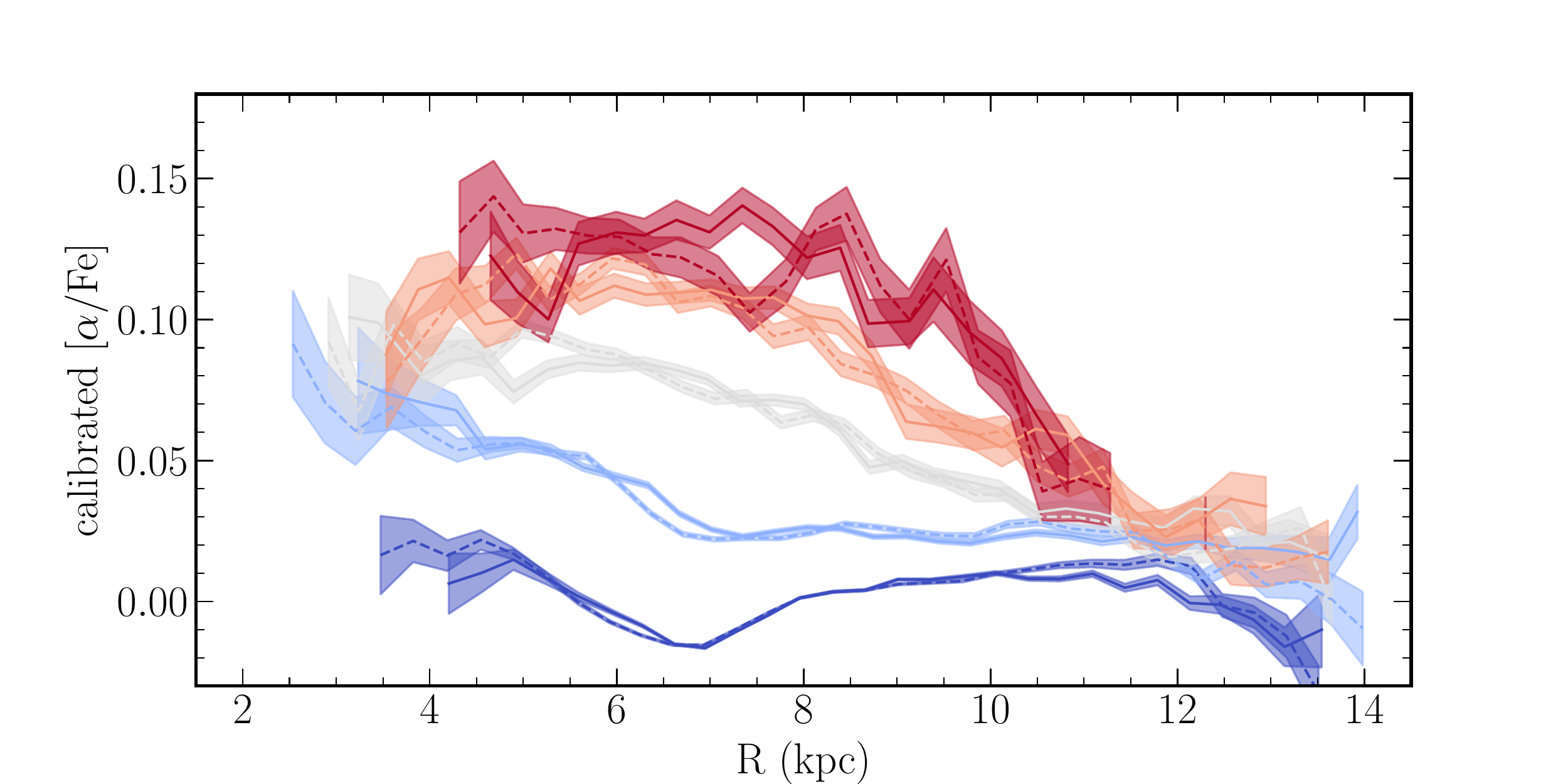}
\caption{Same as Fig.~\ref{fig:radial_gradients}, but selecting only the stars with \g $\le 2$.   }
\label{fig:radial_gradients_RGB}
\end{figure}

\begin{figure*}
\includegraphics[width=\textwidth]{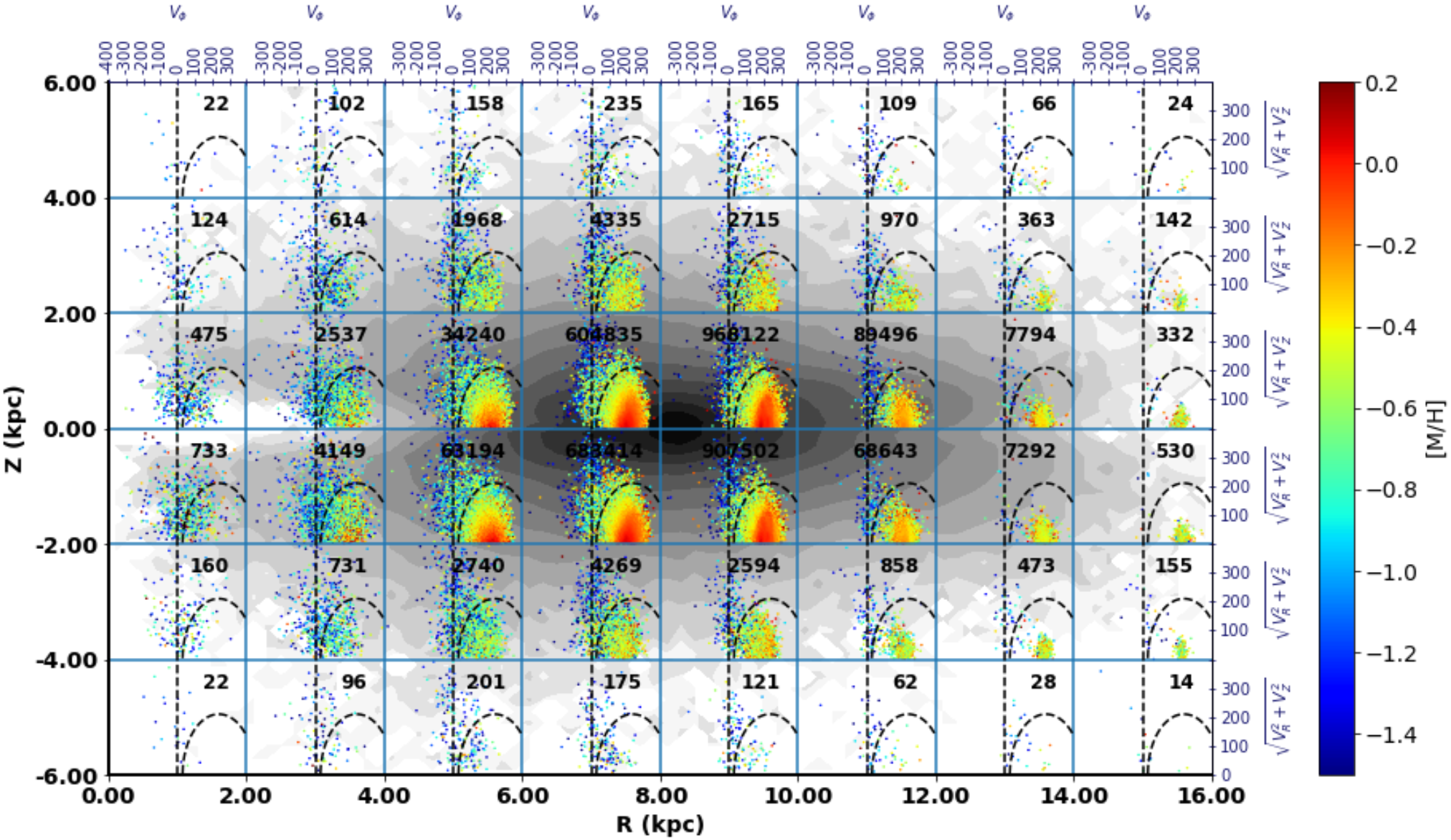}
\caption{Galactic map of the Toomre diagrams colour coded according to \meta. The circular dashed line delimits the regions where the thin- and thick-disc stars are dominant. The vertical dashed line, $V_\phi = 0$, separates prograde and retrograde rotating stars. Halo stars seem preferentially located at positive $V_\phi$ in many panels.}
\label{FigureKinematicsToomre}
\end{figure*}

\begin{figure*}
\includegraphics[width=\textwidth]{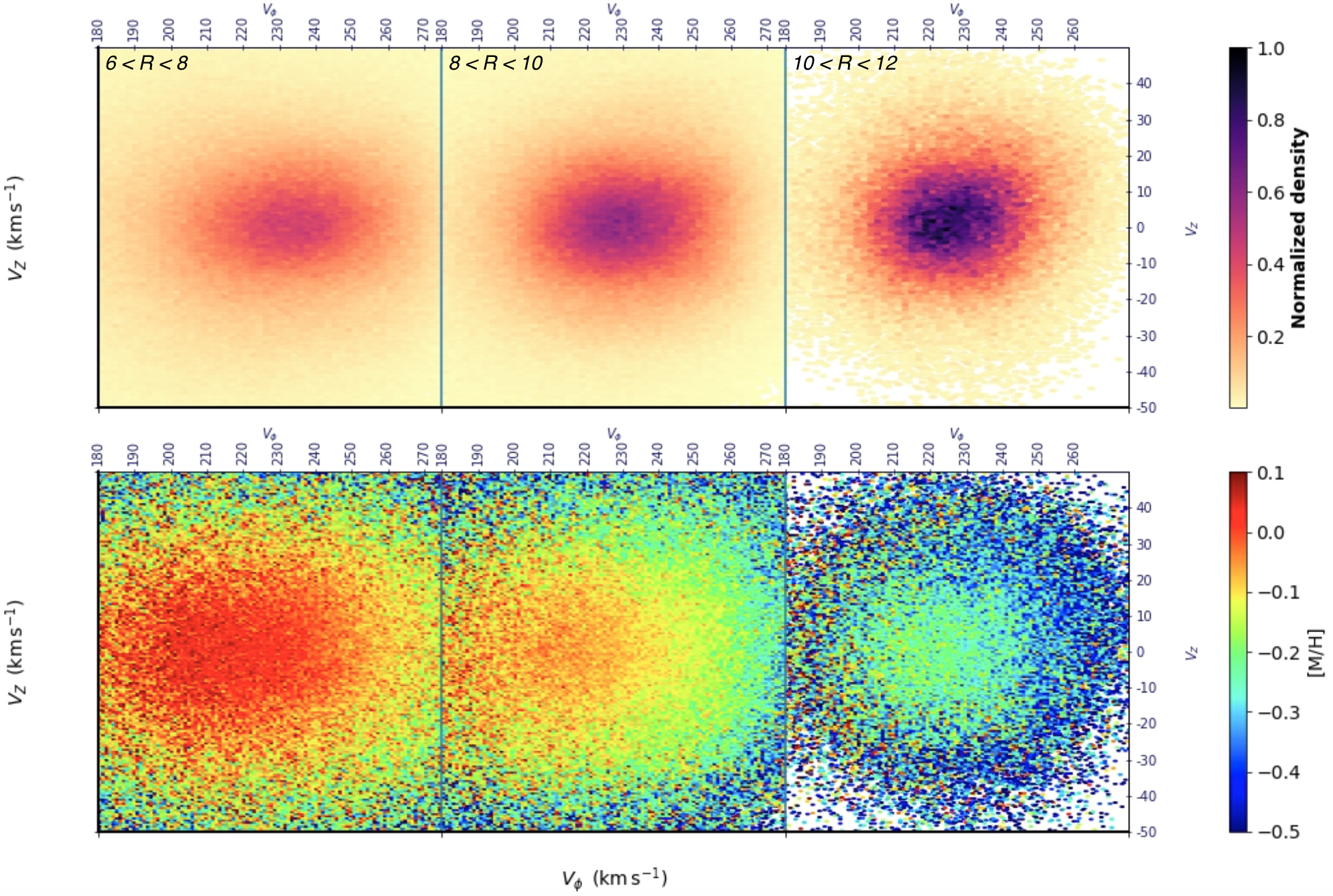}
\caption{Galactic map of the velocity distribution  $V_Z$ vs.\ $V_\phi$ of giant stars (\g $< 3.5$) with $|Z|< 1$ kpc belonging to the Medium Quality sample. The objects are shown, from left to right, in three Galactocentric distance ranges: $6<R<8$~kpc ($883,621$ giants), $8<R<10$~kpc ($867,582$  giants), and $10<R<12$~kpc ($138,208$  giants). Top panel: Normalised density distribution. Bottom panel: Distribution colour coded according to \meta. }
\label{FigureKinematics-Vphi-Vz}
\end{figure*}

\begin{figure*}
\begin{center}
\includegraphics[width=0.90\textwidth]{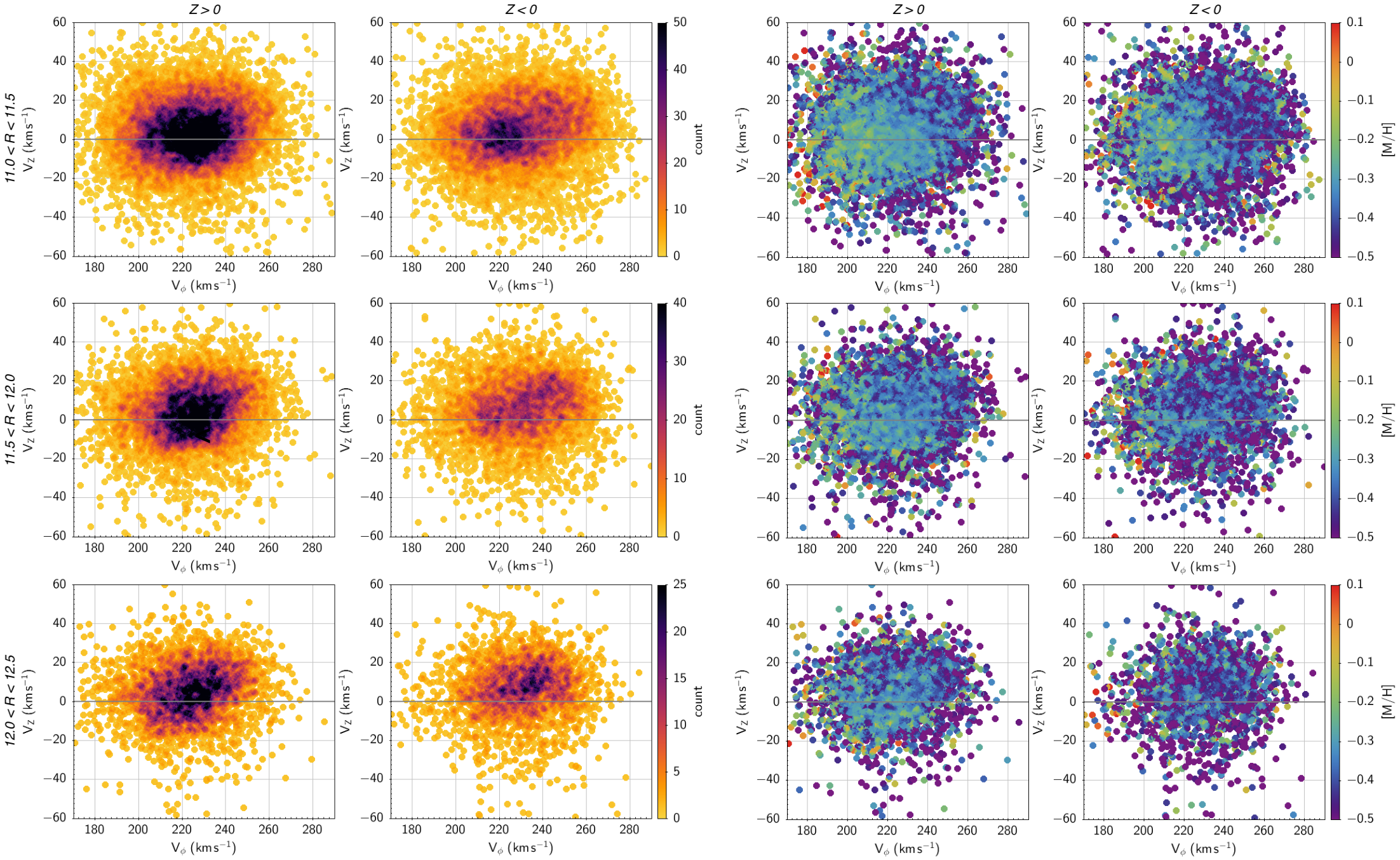}
\end{center}
\caption{Velocity distribution $V_Z$ vs.\ $V_\phi$ of the Medium quality sample for $|Z|< 1$ and $11<R<11.5$ kpc (top), $11.5<R<12$ kpc (middle), $12<R<12.5$ kpc (bottom). The colour coding for stars above and below the plane represents actual counts (first and second columns) and median [M/H] (last two columns), respectively. The bimodality in the velocity distribution observed below the plane and revealed by \cite{Antoja2021} is accompanied by a metallicity bimodality as seen in the rightmost panels.}
\label{FigureKinematics-Vphi-Vz-11-12.5kpc}
\end{figure*}

\begin{figure*}
\begin{center}
\includegraphics[width=0.86\textwidth]{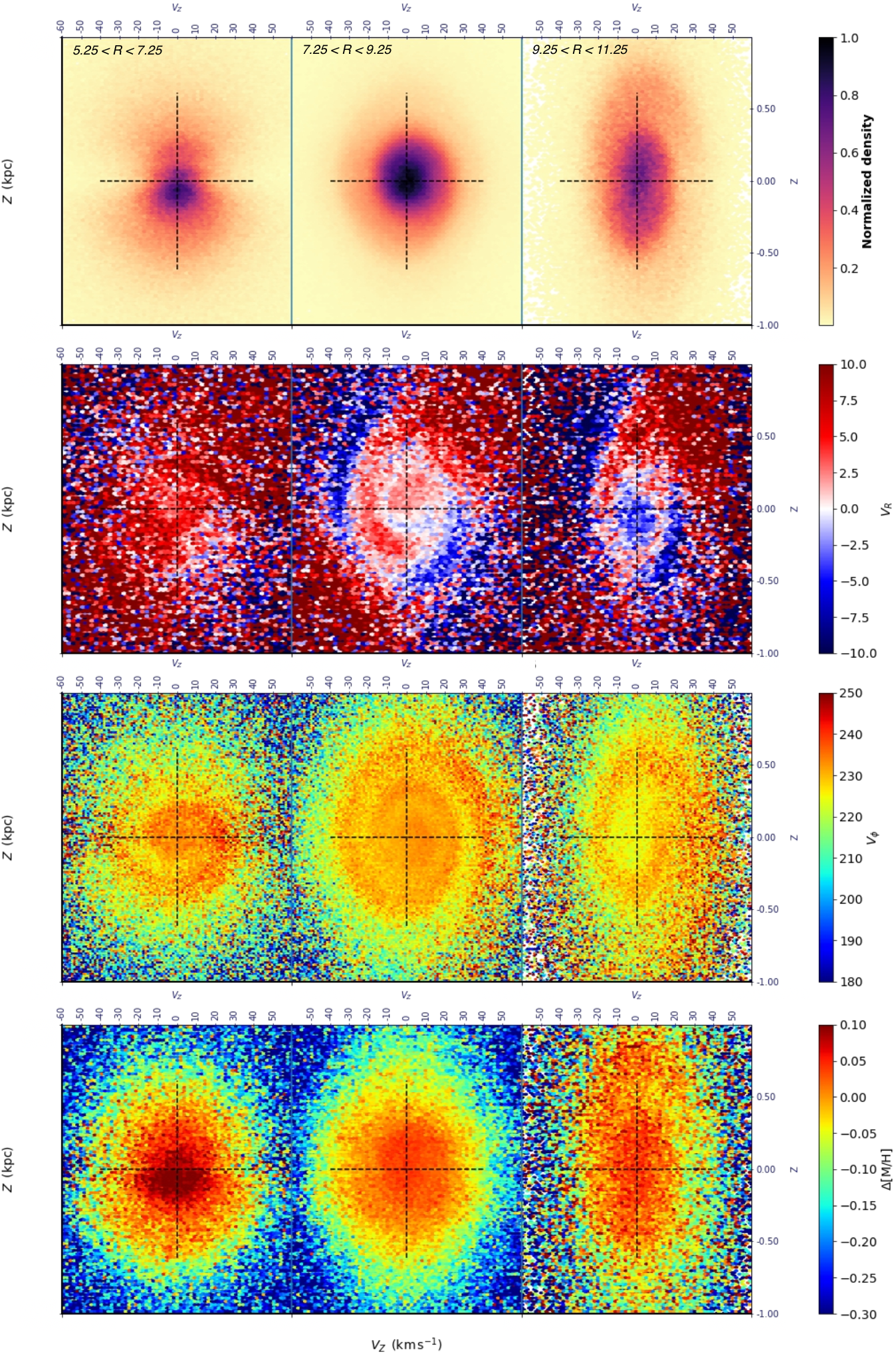}
\end{center}
\caption{Galactic map of the velocity distribution $V_Z$ vs.\ $Z$ of the Medium Quality sample for 5.25 < R < 11.25 kpc. From left to right, the radial range is shown in three rings at R = 6.25, 8.25, and 10.25 kpc, each of  2 kpc in width. From top to bottom: Normalised density distribution, 
$Z-V_Z$ plane coloured as a function of median  $V_R$, $V_\phi$, and $\Delta$\meta.}
\label{FigureKinematics-Vz-Z}
\end{figure*}

Finally, it is worth mentioning that we also find some wiggles in the radial metallicity trends  for $|Z|<0.5$\,kpc and
$R\sim8.5$\,kpc ---both above and below the plane--- that are not related to statistical noise (these bins contain a lot of stars, as they are relatively close to the Sun). 
While we cannot exclude the possibility that these wiggles could partly be due to dynamical effects \citep[e.g. resonances][]{Antoja2018, Kawata:2018}, it is clear that our geometrical selection biases play an important role in producing them.  Indeed, hot TO stars,  that is, relatively young and super-solar metallicity stars, can only be seen relatively close to the solar neighbourhood (see Fig.~\ref{Fig.SelFuncXY1}). This introduces an  age (and metallicity) spatial bias in the sense that the local median metallicity is higher than farther away, simply because TO stars cannot be seen at large distances. Similarly, red clump stars are also lost in the most distant regions, creating a secondary but non-negligible spatial bias. Selecting only giant stars above the red clump (\g\,$\le 2$), which can be seen in a much more unbiased way at all radii, leads to trends without metallicity wiggles (see Fig.~\ref{fig:radial_gradients_RGB}) with gradients that are similar (but slightly flatter by ${\sim} 0.01$ dex\,kpc$^{-1}$)  than the ones quoted above (see Table~\ref{Tab:radial_gradients_giants}). This indicates that the selection function plays an important role in the study of the metallicity gradients.




\begin{figure*}
\includegraphics[width=0.95\textwidth]{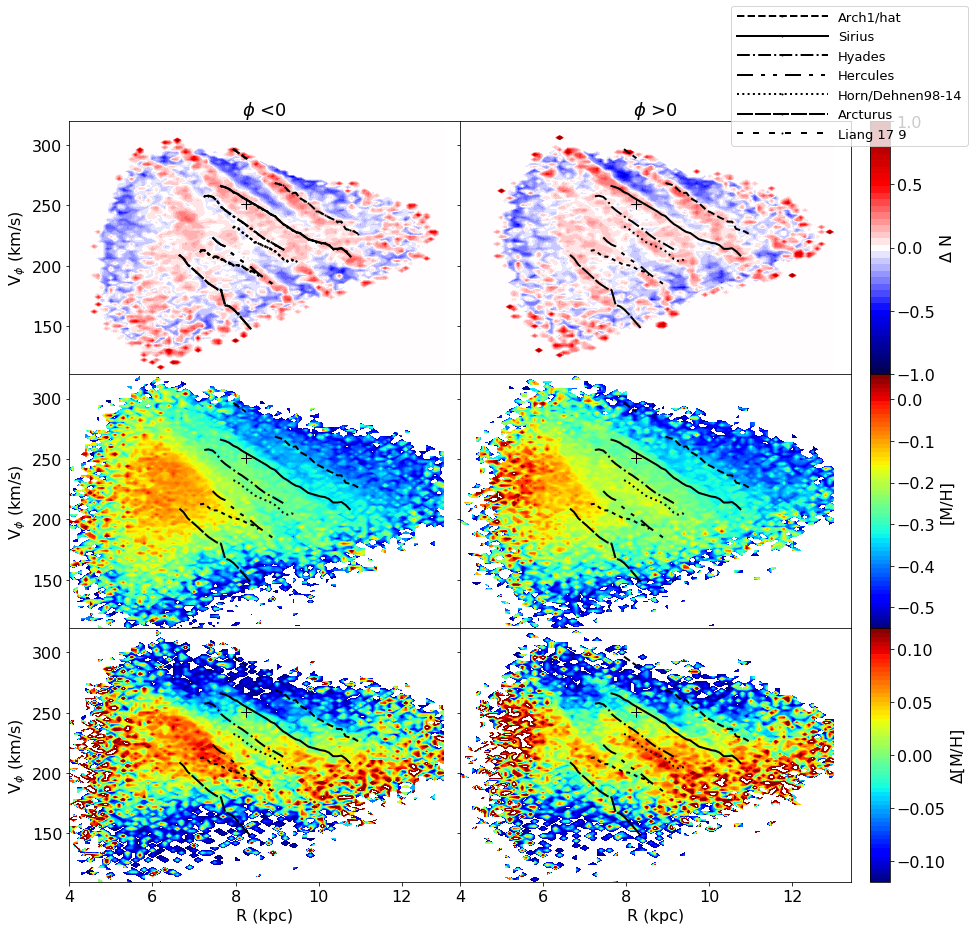}
\caption{
Velocity distribution $V_\phi$ vs.\ $R$ for RGB stars within $|Z|<1$~kpc and $\phi<0$ (left panels) and $\phi> 0$ (right panels).
 Top panels: Distribution of the overdensity $\Delta N$ (see text).  Middle panels: Distribution colour coded according to median metallicity. Bottom panels: Residual metallicity, $\Delta$\meta . Black lines show the known ridges from \citet{Ramos:2018}.
 }
\label{FigureKinematics-Vphi-R}
\end{figure*}

\begin{figure}
\includegraphics[width=0.4\textwidth]{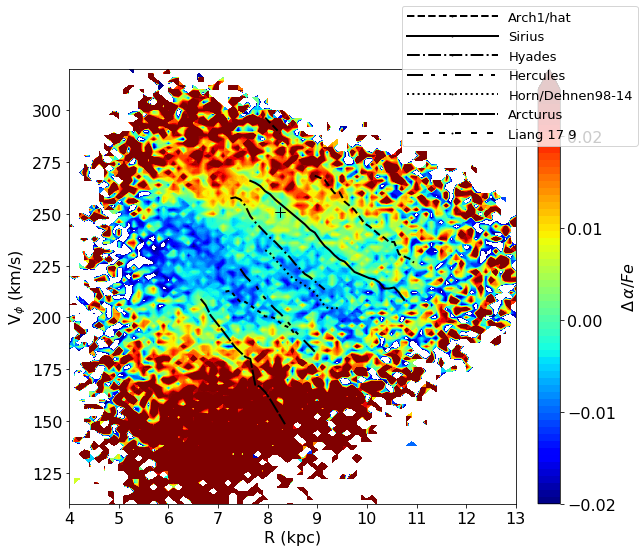}
\caption{
Velocity distribution $V_\phi$ vs.\ $R$ for RGB stars within $|Z|<0.5$~kpc colour coded according to residual \alphaFe, $\Delta$\alphaFe. Black lines show the known ridges from \citet{Ramos:2018}.
 }
\label{FigureKinematics-Vphi-R-alphafe}
\end{figure}

\begin{figure}
\includegraphics[width=0.45\textwidth]{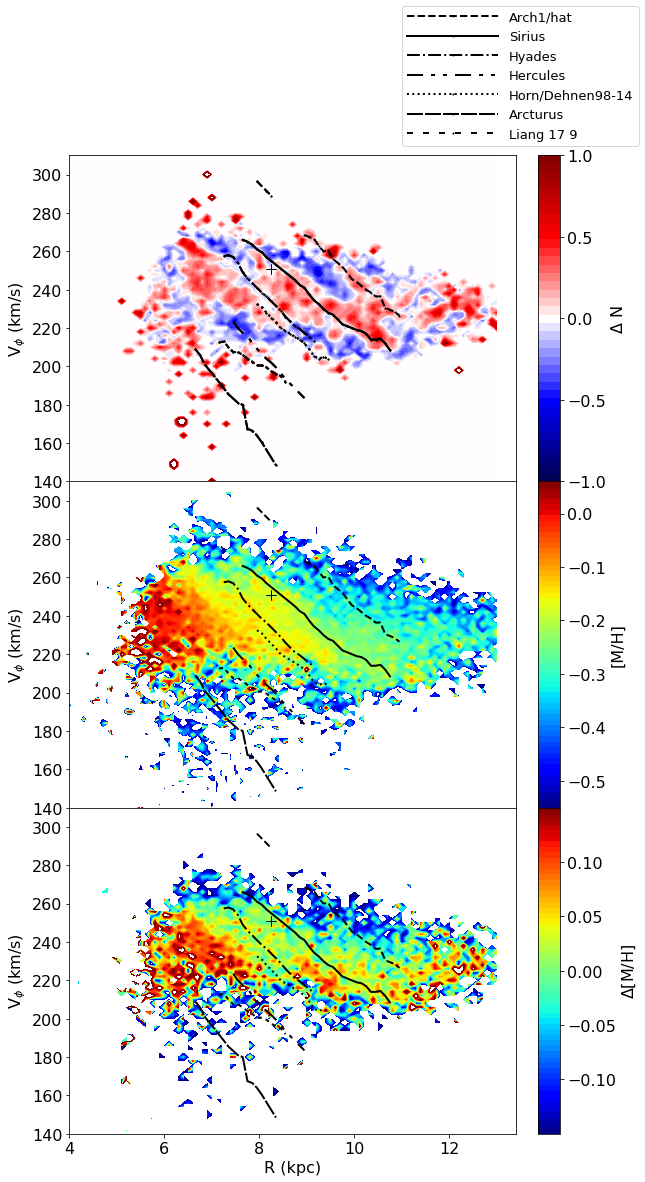}
\caption{Velocity distribution $V_\phi$ vs.\ $R$ for massive giants within $|Z|<1$~kpc.
 Top panel: Density distribution of the sample.  Middle panel: Distribution colour coded according to median metallicity. Bottom panel: Residual metallicity, $\Delta$\meta\ (see text). Black lines shows the known ridges from \citet{Ramos:2018}.
 }
\label{FigureKinematics-Vphi-R-massivestars}
\end{figure}

\begin{figure*}
\includegraphics[width=\textwidth]{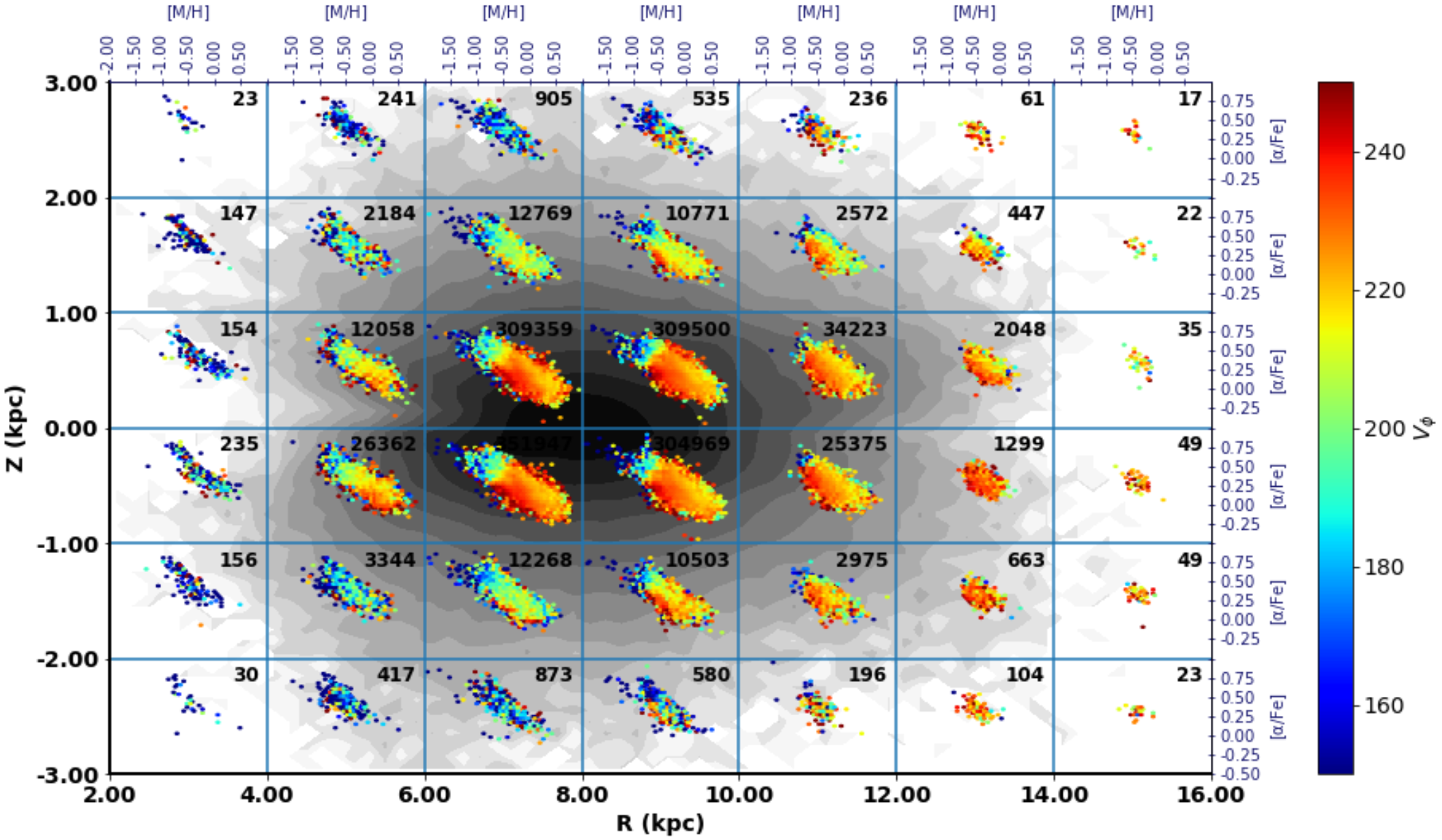}
\caption{Galactic map of the \alphaFe\ vs. \meta\ distribution  of the High Quality  sample colour coded according to median $V_\phi$. }
\label{Fig:alphaFe-MH-ccVPhi}
\end{figure*}

\begin{figure*}
\includegraphics[width=0.99\textwidth]{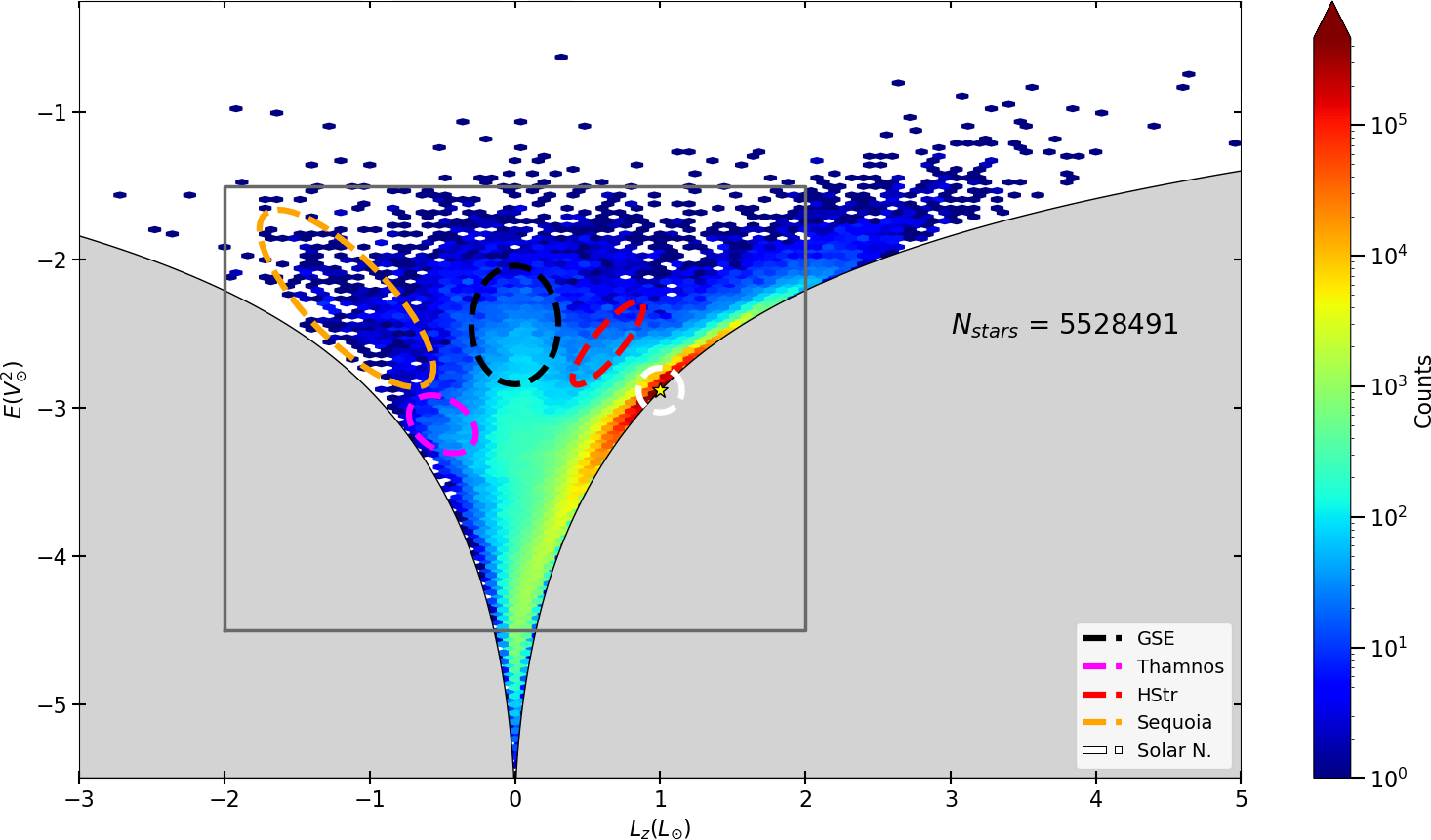}
\caption{Density map in the ($L_Z$, $E$) diagram for the \gspspec\ sample. Dashed ellipses represent the contours of regions discussed in Sect.~\ref{Sec:Dynamics}. The shaded area corresponds to pairs ($L_Z$, $E$) with no physical meaning, in which the upper boundary (solid black line) implies perfect circular orbits. }
\label{Fig_ELz_dens}
\end{figure*}
\begin{figure}[h]
\begin{center}
\includegraphics[width=0.5\textwidth]{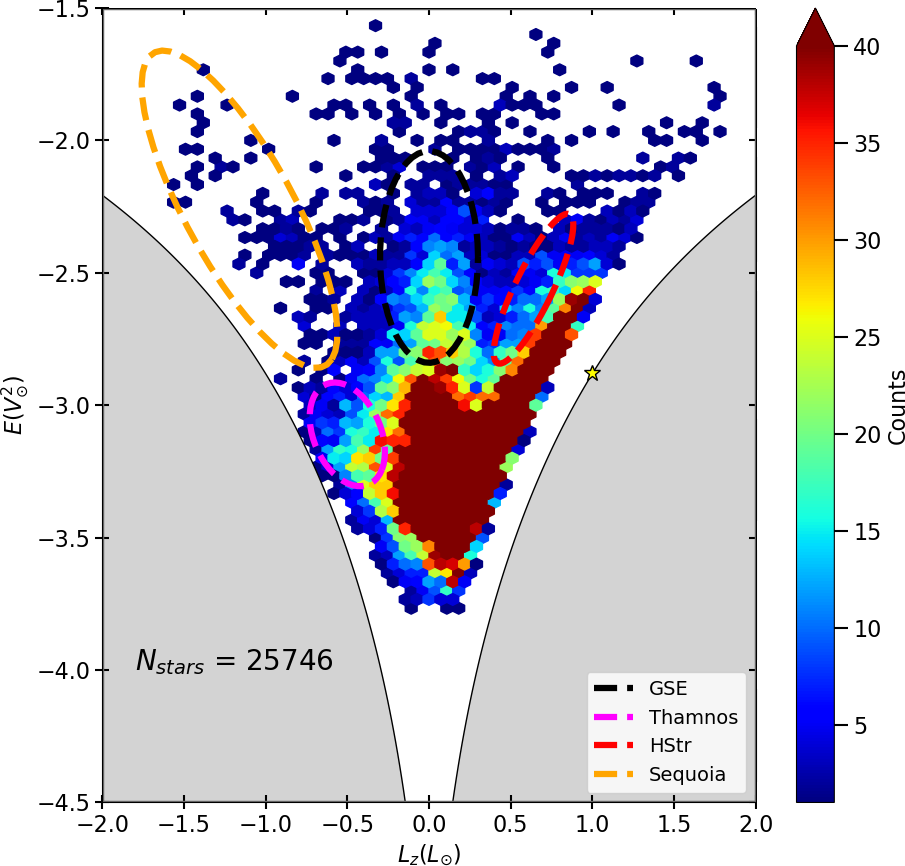}
\caption{Zoomed-in version of the area enclosed by grey rectangles in Fig.~\ref{Fig_ELz_dens} imposing the selection criteria described in \citet{Koppelman_et_al19}. The colour code is saturated in the high-density regions to emphasise the details of the selected areas. No solar neighbourhood area is selected in this figure because it would contain no stars due to the applied selection criteria.}
\label{Fig_ELz_dens_zoom}
\end{center}
\end{figure}
\begin{figure*}[h]
\includegraphics[width=0.99\textwidth]{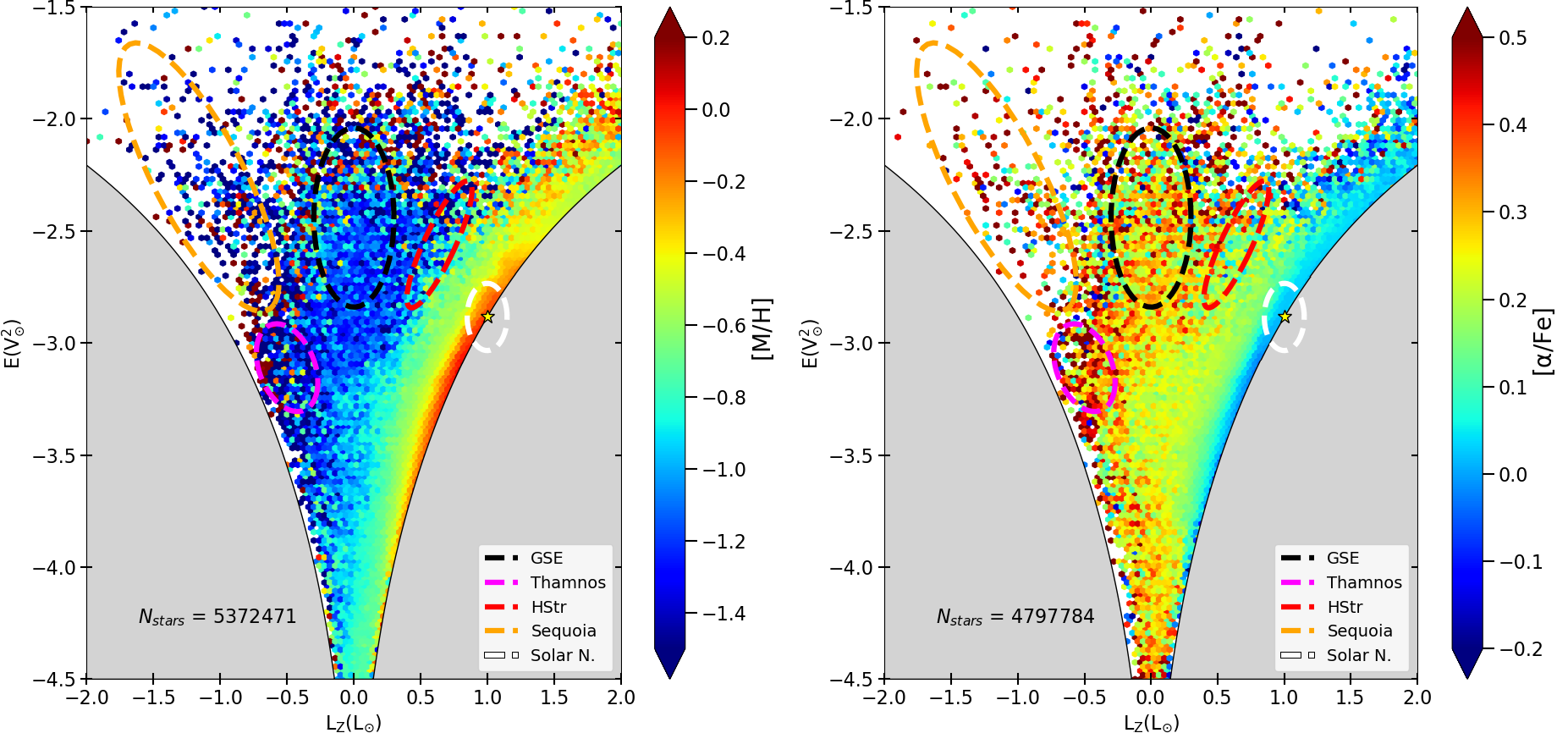}
\caption{Distribution of median metallicities (left panel) and  $\alpha-$element enrichment with respect to iron (right panel) in the energy-angular momentum ($E$, $L_Z$) plane for the General sample stars without those sources with \g<0.5. The coloured ellipses illustrate the selected areas associated with \Gaia-Sausage-Enceladus (black), Thamnos (magenta), the Helmi stream (red), Sequoia (orange), and the solar neighbourhood sample (white).}
\label{Fig_ELz_metalpha_noflags}
\end{figure*}
\begin{figure*}[h]
\begin{center}
	\includegraphics[width=0.95\textwidth]{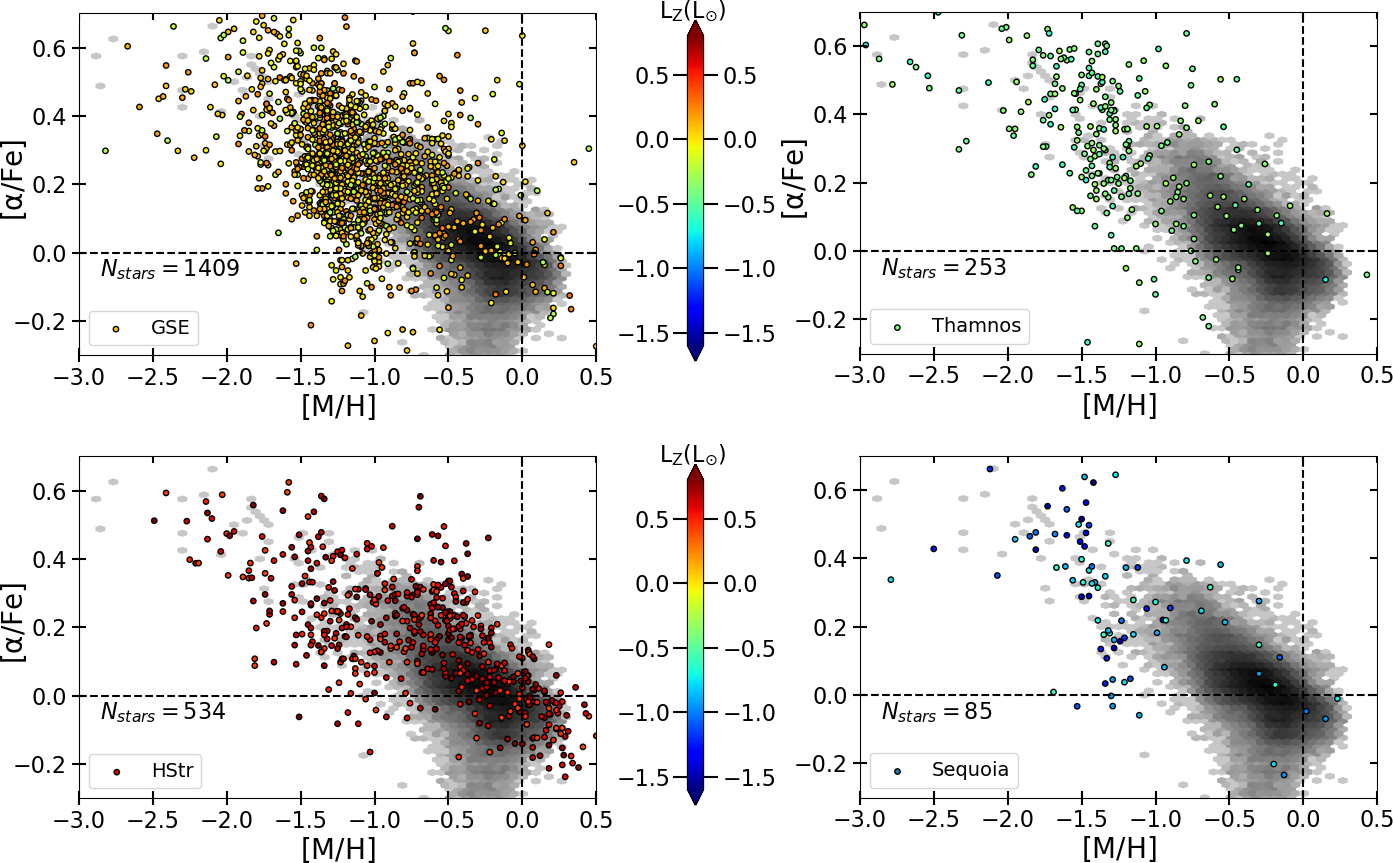}
\caption{\alphaFe\ versus \meta\ diagram for all the stars contained in the regions shown in Fig.~\ref{Fig_ELz_dens} that satisfy the \textit{Medium Quality} selection criteria. The colour code represents the vertical component of the angular momentum $L_Z$. The density plot in the background corresponds to the \textit{Medium Quality} subsample of solar neighbourhood stars.}
\label{Fig_alpha_vs_met_withflags_streams}
\end{center}
\end{figure*}
\begin{figure*}
	\includegraphics[width=0.98\textwidth]{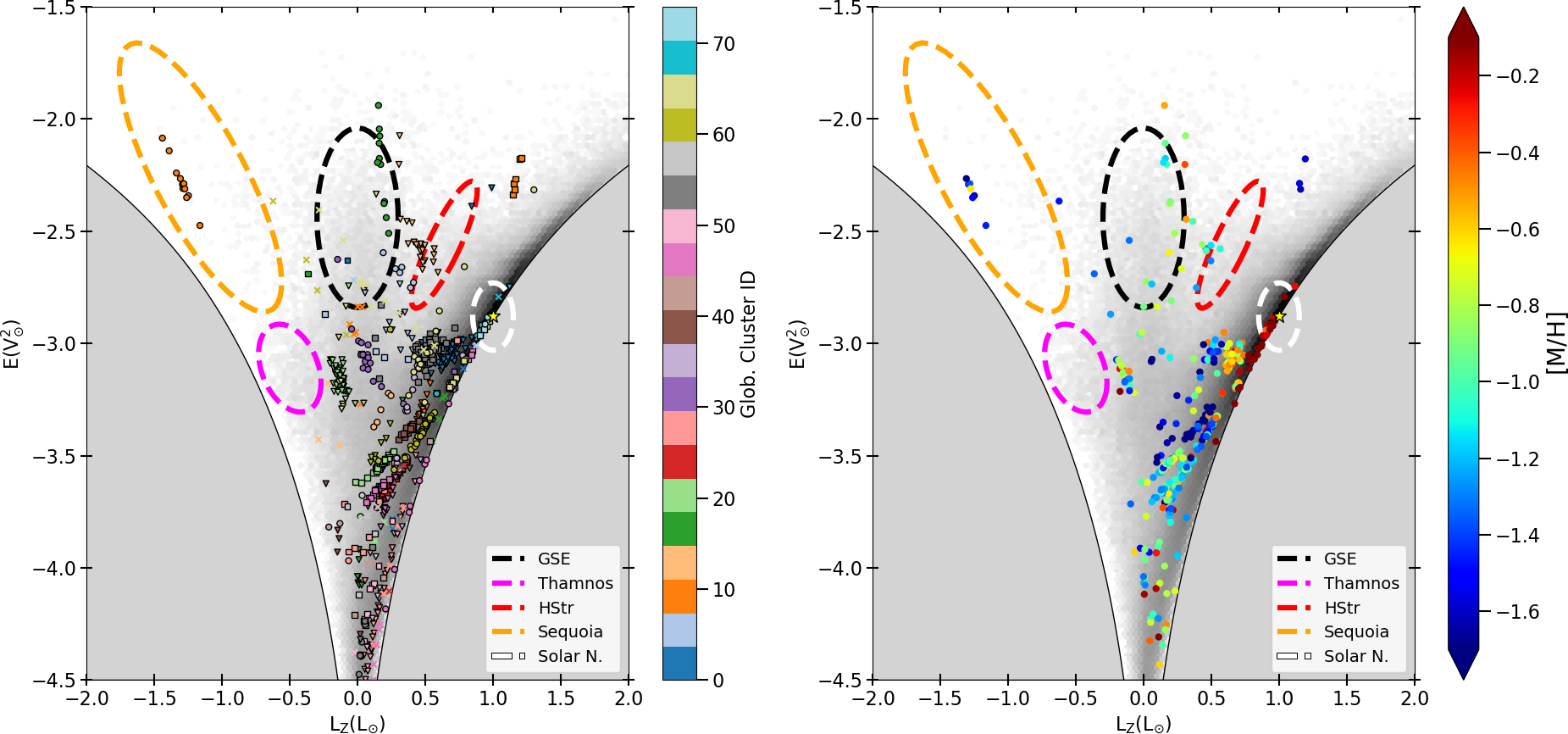}
\caption{Distribution of globular cluster stars in the $E$ vs. $L_Z$ diagram (coloured symbols). In the left panel, each cluster is denoted by a different colour and symbol, while in the right panel the colour code represents the metallicity. The density plot in the background corresponds to the General sample.} 
\label{Fig_GlobClustersELz}
\end{figure*}
\begin{figure*}
\includegraphics[width=0.95\textwidth]{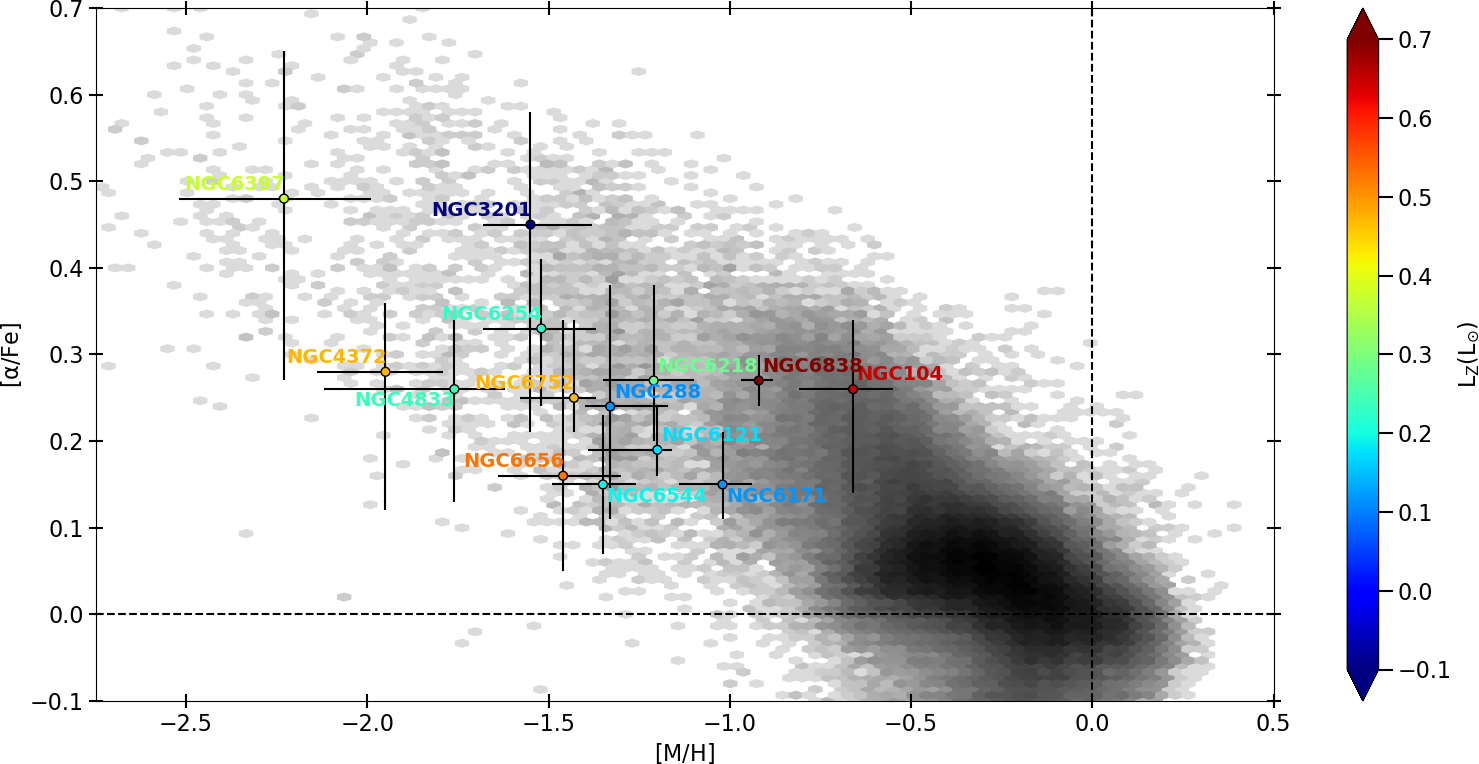}
\caption{
Distribution of \alphaFe\ abundances with respect to \meta\ for the subsample of GCs with more reliable abundances. Error bars are based on the abundance dispersion within the cluster. A metallicity offset of 0.1~dex with respect to the literature is observed and corrected. Globular cluster stars are generally in the low-S/N regime of the \gspspec\ sample and need more deblending corrections of their spectra due to the crowding. The colour code reflects the GC $L_Z$ median values in solar units. In the background, a density plot with the RGB stars sample distribution is presented.}
\label{Fig_GlobClustersAlphaFe}
\end{figure*}
\begin{figure*}[h!]
\begin{center}
\includegraphics[width=0.99\textwidth]{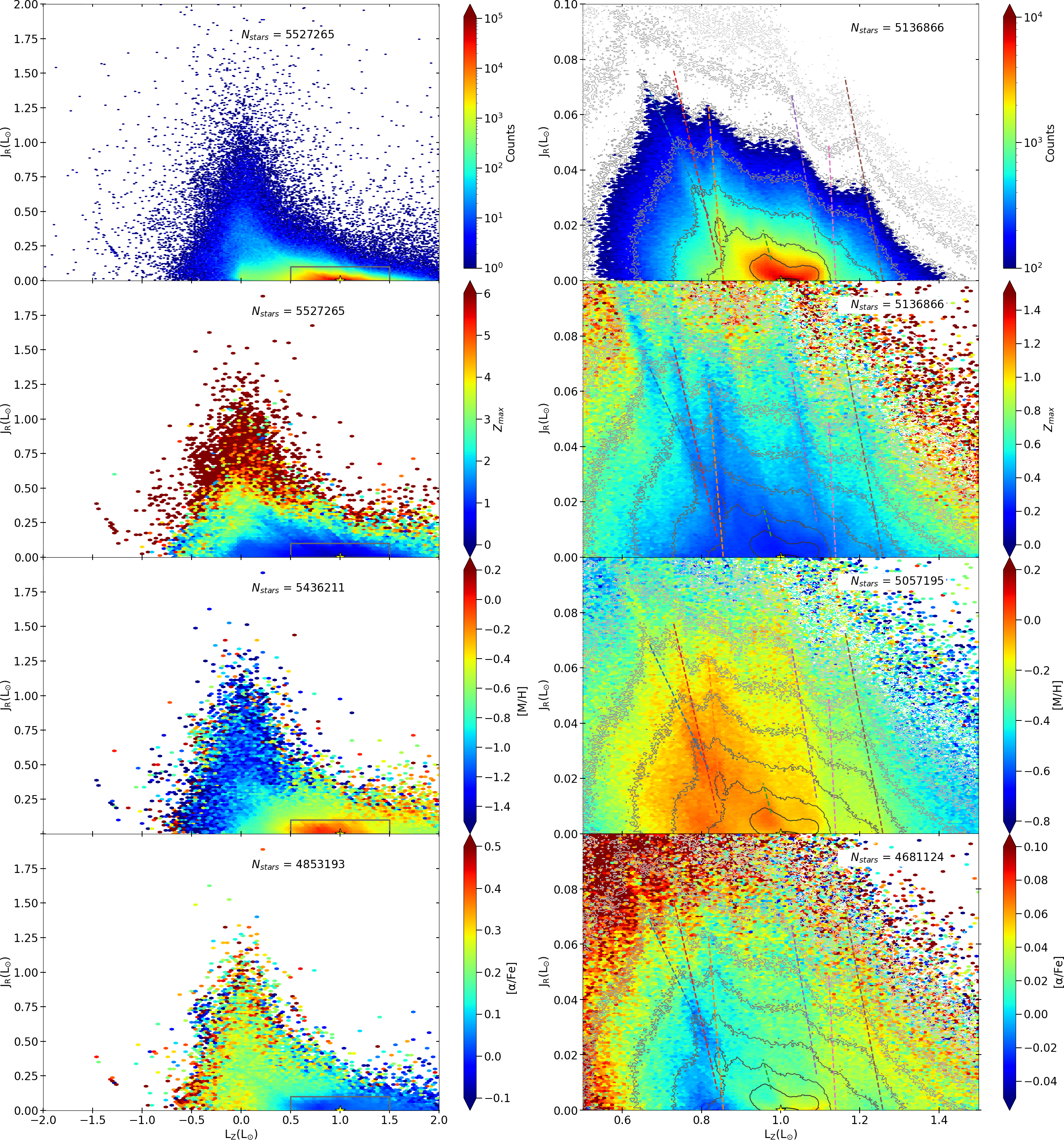}
\caption{Distribution of the General sample in the $J_R-L_Z$ diagram colour coded according to density (upper panels), median maximum distance above the Galactic plane $Z_{\rm max}$ (second panels), median metallicity (third panels), and median \alphaFe\ (lower panels). The grey boxes in the left panels indicate the areas for which a zoomed-in version is shown in the right column. The contour lines of the zoomed density plot are included in the right column as a visual reference, while the tentative positions of the ridges are denoted by the dashed lines. Bins with less than ten stars are omitted in the zoomed density plot to enhance the gradient in the colour code. The position of the LSR at ($J_R$, $L_Z$)$=$($0$, $L_{\odot}$) is denoted by the star symbol.}
\label{Fig_JLz_metalpha_withflags}
\end{center}
\end{figure*}
%

\begin{figure*}[h!]
\includegraphics[width=0.99\textwidth]{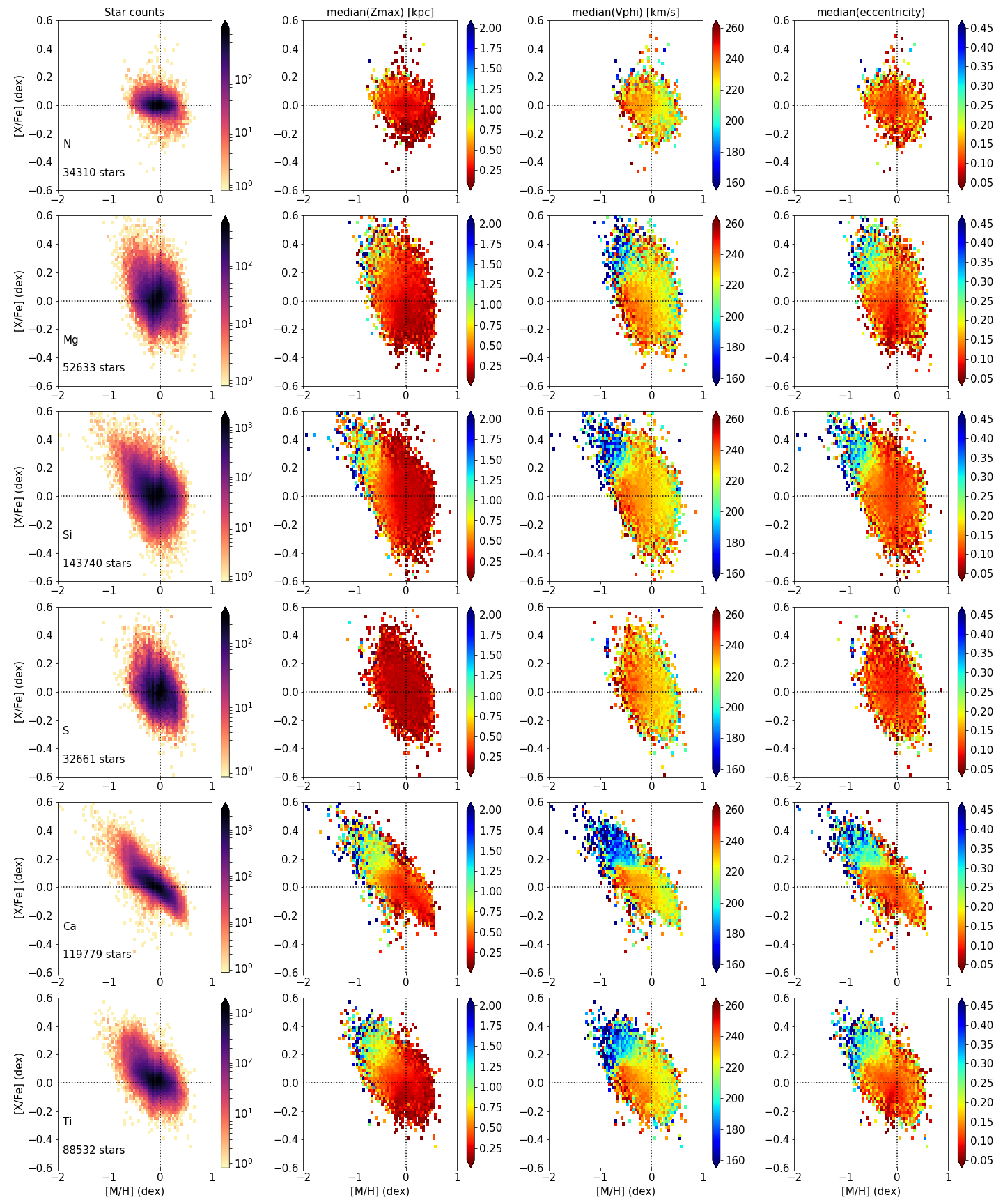}
\caption{Distributions of \NFe, \MgFe, \SiFe, \SFe, \CaFe\ and \TiFe\ (in each row, respectively) as a function of \meta. The colour bars encode: stellar density (col.~1), median maximum vertical distance to the plane (col.~2), median azimuthal velocity (col.~3), and median eccentricity (col.~4), per cell of 0.04~dex in \meta\ and\ per 0.02~dex in \XFe. The number of selected stars is indicated in the lower left corner of the  panels of the first column.}
\label{Fig:F7-1}
\end{figure*}

\begin{figure*}[h!]
\includegraphics[width=0.99\textwidth]{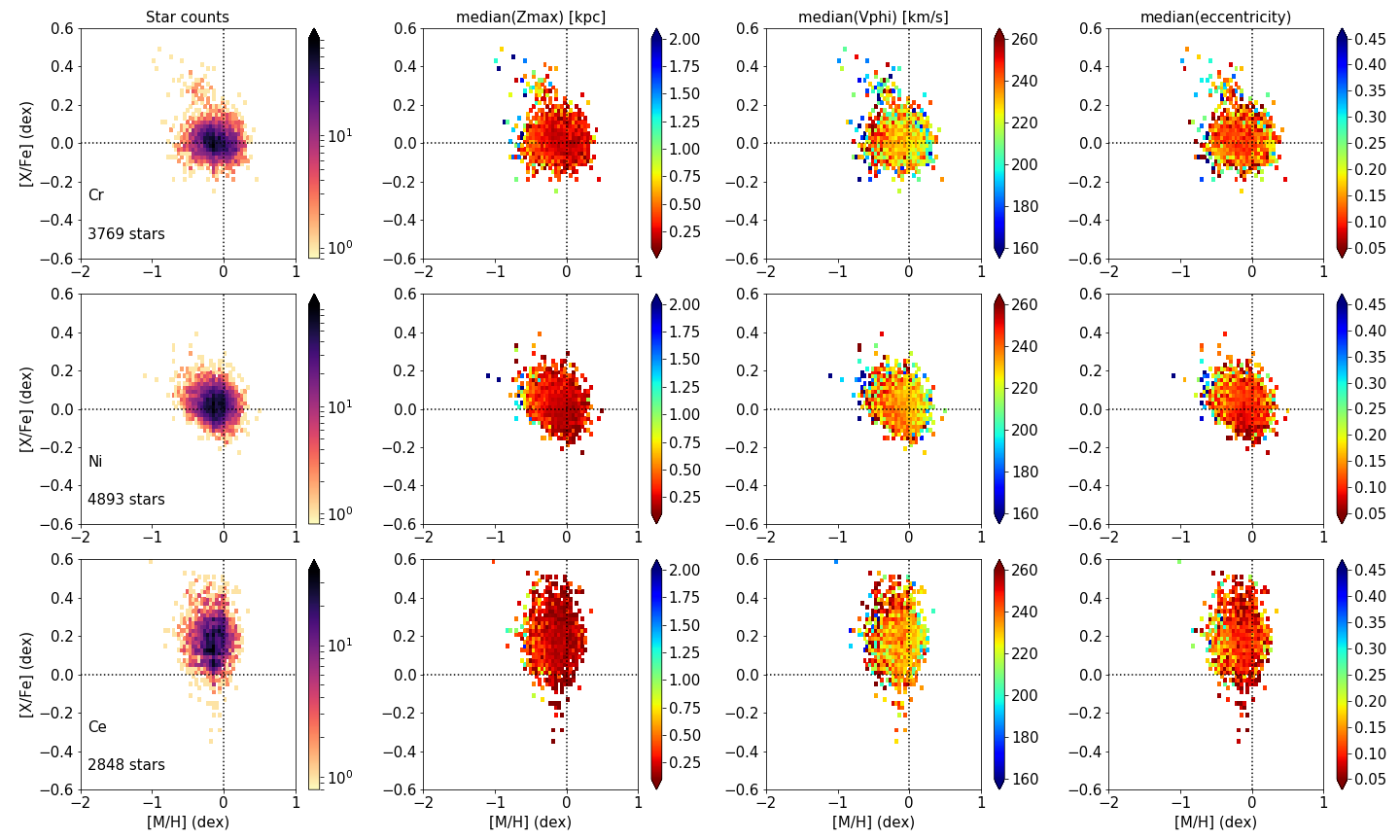}
\caption{Same as Fig.~\ref{Fig:F7-1} but for \CrFe, \NiFe,\ and \CeFe. }
\label{Fig:F7-2}
\end{figure*}



\section{Chemo-kinematical analysis}\label{sec:5}
In this section, we present the large-scale chemo-kinematical properties of the Medium Quality and High Quality  samples presented in Sect.~\ref{Sec:samples}.
We imposed additional conditions on the quality of the astrometric parameters (see also \ref{table:ACSGeoStatistics}) to ensure the use of reliable kinematical velocities. In particular, we selected stars with both five- and six-parameter astrometric solutions and a quality index RUWE $<1.4$. In addition, duplicated sources have been rejected.  
We also performed this analysis by means of the trigonometric distances, $r_\varpi$, from the ACS (see Appendix \ref{Sect:ACS}). No significant difference was detected when using  the trigonometric distances with respect to the chemo-kinematical distributions presented in the following sections. 

The considered spatial range in this section covers the inner disc up to $R=16$~kpc, as well as vertical distances from the Galactic plane in the range $-6 < Z < +6$~kpc.


\subsection{Toomre diagram and Galactic populations}
\label{Sect:Kinematics-Toomre}

Figure~\ref{FigureKinematicsToomre} shows the Toomre diagrams colour coded according to the median \meta, for $3\,468\,272$ stars with the Medium Quality sample within $R < 16$~kpc and $|Z| < 6$~kpc. Each panel presents the distribution $\sqrt{V_R^2 + V_Z^2}$ versus\ $V_\phi$ within bins of $\Delta R=2$~kpc and $\Delta Z=2$~kpc.
The circular dashed line defines the region where the disc stars are dominant:
\begin{equation}
    \sqrt{V_R^2 + V_Z^2 + (V_\phi - V_0)^2} < 210 \,\,\, \mbox{km~s}^{-1}.
\end{equation} 
 Although the mean disc rotation varies as a function of $R$ and $Z$, here we adopt a constant value, $V_0=238.5$~km~s$^{-1}$, corresponding to the velocity of the LSR at the Sun position (see Sect.~\ref{Sec:Orbital_params}).

We note that the chemo-kinematical distributions appear quite symmetrical above and below the plane. Focusing on the  annulus, $4<R<10$~kpc, close to the Galactic plane,  $|Z|< 2$~kpc, we clearly identify the stellar populations belonging to the thin disc with typical solar metallicity $\rm{[M/H]} \approx -0.1$~dex (red colour) and the thick disc with $\rm{[M/H]} \approx -0.6$~dex (green colour). The low metallicity Galactic halo with $\rm{[M/H]} < -1$~dex is also present (blue colours).
As expected, the fraction of disc stars decreases significantly above the plane for $|Z| > 2$~kpc. At $|Z| > 4$~kpc, the main stellar component is the halo formed by \Gaia-Enceladus-Sausage \citep[][]{Belokurov2018,
Haywood2018, Helmi2018, DiMatteo2019, Gallart2019} and heated thick disc stars. 

In the outer disc, $10 < R < 16$~kpc, the  mean metallicity of the disc stars decreases to $\rm{[M/H]} \sim -0.5$~dex (see Sect.~\ref{Sec:Gradients}). In the last panels, $14 < R < 16$~kpc, the thick disc disappears and only thin disc and halo stars are present. 
 Moreover, we still find fast rotating $ V_\phi\approx 200~\rm{km~s}^{-1}$ thin disc stars possibly associated to the flaring of the disc \citep[e.g.][]{Lopez-Corredoira2020, Ding2021} up to $|Z|\sim 4$~kpc (cfr.\ also Fig.\ \ref{Figure_flaring} in Sect.\ref{Sect:Flare}).

In the inner disc, $2<R<4$~kpc, the high-metallicity disc population seems to decrease. In reality, this is  a selection effect produced by the interstellar extinction that reduces the completeness of the sample close to the Galactic plane inside a given $R-Z$ bin. 

\subsection{Velocity distribution  $V_Z$ versus\  $V_\phi$  }\label{sec:VphiVz}


Figure \ref{FigureKinematics-Vphi-Vz} shows the velocity distribution $V_Z$ versus\ $V_\phi$ of giant stars (\g <3.5) with $|Z| < 1$~kpc selected from the Medium Quality sample (with the above described additional conditions on the astrometric parameters). This \g\ cut allows us to select a tracer population with a more homogeneous spatial coverage (cf. Sect.~\ref{Sec:GlobalView}), thus removing dwarf stars to avoid selection function artifacts. However, contrary to the previously defined RGB sample, this selection includes red clump and bottom RGB stars to increase the number statistics, while losing some spatial homogeneity.
The three panels represent the distributions for the subsets within the Galactocentric distance ranges $6<R<8$~kpc (left), $8<R<10$~kpc (middle), and $10<R<12$~kpc (right).

 No significant difference is found between the normalised density distributions above and below the Galactic plane, except for the outer annulus with  $10<R<12$~kpc. Here we find about 80\,000 sources with $Z > 0$~kpc, compared to the approximately 59\,000 sources with $Z < 0$~kpc. Such an effect, which is also present in the {\it General} sample, is possibly the result of greater extinction at negative latitudes than at positive ones.  As for the velocity distributions shown in the upper panels of Fig.~\ref{FigureKinematics-Vphi-Vz}, we notice that they are quite smooth and do not display evidence of substructures, such as those found for $R>10$~kpc towards the Galactic anticentre by \citet[][]{Antoja2021}. In terms of (normalised) density distribution,  such a difference seems associated with the fact that $\sim 80$\% of the stars in our outer region are in the range $10<R<11$ kpc, while the asymmetries found by \citet[][]{Antoja2021} are more evident for $R>11$ kpc (see below for more on this point).

The velocity distributions, $V_Z$ versus\ $V_\phi$, colour coded
according to the median value \meta, are shown in the bottom panels of Fig.~\ref{FigureKinematics-Vphi-Vz}. As expected, the mean metallicity decreases as a function of Galactocentric distance $R$ (cf.\ Sect. 4). Also, stars with higher moduli of vertical velocity, $|V_Z|$, show typically lower \meta\ than stars with smaller vertical velocity dispersions. This effect is consistent with the vertical metallicity gradient. 

Moreover, on average, stars with lower metallicity are rotating faster than those with higher metallicity. This is also an expected result that comes from the {\it negative} rotation--metallicity gradient of the thin disc stellar population   \citep[][]{Kordopatis17, ReFiorentin2019}. This effect is also investigated in Sect.~\ref{Sect:rotation-metallicity}.



Finally, we return to the issue of possible kinematical substructures, such as those found for $R > 10$ kpc towards the Galactic anticentre in \citet[][]{Antoja2021}. Figure \ref{FigureKinematics-Vphi-Vz-11-12.5kpc} shows the $V_Z$ versus\ $V_\phi$ distribution for the Medium Quality sample with $11 < R < 12.5$ kpc and $|Z| < 1$ kpc.
From top to bottom, the panels refer to three Galactocentric bins, each of 0.5 kpc in width. Stars above and below the plane are colour-coded according to their density (first two columns) and their metallicity (last two columns), respectively.

We observe a significant difference in the velocity-space density distribution either side of the Galaxy’s midplane. In particular, below the plane (second column), we see two stellar populations that are similar to those found by these latter authors for $11 < R < 13$ kpc, as seen in their Fig.\ 13. Furthermore, we notice a first group of stars with $V_Z  \simeq 0$ and $V_\phi \simeq 220$ km~s$^{-1}$ that is evident in the range $11<R<11.5$ kpc but decreases in the outer bins, $R=11.5-12$ kpc and $R=12.0-12.5$ kpc, where a second population of stars with $V_Z>0$ and faster rotation, $V_\phi \simeq 240$ km~s$^{-1}$, is more significant. This is clear evidence of the bimodal distribution found by \citet[][]{Antoja2021} and confirmed by McMillan et al.\ (in prep.).

In addition, the chemical information (last column) clearly  shows that the population with lower $V_\phi$ is more metal-rich than the second group of stars with higher $V_\phi$ and positive $V_Z$.  This is an interesting new result that requires further investigation.
This chemo-kinematical signature is confirmed at larger radii, as shown in Figure E.1 which shows the $V_Z$ versus\ $V_\phi$ distribution for a set of radial bins from $R = 10$ kpc to 14 kpc.


\subsection{Vertical velocity oscillations}\label{sec:VzOscillations}

We focus on the $Z$ versus $V_Z$ distribution, where vertical wave-like motions have been detected 
in the form of a snail-like pattern especially 
present when stars are colour-coded according to their azimuthal velocities \citep[][]{Antoja2018}.
Here, we analyse the \Gaia\ DR3 \gspspec\ database selecting  $3\,229\,518$ stars from the Medium Quality sample (with the above-described additional conditions on the astrometric parameters), requiring that the stars be within $|Z|<1$~kpc, $|V_Z| <60~\rm{km~s}^{-1}$, and Galactocentric distance range $5.25 < R < 11.25$~kpc. This region is approximately centred in the smaller volumes of the solar neighbourhood studied previously by \citet[][]{Antoja2018} and other authors such as \citet{ BlandHawthorn19}.

Figure~\ref{FigureKinematics-Vz-Z} shows the projection of phase space in vertical position and velocity, $Z$ versus $V_Z$. The panels refer to three Galactocentric rings, each of 2~kpc in width, with central radius at 
$R = 6.25, 8.25, 10.25$~kpc, respectively. 
The stars are colour-coded according to their 
density (top panels), radial and 
azimuthal velocities (second and third panels from the top, respectively), and residual metallicity 
(bottom panels).

 The density distributions  exhibit structures that are reminiscent of the effects of incomplete phase-mixing. 
Also, the $V_R$ colour-coded  distributions  show strong spiral substructures over the explored Galactocentric range, especially for $R>7$~kpc.

As in \citet{Antoja2018},  
the distribution colour coded  according to $V_\phi$ for the stars near the Sun (central panel of Fig.\ref{FigureKinematics-Vz-Z}) follows a curled, spiral-shaped distribution particularly pronounced  up to 
$|V_Z | \simeq 40~\rm{km~s}^{-1}$. 
In the left panel, that is,\ $5.25 < R < 7.25$~kpc, we still clearly see a  snail-like structure
of fast rotating stars, although with a smaller extension in height  ($|Z|<$ 0.3~kpc) and a vertical velocity amplitude which remains below  $\sim$~30~km~s$^{-1}$  for $V_Z > 0$~km~s$^{-1}$ and reaches $\sim -40$~km~s$^{-1}$ for $V_Z<0$~km~s$^{-1}$.

On the right panel, that is $9.25 < R < 11.25$~kpc, stars with $V_\phi \sim 230~\rm{km~s}^{-1}$ follow a spiral-like pattern with a tail in the region at $V_Z > 20~\rm{km~s}^{-1}$. 
As expected from dynamical considerations \citep[][]{BinneyTremaine}, and as previously reported in the literature using \Gaia\ DR2 data \citep{Laporte19Sag}, the $Z$-axis elongation increases with Galactic radius. Indeed, at large radii, dynamical timescales are longer due to a weaker self-gravity of the disc, inducing a less tightly wound phase-space spiral.

Furthermore, for the first time, we have the possibility to look for chemical signatures associated with the spiral phase-space structures discussed above. These are shown in the bottom panels of  Fig.~\ref{FigureKinematics-Vz-Z}, where the $Z$ versus $V_Z$ distribution is colour coded according to `residual' metallicity $\Delta$\meta, which is calculated as the difference between individual metallicities and the running median over bins of 100~pc in  $R$. The phase space structures seen in the density and kinematic distributions appear to be  linked to \meta\ variations, with more clear correlations in the inner and outer bins. In the inner Galaxy region,  more metal-rich stars tend to be below the plane ($-0.2<Z<$0.0~kpc), where the spiral pattern produces a stellar density excess and lower \vphi\ values. In the external regions, stars with a metallicity excess with respect to the median \meta\ values can be found more than 0.5~kpc above the plane, coinciding again with a stellar density excess and lower rotation. Therefore, in both panels, the rotation velocity $V_\phi$ appears slightly anticorrelated with $\Delta$\meta, in the sense that  stars with higher metallicity have slower rotation and vice versa. This is in agreement with the known \vphi\ $-$ \meta\ correlation in the disc.
 
However, we note that an important warning has to be given with respect to the inner galaxy pattern, where  a significant selection effect due to  an inhomogeneous extinction could be biasing both the stellar density and the metallicity distribution. However, this is not the case  in the external regions, where the metallicity excess correlation with the phase spiral seems more robust.

Finally, it is worth noting that \citet{BlandHawthorn19} used chemical abundances to separate thin-disc ($\alpha$-poor) and thick-disc ($\alpha$-rich) populations, concluding that the phase spiral is confined to the thin disc. In this paper, we explore for the first time the chemical dependencies of the phase spiral within the thin disc at different radii, adding new and more focused constraints.

\subsection{Chemo-kinematical ridges in $V_\phi$ versus\ $R$}\label{sec:Ridges}


Using \Gaia\ DR2, diagonal ridge features were identified in the distribution of stars in the $V_\phi$ versus $R$ space \citep{Antoja2018, Kawata:2018}. Such features might reflect the ongoing phase mixing in the Galactic disc \citep{Fux:2001,Minchev:2009,Gomez:2012}, or, alternatively, be a consequence of resonant orbits induced by the bar and spiral arms \citep[e.g.][]{Dehnen2000,Michtchenko:2018, Fragkoudi19,Hunt2019,Monari2019b}. After their discovery, several properties of the $V_\phi$ versus $R$ distribution were studied to shed light on their origin and dynamical nature \citep[e.g.][]{Ramos:2018, Fragkoudi19, Laporte2019,Antoja2021}. 

In this section, we map $V_\phi$ versus $R$ ridges both in phase space and metallicity space using \Gaia\ DR3. Figure~\ref{FigureKinematics-Vphi-R} shows the diagonal ridges obtained using our sample of RGB stars (see Sect.~\ref{Sec:GlobalView}). The results for the RGB stars at positive (negative) Galactic azimuths $\phi$ are shown in the right (left) column.
Positive $\phi$ values are defined in the direction of Galactic rotation, and as a consequence, are closer to the major axis of the bar.
The upper panels show the ridges in the overdensity $\Delta N = (N - \bar{N})/ \bar{N}$, where $N$ and $\bar{N}$ are the bivariate kernel density estimators adopted in Appendix B1 of \citet{Eloisa21} for the local and mean density, respectively. The adopted bandwidths for the local density are 100~pc in R and 2~km~s$^{-1}$ in $V_{\phi}$, while the corresponding values for the mean density are 700~pc and 14~km~s$^{-1}$. The middle panels in Fig.~\ref{FigureKinematics-Vphi-R} show the median metallicity \meta, while the bottom panels indicate the residual metallicity, $\Delta$\meta, with respect to the median metallicity estimated as a function of Galactocentric distance $R$. Known ridges from \citet{Ramos:2018} are overplotted. As we can see, there is a strong correlation between the ridges found in the overdensity and those observed in \meta\ and $\Delta$\meta. Such a correlation was pointed out in previous studies \citep[e.g.][]{Khanna19}. While some ridges do not exhibit an obvious signature in metallicity (e.g. Arcturus), most of them are typically more metal rich (e.g. Hyades, Horn/Dehnen98). Moreover, some features are markedly different at positive and negative $\phi$. For instance, the Horn/Dehnen98 ridge appears as a metal-rich feature, which is more prominent at $\phi>0$ than at $\phi<0$. Such variations in the metallicity of those ridges are detected here for the first time, taking advantage of the unprecedented coverage of \Gaia\ DR3 chemical measurements. Furthermore, the residual \alphaFe ($\Delta$\alphaFe) in the $V_\phi$ versus $R$ space is shown in Fig.~\ref{FigureKinematics-Vphi-R-alphafe}.
The azimuthal dependency is not shown here in order to increase the statistics and better identify possible features. As we can see, diagonal ridges are apparent when stars are colour coded
according to $\Delta$\alphaFe. In particular, we note that Hercules and Horn/Dehnen exhibit low \alphaFe.

For comparison, Fig.~\ref{FigureKinematics-Vphi-R-massivestars} shows a similar distribution for the sample of massive stars defined in Sect.~\ref{Sec:GlobalView}. Ridges in the young populations were identified by \citet{Wang2019} and \citet{Antoja2021} but here we now present them with an alternative and very robust selection of the young population. Due to the low number statistics, stars are plotted all together, without showing the azimuthal variations. While ridges are also present in this sample, we observe that the signal is somewhat noisier than the RGB sample, presumably due to the low number statistics.

The comparison between the ridges found in the young and old stellar populations potentially contains important clues as to their origin. Ridges in young stellar populations can be generated by  either a recent perturbation (e.g. an interaction with a satellite galaxy) or an ongoing mechanism in the Galactic disc (e.g. resonances). Additionally, they could be related to star formation structures mixing into the disc, but in this case no match between the ridges in the young and old populations would be expected. Due to the low number of massive stars, it is difficult to establish a robust link between the young and old ridges in our data, although ridges such as Sirius and Arch1 seem to be present in both samples. An alternative scenario is that the global disc dynamics and the star formation processes are working so as to produce the observed similarities in the two samples.

The ridges shown in Fig.~\ref{FigureKinematics-Vphi-R} and Fig.~\ref{FigureKinematics-Vphi-R-massivestars} are presumably related to the bumps found in the metallicity gradients as a function of guiding radius (Fig.~\ref{fig:radial_gradients_Zmax_Rguide}; however, it should be noted that similar bumps can also be caused by the selection function when gradients are plotted as a function of observed radius R, as discussed in Section \ref{Sec:Gradients}).

\subsection{Rotation--metallicity correlations}
\label{Sect:rotation-metallicity}

Figure~\ref{Fig:alphaFe-MH-ccVPhi} shows a grid of \alphaFe\ versus\ \meta\ chemical planes colour coded
according to $V_\phi$. The different panels show the distribution across the Milky Way for giants (\g < 3.5) of the High Quality  sample, with the additional conditions on the astrometric parameters defined at the beginning of this section.

In the panels near the Sun and close to the Galactic plane with $6 < R < 10$~kpc and $|Z|< 1$~kpc, red and orange colours represent fast rotating thin disc stars ($V_\phi \sim 220-250$~km~s$^{-1}$) with low \alphaFe.  
We notice that the azimuthal velocity of metal-rich thin-disc stars with 
\meta > 0 is slightly lower than for stars with \meta < 0. This effect is consistent with the well-known {\it negative} rotation--metallicity gradient of the thin-disc population,  
$-20 \lesssim \partial V_\phi/\partial$\meta$ \lesssim -10$~km~s$^{-1}$, as reported by various studies \citep[e.g.][]{Lee2011,AllendePrieto2016,  Kordopatis17, Recio-Blanco14} and is associated with blurring effects in the disc (i.e. stars on higher eccentricity visiting a given radius at their apocentre or pericentre \citep{LyndenBell72, Sellwood2002}.  

The \alphaFe\ abundance from \gspspec\, that mostly traces \CaFe\  does not show a clear gap between the thin- and thick-disc sequences  as in other chemical elements, such as Mg, which is often adopted in the literature because it is mostly produced by high-mass stars, while Ca is also produced by SNe Ia on longer timescales.
Therefore, we derived an approximate estimation of the rotation--metallicity gradient of the thin disc by selecting 763\,845 stars close to the Galactic plane up to $|Z|< 1$~kpc, $8<R<10$~kpc, $-0.4<$\meta$<+0.6$~dex, and $V_\phi > 100$~km~s$^{-1}$. We find a gradient of $-13.9\pm 0.1$~km~s$^{-1}$~dex$^{-1}$, which is consistent with the values from the literature. 

In the panels with $6 < R < 10$~kpc and $|Z|< 1$~kpc, the low metallicity and $\alpha$-enhanced stars belonging to the thick disc and halo are shown with green and blue colours that indicate a slower rotation, $V_\phi < 200$~km~s$^{-1}$. We notice that  the fraction of thin-disc stars decreases significantly in the panels above the plane with  $1< |Z|< 3$~kpc. Here, the thick disc is dominant and the colours provide clear evidence of the {\it positive} rotation--metallicity gradient of this population \citep[][]{Spagna2010, Kordopatis2011, Kordopatis13, Recio-Blanco14, Lee2011}. 
By selecting 1535 sources with $2< |Z|< 3$~kpc, $8<R<10$~kpc, 
$-1<$\meta$<-0.2$~dex, and $V_\phi > 50$~km~s$^{-1}$ to reduce the contamination from thin-disc and halo stars, we measure a gradient  of  $+33\pm 7$~km~s$^{-1}$~dex$^{-1}$ for the thick disc. This result is quite consistent with recent studies based on the chemical classification of  thick-disc stars \citep[e.g.][]{Kordopatis17, ReFiorentin2019} and interpreted as a signature of inside-out formation and gas re-distribution in the primitive disc \citep{Schonrich17}, or as a correlation resulting from the superposition of mono-abundance  subpopulations with negative slopes (as
in the thin disc, \citealt{Minchev19}).

As discussed in Sect.~\ref{Sec:GlobalView} and \ref{Sect:Kinematics-Toomre},  the thin-disc population is incomplete towards the inner disc because of interstellar extinction close to the Galactic plane. 
Instead, the thick disc can be observed and a positive rotation--metallicity gradient for the thick disc is present in the inner annulus, $4<R<6$~kpc, as recently detected also by \citet[][]{ReFiorentin2019}. Finally, because of the shorter radial scale length of the thick disc, the outer disc is mainly formed by thin-disc stars that show the typical negative rotation metallicity gradient until out to $R=12$~kpc.


\section{Chemo-dynamical analysis}
\label{Sec:Dynamics}
\par Here we explore the relation between the orbital parameters and the stellar chemistry inferred by \gspspec\, using the General sample defined in Sect.~\ref{Sec:samples}.  The orbital parameters of the full \gspspec\ sample were computed using the method described in Sect.~\ref{Sec:Orbital_params}. As the potential considered in the determination of the orbital parameters is stationary and axisymmetric \citep[][]{McMillan17}, the vertical component of the angular momentum $L_Z$ and the total energy $E$ are integrals of motion; therefore, their values can be directly estimated from the initial conditions. For sake of simplicity, we adopt the convention $L_Z>0$ for clockwise rotating systems and rescale the units of $L_Z$ and $E$ in terms of $L_{\odot}=R_{\odot}V_{\odot}$ and $V_{\odot}^2$, respectively.

\subsection{Merger debris in the Galactic halo}\label{sec:Mergers}
\par Figure~\ref{Fig_ELz_dens} illustrates the distribution of the \gspspec\ sample in the $E$ versus $L_Z$ plane. 
In this diagram, positive energies correspond to stars not gravitationally bound to the Galaxy, while positive (negative) $L_Z$ values correspond to co-rotating (counter-rotating) systems. As seen, this sample is clearly dominated by stars in the solar neighbourhood ($L_Z=L_{\odot}$, $E\approx -2.88 V_{\odot}^2$) with almost circular orbits ($E$ and $L_Z$ close to the border of the forbidden area). By visual inspection, we can identify the overdensity of stars at low $|L_Z|$ and $-2.8V_{\odot}^2 \lesssim E \lesssim -2.0V_{\odot}^2 $ associated with the \Gaia-Enceladus-Sausage  \citep[GES; ][]{Belokurov18, Myeong18, Helmi2018, Feuillet20, Feuillet21}, as well as an extended structure of counter-rotating stars in the region $L_Z\lesssim -0.80 L_{\odot}$, $E\gtrsim -2.8 V_{\odot}^2$ likely related to the Sequoia stream \citep{Myeong19}. Similarly, we infer the presence of the Thamnos structure \citep{Koppelman_et_al19, Helmi_20} from the extension of the high-density region towards ($L_Z$, $E$)$\approx$($-0.50 L_{\odot}$, $-3.11 V_{\odot}^2$).
\par The comparison of Fig.~\ref{Fig_ELz_dens} with the $E$ versus $L_Z$ diagram of Fig.~2 in \citet[][]{Koppelman_et_al19} motivates the selection of one additional region  associated with the Helmi stream \citep[HStr;][]{Helmi99}.  In Fig.~\ref{Fig_ELz_dens_zoom}, we reproduce the $E-L_Z$ density plot using the same selection criteria as those used by \citet{Koppelman_et_al19} in order to focus on the stars with halo kinematics by imposing $\left|\vec{V}-\vec{V}_{\rm circ}\right|<180$~km~s$^{-1}$, where $\vec{V}_{\rm circ}$ is the circular velocity, and a minimum heliocentric distance of $3$~kpc to exclude nearby sources. It is worthwhile noting that this comparison must be interpreted qualitatively due to the different assumptions made for $R_{\odot}$, $V_{\odot}$, and the values of the solar motion. Therefore, in our Galactic model, the HStr spans an elongated region between ($L_Z$, $E$)$\approx$($0.40 L_{\odot}$, $-2.84 V_{\odot}^2$) and ($0.86 L_{\odot}$, $-2.30 V_{\odot}^2$). 
Finally, we include a control sample of thin disc stars by selecting a circular area of radius $0.15$ around the LSR in the $E$ versus $L_Z$ diagram (white circle in Fig.~\ref{Fig_ELz_dens}). We clean this sample by selecting only the giant stars  ($0.5$~$\leq$\g $<1.8$ dex, $4000$~$\leq$\T $<8000$ K) with good spectral fitting ($log\chi^2<-3.5$). For the sake of visualisation, we focus on a specific area that includes the five selected structures and analyse its chemical distribution. 

\par Figure~\ref{Fig_ELz_metalpha_noflags} illustrates the global distribution of metallicity (left panel) and \alphaFe\ (right panel) of the \textit{Medium Quality} GRVS sample in the $E$ versus $L_Z$ plane. This figure is complemented by the individual \alphaFe\ versus \meta\ diagrams of the four selected structures and shown in the different panels of Fig.~\ref{Fig_alpha_vs_met_withflags_streams}.
As we can see, the Thamnos and Sequoia subsets are dominated by a metal-poor population ($<$\meta$> \sim$-1.3~dex), with lower \alphaFe\ values compared to the disc stars at the same metallicity (density plot in the background, corresponding to the white circle in Fig.~\ref{Fig_ELz_dens} and Fig.~\ref{Fig_ELz_metalpha_noflags}). In contrast, the HStr subset shows a significant contamination by thin disc (\meta$\la$-0.5~dex) and thick disc (\meta$\gtrsim-0.5$~dex, \alphaFe$\gtrsim0.3$~dex) stars due to its proximity to the disc region in the $E-$-$L_Z$ diagram (Fig.~\ref{Fig_ELz_dens}). However, we can infer a low-$\alpha$ population consistent with the stellar streams of accreted satellites in the metal-poor regime that traces a different trend in the \alphaFe\ versus \meta\ diagram from that traced by the disc. Finally, the chemical pattern revealed by the \Gaia-Sausage-Enceladus subset clearly differs from that of the disc despite its large dispersion. An in-depth analysis of accretion debris is beyond the scope of this paper. Nevertheless, the metallicity distributions of the low-\alphaFe\ populations in the selected halo substructures show differences that could be attributed to the mass of the accreted satellites. In particular, Sequoia candidates have a metallicity distribution peaking around \meta$\sim-1.4$~dex for \alphaFe$\sim$0, while \Gaia-Enceladus-Sausage peaks at \meta$\sim-1.2$~dex at the same \alphaFe\ values of $\sim$0. This difference of about 0.2~dex, suggests a more metal-poor knee in the \alphaFe\ versus \meta\ distribution for Sequoia than for \Gaia-Enceladus-Sausage. This is in agreement with literature results indicating a lower mass for Sequoia than for \Gaia-Sausage-Enceladus using ground-based data \citep{Myeong19, Feuillet20, Feuillet21}. In conclusion, Fig.~\ref{Fig_alpha_vs_met_withflags_streams} confirms that the \Gaia\ chemical database possesses the required quality to provide chemical diagnostics of accretion through the abundance ratio of $\alpha$-elements to iron. 

%
%
%
\subsection{Chemo-dynamics of globular clusters}
The globular cluster (GC) system of the  Milky Way has historically driven many Galactic and stellar physics studies. For this reason, we focus here on this particular population, illustrating the quality of the DR3 chemo-dynamical data available for these objects. Figure~\ref{Fig_GlobClustersELz} shows the distribution of individual GC stars in the $E$ versus $L_Z$ diagram (coloured symbols). In the left panel, each cluster is denoted by a different colour and symbol, while in the right panel the colour code represents the \meta\, showing the metallicity coherence of the stars belonging to the same cluster. The density plot in the background corresponds to the General sample. Several clusters associated to accretion overdensities like Gaia-Enceladus \citep{Helmi2018, Myeong18} or Sequoia \citep{Myeong19} are included. GC members have been selected allowing a radial velocity dispersion of 20~km/s with respect to the median value of each GC. It is worth noting that distance uncertainties spread the stars of the same cluster in both $E$ and $L_Z$, as clearly seen for NGC~3201 inside the Sequoia orange ellipse.

GC stars are faint and are therefore generally in the low-S/N regime of the \gspspec\ sample, and require more deblending corrections of their spectra because of the crowding. For this reason, only a subsample of the clusters included in Fig.~\ref{Fig_GlobClustersELz} have reliable \alphaFe\ estimates, reducing the total sample to 14 clusters. Figure~\ref{Fig_GlobClustersAlphaFe} shows the distribution of \alphaFe\ abundances with respect to \meta\ for the subsample of GCs with the most reliable abundances. Error bars are based on the abundance dispersion within each cluster. We also note that GC stars with better quality spectra are often outside the applicability domain of the calibration polynomials in \gspspec. For this reason, we performed a specific calibration with respect to the literature metallicities ; an \meta\ offset of 0.1~dex has been corrected). The colour code reflects the GC $L_Z$ median values in solar units. In the background, a density plot with the RGB star sample distribution is presented as a visual reference. Although,  the \alphaFe\ dispersion is rather high in some cases, which is  possibly due to the data quality and/or membership errors, the global \alphaFe\ versus \meta\  trend follows the expected distribution,
illustrating the potential of Gaia DR3 data for GC studies.
Finally, it is important to note that a more thorough analysis focusing on each cluster individually will certainly improve the results. Moreover, in future Gaia data releases,  GC will clearly benefit from the increase of the RVS spectra S/N, allowing a deeper study of this important Galactic population.

%
%
%
\subsection{Chemo-dynamical signatures in radial actions}
\label{Sec:ChemoRadialAct}
\par The actions ($J_R$, $J_Z$, $L_Z$) in static potentials are integrals of motion that characterise the orbit of the stars. In particular, the action $J_R$ characterises the radial amplitude of the epicyclic orbits while the angular momentum $L_Z$ sets the guiding radius \citep{BinneyTremaine, Trick21}, a more robust estimate of the typical Galactic distance of the star than the present-day Galactocentric radius $R$.
\par The right panels of Fig.~\ref{Fig_JLz_metalpha_withflags} illustrate the distribution of the {\it General  Sample} in the $J_R$ versus $L_Z$ plane with colour coding reflecting stellar density (first row), the maximum orbital distance from the plane $Z_{\rm max}$ (second row), the metallicity (third row), and \alphaFe\ (last row). We observe that the metallicity distribution follows the expected decreasing trend with $Z_{\rm max}$, with the exception of the high angular momentum regime which shows relatively high metallicities. As this sample is biased by nearby stars, most of them have a vertical angular momentum $L_Z\approx L_{\odot}$, the angular momentum of the LSR. These stars show almost circular orbits ($J_R\sim 0$) between around $0.92 \,L_{\odot}$ and $1.10\,L_{\odot}$, which implies a guiding radius of between 7.6 and 9.1~kpc in the assumed Galactic potential. However, the low-density region ($\lesssim 10$~counts) includes a wider variety of substructures.

First, the counter-rotating population ($L_Z<0$), in which we identify the GC NGC~3201 in the elongated structure at $L_Z\lesssim -1.25 L_{\odot}$, $J_R \lesssim 0.5 L_{\odot}$ \citep{Myeong19}. These retrograde stars are predominantly metal poor (\meta<-1.1~dex) as shown in the literature \citep{Carollo07, Morrison09, Kordopatis20}.

Second, a region of almost zero angular momentum that extends towards large values of $J_R$ ($\gtrsim 0.25 L_{\odot}$). This area is characterised by stars in highly eccentric orbits with large vertical displacements from the Galactic plane (second row in Fig. \ref{Fig_JLz_metalpha_withflags}). The high inclination of these orbits and the low metallicity of the stars in this region (third row) are consistent with the halo population.

Third, an extension of the almost solar metallicity region \meta$\gtrsim-0.5\mbox{~dex}$ towards $L_Z\gtrsim 1.2 L_{\odot}$, $J_R\gtrsim 0.1 L_{\odot}$. This area is populated by stars from the outermost part of the Galactic disc ($R\gtrsim 10$~kpc) in relatively eccentric orbits $\epsilon\sim 0.4$. Although these properties are compatible with a radially migrated population from inner regions, further analyses are required to support this interpretation.

\par In addition to these features, a detailed exploration of the most populated region reveals the presence of `ridges' in the distributions of stars (similar to those reported by \citet[][]{Trick2019}), $Z_{\rm max}$ , and metallicity (right panels in Fig.~\ref{Fig_JLz_metalpha_withflags}), while some of them have no counterpart in the \alphaFe\ map. By visual inspection, we trace the configuration of these ridges (dashed lines) to illustrate the comparison of the maps. We find good agreement between the extension of the denser regions (upper right panel), the almost flat orbits (right panel in the second row), and the metal-rich stars (right panel in the third row). We can identify four prominent ridges (first, second, third, and fifth leftmost dashed lines) with negative slope in the $J_R-L_Z$ plane, each of which can be seen  clearly in both the metallicity and $Z_{\rm max}$ distributions. In particular, the fifth ridge delimits a transition in the metallicity map from \meta>-0.2~dex to \meta<-0.2~dex. Similarly, we discern a subtle ridge (fourth dashed line) with minor effects on the chemical pattern but significant imprint on $Z_{\rm max}$. The sixth ridge corresponds to a relative minimum in metallicity and a change in slope in the iso-density contours. Finally, the seventh ridge indicates a weak increase in metallicity as well as another change of slope in the contour curves. In addition to the ridges, two high-metallicity and low-\alphaFe\ areas centred at $(L_Z, J_R)\approx (0.8 L_{\odot}, 0.003 L_{\odot})$ and $(0.8 L_{\odot}, 0.025 L_{\odot})$ can be identified, as well as a third region at $(0.96 L_{\odot}, 0.005 L_{\odot})$ with high metallicity but solar \alphaFe. Although some of the ridges shown in Fig.~\ref{Fig_JLz_metalpha_withflags} are similar to those reported by \citet{Trick2019, Trick22}, who interpret them as signatures of the resonances, we cannot conclude our ridges are produced by the same mechanism, because: (1) we have not explored the influence of the bar and whether it could produce or alter the signatures we see; (2) we do not restrict our study to stars close to the Galactic plane, as in \citet{Trick22}; and (3) some of our ridges have no counterpart in the predictions of \citet{Trick22}. In any case, a complementary and detailed analysis of the orbital frequencies in the ridges would be required to confirm the resonances.

%
%
%
\section{Individual abundances in the solar cylinder}
\label{Sec:SolarCylinder}
We remind the reader that the \gspspec\ module was able to measure the individual abundances of 12 different chemical elements, including iron \citep{GSPspecDR3}. In this section, we turn our attention towards the solar cylinder, here defined by a distance to the Sun projected onto the Galactic plane smaller than 500~pc and no vertical cut. In this volume, we examine the distributions of nine abundance ratios, namely \NFe, \MgFe, \SiFe, \SFe, \CaFe, \TiFe, \CrFe, \NiFe,\ and \CeFe,\ as a function of mean metallicity \meta\ and their dependence on the maximum vertical distance to the plane (\zmax), the azimuthal velocity (\vphi), and the orbital eccentricity (e). \ZrFe\ and \NdFe\ are not presented here, because the number of stars with measured abundances in the solar cylinder is too small.

We use the {\it Individual abundance} subsample (as defined in Sect.~\ref{Sec:samples}), to which we apply some additional filters: \verb|astometric_params_solved = 31 OR 95| (stars with respectively 6- or 5-parameter astrometric solutions), \verb|ruwe < 1.4|, \verb|duplicated_source IS false,| and a projected distance to the Sun of less than 500~pc. The surface gravity range depends on the chemical element, as expected from the presence or not of the related spectral features. Figure~\ref{fig:kiel} shows the Kiel diagrams of the nine individual abundance samples.
We also apply the abundance corrections as a function of surface gravity recommended in \citet{GSPspecDR3} and which are meant to set dwarf and giant stars on the same scale.  Calibrations are provided for nitrogen, $\alpha$-elements, and iron-peak elements, but not for heavy elements \citep[see][for a discussion]{GSPspecDR3} and we therefore apply no correction to the cerium abundances. The samples were restricted to the range in surface gravity where the corrections are applicable.

Figures~\ref{Fig:F7-1} and \ref{Fig:F7-2} show the distributions of the abundance ratios as a function of metallicity for \NFe, \MgFe, \SiFe, \SFe, \CaFe, and \TiFe\ and for \CrFe, \NiFe,\ and \CeFe, respectively. Each row corresponds to one element. The colour bars encode: stellar density (col.~1), median maximum vertical distance to the plane (col.~2), median azimuthal velocity (col.~3), and median eccentricity (col.~4), per cell of 0.04~dex in \meta\ and per 0.02~dex in \XFe.

Nitrogen is produced through the CNO cycle. In Fig.~\ref{Fig:F7-1}, \NFe\ shows a flat trend with \meta. This is in satisfactory agreement with the modest gradient measured by \citet{Ecuvillon2004}, $0.1 \pm 0.05$, or \citet{SuarezAndres2016}, $0.04 \pm 0.07$ and $0.13 \pm 0.04$ (respectively, for their control sample and for a sample of stars with detected planets). It should be noted that \citet{daSilva2015} measured a stronger gradient of $0.28 \pm 0.06$. The proportional enrichment in nitrogen and iron abundances supports the scenario in which nitrogen is produced in intermediate- and low-mass stars, rather than in rotating massive stars \citep[see e.g. the discussion in][]{Ecuvillon2004}. In the solar cylinder sample, nitrogen is mainly measured in giants, including a small proportion of hot-massive giants. Their nitrogen surface abundances may have been enhanced during the dredge-up phases \citep{Masseron2015, Masseron2017, Shetrone2019}. 

We remind the reader that magnesium, silicon, sulfur, calcium, and titanium are $\alpha$-elements. It can be noted that silicon, calcium, and titanium, which have several lines in the RVS wavelength range, have been measured in large sample of stars in the solar cylinder: from more than 80\,000 for \TiFe\ to more than 140\,000 for \SiFe. As shown in Fig.~\ref{Fig:F7-1}, their abundance ratios decrease with metallicity consistently with previous studies in the solar neighbourhood \citep[e.g. among others][]{Adibekyan2012, Recio-Blanco14, Bensby2003, Bensby2014}. This trend is produced by the progressive enrichment in iron of the interstellar medium by thermonuclear supernovae and is well reproduced by Galactic chemical evolution models \citep{Snaith2021, Prantzos18, Koba20, Matteucci21}. As already discussed in Sect.~\ref{Sec:GlobalView}, we again point out the continuous decrease in $\alpha$-element abundances up to supersolar metallicities, as predicted by Galactic evolution models and already revealed by previous studies based on high-resolution spectra \citep[see e.g. ][]{Pablo20, Jeremy21}. A sparsely populated feature is also visible in \MgFe, \CaFe,\ and \TiFe\  in the range \meta\ $\in [-0.5, 0]$~dex and at low abundance ratio values. It is made of a mix of dwarfs and giants in Mg and Ti and of giants in Ca (by construction of the sample). The Massive giants (as defined in Sect. 3) are contained within the feature, although they are not the unique contributors to it. This is due to  the partial overlap between the Massive sample and the contiguous RGB stars in the Kiel diagram, as also seen in Fig~\ref{Fig:TestsCalibs}. The relative contribution of the two populations varies from one element to another and this is why the median values of V$_\phi$ and eccentricity in Fig.~\ref{Fig:F7-1}  are slightly different from one element to another at that abundance locus. It is worth noting that the $\alpha$-element enhancement and the metallicity  compete to regulate the opacity of the stellar atmosphere. Both a higher \alphaFe\ and a higher \meta\ produce a higher opacity and therefore a cooler effective temperature. As a consequence, low-\alphaFe\ stars are hotter than high-\alphaFe\ stars at the same metallicity, but they overlap in \T\ with lower metallicity stars. This is what happens in the Kiel diagram region at the limit of the Massive and RGB samples \citep[see e.g. Fig. 22 in][]{GSPspecDR3}.

Chromium and nickel are iron-peak elements. As such, their abundance ratios are weakly dependent on metallicity, although some authors observe a slow rise in \NiFe\ at metallicities beyond solar \citep[e.g.][]{Adibekyan2012, Bensby2014}. In Fig.~\ref{Fig:F7-2}, as expected, \CrFe\ abundances show no trend with metallicity. \NiFe\ shows a flat trend on the metal-rich side, while on the metal-poor side, the abundance ratios present a small positive offset. 

Cerium is a heavy element mostly produced by the s-process (see Sect.~\ref{Sec:GlobalView}). In Fig.~\ref{Fig:F7-2}, the bulk of solar cylinder stars
show no significant trend with \meta. This is consistent with earlier works using high-resolution spectra \citep[][]{Battistini2016,DelgadoMena2017, Forsberg2019}.  In Fig.~\ref{Fig:F7-2}, the median cerium-to-iron-abundance ratio is around $+0.2$~dex. This is is close to that measured by \citet{Forsberg2019} (for stars with \meta\ $< 0$~dex), but higher than \citet{Battistini2016} and \citet{DelgadoMena2017} whose average \CeFe\ level is around 0~dex or slightly below.

As discussed in Sect.~\ref{Sec:GlobalView}, the possibility of estimating the abundance of a given chemical element depends on the presence of the lines of that element in the related spectra, and therefore it changes with stellar type and luminosity (see also Figure~\ref{fig:kiel}). As a consequence, the tracers of different chemical elements probe different ranges of maximum distance to the Galactic plane (\zmax).  Eight out of the nine elements presented here (Na, Mg, Si, Ca, Ti, Cr, Ni and Ce) can be measured in giant stars. As shown in the second column of Figs.~\ref{Fig:F7-1} and \ref{Fig:F7-2}, on the metal-poor side, the median (\zmax) can reach up to 1 to 2~kpc. As already discussed for Fig.~\ref{Fig.SelFuncXY1}, sulfur is measured in hot F-G dwarf and turn-off stars from the RVS spectra \citep{GSPspecDR3}. It therefore probes a narrower range of \zmax with median (\zmax) being found at around a few hundred parsecs at all metallicities.

In the solar cylinder, the larger the range of \zmax\ probed, the more the thick disc is sampled. At low metallicity, the thick disc is clearly visible in the $\alpha$-elements Mg, Si, Ca, and Ti as a lower median azimuthal velocity and higher median eccentricity structure (green/blue cells in columns 3 and 4). At higher metallicity, the precision of the abundances does not allow us to see the gap between the high- and low-$\alpha$ sequences in stellar density space (col. 1). Nevertheless, in the azimuthal velocity space (col. 3), looking at Mg and Ti, the high-$\alpha$ sequence rotates a few tens of km~s$^{-1}$ slower than the low-$\alpha$ sequence, as expected from previous studies \citep[e.g.][and Sect.~\ref{Sect:rotation-metallicity}]{Adibekyan2013, Haywood2013, Recio-Blanco14, AllendePrieto2016, Kordopatis17, ReFiorentin2019}. This can be used to separate the thick and the thin disc in chemical space for these two elements. In the solar cylinder sample, sulfur is dominated by thin-disc stars where most hot dwarfs lie. It is interesting to note that the thin-disc negative azimuthal velocity gradient as a function of metallicity reported by \citet{Adibekyan2013, Haywood2013, AllendePrieto2016, Kordopatis17, ReFiorentin2019} for example and also measured in Sect.~\ref{Sect:rotation-metallicity} is visible in the third column as a transition from red to yellow-green (from metal-poor to metal-rich).

In summary,  the \Gaia\ solar cylinder individual abundance distributions reveal the expected chemo-dynamical patterns of thin and thick disc stellar populations for a variety of chemical elements, highlighting the richness of the \Gaia\ chemical space.

%

\section{Chemo-kinematical cartography of the open clusters population}
\label{Sec:clusters}
Open stellar clusters are groups of stars that formed together at the same time and from the same material. For this reason, their ages, distances, kinematics, and chemical patterns can be estimated easily and with very high precision. Given these properties, open clusters are widely considered excellent tracers of the Galactic disc, both in space and time \citep[e.g.][]{Friel95,Yong12,Cunha16,Donor20,Casamiquela21,Spina21}. In this section, we investigate the distribution of metals (\meta) traced by open clusters across the Milky Way disc.

To date, around 2600 open clusters spread throughout the Milky Way disc have been identified in the \Gaia\ data set. Cluster ages, parallaxes, proper motions, and radial velocities have been derived by \citet{CantatGaudin20,CastroGinard21,Tarricq21} from stars with membership probability $\geq$0.7. These parameters have then been employed to derive the Galactocentric positions, velocities, and orbital parameters for 2162 open clusters through the same procedure described in Section~\ref{Sec:Orbital_params}. It was not possible to derive these quantities for the full sample of known open clusters because a few hundred of them have no radial velocity estimates. 

As a result of the cluster membership analysis based on the \Gaia\ astrometric solutions, \citet{CantatGaudin20,CastroGinard21,Tarricq21} also provide the list of stars that are possibly associated to these open clusters along with their membership probabilities. Of all these stars with membership probability $\geq$0.7, 6861 have \texttt{mh\_gspspec} determinations. Among these latter, we select only those belonging to the \textit{High-Quality} subsample (Sect.~\ref{Sec:samples}). From this sample, we further discard all sources with \texttt{teff\_gspspec}$\geq$8000 and \texttt{rv\_expected\_sig\_to\_noise}$\geq$40. These two additional conditions further improve the quality of our final data set which is composed of 1613 stars. These objects are members of 597 open clusters for which we derive median metallicities \meta\ and relative uncertainties $\Delta$~\meta. This latter is defined for each open cluster as the standard deviation of the \texttt{mh\_gspspec} values of its members. These open clusters have a number of members with \gspspec\ \meta\ ranging between 1 and 55, with an average of three members. When only one member is available, $\Delta$\meta\ is assumed to be equal to the \texttt{[M/H]\_unc} of the single star.

\subsection{Galactic radial metallicity and \alphaFe\ gradients}\label{sec:OC_globalGradient}

\begin{figure}
\includegraphics[width=0.5\textwidth]{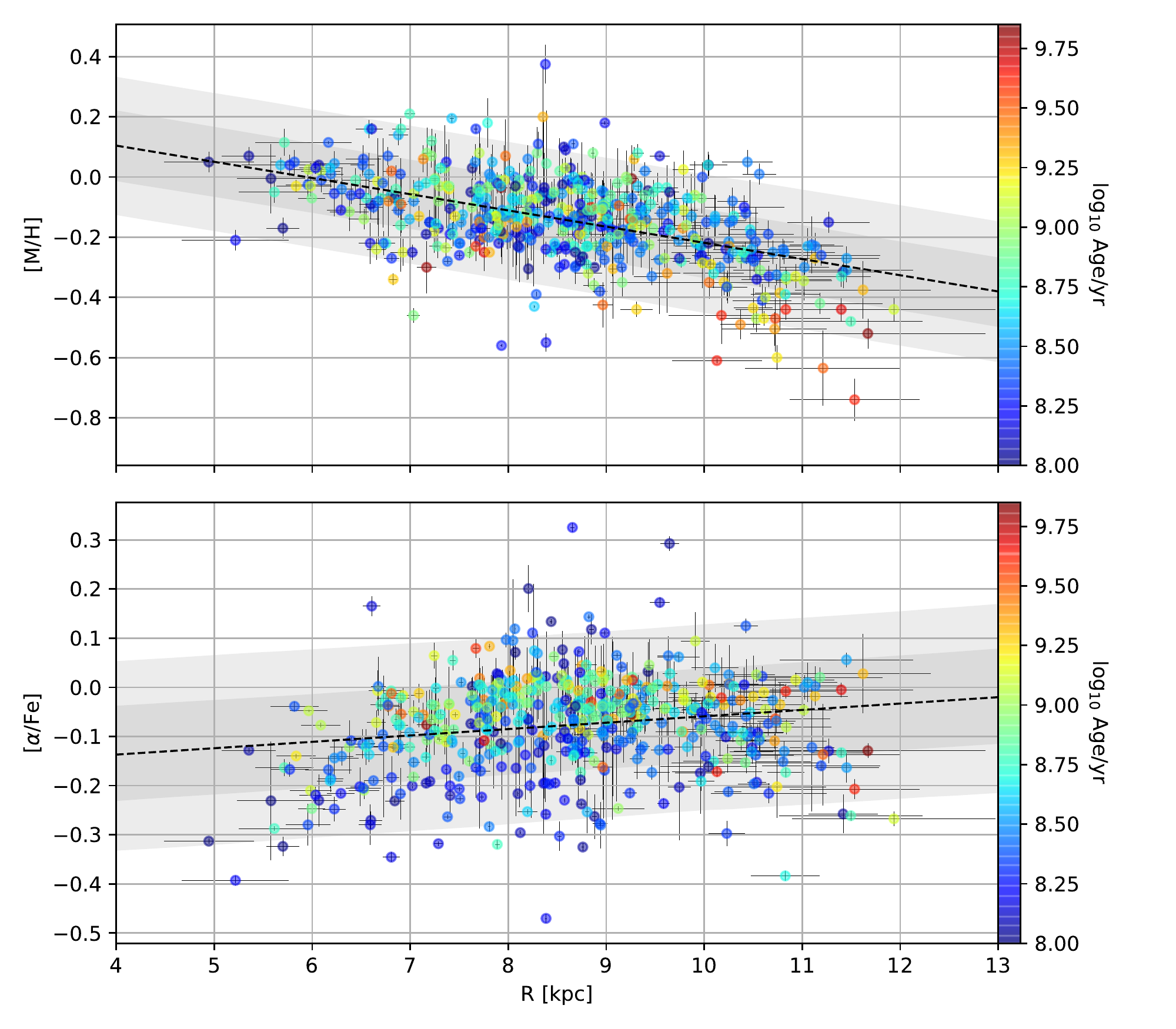}
\caption{ Galactic radial metallicity and [$\alpha$/Fe] gradients traced by open clusters. Top: Cluster \meta\ values as a function of their Galactocentric distance $R$. Clusters are marked by  circles colour coded
according to their age. The grey shaded areas represent the 68\%\ and 95$\%$ confidence intervals of the linear models resulting from the Bayesian regression, while the black dashed line traces the most probable model. Bottom: Cluster [$\alpha$/Fe] values as a function of their Galactocentric distance $R$. Symbol colours and shaded areas are the same as those in the top panel.}
\label{Fig:OCs_gradient}
\end{figure}

Here we use the data set described above to investigate the radial metallicity gradient traced by open clusters. The top panel of Fig.~\ref{Fig:OCs_gradient} shows the \meta\ abundance of the 503 open clusters older than 100~Myr and with $R\leq 12$~kpc. 

The cut in age was necessary because the spectroscopic analysis of very young stars is considerably complicated by high rotational velocities and by the effects of chromospheric activity which make stars appear more metal-poor than they are in reality \citep{Galarza19,Baratella20,Spina20,Zhang21}. Instead, the cut in $R$ allows us to include only the open clusters located before the break in the radial metallicity profile which is located at $R\sim 12-14$~kpc \citep[for a recent study, see e.g.][and literature therein]{Donor20}. As our sample does not include open clusters located at $R>14$~kpc, the analysis of this section is focused on the metallicity distribution of the inner disc. Symbols in the top panel of Fig.~\ref{Fig:OCs_gradient} are colour coded as a function of age. The open cluster distribution is modelled with a Bayesian regression using the same procedure adopted by \citet{Spina21}. Namely, we employ a linear model $y_i = \alpha \times x_i + \beta$, where x$_i$ and y$_i$ are normal distributions centred on the $R$ and \meta\ values of the $i^{\rm th}$ cluster. The x$_i$ normal distributions have standard deviations equal to $\sqrt{\Delta[M/H]^2 + \epsilon^2}$. The $\epsilon$ parameter accounts for the intrinsic chemical scatter between clusters at fixed Galactic radius. The simulation is run with 10\,000 samples, half of which are used for burn in, and a No-U-Turn Sampler \citep{Hoffman11}. The script is written in Python using the $\tt{pymc3}$ package \citep{Salvatier16}. The 68\% and 95$\%$ confidence intervals of the models resulting from the posteriors are represented in Fig.~\ref{Fig:OCs_gradient} with grey shaded areas. The results are also listed in Table~\ref{Tab:posteriors_tab}, which reports the mean, standard deviation, and 95$\%$ confidence interval of each posterior distribution\footnote{Note that the $\alpha$ and $\beta$ posteriors are highly correlated with a Pearson coefficient of 0.98.}. 

Interestingly, the radial metallicity gradient $\alpha$=-0.054$\pm$0.008~dex~kpc$^{-1}$ derived from our sample of 503 open clusters is consistent with the value given by \citet{Casamiquela19}, that is $-$0.056$\pm$0.011~dex~kpc$^{-1}$ (18 open clusters), and that of \citet{Spina22}, namely $-$0.064$\pm$0.007~dex~kpc$^{-1}$ (175 open clusters). On the other hand, our slope estimation is higher than those provided by other recent works in the literature: for example  $-$0.10$\pm$0.02~dex~kpc$^{-1}$ given by  \citet{Jacobson16} for 12 open clusters;  $-0.086\pm0.009$  by \citet{Netopil16} for 88 open clusters); $-$0.077$\pm$0.007~dex~kpc$^{-1}$ by  \citet{Carrera19} for 90 open clusters; $-$0.068$\pm$0.004~dex~kpc$^{-1}$ by  \citet{Donor20} for 71 open clusters;  $-$0.076$\pm$0.009~dex~kpc$^{-1}$ by \citet{Spina21} for 134 open clusters; and $-$0.066$\pm$0.005~dex~kpc$^{-1}$ by  \citet{Zhang21} for 157 open clusters.

The great diversity of these estimations of the radial metallicity gradient highlights how studies based on open clusters are strongly affected by the samples employed in their analysis. In this regard, our study represents a significant step forward in the field as our data set is a few times larger than those employed in the past.

The full analysis is repeated for the distribution of open clusters in the \meta--r$_{\rm guid}$ plane, where r$_{\rm guid}$ is the guiding radius defined in Sect.~\ref{Sec:Gradients} as $(R_{\rm apo}+R_{\rm peri})/2$. The resulting posteriors are listed in Table~\ref{Tab:posteriors_tab} and are consistent within the uncertainties with those obtained from the \meta--R$_{\rm Gal}$ distribution. This is due to the fact that, with some rare exception, the vast majority of the open clusters in our sample are on almost circular orbits. 

Finally, we analysed the radial distribution of [$\alpha$/Fe] traced by open clusters. The resulting gradient is slightly positive, in agreement with recent studies \citep[e.g.]{Donor20,Spina21}. The results are plotted in the bottom panel of Fig.~\ref{Fig:OCs_gradient}. The posteriors are listed in Table~\ref{Tab:posteriors_tab}.

\begin{table}
\caption{Posteriors linear regression}
\centering
\label{Tab:posteriors_tab}
{\small
\begin{tabular}{cccc}
\hline
\multicolumn{4}{c}{[M/H] versus $R$} \\ \hline \hline
Parameter & Mean & $\sigma$ & 95$\%$ C.I.  \\ \hline
$\alpha$ [dex kpc$^{-1}$ ] & $-$0.054 & 0.008 & $-$0.062 - $-$0.046\\
$\beta$ [dex] & 0.32 & 0.07 & 0.25 - 0.39\\
$\epsilon$ [dex] & 0.116 & 0.008 & 0.108 - 0.124 \\ \hline \hline

\multicolumn{4}{c}{[M/H] versus r$_{\rm guid}$} \\ \hline
Parameter & Mean & $\sigma$ & 95$\%$ C.I.  \\ \hline 
$\alpha$ [dex kpc$^{-1}$ ] & $-$0.063 & 0.008 & $-$0.071 - $-$0.055\\
$\beta$ [dex] & 0.38 & 0.07 & 0.31 - 0.45\\
$\epsilon$ [dex] & 0.110 & 0.008 & 0.102 - 0.118\\ \hline \hline

\multicolumn{4}{c}{[$\alpha$/Fe] versus $R$} \\ \hline
Parameter & Mean & $\sigma$ & 95$\%$ C.I.  \\ \hline 
$\alpha$ [dex kpc$^{-1}$ ] & 0.013 & 0.007 & 0.006 - 0.020\\
$\beta$ [dex] & $-$0.19 & 0.06 & $-$0.24 - $-$0.13\\
$\epsilon$ [dex] & 0.096 & 0.007 & 0.089 - 0.102\\ \hline \hline

\multicolumn{4}{c}{[$\alpha$/Fe] versus r$_{\rm guid}$} \\ \hline
Parameter & Mean & $\sigma$ & 95$\%$ C.I.  \\ \hline 
$\alpha$ [dex kpc$^{-1}$ ] & 0.017 & 0.008 & 0.009 - 0.024\\
$\beta$ [dex] & $-$0.21 & 0.06 & $-$0.28 - $-$0.15\\
$\epsilon$ [dex] & 0.095 & 0.006 & 0.089 - 0.102\\ \hline

\end{tabular}
}
\end{table}

\subsection{Temporal evolution of the Galactic radial metallicity gradients}
\label{Sec:OCs_gradients}

The temporal evolution of Galactic metallicity gradients provides information about how the interstellar medium has been progressively enriched in metals, but also about how the metallicity tracers migrate across the Galactic disc. For instance, it is well known that the radial metallicity gradient traced by field stars gets flatter for older stars \citep[e.g.][]{Casagrande11,Anders17}. This behaviour is due to radial migration which is expected to smooth the gradient with time \citep[e.g.][]{Schoenrich09, Minchev18}. In contrast to what is observed for field stars, several works have shown that the radial metallicity gradient traced by open clusters flattens with decreasing age \citep[e.g.][]{Friel02,Donor20,Spina21,Spina22}. However, the literature is not in complete consensus \citep[see][]{Salaris04,Netopil22}.

\begin{figure}
\includegraphics[width=0.5\textwidth]{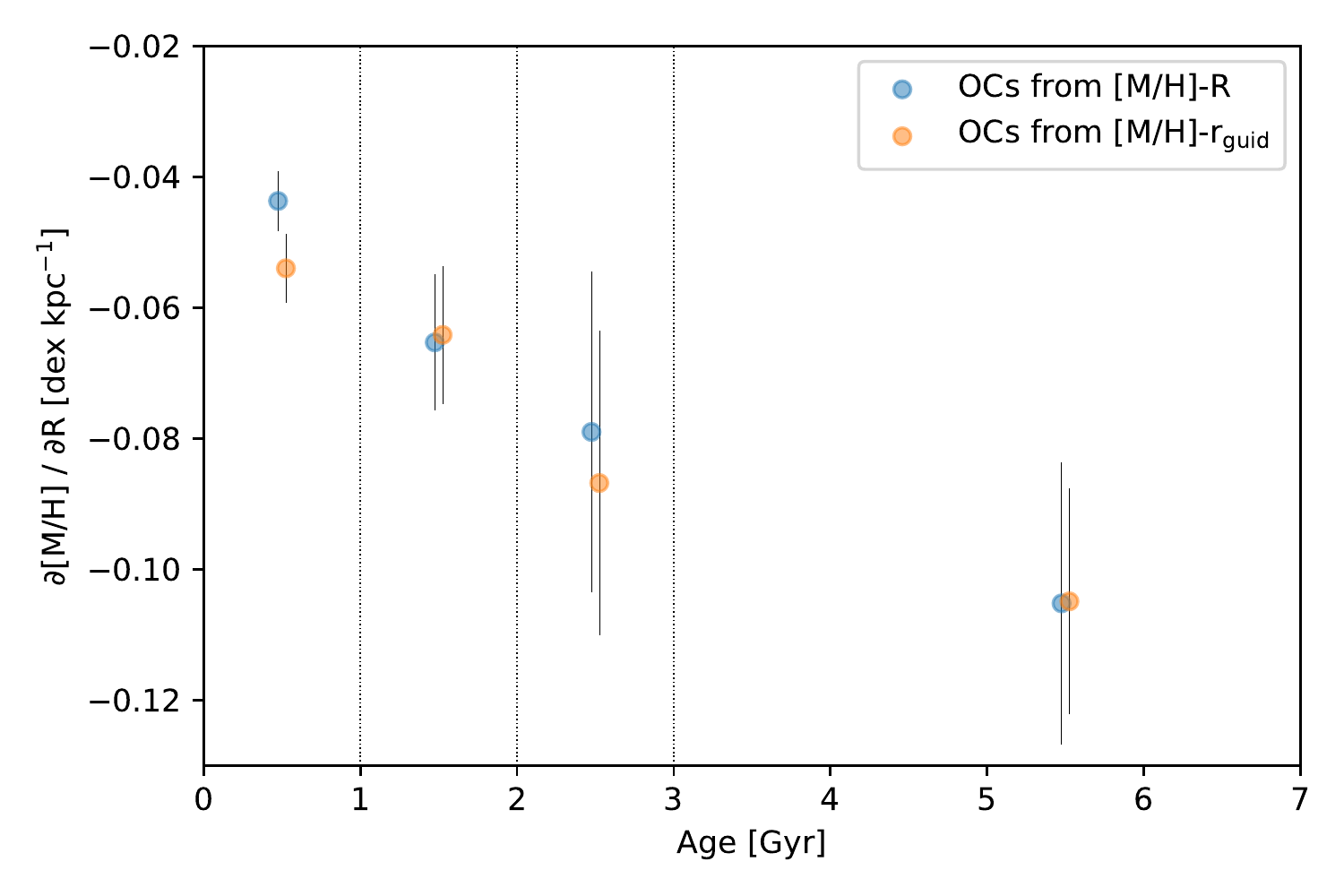}
\caption{Age dependence of the Galactic metallicity gradient traced by open clusters in the [M/H]--$R$ (blue dots) and [M/H]--r$_{\rm guid}$ (red dots) diagrams.}
\label{Fig:gradient_evolution_MCMC}
\end{figure}

We split the sample of open clusters into four groups depending on their ages: age $<$1 Gyr,  1$\leq$ age $<$2, 2$\leq$ age $<$3, and age $\geq$3 Gyr. These bins are arbitrarily chosen to contain a balanced number of clusters. Figure~\ref{Fig:gradient_evolution_MCMC} shows the metallicity gradients calculated with the same method outlined above. Based on our analysis, we confirm that both the [M/H]--$R$ and [M/H]--r$_{\rm guid}$ gradients evolve with age in the same way, with the younger cluster gradients being flatter.

Interestingly, literature results for field stars seem to indicate an opposite temporal evolution of the gradient between open clusters and field stars \citep[e.g.][]{Casagrande11,Anders17,Minchev18}. This opposite behaviour could be due to a bias imposed by the Galaxy on the current open cluster population \citep{Anders17,Spina21}. In fact, while stars are free to migrate within the Galactic disc, the metal-rich clusters formed in the inner disc can survive only if they migrate outward where the Galactic potentials (e.g. spirals, bar, giant molecular clouds) are weaker and less destructive. In contrast, the metal-rich clusters migrating inward are quickly disrupted. The direct consequence of this Galactic bias is that the gradient traced by old clusters is steeper than the one seen for young clusters and field stars. 

Nevertheless, our analysis of the young field stars in the Massive sample presented in Sect.~\ref{Sec:Gradients} suggests a flattening of the gradient in recent times (a slope of -0.036$\pm$0.002~dex kpc$^{-1}$ is obtained for the Massive sample, while it is of -0.055$\pm$0.007~dex kpc$^{-1}$ for the entire field population considered in the $Gradient$ sample). 
Therefore, from purely \Gaia\ data, both open clusters and field stars seem to show a flattening of their radial metallicity gradient in the more recent epochs of Milky Way evolution. However, a careful analysis of stellar ages is necessary to put this result in robust context and to allow a significant comparison with the temporal evolution of the gradient traced by open clusters.

\subsection{Kinematic features in the radial metallicity distribution}
\label{Sec:OCs_waves}

Similarly to what is observed for field stars, it is expected that the complex dynamical substructure of the Galactic disc discussed in Section~\ref{Sect:Kinematics-Toomre} would produce visible signatures in the chemical distribution of elements traced by open clusters.  In fact, based on a sample of 1382 open clusters, \citet{Tarricq21} recently showed that they seem to follow the main diagonal ridges in the Galactic radial distribution of azimuthal velocities outlined by field stars. 

\begin{figure}
\includegraphics[width=0.5\textwidth]{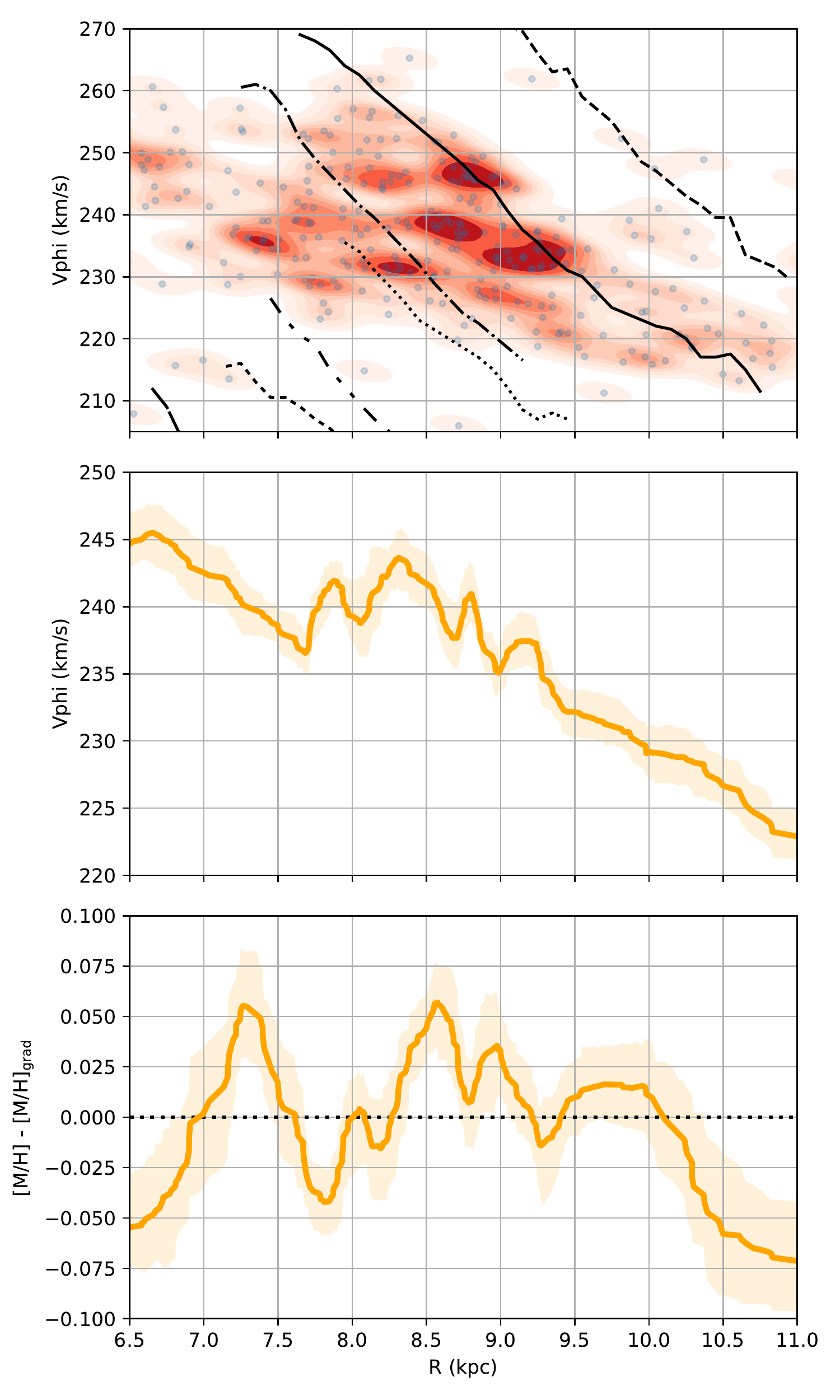}
\caption{ Kinematic features in the radial metallicity distribution as traced by open clusters. Top: Distribution of open clusters in the V$_{\rm \phi}$--$R$ diagram. The 332 open clusters with [M/H] estimation are shown as grey dots. The red area shows the kernel density estimate of that sample. The Galactic ridges identified by \citet{Ramos:2018} are plotted as in Fig. \ref{FigureKinematics-Vphi-R}. Middle: Locally weighted scatterplot smoothing (LOWESS) regression of cluster V$_{\rm \phi}$ as a function of $R$ (solid line). The shaded area shows the 90$\%$ confidence interval of the LOWESS functions obtained by sampling within the uncertainties of the two variables. Bottom: Residuals between the cluster [M/H] values and the metallicities predicted by the radial metallicity gradient as in Table~\ref{Tab:posteriors_tab} as a function of $R$. As in the middle panel, the solid line shows the LOWESS function and the shaded area shows its 90$\%$ confidence interval.}
\label{Fig:OCs_waves}
\end{figure}

The top panel of Fig.~\ref{Fig:OCs_waves} shows the distribution of open clusters in the V$_{\rm \phi}$--R$_{\rm Gal}$ diagram (grey dots). Here we consider only the 322 open clusters with |$\phi$-$\phi_{\odot}$|$<$10$^{\circ}$ older than 100 Gyr and with a metallicity estimation. We also plot the kernel density estimate of the open cluster distribution (in red) and the ridges (black lines) identified for field stars by \citet{Ramos:2018}. Similarly to what is observed by \citet{Tarricq21}, here we notice a steep decrease in the number of open clusters on the right of the `Sirius' ridge (solid line). This could be due to a number of factors, such as the steep decrease in the number of open clusters beyond 9.5-10 kpc, and the fact that a significant fraction of our outermost clusters are older than a few gigayears in contrast to the rest of the sample. In any case,
 other than that, the density features outlined by open clusters do not perfectly match the locations of the other ridges. This may be related to the complex and unexplored selection function of the known open cluster sample which is presumably different from that of field stars. Nevertheless, what we want to highlight here is that the distribution of open clusters shows complicated patterns characterised by different overdensities and a V$_{\rm \phi}$ range which is very restricted compared to that of field stars.

The complexity of this pattern is also shown in the middle
panel of Fig.~\ref{Fig:OCs_waves}, where the solid orange line represents the locally weighted scatterplot smoothing (LOWESS) regression of the V$_{\rm \phi}$ of clusters as a function of R$_{\rm Gal}$. The shaded area shows the 90$\%$ confidence interval of the LOWESS functions obtained by sampling within the uncertainties of the two variables. As expected, the innermost and the outermost clusters have the highest and the lowest V$_{\rm \phi}$ values, respectively. However, in between these extremes, the LOWESS function shows a noticeably wavy pattern which is particularly evident within 7 and 9.5 kpc. A similar wavy pattern was also noticed by \citet{Tarricq21} who identified a significant dip at 7.5 kpc in the V$_{\rm \phi}$--$R$ diagram with a departure of more than 2$\sigma$ from the Galactic rotation curve derived analytically from a simple axisymmetric potential.

Interestingly, this wavy pattern seems to be linked to the metallicity distribution as shown in the bottom panel of Figure~\ref{Fig:OCs_waves}. Here, we plot the LOWESS regression of [M/H]--[M/H]$_{\rm grad}$ as a function of R$_{\rm Gal}$, where [M/H]$_{\rm grad}$ is the metallicity predicted by the linear model of Fig.~\ref{Fig:gradient_evolution_MCMC} (top panel). As in the V$_{\rm \phi}$--R$_{\rm Gal}$ diagram, this LOWESS function also shows a wavy behaviour with the residual metallicity [M/H]--[M/H]$_{\rm grad}$ swinging around the zero value. We note that each crest in the [M/H]--[M/H]$_{\rm grad}$ versus $R$ diagram is located near a trough in the V$_{\rm \phi}$--R$_{\rm Gal}$ plot. 

This relation between the waves seen in the two LOWESS functions can be explained as the natural consequence of open cluster migration across the Milky Way disc. More specifically, if we consider only the type of migration that conserves angular momentum (i.e. \textit{blurring}), the open clusters coming from the outer disc have higher V$_{\rm \phi}$ and lower residual metallicity than adjacent clusters at a given $R$. On the other hand, open clusters coming from the inner disc have lower V$_{\rm \phi}$ and higher residual metallicity than adjacent clusters. 

However, we still need to explain the origin of the wavy V$_{\rm \phi}$--$R$ distribution in the first place. It is likely that the V$_{\rm \phi}$ oscillations are somehow linked to the orbital resonances and perturbations observed from field stars. Although the top panel of Fig.~\ref{Fig:OCs_waves} does not provide strong evidence in that sense, the waves are likely the consequence of the same process that produced the diagonal ridges. A deeper analysis of the open cluster population, of their orbital properties, and knowledge of their selection function are necessary to provide additional clues on this topic.

\section{Summary: The \Gaia\ chemical cartography of the Milky Way}\label{Sec:Summary}
The previous sections reveal a number of chemo-dynamical features observed in the RVS \gspspec\ database. Some of these are the same underlying Galactic characteristics  seen from different perspectives; this section summarises these characteristics, providing a more global view. 

\subsection{Chemical markers of Milky Way structure}
\label{Sec:MWstruct}
The large radial and vertical coverage of the database, together with its substantial azimuthal coverage, allow us to explore the detail of the structure of the Milky Way  within an unprecedentedly large volume.
Sections \ref{Sec:GlobalView}, \ref{Sec:Gradients}, and \ref{sec:5} explore the Galaxy's spatial symmetry through chemo-kinematical considerations. While radial and azimuthal substructures are detected (cf. Sect.~\ref{Sec:ChemSeis}), the Galaxy shows an important degree of vertical symmetry (agreement within 0.05 dex in \meta, see for instance Figs.~\ref{fig:radial_gradients} and ~\ref{fig:vertical_gradients}) in the region of radial distances of between 5 and 10 kpc from the centre, and within 2 kpc of the plane. Outside this region, selection functions and statistical fluctuations perturb the diagnostics. This vertical symmetry imposes an important constraint on simulations, in the context of the rich accretion history of the Milky Way. In particular, further detailed studies of chemo-dynamical correlations including halo stars would help us to explore the possible existence of a preferential axis of accretion, extending recent literature analyses hampered by low number statistics \citep[e.g.][]{RecioBlanco21}.

Moreover, stars with typical chemo-dynamical characteristics of the disc population (fast rotation associated with metallicities higher than $\sim$-0.75~dex) are detected at large distances above and below the plane. The maximum height at which disc stars are detected grows with increasing radial distance. This behaviour is typical of a flared disc, as already detected in the Milky Way for both the stellar population and the gas \citep[e.g.][]{Grabelsky87, Feast2014, LopezCorredoira2014, Yazan2006, Reyle2009, Li2019, Mosenkov21}. In the external regions, the disc population is detected up to 4~kpc from the plane (cf.~\ref{Figure_flaring}), suggesting a higher scale height of the flare than previously reported. We note however that an exact quantification of the disc shape is out of the scope of this paper. In addition, a certain contribution of the disc warp  \citep[e.g.][]{PoggioWarp, Reyle2009, Lopez-Corredoira2020} in the external regions cannot be excluded.

Finally, the identification of massive stars (blue loop and other variable objects) in the \gspspec\ Kiel diagram has allowed us to detect the disc spiral structure. The spiral pattern traced by this population is in agreement with \cite{Eloisa21}, who used upper main sequence stars, open clusters, and classical Cepheids. A more detailed analysis of the spiral arms can be found in \cite{DPACP75}. In the present paper, the metallicity distribution of the population of massive stars  shows the expected radial gradient, although azimuthal variations also exist. In particular, at the solar Galactic radius,  we find that stars closer in azimuth than the Sun to the major axis of the bar have higher metallicities (Sect.~\ref{sec:5}, Fig.~\ref{Fig.SelFuncXY3}). A detailed characterisation of these chemical inhomogeneities and their link to the spiral arms remains to be performed.   

\subsection{Chemical markers of disc kinematic disturbances }
\label{Sec:ChemSeis}
One of the most evident results of this analysis is that the known kinematic disturbances of the disc, including different manifestations of phase space correlations and kinematic substructure, are clearly associated with chemical patterns, acting as markers of the perturbations. From the observational point of view, it is important to highlight that chemistry allows a completely independent diagnostic from the kinematical one.

First, the $V_Z$ versus \vphi\ distribution in the outer Galactic disc shows density substructures associated to vertical asymmetries and metallicity patterns (cf. Sect.~\ref{sec:VphiVz}).
Second, the phase spiral in the $Z$ versus $V_Z$ plane \citep[][]{Antoja2018}, already associated with $V_R$ and \vphi\ patterns, appears to be linked to \meta\ variations 
(Fig.~\ref{FigureKinematics-Vz-Z}, bottom panels). In particular, in the external region,  $R>$ 9.25~kpc,  where no significant bias from differential extinction exists, stars with metallicity excess with respect to the median \meta\ values clearly trace the observed phase-spiral pattern. Stars with metallicity excess are found more than 0.5 kpc above the plane, coinciding  with stellar overdensity and lower rotation values (which seem to follow the expected metallicity trend with mean azimuthal velocities in the disc).

Third, the identified ridges in the \vphi\ versus R space are typically more metal-rich than their phase space surroundings and show azimuthal dependencies (Fig.~\ref{FigureKinematics-Vphi-R}). In addition, although the lower number statistics perturb the interpretation of results, substructures in the \vphi\ versus R  plane are also seen in the young field stars tracing the spiral structure (Fig.~\ref{FigureKinematics-Vphi-R-massivestars}) and in the open cluster population (Fig.~\ref{Fig:OCs_waves}), possibly associated to metallicity fluctuations.

Additional evidence of chemical fluctuations associated with the {\it seismology} of the  Milky Way comes, for the first time, from the orbital space.  The action plane defined by $J_R$ versus $L_Z$ (Fig.~\ref{Fig_JLz_metalpha_withflags}) shows clear ridges of higher stellar density characterised by stars in low $Z_{max}$ orbits with metallicities higher than the surrounding median values. 

In summary, these different  diagnostics indicate that stars tracing disc kinematic disturbances have higher metallicities. This is true both for phase space correlation patterns and kinematic substructure.
Although results from ground-based surveys \citep{Khanna19} already identified solar metallicity values in the local \vphi\ versus R ridges, this large-scale analysis of \Gaia\ \gspspec\ data shows that stars in kinematic substructures are more metal-rich than the typical metallicity at a given radius, irrespective of the metallicity regime.
The best chemical marker of disc kinematic disturbances is in fact the differential metallicity of the stars with respect to the median \meta\ value at each Galactic radius (as provided by the radial metallicity gradient).

To explain the higher metallicities tracing phase-spiral patterns and ridges, a younger age or an inner Galaxy origin could be invoked. Younger stars are indeed formed from a more chemically evolved (more metal-rich) interstellar medium. In addition, they are expected to be dynamically cooler, and therefore react more strongly than older dynamically hotter components when subject to either internal perturbations
\citep[][]{Fragkoudi19} or to satellite impacts \citep[e.g][]{Checkers18}. A literature analysis using \Gaia\ DR2 data and a heterogeneous compilation of stellar ages \citep{Laporte20} report slightly younger ages for stars in kinematical substructures. Extending previous results of \cite{Wang2019} and \cite{Antoja2021}, we show that objects formed only a few hundred million years ago (such as the massive population defined in Sect.~\ref{Sec:GlobalView} tracing the spiral arms or the open cluster population), also show kinematical substructure. On one hand, this is in agreement with N-body simulations of a recent \citep[500-800 Myr, e.g.][]{Laporte19Sag} or a 1-2 Gyr old \citep[e.g. ][]{Jos2021} and repeated interaction of the Milky Way with a massive, Sagittarius-like satellite. On the other hand, internal mechanisms of disc perturbation such as the buckling of a stellar bar can also generate long-living phase-spirals \citep{Khoperskov20} affecting stars of different ages, in agreement with observations. 

It is also worth mentioning that the observed azimuthal dependencies in  kinematic substructure (cf. Fig.~\ref{FigureKinematics-Vphi-R}) already detected by previous studies for some ridges \citep[e.g.][for Hyades and Hercules]{Ramos:2018, Monari19} imply azimuthal metallicity variations due to the corresponding chemo-kinematical correlation. Moreover, azimuthal dependencies are also present in the metallicity distribution of disc RGB and young stars (cf. Sect.~\ref{Sec:GlobalView}), with stars in azimuthal angles closer to the major axis of the bar being slightly more metal rich. Recent studies of the impact of Galactic disc asymmetries on azimuthal chemical variations 
\citep{Spitoni19} predict stronger azimuthal fluctuations near the spiral arm co-rotation radius. A detailed spatial analysis of chemical fluctuations is nevertheless out of the scope of this performance verification paper.

\subsection{Chemical markers of satellite accretion.}\label{Sec:ChemMergers}
The large spatial coverage and large number statistics of the \Gaia\ \gspspec\ database allow a chemo-dynamical study of halo substructure (cf. Sect.~\ref{Sec:Dynamics}). From this analysis, which is not intended to be exhaustive, three main conclusions emerge. 

First of all, halo substructure in the orbital space (cf. Figs.~\ref{Fig_ELz_dens_zoom} and \ref{Fig_JLz_metalpha_withflags}) is clearly observed, extending previous observations \citep[cf.][]{Koppelman_et_al19} using \Gaia\ DR2 and ground-based data. Second, the new \Gaia\ chemical data are of sufficient quality to provide chemical diagnostics of accretion through the abundance ratio of $\alpha$-elements to iron (cf. Fig.~\ref{Fig_alpha_vs_met_withflags_streams}). This opens new horizons for in-depth analysis of accretion debris and stellar streams.
In addition, the present study clearly detects stars with chemical patterns typical of the disc, but on halo orbits. The number of these puffed-up  disc stars increases as we move towards the disc region in the energy versus $L_Z$ plane. Finally, the chemo-dynamical properties of the Milky Way system of GCs can also be explored.


\subsection{Chemo-dynamics of the most recent billion years.}\label{Sec:1Gyr}
The \Gaia\ chemical database includes 687 open clusters (c.f Sect.~\ref{Sec:clusters}) and about 30~000 young stars tracing the spiral arms (cf. Sect.~\ref{Sec:GlobalView}). These encode the evolution of the last billion years, anchoring the long Galactic history to the present.

On one hand, age estimates of Galactic open clusters allow us to assess the temporal evolution of the metallicity radial gradient of the  cluster.  Flatter gradients are progressively observed in the younger open cluster populations (cf. Fig.~\ref{Fig:gradient_evolution_MCMC}), in contrast to the opposite trend reported in the literature for field stars \citep[][]{Casagrande11, Anders17, Minchev18}. This could be due to differential effects of radial migration between the clusters and the field population \citep[e.g.][]{Anders17,Spina21}. 

On the other hand, interestingly, the young field population that is traced here by massive F- and G-type stars shows a chemical impoverishment with respect to older populations. This impoverishment is observed not only in the global metallicity, but also in the abundance ratio of several chemical species with respect to iron, including $\alpha$-elements (cf. Fig.~\ref{Fig.SelFuncXY3}) and heavy elements like Cerium (cf. Fig.~\ref{Fig.SelFuncCe}). Explaining this metallicity decrease requires a careful study with chemical evolution models. However, its link to the coeval open cluster population could be key to solving the chemo-dynamical puzzle of the most recent epochs of Galactic evolution.

\subsection{The Sun in its Galactic environment}\label{sec:Sun}
The results discussed above bring to the fore the question of the position of our Sun  in the global chemo-dynamical evolution of the Milky Way. Spatially, the solar position in the local void outside the spiral arms is clearly visible in Fig.~\ref{Fig.SelFuncXY3}.
From a chemical point of view, when the Sun is considered within the global stellar population in the solar cylinder (cf. Figs.~\ref{Fig:F7-1} and \ref{Fig:F7-2}), it has a chemical pattern and chemo-dynamical characteristics typical of a thin disc star.
Interestingly, it has a higher metallicity than the corresponding median value at its Galactic position (assumed here to be R$=$8.249 kpc) from the measured radial gradient ($\Delta$\meta$=0.04$~dex from the {\it Gradient analysis} sample and $\Delta$\meta$=0.16$~dex from the more homogeneous RGB sample).
In addition, despite being born 4.6~Gyr ago, the Sun is also more chemically enriched than the young stars in the solar neighbourhood, even if they can be more  than $\sim$4~Gyr younger. These two facts seem to invoke either a discontinuity in the chemical evolution or a change in the Sun's orbital parameters (migration) or both phenomena at the same time. 

From a kinematical point of view, the solar $V_Z$ and $Z$ values place the Sun near the central grid point of the phase spiral (medium panels  Fig.~\ref{FigureKinematics-Vz-Z}). In addition, the Sun has a negative $V_R$ velocity and a positive $\Delta$\meta, which is coherent with the typical values observed in its phase space surroundings. Moreover, as indicated by the solar position in the \vphi\ versus $R$ plane (identified by a cross in Fig.~\ref{FigureKinematics-Vphi-R}), the Sun seems to lay near the Sirius ridge.   In the action space, the solar position near (1,0) in Fig.~\ref{Fig_JLz_metalpha_withflags} (in units of $L_{\odot}$), together with its low $Z_{\rm max}$ value of 0.12~kpc, again place the Sun near one of the identified ridges (the fifth one from the left in Fig. 31). 
The above information will be crucial in determining whether or not the Sun has been affected by the different perturbations at play.

\section{Conclusions}\label{Sec:Conclusions}
This article explores the scientific quality of the \Gaia\ chemical database produced by the \gspspec\ module from RVS spectra and published as part of the third \Gaia\ data release (DR3).
Combined with \Gaia\ EDR3 astrometric data and DR3 radial velocities, the all-sky \Gaia\ chemical cartography allows a powerful and precise chemo-dynamical view of the Milky Way with unprecedented spatial coverage. 

In this framework, chemical abundances carry pivotal information on the origins of the stars. With the exception of some elements that suffer from internal mixing in certain stellar evolutionary phases, the chemical composition of a star shows a preserved pattern of element abundances encoding the conditions of the interstellar medium at the time and place of its formation. Therefore, chemical markers are crucial for validating kinematical and dynamical diagnostics, providing constraints on the properties of matter and their temporal evolution in theoretical modelling and analysis.

In addition, 
understanding the history of the Milky Way requires the exploration of a complex physico-chemical space that only huge data sets such as the \Gaia\ catalogue can appropriately probe. Large and precise databases not only improve the confidence level of the derived general trends, but also allow researchers to reveal the crucial but rare non-standard populations in distribution tails. Adequately resolving the natural scatter in chemo-dynamical properties of Galactic stellar populations  is essential for uncovering the imprint of physical processes of evolution. 
This scatter is indeed the consequence of star and gas dynamics, of discontinuous star formation histories, and of galaxy accretion, among other mechanisms.

The results of the present analysis illustrate how this chemo-dynamical distribution tail allows us (i) to unveil the pronouncedly flared structure of the disc, (ii) to robustly assess the Galaxy vertical symmetry, (iii) to chemically trace the large-scale motions and seismology of the disc, (iv) to uncover galaxy accretion debris and heated disc stars on halo orbits, and (v) to anchor the Galactic history to its present through the young but scarce and ephemeral massive stars and open clusters.

\Gaia\ DR3 chemo-dynamical diagnostics open new horizons before the era of ground-based wide-field high-resolution spectroscopic surveys like WEAVE \citep{Dalton20} or 4MOST \citep{deJong19}  which will provide detailed abundances for millions of  Milky Way stars. Moreover, the \Gaia\ satellite allows all-sky, continuous data collection over timescales measurable in years, which is impossible to achieve from the ground. This  results in an extraordinary homogeneity of data, and allows \Gaia\ to include detailed bias modelling in its analysis and data-treatment procedures, which is at the core of its precision and observational success. A beautiful illustration of this are the \Gaia\ DR3 heavy element abundances, which at first appeared to be out of reach for the RVS instrumental characteristics. 

Finally, the results of the present work reinforce the vision of a heterogeneous Milky Way in permanent interaction with its environment, the outcome of an eventful history that continues to shape our Galaxy to the present day.

\begin{acknowledgements}
This work presents results from the European Space Agency (ESA) space mission \Gaia\  (https://www.cosmos.esa.int/gaia). \Gaia\ data are processed by the \Gaia\ Data Processing and Analysis Consortium (DPAC). Funding for the DPAC is provided by national institutions, in particular the institutions participating in the \Gaia\ MultiLateral Agreement (MLA). The \Gaia\ archive website is \url{https://archives.esac.esa.int/gaia}. Acknowledgments from the financial institutions are given in Appendix~\ref{sec_acknowledgements}.

Part of the calculations used for this article have been performed with the high-performance computing facility SIGAMM, hosted by the Observatoire de la Côte d'Azur.

\end{acknowledgements}

\bibliographystyle{aa}  
\bibliography{biblio} 

\begin{appendix}

\section{The astrometric control sample}
\label{Sect:ACS}

The astrometric control sample (ACS) is a subset of \gspspec\ 
whose distances are derived directly from the \Gaia\ DR3 parallaxes,
$\varpi$, after correcting for the individual parallax zero-point,
$\Delta_\varpi$, calculated according to
\citet[][]{Lindegren2021}, that is, using the corrected parallaxes 
${\varpi}' = {\varpi - \Delta_\varpi}$, and after including 
only sources with high-quality astrometric parameters, full three-band photometry, and radial velocities.
In short, the ACS selection criteria are:\\ \noindent
\begin{verbatim}
astrometric_params_solved = 31 AND 
ruwe < 1.4 AND
parallax_over_error > 5 AND
duplicated_source IS false AND
radial_velocity IS NOT null  AND
phot_g_mean_mag IS NOT null AND 
phot_bp_mean_mag IS NOT null AND
phot_rp_mean_mag IS NOT null 

\end{verbatim} 

\noindent
We note that {\it astrometric$_{-}$params$_{-}$solved = 31} means that only five-parameter astrometric solutions are retained, and the criteria {\it parallax$_{-}$over$_{-}$error > 5} immediately eliminates all negative parallaxes. This has no statistical consequence given the nature of the ACS sample and its use in this work.  
Trigonometric distances are then 
\begin{equation}
    r_\varpi = \frac{1}{\varpi'},
    \label{eq:r_plx}
\end{equation}
and, as is often the case in the astronomical literature, the {\it lower} and {\it upper} distance errors are, respectively:
\begin{equation}
    \sigma_ {r_\varpi}^- =  r_\varpi \left(\frac{\alpha}{1+\alpha}\right)
    \,\,\,\,\,\,\,\,
       \sigma_ {r_\varpi}^+ =  r_\varpi \left(\frac{\alpha}{1-\alpha}\right),
\label{distance-errors}
\end{equation}
where $\alpha = \sigma_\varpi/\varpi$ is the parallax relative error.
These distance errors 
correspond to the 1-$\sigma$ parallax errors $\sigma_\varpi$ with the asymmetry, $\sigma_ {r_\varpi}^- / \sigma_ {r_\varpi}^+ = (1-\alpha) / (1+\alpha)$, decreasing with $\alpha$.

Table \ref{table:ACSGeoStatistics} reports the basic statistics of the complete ACS and of the subsets ACS$_{\rm MQ}$ and ACS$_{\rm HQ}$, that is, the lists with only {\it Medium} and High Quality  samples, respectively (see Sect. \ref{Sec:samples}).
The full ACS includes 4\,397\,814 stars of the  {\it General} sample, about half of which (53\%) are astrometric primary flagged 
({\it astrometric$_{-}$primary$_{-}$flag = True}).
It reaches distances of up to $\sim$ 10 kpc from the Sun with a median of 1.3 kpc and 90\% of its stars within $\simeq$ 3.7 kpc. The parallax relative error $\alpha$ 
is below 7\% for the 90\% quantile.
The ACS$_{\rm MQ}$ includes 3\,362\,012
 stars up to 10 kpc with median distance $r_\varpi = 1.5$ kpc and mostly (90\%) within $\simeq$ 2.9 kpc. For this sample, $\alpha$ is below 5\% for 90\% of the sources. Finally, the ACS$_{\rm HQ}$ is formed by 1\,815\,072 stars up to 6 kpc with median distance $r_\varpi = 1.5$ kpc, mostly (90\%) within $\simeq$ 3.4 kpc, and  $\alpha$ under 6\% for 90\% of the cases. 


Table \ref{table:ACSGeoStatistics} also includes the analogous statistics for the corresponding samples with Geo distances, $r_{\rm Geo}$, from  \citet[][]{Bailer-Jones21}. These Geo samples contain  sources with five- and six-parameter astrometric solutions, a quality index RUWE $<1.4$, and no duplicated sources. 
The full Geo sample includes 4\,654\,150 stars, of which only 130\,114 (3\%) are from six-parameter solutions. The {\it Medium} and High Quality  samples, Geo$_{\rm MQ}$ and Geo$_{\rm HQ}$, list 3\,469\,143 and 1\,894\,809 sources, respectively.
These Geo samples appear to be comparable to their corresponding astrometric control samples in terms of number of objects, spatial extension, and astrometric quality.

\begin{table*}
\caption{Statistics of the ACS and Geo samples.}          
\label{table:1new}      
\centering                       
\begin{tabular}{l c c c c c c}          
\hline\hline                     
   & ACS & ACS$_{\rm MQ}$ & ACS$_{\rm HQ}$ &  Geo & Geo$_{\rm MQ}$ & Geo$_{\rm HQ}$ \\     
\hline  
N. sources         & 4,397,814 & 3,362,012 & 1,815,072 & 4,654,150 & 3,469,143 & 1,894,809\\
Fraction of primaries &       53\% & 57\% & 53\% &  50\% & 55\% & 51\% \\

 distance (median) & 1.3 kpc & 1.3 kpc & 1.5 kpc &   1.3 kpc  & 1.2 kpc & 1.5 kpc \\
 distance (90$^{\rm th}$ quantile)  &  3.7 kpc & 2.9 kpc & 3.4 kpc & 4.1 kpc & 2.9 kpc & 3.4 kpc \\
$\sigma_\varpi/\varpi$ (median)  & 1.9\% & 1.8\% & 2.3\% & 1.9\% & 1.8\% & 2.2\%  \\
$\sigma_\varpi/\varpi$ (90$^{\rm th}$ quantile)   & 6.9\%  &  4.7\% & 6.0\% & 8.3\% & 4.7\% &  6.1\% \\
\hline
\end{tabular}
\label{table:ACSGeoStatistics}
\end{table*}


In order to compare trigonometric and Geo distances, we consider the fractional residuals, 
$(r_{\rm Geo} - r_{\varpi})/r_\varpi$, for the common objects. 
Median residuals over the whole distance range are $0.04$\% for both the ACS and ACS$_{\rm MQ}$, while for the high-quality sample, ACS$_{\rm HQ}$, the median residual with respect to Geo$_{\rm HQ}$ is $0.07$\%. The median absolute deviation is 0.2\% for the three data sets.

\begin{figure*}
\includegraphics[width=0.5\textwidth]{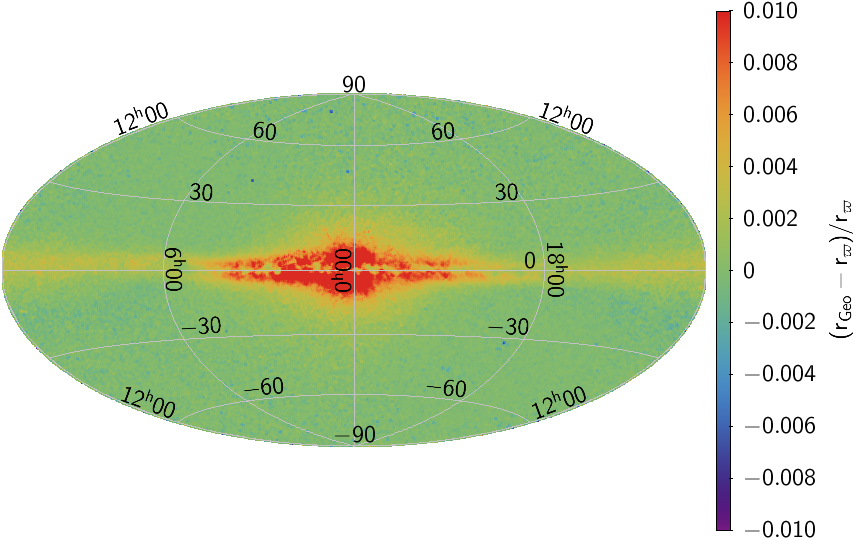}
\includegraphics[width=0.5\textwidth]{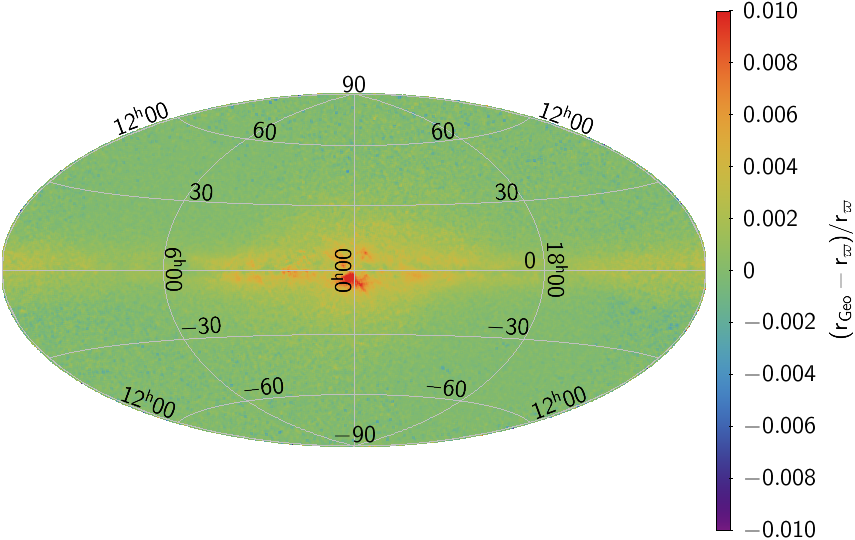}
\caption{Galactic map of the fractional residuals between trigonometric and Bayesian distances. {\it Left panel:} full ACS. {\it Right panel:} the same but for the Medium Quality  chemical abundances, ACS$_{\rm MQ}$, subsample.}
\label{ACSfig1}
\end{figure*}

The Galactic distribution of the mean fractional residuals is shown in Figure~\ref{ACSfig1} for the ACS and ACS$_{\rm MQ}$ on the left- and right-hand panel, respectively. The predominance of the colour green, with only traces of blue, indicates general good agreement ($\lesssim$~0.2\%) between the two distance estimates away from the Galactic plane. The yellow-orange-red colour of the in-plane regions
signifies that Bayesian and Geo distances are $\lesssim$ 4\% larger than trigonometric parallax distances.

\begin{figure*}
\includegraphics[width=0.49\textwidth]{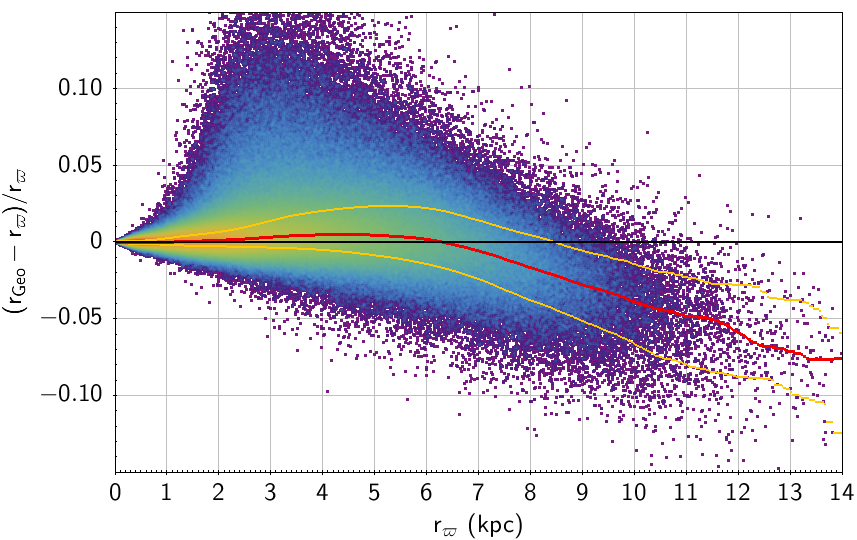}
\includegraphics[width=0.49\textwidth]{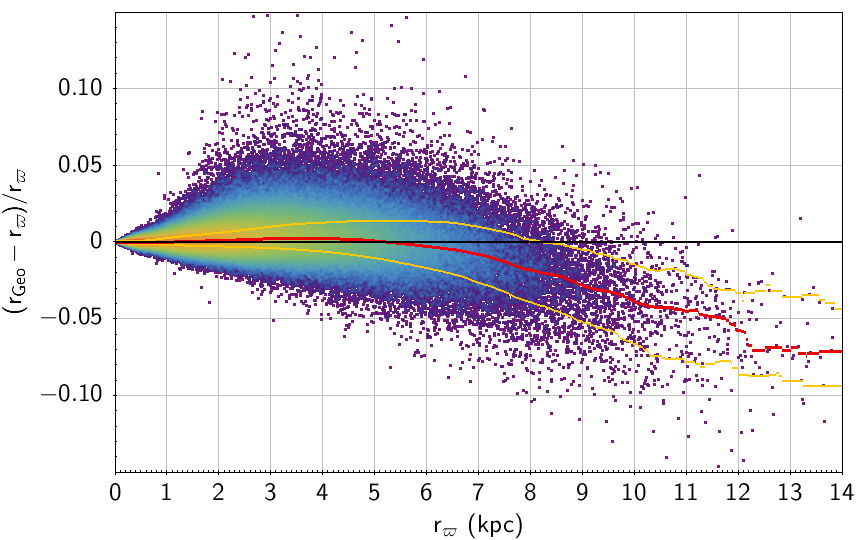}
\caption{Fractional residuals as a function of trigonometric distances ($r_\varpi$). The red line shows the median of the residuals, while the yellow curves trace the 16\% and 84\% quantiles. 
{\it Left panel:} Full ACS. {\it Right panel:}  ACS$_{\rm MQ}$.}
\label{ACSfig2}
\end{figure*}

Figure \ref{ACSfig2} shows the fractional residuals  as a function of $r_\varpi$. The median is shown by the red line, while the yellow curves trace the 16\% and 84\% quantiles.
The left panel shows the full ACS extending up to 10 kpc.
At larger distances ($>6.5$ kpc), trigonometric distances are larger than the Geo ones;
conversely, below 6.5 kpc, $r_\varpi$ is slightly smaller, that is by $\sim 1$\%. This systematic residual is produced by stars close to the Galactic plane (cf.\ Fig.\ \ref{ACSfig1}).
Although better agreement is found between the cross-matched ACS$_{\rm MQ}$ and the Geo$_{\rm MQ}$ (right panels of Figs.\ \ref{ACSfig1}-\ref{ACSfig2}), similar systematic effects, such as a function of distance and Galactic latitude, are still present. These relatively small differences are consistent with the intrinsic biases expected in the trigonometric distances as a function of the parallax relative error \citep[see e.g. ][]{Kovalevsky1998}, 
and the interplay with analogous effects in the Geo distances.

As for possible consequences of those differences, we recall that the chemo-kinematical distributions discussed in Sect.~\ref{sec:5} were scrutinised using both
$r_\varpi$ and $r_{\rm Geo}$ distances and were found to be statistically indistinguishable. Therefore,  
the present comparison allows us to conclude that the above-mentioned biases do not affect the different analysis of this paper nor its conclusions.

\section{Table queries of data samples}
\label{Appendix:queries}

The following ADQL queries to the $Astrophysical\_parameters$ table can be used to retrieve the data samples employed in this work. We first present the definition of the different working samples and then provide the ADQL queries at the end of this Appendix. 
\smallskip

\noindent General sample (5,527,090 stars, Sects.~\ref{Sec:GlobalView} and \ref{Sec:Dynamics}): 

\noindent \begin{verbatim} 
From the complete \gspspec\ sample, we have removed: 
 (i) very cool dwarfs defined by (\T<3800K \& \g>3.5)
 (ii) the {\it KM-extended flag}-stars defined by 
(\T<4150K \& 2.4<\g<3.6).
\end{verbatim}
From this query, objects without $r_{\rm Geo}$ distances were also removed.

\bigskip
\noindent{\it Medium quality sample} (4,140,759 stars, Sects.~\ref{Sec:GlobalView}, ~\ref{sec:5}, and ~\ref{Sec:Dynamics}): 

\noindent \begin{verbatim}
General sample & Teff>3500 & 0<log(g)<5 &
Teff_unc<375 & log(g)_unc<0.5 & [M/H]_unc<0.25 &
vbroadT<2 & vbroadG<2 & vbroadM<2 &
vradT<2 & vradG<2 & vradM<2 &
fluxNoise<4 & extrapol<3 & KMtypestars<2
\end{verbatim}

\bigskip
\noindent Gradient analysis sample (2,762,809 stars, Sect.~\ref{Sec:Gradients}): 

\begin{verbatim}
vbroadT<2 & vbroadG<2 & vbroadM<2 &
vradT<2 & vradG<2 & vradM<2 &
fluxNoise<2 & extrapol=0 & KMtypestars=0 &
NegFlux=0 & nanFlux=0 & emission=0 &
nullFluxErr=0 & RUWE<=1.4 & Teff>4000
\end{verbatim}

\medskip
\noindent High Quality sample (2,218,573 stars, Sect.~\ref{sec:5}): 

\noindent 
\begin{verbatim}
Teff>3500 & 0<log(g)<5 AND
vbroadT=0 & vbroadG=0 & vbroadM=0 AND
vradT=0 & vradG=0 & vradM=0 AND 
fluxNoise=0 & extrapol<3 & KMtypestars<2 AND
NegFlux=0 & nanFlux=0 & emission=0 & nullFluxErr=0 
\end{verbatim}

\medskip
\noindent Individual abundances samples (Sects.~\ref{Sec:GlobalView} and \ref{Sec:SolarCylinder}):

\noindent
\begin{verbatim}
High Quality & Teff<7000 AND 
[X/Fe]_unc<0.2 & XUpLim<2 & XUncer=0 AND
IF [Ca/Fe]: 
cafe_gspspec_nlines>=3 AND
cafe_gspspec_linescatter<0.1 AND
(logg_gspspec < 3.5) AND
((teff_gspspec != 4250) | (logg_gspspec != 1.5))
IF [N/Fe]: 
nfe_gspspec_nlines>=2 AND
nfe_gspspec_linescatter<0.1 AND
[N/Fe]_unc<0.15 
IF [Cr/Fe]: teff_gspspec > 4100 & logg_gspspec < 2.5
IF [Ni/Fe]: nife_gspspec_unc < 0.1 & logg_gspspec < 3
IF [Ce/Fe]: CeUpLim<=2 AND
logg_gspspec >= 1.21 & logg_gspspec <= 2.46 AND
((teff_gspspec != 4250 | logg_gspspec != 1.5))
\end{verbatim}

\textbf{\texttt{ADQL} queries for the working samples from the \Gaia\ Archive}

\lstset{language=SQL}

\begin{lstlisting}[caption={\texttt{ADQL} query for the \textit{General sample}.},captionpos=b]
SELECT source_id
FROM user_dr3int6.astrophysical_parameters
WHERE ((teff_gspspec>=3800) OR (logg_gspspec>=3.5)) AND ((teff_gspspec>=4150) OR (logg_gspspec>=3.6) OR (logg_gspspec<=2.4))
\end{lstlisting}

\begin{lstlisting}[caption={\texttt{ADQL} query for the \textit{Medium quality sample}.},captionpos=b]
SELECT source_id 
FROM user_dr3int6.astrophysical_parameters 
WHERE  (teff_gspspec>3500) AND (logg_gspspec>0) AND (logg_gspspec<5) AND ((teff_gspspec_upper-teff_gspspec_lower)<750) AND ((logg_gspspec_upper-logg_gspspec_lower)<1.) AND ((mh_gspspec_upper-mh_gspspec_lower)<.5) AND (teff_gspspec>=3800 OR logg_gspspec<=3.5) AND (teff_gspspec>=4150 OR logg_gspspec<=2.4 OR logg_gspspec>=3.6 ) AND  ((flags_gspspec LIKE "____________0%") OR (flags_gspspec LIKE "____________1%")) AND ((flags_gspspec LIKE "0%") OR (flags_gspspec LIKE "1%")) AND  ((flags_gspspec LIKE "_0%") OR (flags_gspspec LIKE "_1%")) AND  ((flags_gspspec LIKE "__0%") OR (flags_gspspec LIKE "__1%")) AND  ((flags_gspspec LIKE "___0%") OR (flags_gspspec LIKE "___1%")) AND  ((flags_gspspec LIKE "____0%") OR (flags_gspspec LIKE "____1%")) AND  ((flags_gspspec LIKE "_____0%") OR (flags_gspspec LIKE "_____1%")) AND  ((flags_gspspec LIKE "______0%") OR (flags_gspspec LIKE "______1%") OR (flags_gspspec LIKE "______2%") OR (flags_gspspec LIKE "______3%")) AND  ((flags_gspspec LIKE "_______0%") OR (flags_gspspec LIKE "_______1%") OR (flags_gspspec LIKE "_______2%"))
\end{lstlisting}

\begin{lstlisting}[caption={\texttt{ADQL} query for the \textit{Gradient analysis sample}.},captionpos=b]
SELECT gaia.source_id
FROM  user_dr3int6.gaia_source as gaia INNER JOIN (
 SELECT source_id
 FROM user_dr3int6.astrophysical_parameters
 WHERE ((flags_gspspec LIKE "0%") OR (flags_gspspec LIKE "1%")) AND  ((flags_gspspec LIKE "_0%") OR (flags_gspspec LIKE "_1%")) AND  ((flags_gspspec LIKE "__0%") OR (flags_gspspec LIKE "__1%")) AND  ((flags_gspspec LIKE "___0%") OR (flags_gspspec LIKE "___1%")) AND  ((flags_gspspec LIKE "____0%") OR (flags_gspspec LIKE "____1%")) AND  ((flags_gspspec LIKE "_____0%") OR (flags_gspspec LIKE "_____1%")) AND ((flags_gspspec LIKE "______0%") OR (flags_gspspec LIKE "______1%")) AND  (flags_gspspec LIKE "_______0%") AND  (flags_gspspec LIKE "________0%") AND  (flags_gspspec LIKE "_________0%") AND  (flags_gspspec LIKE "__________0%") AND  (flags_gspspec LIKE "___________0%") AND (flags_gspspec LIKE "____________0%") AND (teff_gspspec>4000) ) as aux USING (source_id) 
WHERE(gaia.ruwe<=1.4)
\end{lstlisting}

\begin{lstlisting}[caption={\texttt{ADQL} query for the \textit{High quality sample}.},captionpos=b]
SELECT source_id 
FROM user_dr3int6.astrophysical_parameters 
WHERE (teff_gspspec>3500) AND (logg_gspspec>0 AND logg_gspspec<5) AND (flags_gspspec LIKE "0%") AND (flags_gspspec LIKE "_0%") AND (flags_gspspec LIKE "__0%") AND (flags_gspspec LIKE "___0%") AND (flags_gspspec LIKE "____0%") AND (flags_gspspec LIKE "_____0%") AND (flags_gspspec LIKE "______0%") AND ((flags_gspspec LIKE "_______0%") OR (flags_gspspec LIKE "_______1%") OR (flags_gspspec LIKE "_______2%") )  AND ((flags_gspspec LIKE "____________0%") OR (flags_gspspec LIKE "____________1%") ) AND (flags_gspspec LIKE "________0%") AND (flags_gspspec LIKE "_________0%") AND (flags_gspspec LIKE "__________0%") AND (flags_gspspec LIKE "___________0%")
\end{lstlisting}

\begin{lstlisting}[caption={\texttt{ADQL} query for the \textit{Calcium abundance sample}.},captionpos=b]
SELECT source_id 
FROM user_dr3int6.astrophysical_parameters 
WHERE (teff_gspspec>3500 AND teff_gspspec<7000) AND (logg_gspspec>0 AND logg_gspspec<5) AND (flags_gspspec LIKE "0%") AND (flags_gspspec LIKE "_0%") AND (flags_gspspec LIKE "__0%") AND (flags_gspspec LIKE "___0%") AND (flags_gspspec LIKE "____0%") AND (flags_gspspec LIKE "_____0%") AND (flags_gspspec LIKE "______0%") AND ((flags_gspspec LIKE "_______0%") OR (flags_gspspec LIKE "_______1%") OR (flags_gspspec LIKE "_______2%") )  AND ((flags_gspspec LIKE "____________0%") OR (flags_gspspec LIKE "____________1%") ) AND (flags_gspspec LIKE "________0%")  AND (flags_gspspec LIKE "_________0%")  AND (flags_gspspec LIKE "__________0%")  AND (flags_gspspec LIKE "___________0%")    AND ((cafe_gspspec_upper-cafe_gspspec_lower)<0.4)  AND ((flags_gspspec LIKE "_____________________0%") OR (flags_gspspec LIKE "_____________________1%")) AND (flags_gspspec LIKE "______________________0%")  AND (cafe_gspspec_nlines>=3) AND (cafe_gspspec_linescatter<0.1) AND (logg_gspspec<3.5) AND ((teff_gspspec!=4250) OR (logg_gspspec!=1.5))
\end{lstlisting}

\begin{lstlisting}[caption={\texttt{ADQL} query for the \textit{Nitrogen abundance sample}.},captionpos=b]
SELECT source_id 
FROM user_dr3int6.astrophysical_parameters 
WHERE (teff_gspspec>3500 AND teff_gspspec<7000) AND (logg_gspspec>0 AND logg_gspspec<5) AND (flags_gspspec LIKE "0%") AND (flags_gspspec LIKE "_0%") AND (flags_gspspec LIKE "__0%") AND (flags_gspspec LIKE "___0%") AND (flags_gspspec LIKE "____0%") AND (flags_gspspec LIKE "_____0%") AND (flags_gspspec LIKE "______0%") AND ((flags_gspspec LIKE "_______0%") OR (flags_gspspec LIKE "_______1%") OR (flags_gspspec LIKE "_______2%") )  AND ((flags_gspspec LIKE "____________0%") OR (flags_gspspec LIKE "____________1%") ) AND (flags_gspspec LIKE "________0%")  AND (flags_gspspec LIKE "_________0%")  AND (flags_gspspec LIKE "__________0%")  AND (flags_gspspec LIKE "___________0%")  AND ((nfe_gspspec_upper-nfe_gspspec_lower)<0.30)  AND ((flags_gspspec LIKE "_____________0%") OR (flags_gspspec LIKE "_____________1%")) AND (flags_gspspec LIKE "______________0%") AND (nfe_gspspec_nlines>=2) AND (nfe_gspspec_linescatter<0.1)  
\end{lstlisting}

\begin{lstlisting}[caption={\texttt{ADQL} query for the \textit{Chromium abundance sample}.},captionpos=b]
SELECT source_id 
FROM user_dr3int6.astrophysical_parameters 
WHERE (teff_gspspec>3500 AND teff_gspspec<7000) AND (logg_gspspec>0 AND logg_gspspec<5) AND (flags_gspspec LIKE "0%") AND (flags_gspspec LIKE "_0%") AND (flags_gspspec LIKE "__0%") AND (flags_gspspec LIKE "___0%") AND (flags_gspspec LIKE "____0%") AND (flags_gspspec LIKE "_____0%") AND (flags_gspspec LIKE "______0%") AND ((flags_gspspec LIKE "_______0%") OR (flags_gspspec LIKE "_______1%") OR (flags_gspspec LIKE "_______2%") ) AND ((flags_gspspec LIKE "____________0%") OR (flags_gspspec LIKE "____________1%") ) AND (flags_gspspec LIKE "________0%") AND (flags_gspspec LIKE "_________0%") AND (flags_gspspec LIKE "__________0%") AND (flags_gspspec LIKE "___________0%")   AND ((crfe_gspspec_upper-crfe_gspspec_lower)<0.4) AND ((flags_gspspec LIKE "_________________________0%") OR (flags_gspspec LIKE "_________________________1%")) AND (flags_gspspec LIKE "_________________________0%") AND ((teff_gspspec>4100) AND (logg_gspspec<2.5)) 
\end{lstlisting}

\begin{lstlisting}[caption={\texttt{ADQL} query for the \textit{Nickel abundance sample}.},captionpos=b]
SELECT source_id 
FROM user_dr3int6.astrophysical_parameters 
WHERE (teff_gspspec>3500 AND teff_gspspec<7000) AND (logg_gspspec>0 AND logg_gspspec<5) AND (flags_gspspec LIKE "0%") AND (flags_gspspec LIKE "_0%") AND (flags_gspspec LIKE "__0%") AND (flags_gspspec LIKE "___0%") AND (flags_gspspec LIKE "____0%") AND (flags_gspspec LIKE "_____0%") AND (flags_gspspec LIKE "______0%") AND ((flags_gspspec LIKE "_______0%") OR (flags_gspspec LIKE "_______1%") OR (flags_gspspec LIKE "_______2%") ) AND ((flags_gspspec LIKE "____________0%") OR (flags_gspspec LIKE "____________1%") ) AND (flags_gspspec LIKE "________0%") AND (flags_gspspec LIKE "_________0%") AND (flags_gspspec LIKE "__________0%") AND (flags_gspspec LIKE "___________0%")  AND ((nife_gspspec_upper-nife_gspspec_lower)<0.2) AND ((flags_gspspec LIKE "%0_________") OR (flags_gspspec LIKE "%1_________")) AND (flags_gspspec LIKE "%0________") AND (logg_gspspec<3)
\end{lstlisting}

\begin{lstlisting}[caption={\texttt{ADQL} query for the \textit{Cerium abundance sample}.},captionpos=b]
SELECT source_id 
FROM user_dr3int6.astrophysical_parameters 
WHERE (teff_gspspec>3500 AND teff_gspspec<7000) AND (logg_gspspec>0 AND logg_gspspec<5) AND (flags_gspspec LIKE "0%") AND (flags_gspspec LIKE "_0%") AND (flags_gspspec LIKE "__0%") AND (flags_gspspec LIKE "___0%") AND (flags_gspspec LIKE "____0%") AND (flags_gspspec LIKE "_____0%") AND (flags_gspspec LIKE "______0%") AND ((flags_gspspec LIKE "_______0%") OR (flags_gspspec LIKE "_______1%") OR (flags_gspspec LIKE "_______2%") )  AND ((flags_gspspec LIKE "____________0%") OR (flags_gspspec LIKE "____________1%") ) AND (flags_gspspec LIKE "________0%") AND (flags_gspspec LIKE "_________0%") AND (flags_gspspec LIKE "__________0%") AND (flags_gspspec LIKE "___________0%")  AND ((cefe_gspspec_upper-cefe_gspspec_lower)<0.4) AND ((flags_gspspec LIKE "%0_____") OR (flags_gspspec LIKE "%1_____") OR (flags_gspspec LIKE "%2_____")) AND (flags_gspspec LIKE "%0____")  AND ((logg_gspspec>=1.21) AND (logg_gspspec<=2.46)) AND ((teff_gspspec!=4250) OR (logg_gspspec!=1.5))
\end{lstlisting}

\section{The Milky Way in \alphaFe\ for different distance intervals}
Here, we provide a complementary figure to Fig.~\ref{Fig.lb2} in Sect. \ref{Sec:GlobalView},  showing the \alphaFe\ content of the Milky Way, cut in various distance intervals.
\begin{figure*}[h]
\includegraphics[width=0.49\textwidth]{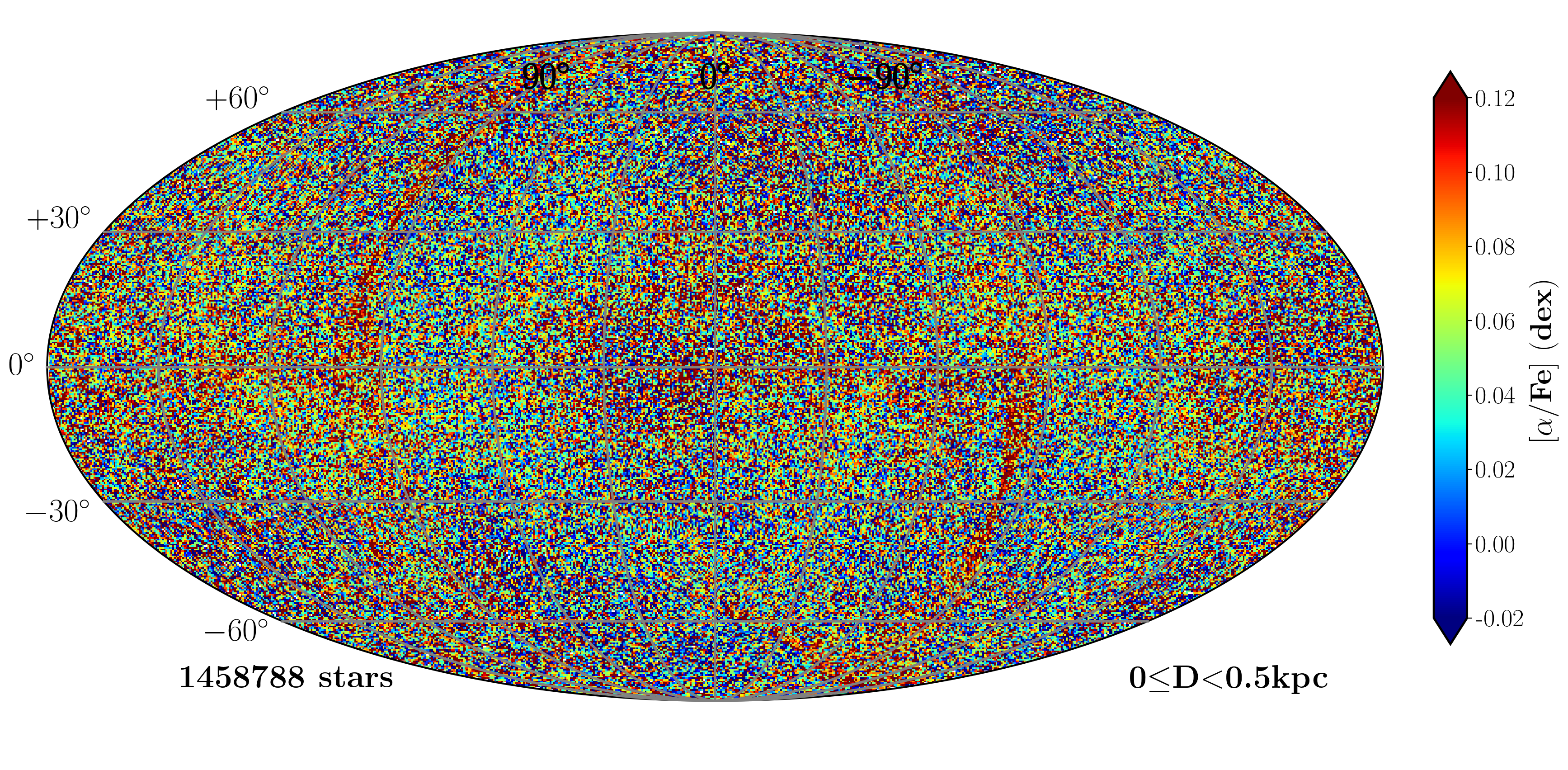}
\includegraphics[width=0.49\textwidth]{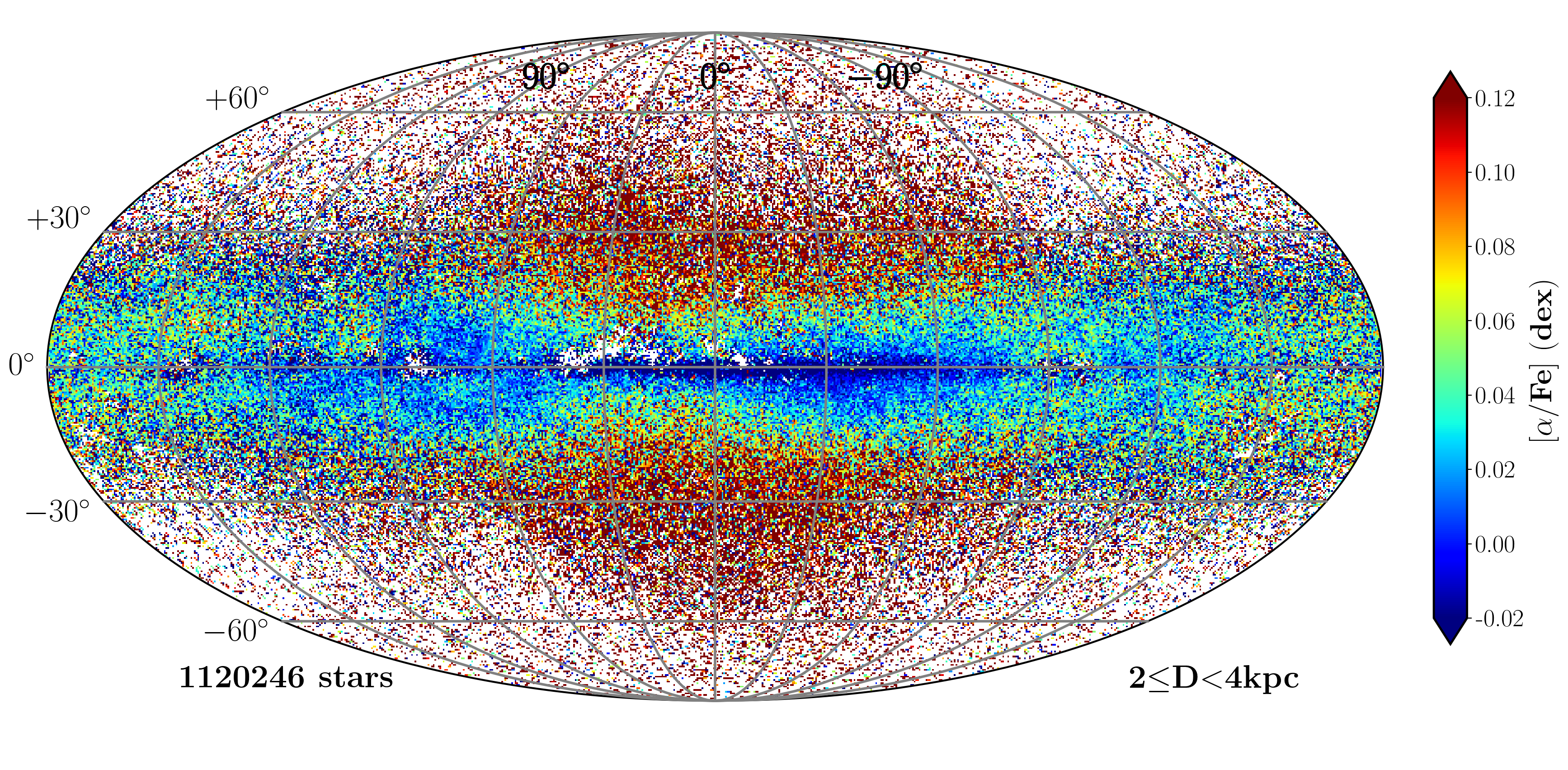}
\includegraphics[width=0.49\textwidth]{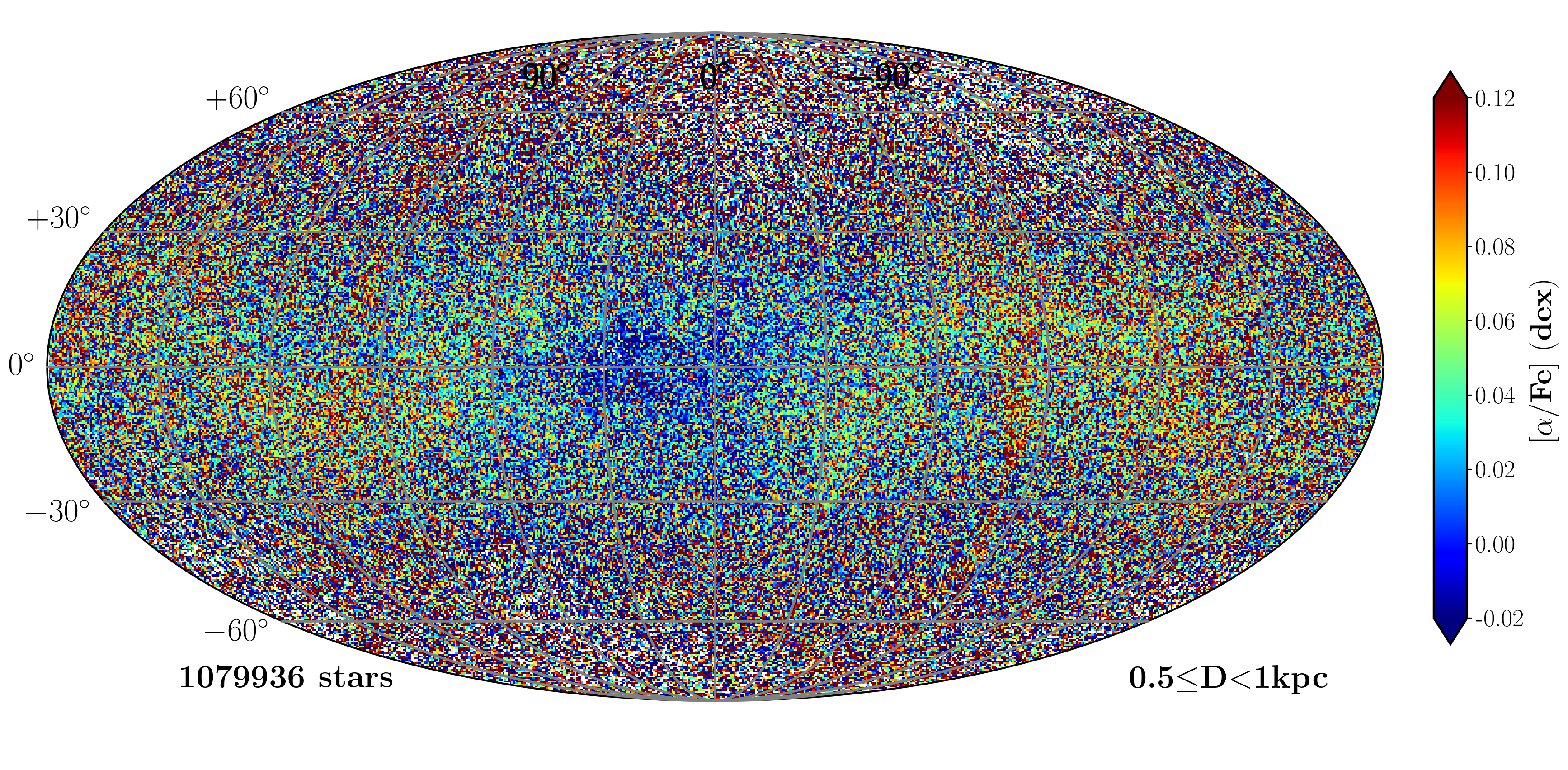}
\includegraphics[width=0.49\textwidth]{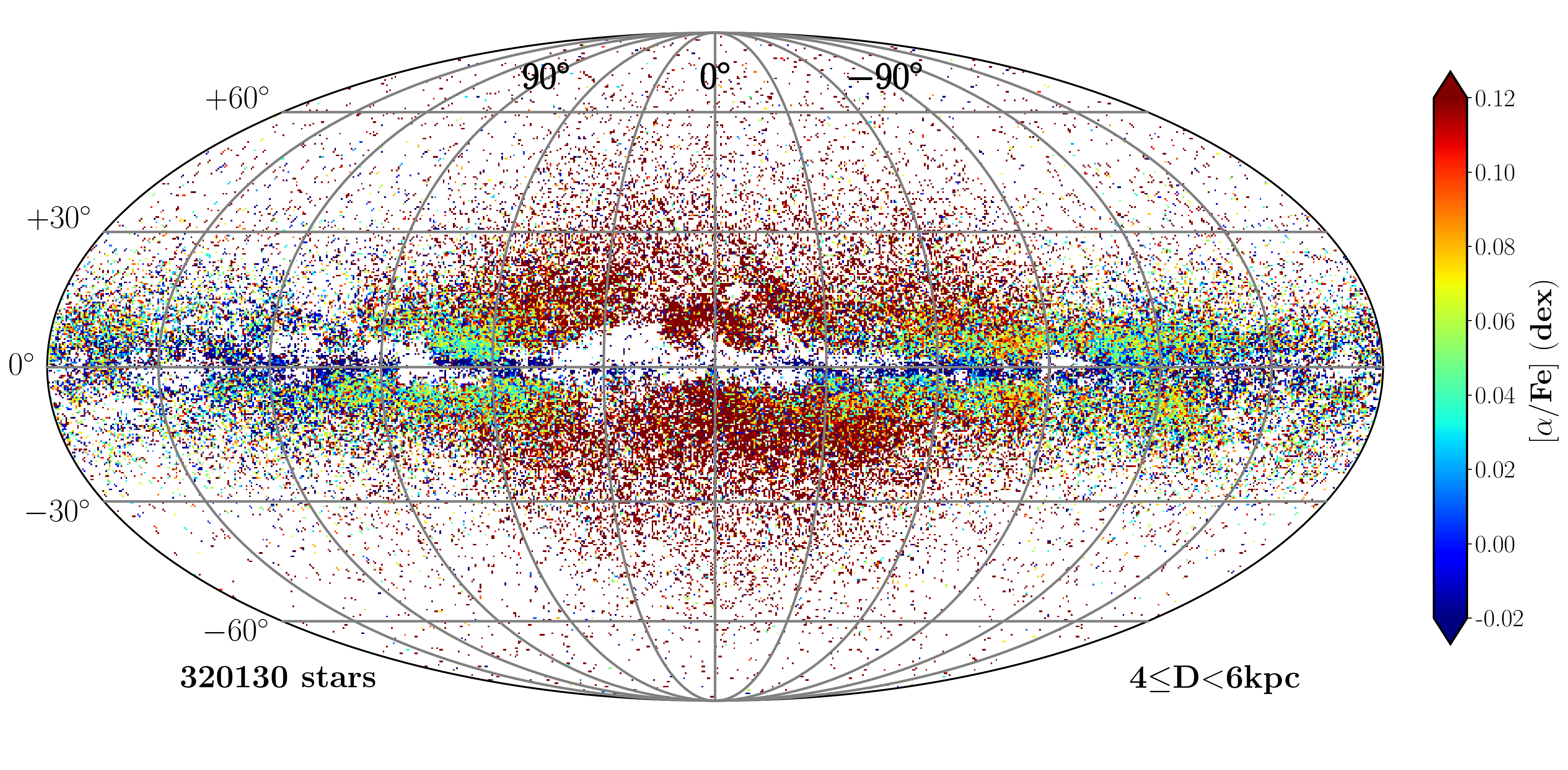}
\includegraphics[width=0.49\textwidth]{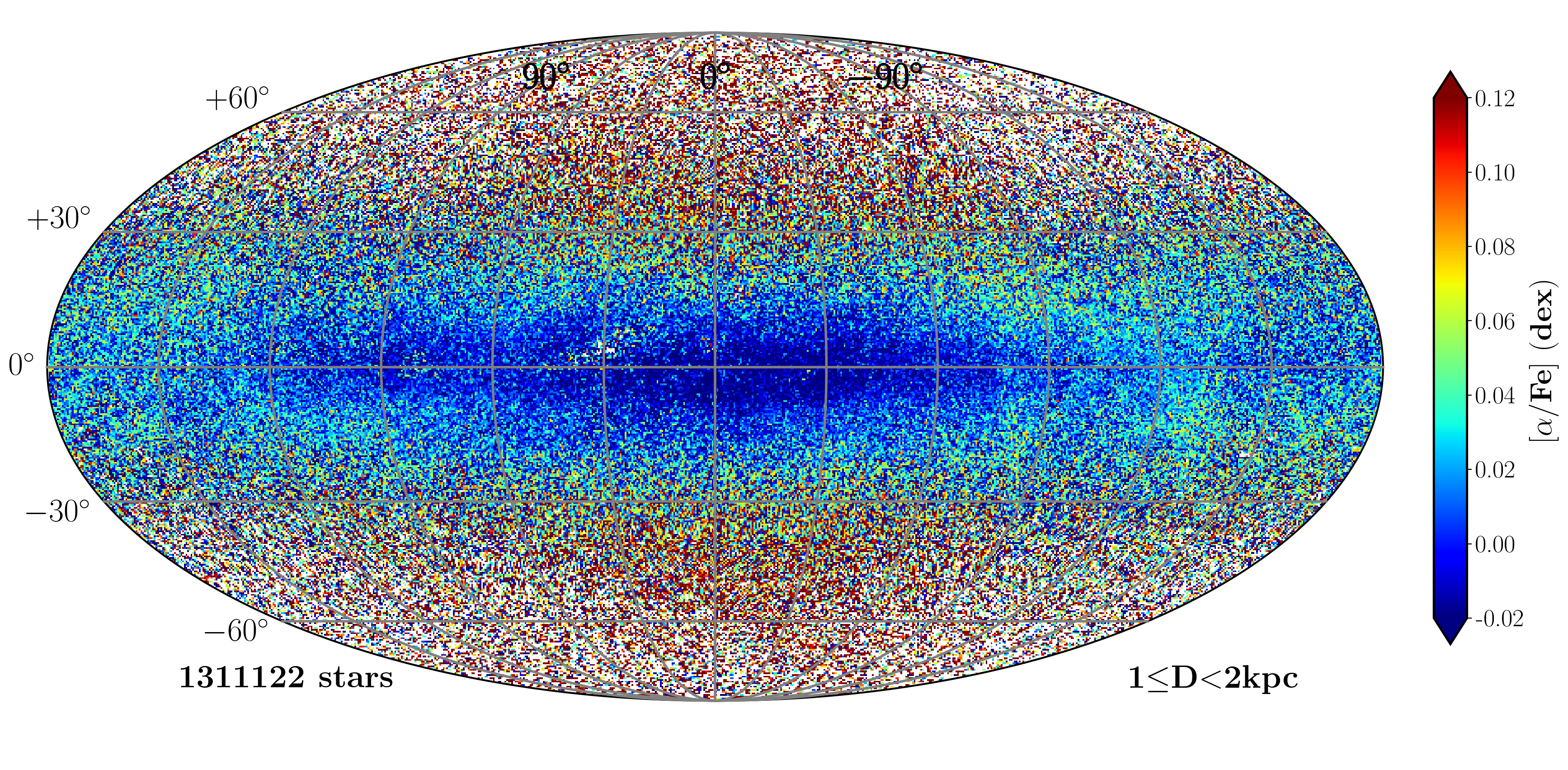} 
\hspace{0.2cm}
\includegraphics[width=0.49\textwidth]{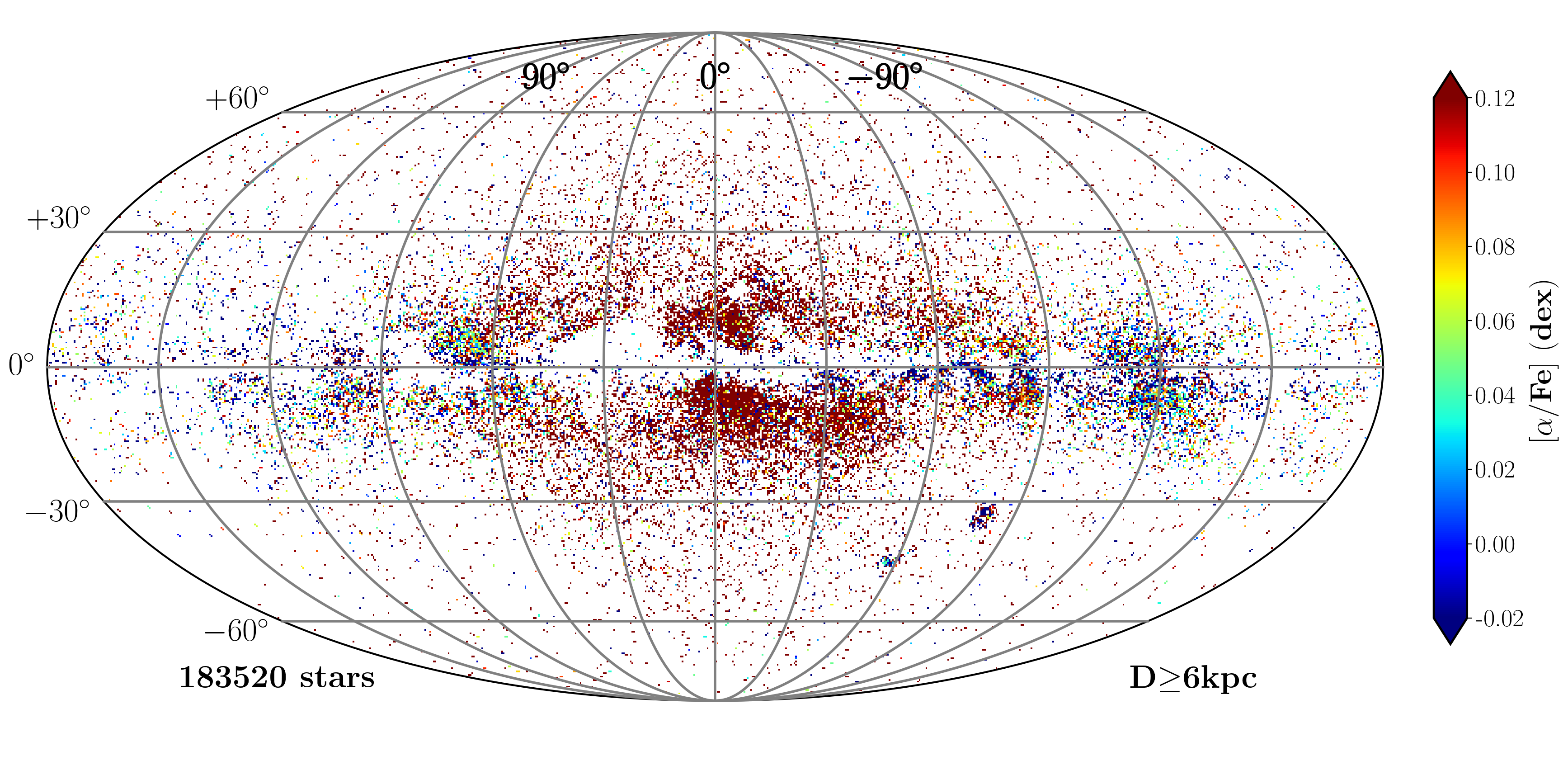}
\caption{Same as Fig.~\ref{Fig.lb2} but for \alphaFe .}
\label{Fig.lb3}
\end{figure*}

\section{Impact of data calibrations on \alphaFe\ versus \meta\ trends}
\label{Sec:calibsTests}

\begin{figure*}
\begin{tabular}{cc}
\includegraphics[width=0.49\textwidth]{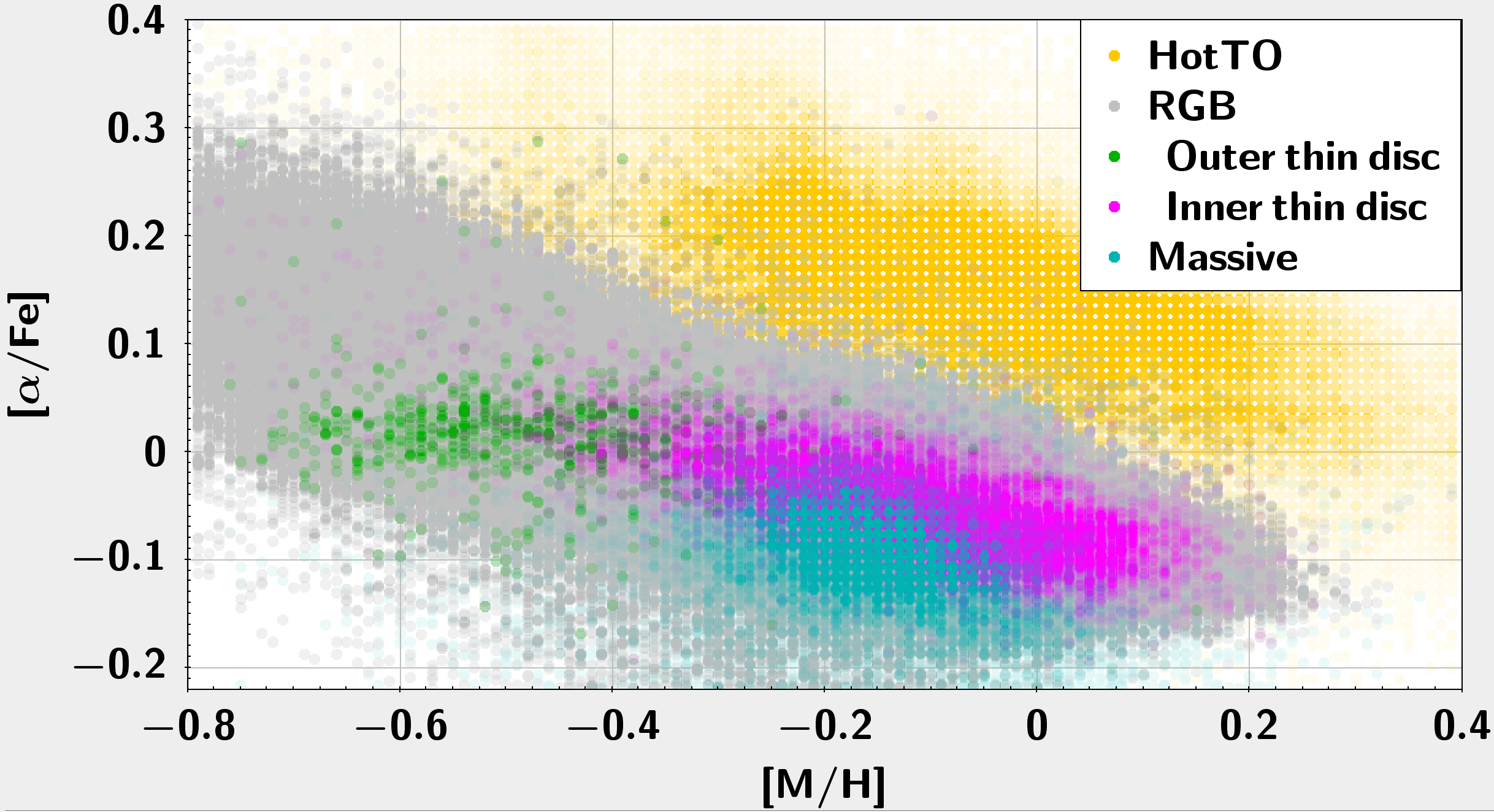} & \\
\includegraphics[width=0.49\textwidth]{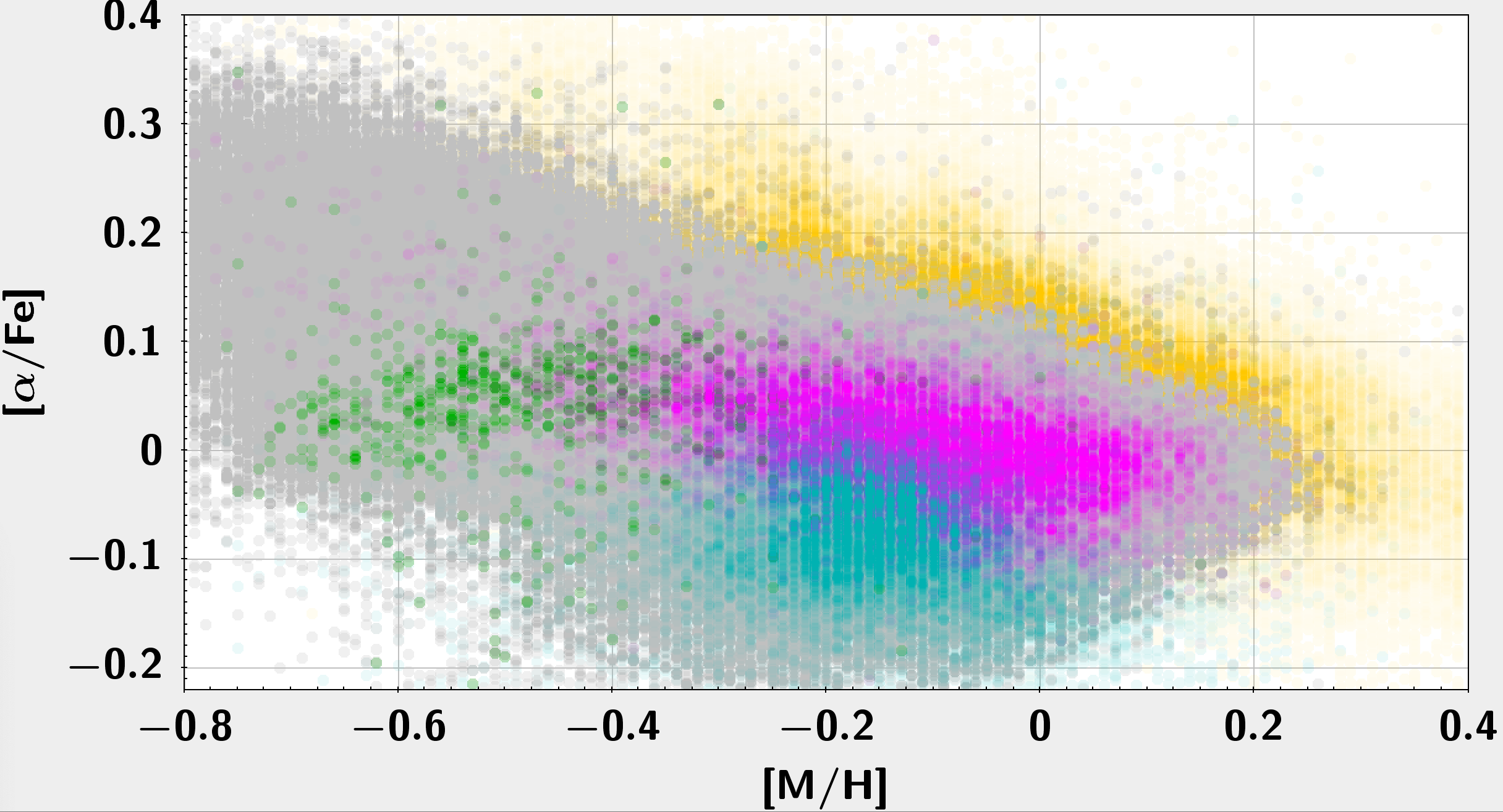}& \includegraphics[width=0.49\textwidth]{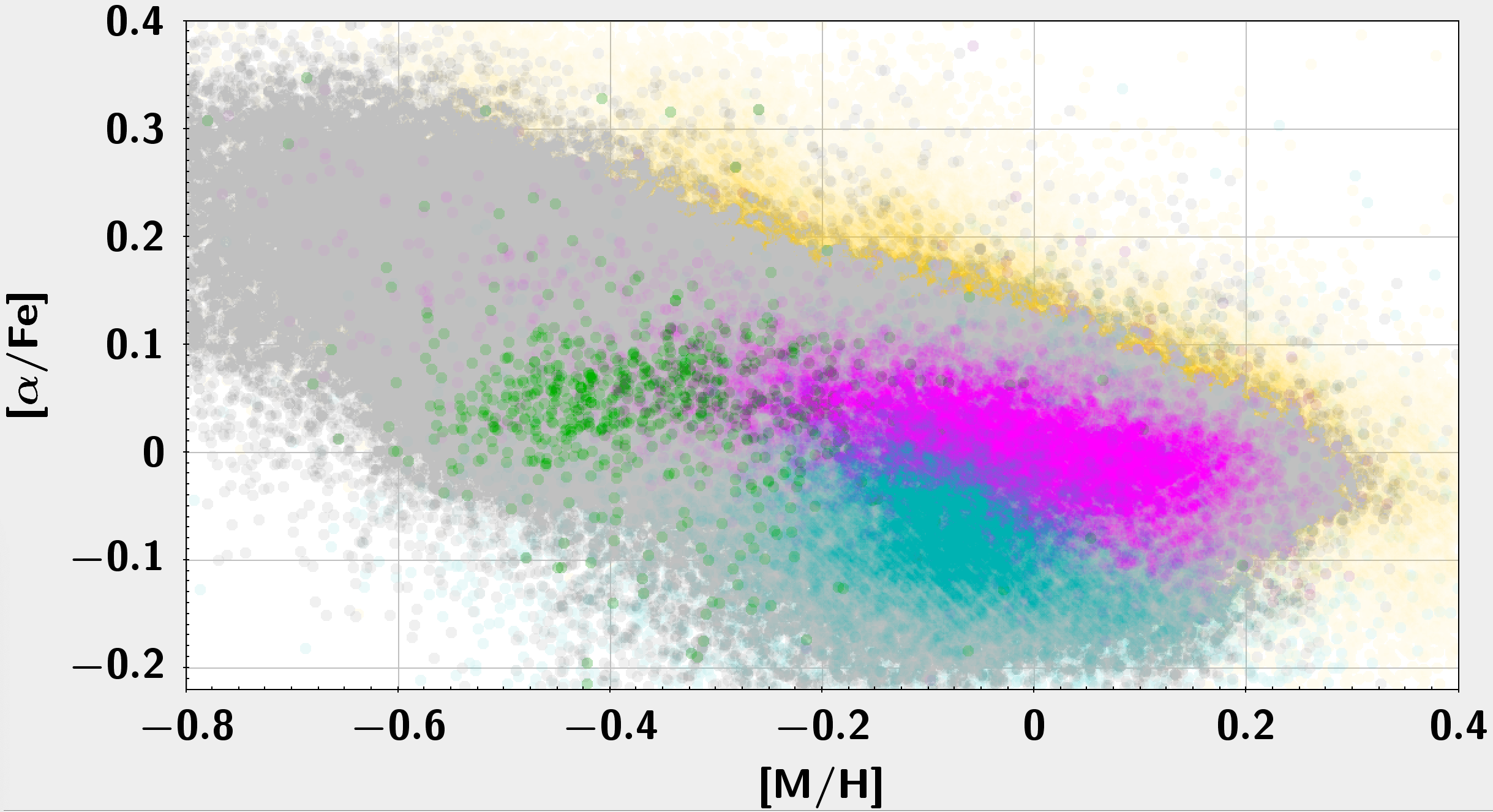}\\
\includegraphics[width=0.49\textwidth]{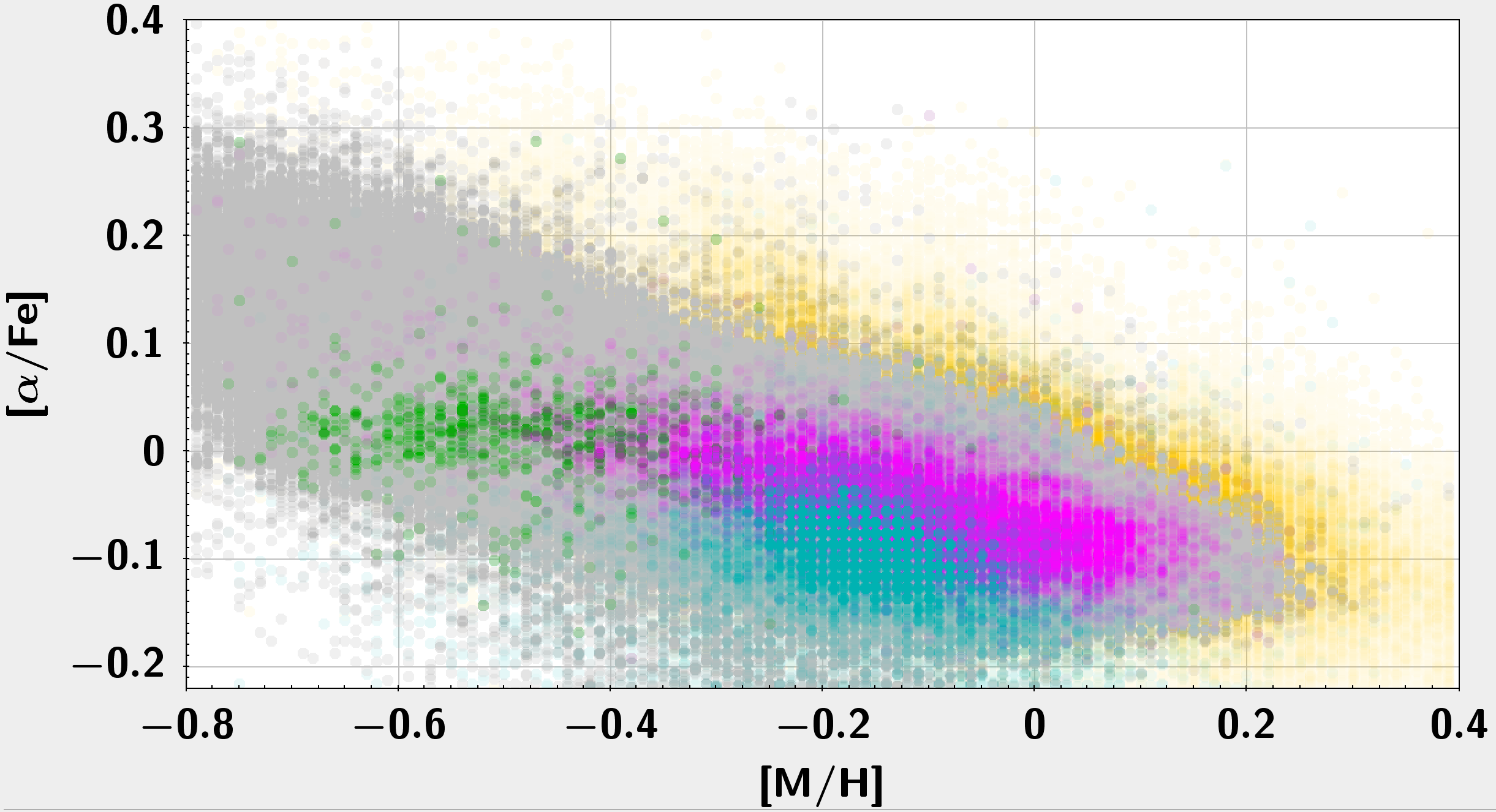} &\includegraphics[width=0.49\textwidth]{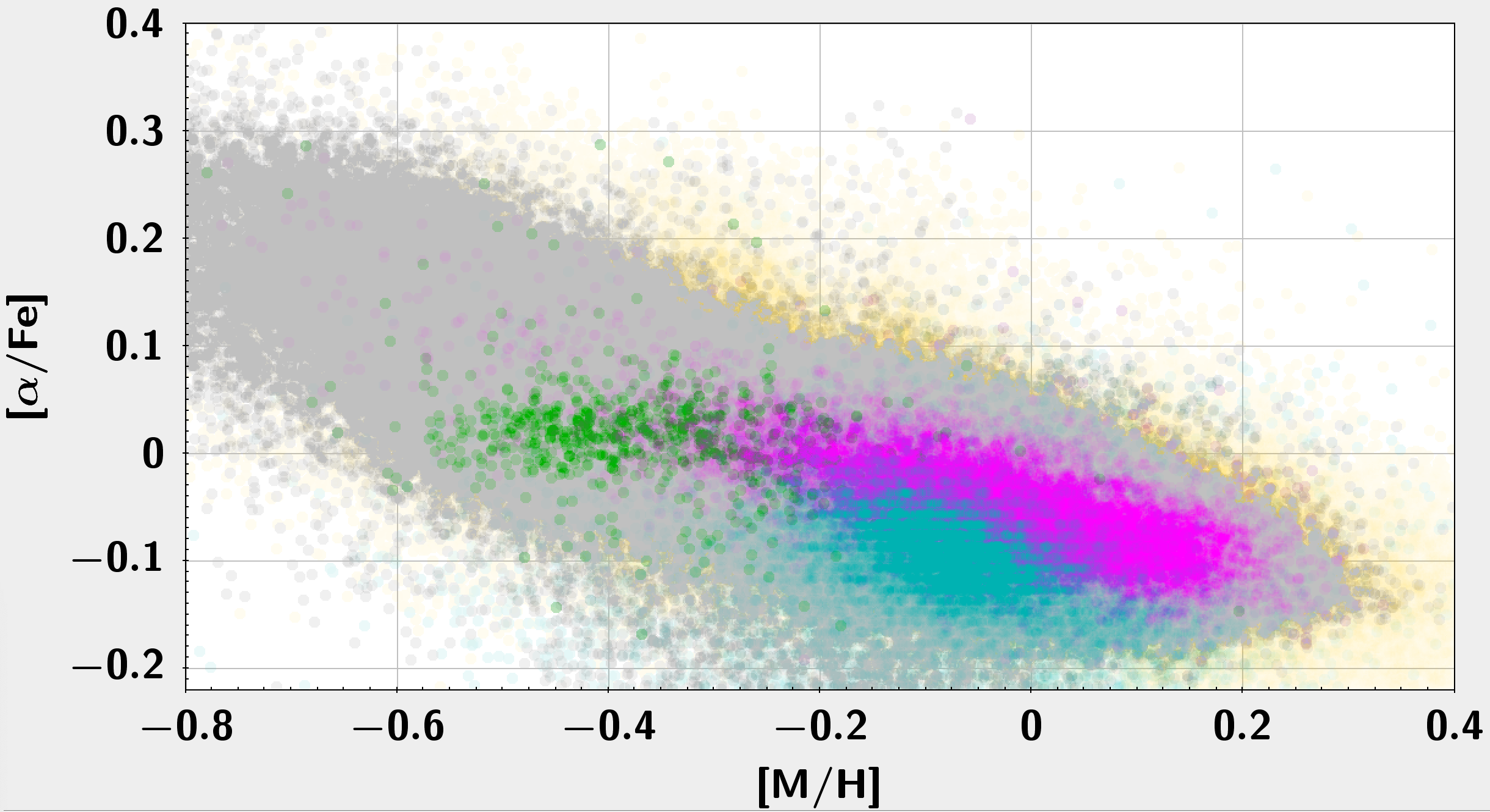}\\
\end{tabular}
\caption{Illustration of the impact of different calibrations on the observed \alphaFe\ vs. \meta\ trends for the samples defined in Sect.~\ref{Sec:selFunc}: HotTO (orange), RGB (grey, with the inner and outer thin disc in pink and green, respectively), and Massive (turquoise) samples. Upper panel: Raw data (no calibration). Left middle panel: Calibrated \alphaFe\ based on \g\ and no \meta\ calibration. This is the solution used throughout the different sections of this article. Right middle panel:  Calibrated \alphaFe\ and \meta\ based on \g. Left lower panel: Calibrated \alphaFe\ based on \T\ and no \meta\ calibration. Right left panel: calibrated \alphaFe\ and \meta\ based on \T.}
\label{Fig:TestsCalibs}
\end{figure*}

Discussing the different abundance patterns observed in the investigated Galactic stellar populations as presented in Sect.\ref{Sec:selFunc} requires knowledge of the used abundance calibrations and their relative impact on the studied samples. This is presented in Fig.~\ref{Fig:TestsCalibs} for the observed distributions in the \alphaFe\ versus \meta\ plane for (i) the raw uncalibrated data (upper panel), (ii) an \alphaFe\ calibration based on \g\ (left middle panel) or \T (left lower panel), and (iii) these two \alphaFe\ calibration flavours combined with a \meta\ calibration based on \g\ (middle and lower right panels). The calibration polynomial are taken from Tables~3 and 4 in \cite{GSPspecDR3}. In addition, the figure allows to compare the three samples defined in  \ref{Sec:selFunc} (HotTO in orange,  RGB in grey and Massive in turquoise) and how their relative positions in the \alphaFe\ versus \meta\ plane depend on the calibrations. To help this comparison, the locus of the thin-disc sequence in the RGB sample is highlighted, distinguishing an outer (green points) and an inner (pink) thin disc, as selected through the stellar orbit apocentres ($<$7.5~kpc and $>$12.5~kpc for the inner and outer disc respectively) and Zmax values ($<$0.5~kpc).

Figure~\ref{Fig:TestsCalibs} shows that the Massive sample is, in all the combinations of calibration and no calibration flavours, impoverished in \meta\ with respect to the thin disc sequence. In addition,
the maximum metallicity reached by the Massive sample is also lower than that of the RGB and HotTO samples, independently of the chosen calibration procedure. 

\section{Disc flaring}
\label{Sect:Flare}
\begin{figure*}
\includegraphics[width=0.49\textwidth]{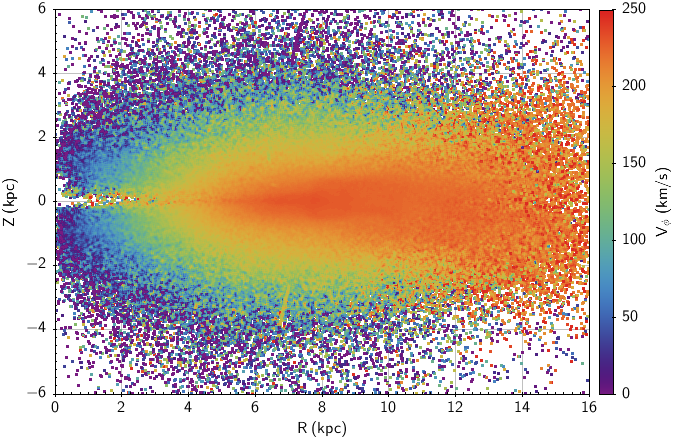}
\includegraphics[width=0.49\textwidth]{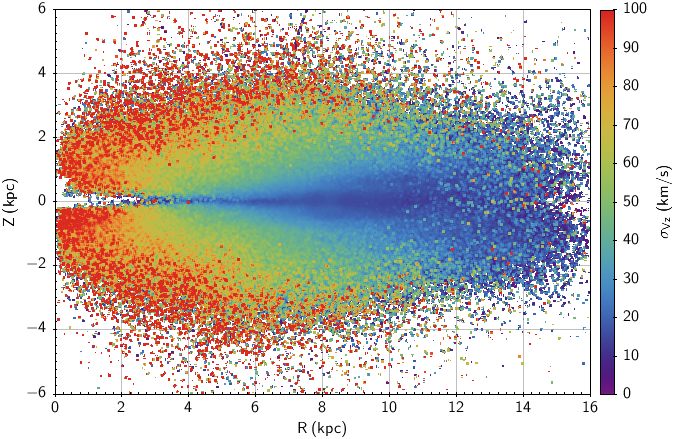}
\caption{Galactic distribution $(R,Z)$ of 4\,682\,659 stars of the General sample with good astrometric parameters, i.e. not duplicated sources with RUWE $ < 1.4$.  Sources are colour coded according to the mean $V_\phi$ (left panel) and $\sigma_{Vz}$ (right panel).}
\label{Figure_flaring}
\end{figure*}

Figure \ref{Figure_flaring} presents a kinematical evidence of the disc flaring discussed in Sect. \ref{Sect:Kinematics-Toomre}.

\section{Chemo-kinematical asymmetries in the outer disc}
\label{Appendix:OuterDisc}
\begin{figure*}
\begin{center}
\includegraphics[width=0.75\textwidth]{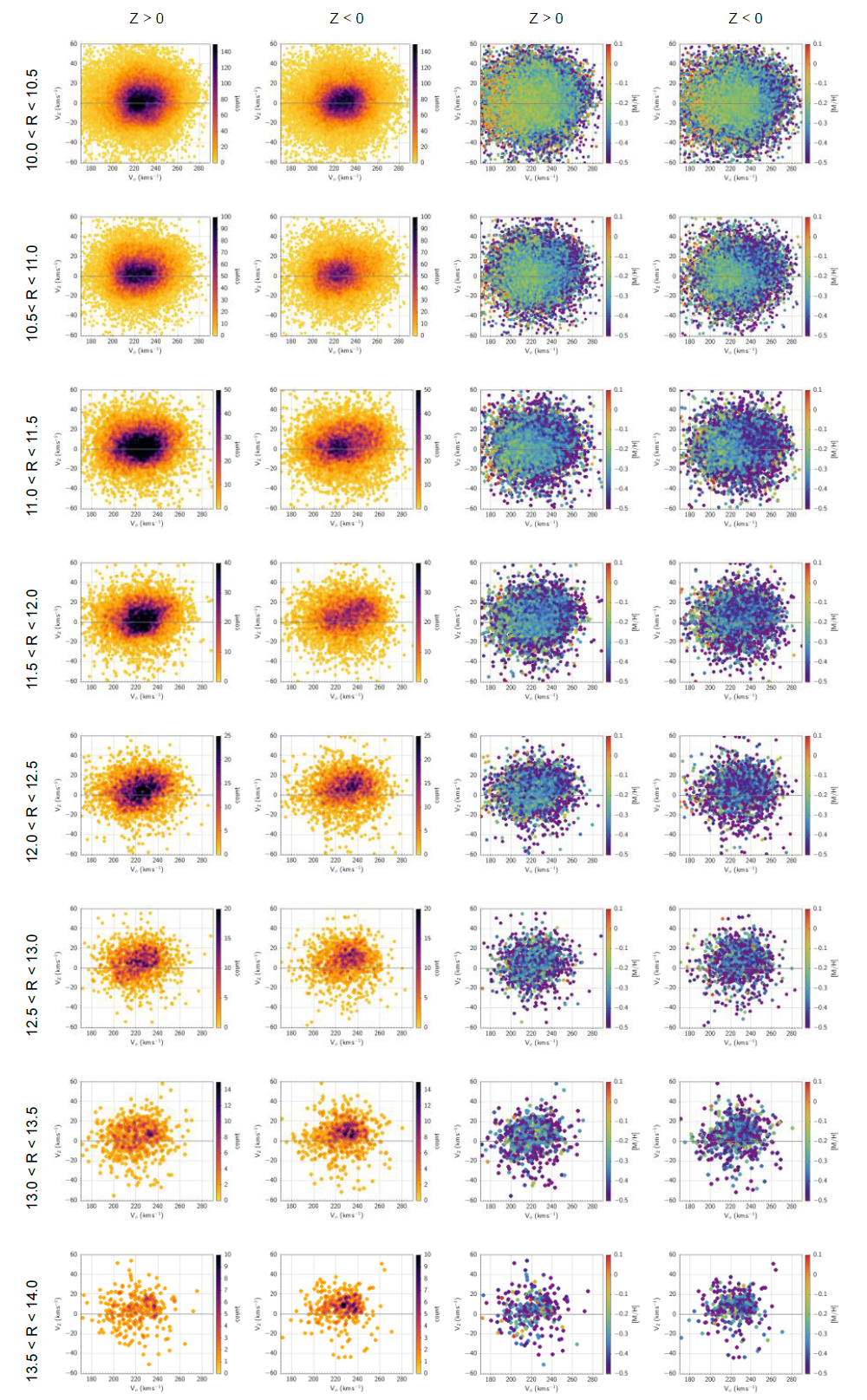}
\end{center}
\caption{Velocity distribution $V_Z$ vs.\ $V_\phi$ for the Medium Quality stars with good astrometric parameters, $10<R<14$ kpc and $|Z|< 1$ kpc. The panels in the first and second columns show the velocity distribution for $Z>0$ and $Z<0$, respectively. The third and fourth columns show the corresponding distributions colour coded according to the median {\meta}.}
\label{FigureKinematics-Vphi-Vz-10-14kpc}
\end{figure*}

Figure \ref{FigureKinematics-Vphi-Vz-10-14kpc} 
 shows the $V_Z$ versus\ $V_\phi$ distributions from $R=10$ kpc to 14.5 kpc and $|Z|< 1$ kpc for the Medium Quality stars with good astrometric parameters, i.e.\ not duplicated sources with RUWE $ < 1.4$.  
 These plots are consistent with the asymmetric velocity distributions found in \citet[][]{Antoja2021} towards the Galactic anticenter.

\section{Kiel diagrams of individual abundance samples in the solar cylinder}

\begin{figure*}[h!]
\includegraphics[width=0.99\textwidth]{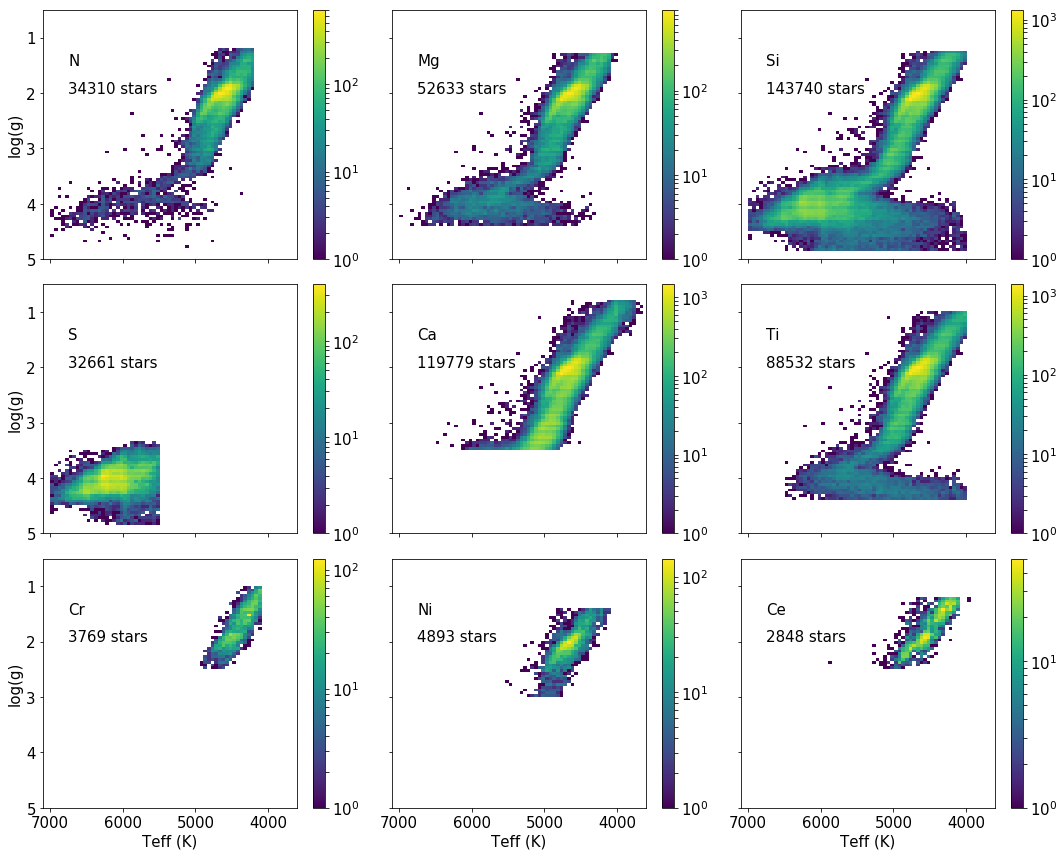}
\caption{Kiel diagrams of the individual abundances samples. The colour bars show the number of stars per 50~K by 0.05~dex bin.}
\label{fig:kiel}
\end{figure*}

For a better understanding of the abundance plots in the solar cylinder presented in Section~\ref{Sec:SolarCylinder}, Figure~\ref{fig:kiel} shows the Kiel diagrams of the analysed individual abundance samples. The differences in the underlying selection function are clearly illustrated with this figure. They are imposed by the dectectability of the spectral lines of each element in the RVS spectra of the stars, which depends on the stellar parameters, the element abundance and the S/N.

\section{Acknowledgements \label{sec_acknowledgements}}
The \Gaia\ mission and data processing have financially been supported by, in alphabetical order by country: \\

      -- the Algerian Centre de Recherche en Astronomie, Astrophysique et G\'{e}ophysique of Bouzareah Observatory;
\\ -- the Austrian Fonds zur F\"{o}rderung der wissenschaftlichen Forschung (FWF) Hertha Firnberg Programme through grants T359, P20046, and P23737;
\\ -- the BELgian federal Science Policy Office (BELSPO) through various PROgramme de D\'{e}veloppement d'Exp\'{e}riences scientifiques (PRODEX)
      grants, the Research Foundation Flanders (Fonds Wetenschappelijk Onderzoek) through grant VS.091.16N,
      the Fonds de la Recherche Scientifique (FNRS), and the Research Council of Katholieke Universiteit (KU) Leuven through
      grant C16/18/005 (Pushing AsteRoseismology to the next level with TESS, GaiA, and the Sloan DIgital Sky SurvEy -- PARADISE);  
\\ -- the Brazil-France exchange programmes Funda\c{c}\~{a}o de Amparo \`{a} Pesquisa do Estado de S\~{a}o Paulo (FAPESP) and Coordena\c{c}\~{a}o de Aperfeicoamento de Pessoal de N\'{\i}vel Superior (CAPES) - Comit\'{e} Fran\c{c}ais d'Evaluation de la Coop\'{e}ration Universitaire et Scientifique avec le Br\'{e}sil (COFECUB);
\\ -- the Chilean Agencia Nacional de Investigaci\'{o}n y Desarrollo (ANID) through Fondo Nacional de Desarrollo Cient\'{\i}fico y Tecnol\'{o}gico (FONDECYT) Regular Project 1210992 (L.~Chemin);
\\ -- the National Natural Science Foundation of China (NSFC) through grants 11573054, 11703065, and 12173069, the China Scholarship Council through grant 201806040200, and the Natural Science Foundation of Shanghai through grant 21ZR1474100;  
\\ -- the Tenure Track Pilot Programme of the Croatian Science Foundation and the \'{E}cole Polytechnique F\'{e}d\'{e}rale de Lausanne and the project TTP-2018-07-1171 `Mining the Variable Sky', with the funds of the Croatian-Swiss Research Programme;
\\ -- the Czech-Republic Ministry of Education, Youth, and Sports through grant LG 15010 and INTER-EXCELLENCE grant LTAUSA18093, and the Czech Space Office through ESA PECS contract 98058;
\\ -- the Danish Ministry of Science;
\\ -- the Estonian Ministry of Education and Research through grant IUT40-1;
\\ -- the European Commission’s Sixth Framework Programme through the European Leadership in Space Astrometry (\href{https://www.cosmos.esa.int/web/gaia/elsa-rtn-programme}{ELSA}) Marie Curie Research Training Network (MRTN-CT-2006-033481), through Marie Curie project PIOF-GA-2009-255267 (Space AsteroSeismology \& RR Lyrae stars, SAS-RRL), and through a Marie Curie Transfer-of-Knowledge (ToK) fellowship (MTKD-CT-2004-014188); the European Commission's Seventh Framework Programme through grant FP7-606740 (FP7-SPACE-2013-1) for the \Gaia\ European Network for Improved data User Services (\href{https://gaia.ub.edu/twiki/do/view/GENIUS/}{GENIUS}) and through grant 264895 for the \Gaia\ Research for European Astronomy Training (\href{https://www.cosmos.esa.int/web/gaia/great-programme}{GREAT-ITN}) network;
\\ -- the European Cooperation in Science and Technology (COST) through COST Action CA18104 `Revealing the Milky Way with \Gaia\ (MW-Gaia)';
\\ -- the European Research Council (ERC) through grants 320360, 647208, and 834148 and through the European Union’s Horizon 2020 research and innovation and excellent science programmes through Marie Sk{\l}odowska-Curie grant 745617 (Our Galaxy at full HD -- Gal-HD) and 895174 (The build-up and fate of self-gravitating systems in the Universe) as well as grants 687378 (Small Bodies: Near and Far), 682115 (Using the Magellanic Clouds to Understand the Interaction of Galaxies), 695099 (A sub-percent distance scale from binaries and Cepheids -- CepBin), 716155 (Structured ACCREtion Disks -- SACCRED), 951549 (Sub-percent calibration of the extragalactic distance scale in the era of big surveys -- UniverScale), and 101004214 (Innovative Scientific Data Exploration and Exploitation Applications for Space Sciences -- EXPLORE);
\\ -- the European Science Foundation (ESF), in the framework of the \Gaia\ Research for European Astronomy Training Research Network Programme (\href{https://www.cosmos.esa.int/web/gaia/great-programme}{GREAT-ESF});
\\ -- the European Space Agency (ESA) in the framework of the \Gaia\ project, through the Plan for European Cooperating States (PECS) programme through contracts C98090 and 4000106398/12/NL/KML for Hungary, through contract 4000115263/15/NL/IB for Germany, and through PROgramme de D\'{e}veloppement d'Exp\'{e}riences scientifiques (PRODEX) grant 4000127986 for Slovenia;  
\\ -- the Academy of Finland through grants 299543, 307157, 325805, 328654, 336546, and 345115 and the Magnus Ehrnrooth Foundation;
\\ -- the French Centre National d’\'{E}tudes Spatiales (CNES), the Agence Nationale de la Recherche (ANR) through grant ANR-10-IDEX-0001-02 for the `Investissements d'avenir' programme, through grant ANR-15-CE31-0007 for project `Modelling the Milky Way in the \Gaia\ era’ (MOD4Gaia), through grant ANR-14-CE33-0014-01 for project `The Milky Way disc formation in the \Gaia\ era’ (ARCHEOGAL), through grant ANR-15-CE31-0012-01 for project `Unlocking the potential of Cepheids as primary distance calibrators’ (UnlockCepheids), through grant ANR-19-CE31-0017 for project `Secular evolution of galaxies' (SEGAL), and through grant ANR-18-CE31-0006 for project `Galactic Dark Matter' (GaDaMa), the Centre National de la Recherche Scientifique (CNRS) and its SNO \Gaia\ of the Institut des Sciences de l’Univers (INSU), its Programmes Nationaux: Cosmologie et Galaxies (PNCG), Gravitation R\'{e}f\'{e}rences Astronomie M\'{e}trologie (PNGRAM), Plan\'{e}tologie (PNP), Physique et Chimie du Milieu Interstellaire (PCMI), and Physique Stellaire (PNPS), the `Action F\'{e}d\'{e}ratrice \Gaia ' of the Observatoire de Paris, the R\'{e}gion de Franche-Comt\'{e}, the Institut National Polytechnique (INP) and the Institut National de Physique nucl\'{e}aire et de Physique des Particules (IN2P3) co-funded by CNES;
\\ -- the German Aerospace Agency (Deutsches Zentrum f\"{u}r Luft- und Raumfahrt e.V., DLR) through grants 50QG0501, 50QG0601, 50QG0602, 50QG0701, 50QG0901, 50QG1001, 50QG1101, 50\-QG1401, 50QG1402, 50QG1403, 50QG1404, 50QG1904, 50QG2101, 50QG2102, and 50QG2202, and the Centre for Information Services and High Performance Computing (ZIH) at the Technische Universit\"{a}t Dresden for generous allocations of computer time;
\\ -- the Hungarian Academy of Sciences through the Lend\"{u}let Programme grants LP2014-17 and LP2018-7 and the Hungarian National Research, Development, and Innovation Office (NKFIH) through grant KKP-137523 (`SeismoLab');
\\ -- the Science Foundation Ireland (SFI) through a Royal Society - SFI University Research Fellowship (M.~Fraser);
\\ -- the Israel Ministry of Science and Technology through grant 3-18143 and the Tel Aviv University Center for Artificial Intelligence and Data Science (TAD) through a grant;
\\ -- the Agenzia Spaziale Italiana (ASI) through contracts I/037/08/0, I/058/10/0, 2014-025-R.0, 2014-025-R.1.2015, and 2018-24-HH.0 to the Italian Istituto Nazionale di Astrofisica (INAF), contract 2014-049-R.0/1/2 to INAF for the Space Science Data Centre (SSDC, formerly known as the ASI Science Data Center, ASDC), contracts I/008/10/0, 2013/030/I.0, 2013-030-I.0.1-2015, and 2016-17-I.0 to the Aerospace Logistics Technology Engineering Company (ALTEC S.p.A.), INAF, and the Italian Ministry of Education, University, and Research (Ministero dell'Istruzione, dell'Universit\`{a} e della Ricerca) through the Premiale project `MIning The Cosmos Big Data and Innovative Italian Technology for Frontier Astrophysics and Cosmology' (MITiC);
\\ -- the Netherlands Organisation for Scientific Research (NWO) through grant NWO-M-614.061.414, through a VICI grant (A.~Helmi), and through a Spinoza prize (A.~Helmi), and the Netherlands Research School for Astronomy (NOVA);
\\ -- the Polish National Science Centre through HARMONIA grant 2018/30/M/ST9/00311 and DAINA grant 2017/27/L/ST9/03221 and the Ministry of Science and Higher Education (MNiSW) through grant DIR/WK/2018/12;
\\ -- the Portuguese Funda\c{c}\~{a}o para a Ci\^{e}ncia e a Tecnologia (FCT) through national funds, grants SFRH/\-BD/128840/2017 and PTDC/FIS-AST/30389/2017, and work contract DL 57/2016/CP1364/CT0006, the Fundo Europeu de Desenvolvimento Regional (FEDER) through grant POCI-01-0145-FEDER-030389 and its Programa Operacional Competitividade e Internacionaliza\c{c}\~{a}o (COMPETE2020) through grants UIDB/04434/2020 and UIDP/04434/2020, and the Strategic Programme UIDB/\-00099/2020 for the Centro de Astrof\'{\i}sica e Gravita\c{c}\~{a}o (CENTRA);  
\\ -- the Slovenian Research Agency through grant P1-0188;
\\ -- the Spanish Ministry of Economy (MINECO/FEDER, UE), the Spanish Ministry of Science and Innovation (MICIN), the Spanish Ministry of Education, Culture, and Sports, and the Spanish Government through grants BES-2016-078499, BES-2017-083126, BES-C-2017-0085, ESP2016-80079-C2-1-R, ESP2016-80079-C2-2-R, FPU16/03827, PDC2021-121059-C22, RTI2018-095076-B-C22, and TIN2015-65316-P (`Computaci\'{o}n de Altas Prestaciones VII'), the Juan de la Cierva Incorporaci\'{o}n Programme (FJCI-2015-2671 and IJC2019-04862-I for F.~Anders), the Severo Ochoa Centre of Excellence Programme (SEV2015-0493), and MICIN/AEI/10.13039/501100011033 (and the European Union through European Regional Development Fund `A way of making Europe') through grant RTI2018-095076-B-C21, the Institute of Cosmos Sciences University of Barcelona (ICCUB, Unidad de Excelencia `Mar\'{\i}a de Maeztu’) through grant CEX2019-000918-M, the University of Barcelona's official doctoral programme for the development of an R+D+i project through an Ajuts de Personal Investigador en Formaci\'{o} (APIF) grant, the Spanish Virtual Observatory through project AyA2017-84089, the Galician Regional Government, Xunta de Galicia, through grants ED431B-2021/36, ED481A-2019/155, and ED481A-2021/296, the Centro de Investigaci\'{o}n en Tecnolog\'{\i}as de la Informaci\'{o}n y las Comunicaciones (CITIC), funded by the Xunta de Galicia and the European Union (European Regional Development Fund -- Galicia 2014-2020 Programme), through grant ED431G-2019/01, the Red Espa\~{n}ola de Supercomputaci\'{o}n (RES) computer resources at MareNostrum, the Barcelona Supercomputing Centre - Centro Nacional de Supercomputaci\'{o}n (BSC-CNS) through activities AECT-2017-2-0002, AECT-2017-3-0006, AECT-2018-1-0017, AECT-2018-2-0013, AECT-2018-3-0011, AECT-2019-1-0010, AECT-2019-2-0014, AECT-2019-3-0003, AECT-2020-1-0004, and DATA-2020-1-0010, the Departament d'Innovaci\'{o}, Universitats i Empresa de la Generalitat de Catalunya through grant 2014-SGR-1051 for project `Models de Programaci\'{o} i Entorns d'Execuci\'{o} Parallels' (MPEXPAR), and Ramon y Cajal Fellowship RYC2018-025968-I funded by MICIN/AEI/10.13039/501100011033 and the European Science Foundation (`Investing in your future');
\\ -- the Swedish National Space Agency (SNSA/Rymdstyrelsen);
\\ -- the Swiss State Secretariat for Education, Research, and Innovation through the Swiss Activit\'{e}s Nationales Compl\'{e}mentaires and the Swiss National Science Foundation through an Eccellenza Professorial Fellowship (award PCEFP2\_194638 for R.~Anderson);
\\ -- the United Kingdom Particle Physics and Astronomy Research Council (PPARC), the United Kingdom Science and Technology Facilities Council (STFC), and the United Kingdom Space Agency (UKSA) through the following grants to the University of Bristol, the University of Cambridge, the University of Edinburgh, the University of Leicester, the Mullard Space Sciences Laboratory of University College London, and the United Kingdom Rutherford Appleton Laboratory (RAL): PP/D006511/1, PP/D006546/1, PP/D006570/1, ST/I000852/1, ST/J005045/1, ST/K00056X/1, ST/\-K000209/1, ST/K000756/1, ST/L006561/1, ST/N000595/1, ST/N000641/1, ST/N000978/1, ST/\-N001117/1, ST/S000089/1, ST/S000976/1, ST/S000984/1, ST/S001123/1, ST/S001948/1, ST/\-S001980/1, ST/S002103/1, ST/V000969/1, ST/W002469/1, ST/W002493/1, ST/W002671/1, ST/W002809/1, and EP/V520342/1.

The GBOT programme  uses observations collected at (i) the European Organisation for Astronomical Research in the Southern Hemisphere (ESO) with the VLT Survey Telescope (VST), under ESO programmes
092.B-0165,
093.B-0236,
094.B-0181,
095.B-0046,
096.B-0162,
097.B-0304,
098.B-0030,
099.B-0034,
0100.B-0131,
0101.B-0156,
0102.B-0174, and
0103.B-0165;
%
%
and (ii) the Liverpool Telescope, which is operated on the island of La Palma by Liverpool John Moores University in the Spanish Observatorio del Roque de los Muchachos of the Instituto de Astrof\'{\i}sica de Canarias with financial support from the United Kingdom Science and Technology Facilities Council, and (iii) telescopes of the Las Cumbres Observatory Global Telescope Network.\\

\end{appendix}

\end{document}

%% file: authors_d_nosp.tex
\author{
{\it Gaia} Collaboration
\and         A.~Recio-Blanco\orcit{0000-0002-6550-7377}\inst{\ref{inst:0001}}
\and         G.~Kordopatis\orcit{0000-0002-9035-3920}\inst{\ref{inst:0001}}
\and         P.~de Laverny\orcit{0000-0002-2817-4104}\inst{\ref{inst:0001}}
\and       P.A.~Palicio\orcit{0000-0002-7432-8709}\inst{\ref{inst:0001}}
\and         A.~Spagna\orcit{0000-0003-1732-2412}\inst{\ref{inst:0005}}
\and         L.~Spina\orcit{0000-0002-9760-6249}\inst{\ref{inst:0006}}
\and         D.~Katz\orcit{0000-0001-7986-3164}\inst{\ref{inst:0007}}
\and         P.~Re Fiorentin\orcit{0000-0002-4995-0475}\inst{\ref{inst:0005}}
\and         E.~Poggio\orcit{0000-0003-3793-8505}\inst{\ref{inst:0001},\ref{inst:0005}}
\and       P.J.~McMillan\orcit{0000-0002-8861-2620}\inst{\ref{inst:0011}}
\and         A.~Vallenari\orcit{0000-0003-0014-519X}\inst{\ref{inst:0006}}
\and       M.G.~Lattanzi\orcit{0000-0003-0429-7748}\inst{\ref{inst:0005},\ref{inst:0014}}
\and       G.M.~Seabroke\orcit{0000-0003-4072-9536}\inst{\ref{inst:0015}}
\and         L.~Casamiquela\orcit{0000-0001-5238-8674}\inst{\ref{inst:0016},\ref{inst:0007}}
\and         A.~Bragaglia\orcit{0000-0002-0338-7883}\inst{\ref{inst:0018}}
\and         T.~Antoja\orcit{0000-0003-2595-5148}\inst{\ref{inst:0019}}
\and     C.A.L.~Bailer-Jones\inst{\ref{inst:0020}}
\and       M.~Schultheis\orcit{0000-0002-6590-1657}\inst{\ref{inst:0001}}
\and         R.~Andrae\orcit{0000-0001-8006-6365}\inst{\ref{inst:0020}}
\and         M.~Fouesneau\orcit{0000-0001-9256-5516}\inst{\ref{inst:0020}}
\and         M.~Cropper\orcit{0000-0003-4571-9468}\inst{\ref{inst:0015}}
\and         T.~Cantat-Gaudin\orcit{0000-0001-8726-2588}\inst{\ref{inst:0019},\ref{inst:0020}}
\and         A.~Bijaoui\inst{\ref{inst:0001}}
\and         U.~Heiter\orcit{0000-0001-6825-1066}\inst{\ref{inst:0026}}
\and     A.G.A.~Brown\orcit{0000-0002-7419-9679}\inst{\ref{inst:0028}}
\and         T.~Prusti\orcit{0000-0003-3120-7867}\inst{\ref{inst:0029}}
\and     J.H.J.~de Bruijne\orcit{0000-0001-6459-8599}\inst{\ref{inst:0029}}
\and         F.~Arenou\orcit{0000-0003-2837-3899}\inst{\ref{inst:0007}}
\and         C.~Babusiaux\orcit{0000-0002-7631-348X}\inst{\ref{inst:0032},\ref{inst:0007}}
\and         M.~Biermann\inst{\ref{inst:0034}}
\and       O.L.~Creevey\orcit{0000-0003-1853-6631}\inst{\ref{inst:0001}}
\and         C.~Ducourant\orcit{0000-0003-4843-8979}\inst{\ref{inst:0016}}
\and       D.W.~Evans\orcit{0000-0002-6685-5998}\inst{\ref{inst:0037}}
\and         L.~Eyer\orcit{0000-0002-0182-8040}\inst{\ref{inst:0038}}
\and         R.~Guerra\orcit{0000-0002-9850-8982}\inst{\ref{inst:0039}}
\and         A.~Hutton\inst{\ref{inst:0040}}
\and         C.~Jordi\orcit{0000-0001-5495-9602}\inst{\ref{inst:0019}}
\and       S.A.~Klioner\orcit{0000-0003-4682-7831}\inst{\ref{inst:0042}}
\and       U.L.~Lammers\orcit{0000-0001-8309-3801}\inst{\ref{inst:0039}}
\and         L.~Lindegren\orcit{0000-0002-5443-3026}\inst{\ref{inst:0011}}
\and         X.~Luri\orcit{0000-0001-5428-9397}\inst{\ref{inst:0019}}
\and         F.~Mignard\inst{\ref{inst:0001}}
\and         C.~Panem\inst{\ref{inst:0047}}
\and         D.~Pourbaix$^\dagger$\orcit{0000-0002-3020-1837}\inst{\ref{inst:0048},\ref{inst:0049}}
\and         S.~Randich\orcit{0000-0003-2438-0899}\inst{\ref{inst:0050}}
\and         P.~Sartoretti\inst{\ref{inst:0007}}
\and         C.~Soubiran\orcit{0000-0003-3304-8134}\inst{\ref{inst:0016}}
\and         P.~Tanga\orcit{0000-0002-2718-997X}\inst{\ref{inst:0001}}
\and       N.A.~Walton\orcit{0000-0003-3983-8778}\inst{\ref{inst:0037}}
\and         U.~Bastian\orcit{0000-0002-8667-1715}\inst{\ref{inst:0034}}
\and         R.~Drimmel\orcit{0000-0002-1777-5502}\inst{\ref{inst:0005}}
\and         F.~Jansen\inst{\ref{inst:0057}}
\and         F.~van Leeuwen\inst{\ref{inst:0037}}
\and         J.~Bakker\inst{\ref{inst:0039}}
\and         C.~Cacciari\orcit{0000-0001-5174-3179}\inst{\ref{inst:0018}}
\and         J.~Casta\~{n}eda\orcit{0000-0001-7820-946X}\inst{\ref{inst:0061}}
\and         F.~De Angeli\orcit{0000-0003-1879-0488}\inst{\ref{inst:0037}}
\and         C.~Fabricius\orcit{0000-0003-2639-1372}\inst{\ref{inst:0019}}
\and         Y.~Fr\'{e}mat\orcit{0000-0002-4645-6017}\inst{\ref{inst:0064}}
\and         L.~Galluccio\orcit{0000-0002-8541-0476}\inst{\ref{inst:0001}}
\and         A.~Guerrier\inst{\ref{inst:0047}}
\and         E.~Masana\orcit{0000-0002-4819-329X}\inst{\ref{inst:0019}}
\and         R.~Messineo\inst{\ref{inst:0068}}
\and         N.~Mowlavi\orcit{0000-0003-1578-6993}\inst{\ref{inst:0038}}
\and         C.~Nicolas\inst{\ref{inst:0047}}
\and         K.~Nienartowicz\orcit{0000-0001-5415-0547}\inst{\ref{inst:0071},\ref{inst:0072}}
\and         F.~Pailler\orcit{0000-0002-4834-481X}\inst{\ref{inst:0047}}
\and         P.~Panuzzo\orcit{0000-0002-0016-8271}\inst{\ref{inst:0007}}
\and         F.~Riclet\inst{\ref{inst:0047}}
\and         W.~Roux\orcit{0000-0002-7816-1950}\inst{\ref{inst:0047}}
\and         R.~Sordo\orcit{0000-0003-4979-0659}\inst{\ref{inst:0006}}
\and         F.~Th\'{e}venin\inst{\ref{inst:0001}}
\and         G.~Gracia-Abril\inst{\ref{inst:0079},\ref{inst:0034}}
\and         J.~Portell\orcit{0000-0002-8886-8925}\inst{\ref{inst:0019}}
\and         D.~Teyssier\orcit{0000-0002-6261-5292}\inst{\ref{inst:0082}}
\and         M.~Altmann\orcit{0000-0002-0530-0913}\inst{\ref{inst:0034},\ref{inst:0084}}
\and         M.~Audard\orcit{0000-0003-4721-034X}\inst{\ref{inst:0038},\ref{inst:0072}}
\and         I.~Bellas-Velidis\inst{\ref{inst:0087}}
\and         K.~Benson\inst{\ref{inst:0015}}
\and         J.~Berthier\orcit{0000-0003-1846-6485}\inst{\ref{inst:0089}}
\and         R.~Blomme\orcit{0000-0002-2526-346X}\inst{\ref{inst:0064}}
\and       P.W.~Burgess\inst{\ref{inst:0037}}
\and         D.~Busonero\orcit{0000-0002-3903-7076}\inst{\ref{inst:0005}}
\and         G.~Busso\orcit{0000-0003-0937-9849}\inst{\ref{inst:0037}}
\and         H.~C\'{a}novas\orcit{0000-0001-7668-8022}\inst{\ref{inst:0082}}
\and         B.~Carry\orcit{0000-0001-5242-3089}\inst{\ref{inst:0001}}
\and         A.~Cellino\orcit{0000-0002-6645-334X}\inst{\ref{inst:0005}}
\and         N.~Cheek\inst{\ref{inst:0097}}
\and         G.~Clementini\orcit{0000-0001-9206-9723}\inst{\ref{inst:0018}}
\and         Y.~Damerdji\orcit{0000-0002-3107-4024}\inst{\ref{inst:0099},\ref{inst:0100}}
\and         M.~Davidson\inst{\ref{inst:0101}}
\and         P.~de Teodoro\inst{\ref{inst:0039}}
\and         M.~Nu\~{n}ez Campos\inst{\ref{inst:0040}}
\and         L.~Delchambre\orcit{0000-0003-2559-408X}\inst{\ref{inst:0099}}
\and         A.~Dell'Oro\orcit{0000-0003-1561-9685}\inst{\ref{inst:0050}}
\and         P.~Esquej\orcit{0000-0001-8195-628X}\inst{\ref{inst:0106}}
\and         J.~Fern\'{a}ndez-Hern\'{a}ndez\inst{\ref{inst:0107}}
\and         E.~Fraile\inst{\ref{inst:0106}}
\and         D.~Garabato\orcit{0000-0002-7133-6623}\inst{\ref{inst:0109}}
\and         P.~Garc\'{i}a-Lario\orcit{0000-0003-4039-8212}\inst{\ref{inst:0039}}
\and         E.~Gosset\inst{\ref{inst:0099},\ref{inst:0049}}
\and         R.~Haigron\inst{\ref{inst:0007}}
\and      J.-L.~Halbwachs\orcit{0000-0003-2968-6395}\inst{\ref{inst:0114}}
\and       N.C.~Hambly\orcit{0000-0002-9901-9064}\inst{\ref{inst:0101}}
\and       D.L.~Harrison\orcit{0000-0001-8687-6588}\inst{\ref{inst:0037},\ref{inst:0117}}
\and         J.~Hern\'{a}ndez\orcit{0000-0002-0361-4994}\inst{\ref{inst:0039}}
\and         D.~Hestroffer\orcit{0000-0003-0472-9459}\inst{\ref{inst:0089}}
\and       S.T.~Hodgkin\orcit{0000-0002-5470-3962}\inst{\ref{inst:0037}}
\and         B.~Holl\orcit{0000-0001-6220-3266}\inst{\ref{inst:0038},\ref{inst:0072}}
\and         K.~Jan{\ss}en\orcit{0000-0002-8163-2493}\inst{\ref{inst:0123}}
\and         G.~Jevardat de Fombelle\inst{\ref{inst:0038}}
\and         S.~Jordan\orcit{0000-0001-6316-6831}\inst{\ref{inst:0034}}
\and         A.~Krone-Martins\orcit{0000-0002-2308-6623}\inst{\ref{inst:0126},\ref{inst:0127}}
\and       A.C.~Lanzafame\orcit{0000-0002-2697-3607}\inst{\ref{inst:0128},\ref{inst:0129}}
\and         W.~L\"{ o}ffler\inst{\ref{inst:0034}}
\and         O.~Marchal\orcit{ 0000-0001-7461-892}\inst{\ref{inst:0114}}
\and       P.M.~Marrese\orcit{0000-0002-8162-3810}\inst{\ref{inst:0132},\ref{inst:0133}}
\and         A.~Moitinho\orcit{0000-0003-0822-5995}\inst{\ref{inst:0126}}
\and         K.~Muinonen\orcit{0000-0001-8058-2642}\inst{\ref{inst:0135},\ref{inst:0136}}
\and         P.~Osborne\inst{\ref{inst:0037}}
\and         E.~Pancino\orcit{0000-0003-0788-5879}\inst{\ref{inst:0050},\ref{inst:0133}}
\and         T.~Pauwels\inst{\ref{inst:0064}}
\and         C.~Reyl\'{e}\orcit{0000-0003-2258-2403}\inst{\ref{inst:0141}}
\and         M.~Riello\orcit{0000-0002-3134-0935}\inst{\ref{inst:0037}}
\and         L.~Rimoldini\orcit{0000-0002-0306-585X}\inst{\ref{inst:0072}}
\and         T.~Roegiers\orcit{0000-0002-1231-4440}\inst{\ref{inst:0144}}
\and         J.~Rybizki\orcit{0000-0002-0993-6089}\inst{\ref{inst:0020}}
\and       L.M.~Sarro\orcit{0000-0002-5622-5191}\inst{\ref{inst:0146}}
\and         C.~Siopis\orcit{0000-0002-6267-2924}\inst{\ref{inst:0048}}
\and         M.~Smith\inst{\ref{inst:0015}}
\and         A.~Sozzetti\orcit{0000-0002-7504-365X}\inst{\ref{inst:0005}}
\and         E.~Utrilla\inst{\ref{inst:0040}}
\and         M.~van Leeuwen\orcit{0000-0001-9698-2392}\inst{\ref{inst:0037}}
\and         U.~Abbas\orcit{0000-0002-5076-766X}\inst{\ref{inst:0005}}
\and         P.~\'{A}brah\'{a}m\orcit{0000-0001-6015-646X}\inst{\ref{inst:0153},\ref{inst:0154}}
\and         A.~Abreu Aramburu\inst{\ref{inst:0107}}
\and         C.~Aerts\orcit{0000-0003-1822-7126}\inst{\ref{inst:0156},\ref{inst:0157},\ref{inst:0020}}
\and       J.J.~Aguado\inst{\ref{inst:0146}}
\and         M.~Ajaj\inst{\ref{inst:0007}}
\and         F.~Aldea-Montero\inst{\ref{inst:0039}}
\and         G.~Altavilla\orcit{0000-0002-9934-1352}\inst{\ref{inst:0132},\ref{inst:0133}}
\and       M.A.~\'{A}lvarez\orcit{0000-0002-6786-2620}\inst{\ref{inst:0109}}
\and         J.~Alves\orcit{0000-0002-4355-0921}\inst{\ref{inst:0165}}
\and         F.~Anders\inst{\ref{inst:0019}}
\and       R.I.~Anderson\orcit{0000-0001-8089-4419}\inst{\ref{inst:0167}}
\and         E.~Anglada Varela\orcit{0000-0001-7563-0689}\inst{\ref{inst:0107}}
\and         D.~Baines\orcit{0000-0002-6923-3756}\inst{\ref{inst:0082}}
\and       S.G.~Baker\orcit{0000-0002-6436-1257}\inst{\ref{inst:0015}}
\and         L.~Balaguer-N\'{u}\~{n}ez\orcit{0000-0001-9789-7069}\inst{\ref{inst:0019}}
\and         E.~Balbinot\orcit{0000-0002-1322-3153}\inst{\ref{inst:0172}}
\and         Z.~Balog\orcit{0000-0003-1748-2926}\inst{\ref{inst:0034},\ref{inst:0020}}
\and         C.~Barache\inst{\ref{inst:0084}}
\and         D.~Barbato\inst{\ref{inst:0038},\ref{inst:0005}}
\and         M.~Barros\orcit{0000-0002-9728-9618}\inst{\ref{inst:0126}}
\and       M.A.~Barstow\orcit{0000-0002-7116-3259}\inst{\ref{inst:0179}}
\and         S.~Bartolom\'{e}\orcit{0000-0002-6290-6030}\inst{\ref{inst:0019}}
\and      J.-L.~Bassilana\inst{\ref{inst:0181}}
\and         N.~Bauchet\inst{\ref{inst:0007}}
\and         U.~Becciani\orcit{0000-0002-4389-8688}\inst{\ref{inst:0128}}
\and         M.~Bellazzini\orcit{0000-0001-8200-810X}\inst{\ref{inst:0018}}
\and         A.~Berihuete\orcit{0000-0002-8589-4423}\inst{\ref{inst:0185}}
\and         M.~Bernet\orcit{0000-0001-7503-1010}\inst{\ref{inst:0019}}
\and         S.~Bertone\orcit{0000-0001-9885-8440}\inst{\ref{inst:0187},\ref{inst:0188},\ref{inst:0005}}
\and         L.~Bianchi\orcit{0000-0002-7999-4372}\inst{\ref{inst:0190}}
\and         A.~Binnenfeld\orcit{0000-0002-9319-3838}\inst{\ref{inst:0191}}
\and         S.~Blanco-Cuaresma\orcit{0000-0002-1584-0171}\inst{\ref{inst:0192}}
\and         T.~Boch\orcit{0000-0001-5818-2781}\inst{\ref{inst:0114}}
\and         A.~Bombrun\inst{\ref{inst:0194}}
\and         D.~Bossini\orcit{0000-0002-9480-8400}\inst{\ref{inst:0195}}
\and         S.~Bouquillon\inst{\ref{inst:0084},\ref{inst:0197}}
\and         L.~Bramante\inst{\ref{inst:0068}}
\and         E.~Breedt\orcit{0000-0001-6180-3438}\inst{\ref{inst:0037}}
\and         A.~Bressan\orcit{0000-0002-7922-8440}\inst{\ref{inst:0200}}
\and         N.~Brouillet\orcit{0000-0002-3274-7024}\inst{\ref{inst:0016}}
\and         E.~Brugaletta\orcit{0000-0003-2598-6737}\inst{\ref{inst:0128}}
\and         B.~Bucciarelli\orcit{0000-0002-5303-0268}\inst{\ref{inst:0005},\ref{inst:0014}}
\and         A.~Burlacu\inst{\ref{inst:0205}}
\and       A.G.~Butkevich\orcit{0000-0002-4098-3588}\inst{\ref{inst:0005}}
\and         R.~Buzzi\orcit{0000-0001-9389-5701}\inst{\ref{inst:0005}}
\and         E.~Caffau\orcit{0000-0001-6011-6134}\inst{\ref{inst:0007}}
\and         R.~Cancelliere\orcit{0000-0002-9120-3799}\inst{\ref{inst:0209}}
\and         R.~Carballo\orcit{0000-0001-7412-2498}\inst{\ref{inst:0210}}
\and         T.~Carlucci\inst{\ref{inst:0084}}
\and       M.I.~Carnerero\orcit{0000-0001-5843-5515}\inst{\ref{inst:0005}}
\and       J.M.~Carrasco\orcit{0000-0002-3029-5853}\inst{\ref{inst:0019}}
\and         M.~Castellani\orcit{0000-0002-7650-7428}\inst{\ref{inst:0132}}
\and         A.~Castro-Ginard\orcit{0000-0002-9419-3725}\inst{\ref{inst:0028}}
\and         L.~Chaoul\inst{\ref{inst:0047}}
\and         P.~Charlot\orcit{0000-0002-9142-716X}\inst{\ref{inst:0016}}
\and         L.~Chemin\orcit{0000-0002-3834-7937}\inst{\ref{inst:0218}}
\and         V.~Chiaramida\inst{\ref{inst:0068}}
\and         A.~Chiavassa\orcit{0000-0003-3891-7554}\inst{\ref{inst:0001}}
\and         N.~Chornay\orcit{0000-0002-8767-3907}\inst{\ref{inst:0037}}
\and         G.~Comoretto\inst{\ref{inst:0082},\ref{inst:0223}}
\and         G.~Contursi\orcit{0000-0001-5370-1511}\inst{\ref{inst:0001}}
\and       W.J.~Cooper\orcit{0000-0003-3501-8967}\inst{\ref{inst:0225},\ref{inst:0005}}
\and         T.~Cornez\inst{\ref{inst:0181}}
\and         S.~Cowell\inst{\ref{inst:0037}}
\and         F.~Crifo\inst{\ref{inst:0007}}
\and         M.~Crosta\orcit{0000-0003-4369-3786}\inst{\ref{inst:0005},\ref{inst:0231}}
\and         C.~Crowley\inst{\ref{inst:0194}}
\and         C.~Dafonte\orcit{0000-0003-4693-7555}\inst{\ref{inst:0109}}
\and         A.~Dapergolas\inst{\ref{inst:0087}}
\and         P.~David\inst{\ref{inst:0089}}
\and         F.~De Luise\orcit{0000-0002-6570-8208}\inst{\ref{inst:0236}}
\and         R.~De March\orcit{0000-0003-0567-842X}\inst{\ref{inst:0068}}
\and         J.~De Ridder\orcit{0000-0001-6726-2863}\inst{\ref{inst:0156}}
\and         R.~de Souza\inst{\ref{inst:0239}}
\and         A.~de Torres\inst{\ref{inst:0194}}
\and       E.F.~del Peloso\inst{\ref{inst:0034}}
\and         E.~del Pozo\inst{\ref{inst:0040}}
\and         M.~Delbo\orcit{0000-0002-8963-2404}\inst{\ref{inst:0001}}
\and         A.~Delgado\inst{\ref{inst:0106}}
\and      J.-B.~Delisle\orcit{0000-0001-5844-9888}\inst{\ref{inst:0038}}
\and         C.~Demouchy\inst{\ref{inst:0246}}
\and       T.E.~Dharmawardena\orcit{0000-0002-9583-5216}\inst{\ref{inst:0020}}
\and         P.~Di Matteo\inst{\ref{inst:0007}}
\and         S.~Diakite\inst{\ref{inst:0249}}
\and         C.~Diener\inst{\ref{inst:0037}}
\and         E.~Distefano\orcit{0000-0002-2448-2513}\inst{\ref{inst:0128}}
\and         C.~Dolding\inst{\ref{inst:0015}}
\and         B.~Edvardsson\inst{\ref{inst:0253}}
\and         H.~Enke\orcit{0000-0002-2366-8316}\inst{\ref{inst:0123}}
\and         C.~Fabre\inst{\ref{inst:0255}}
\and         M.~Fabrizio\orcit{0000-0001-5829-111X}\inst{\ref{inst:0132},\ref{inst:0133}}
\and         S.~Faigler\orcit{0000-0002-8368-5724}\inst{\ref{inst:0258}}
\and         G.~Fedorets\orcit{0000-0002-8418-4809}\inst{\ref{inst:0135},\ref{inst:0260}}
\and         P.~Fernique\orcit{0000-0002-3304-2923}\inst{\ref{inst:0114},\ref{inst:0262}}
\and         F.~Figueras\orcit{0000-0002-3393-0007}\inst{\ref{inst:0019}}
\and         Y.~Fournier\orcit{0000-0002-6633-9088}\inst{\ref{inst:0123}}
\and         C.~Fouron\inst{\ref{inst:0205}}
\and         F.~Fragkoudi\orcit{0000-0002-0897-3013}\inst{\ref{inst:0266},\ref{inst:0267},\ref{inst:0268}}
\and         M.~Gai\orcit{0000-0001-9008-134X}\inst{\ref{inst:0005}}
\and         A.~Garcia-Gutierrez\inst{\ref{inst:0019}}
\and         M.~Garcia-Reinaldos\inst{\ref{inst:0039}}
\and         M.~Garc\'{i}a-Torres\orcit{0000-0002-6867-7080}\inst{\ref{inst:0272}}
\and         A.~Garofalo\orcit{0000-0002-5907-0375}\inst{\ref{inst:0018}}
\and         A.~Gavel\orcit{0000-0002-2963-722X}\inst{\ref{inst:0026}}
\and         P.~Gavras\orcit{0000-0002-4383-4836}\inst{\ref{inst:0106}}
\and         E.~Gerlach\orcit{0000-0002-9533-2168}\inst{\ref{inst:0042}}
\and         R.~Geyer\orcit{0000-0001-6967-8707}\inst{\ref{inst:0042}}
\and         P.~Giacobbe\orcit{0000-0001-7034-7024}\inst{\ref{inst:0005}}
\and         G.~Gilmore\orcit{0000-0003-4632-0213}\inst{\ref{inst:0037}}
\and         S.~Girona\orcit{0000-0002-1975-1918}\inst{\ref{inst:0280}}
\and         G.~Giuffrida\inst{\ref{inst:0132}}
\and         R.~Gomel\inst{\ref{inst:0258}}
\and         A.~Gomez\orcit{0000-0002-3796-3690}\inst{\ref{inst:0109}}
\and         J.~Gonz\'{a}lez-N\'{u}\~{n}ez\orcit{0000-0001-5311-5555}\inst{\ref{inst:0097},\ref{inst:0285}}
\and         I.~Gonz\'{a}lez-Santamar\'{i}a\orcit{0000-0002-8537-9384}\inst{\ref{inst:0109}}
\and       J.J.~Gonz\'{a}lez-Vidal\inst{\ref{inst:0019}}
\and         M.~Granvik\orcit{0000-0002-5624-1888}\inst{\ref{inst:0135},\ref{inst:0289}}
\and         P.~Guillout\inst{\ref{inst:0114}}
\and         J.~Guiraud\inst{\ref{inst:0047}}
\and         R.~Guti\'{e}rrez-S\'{a}nchez\inst{\ref{inst:0082}}
\and       L.P.~Guy\orcit{0000-0003-0800-8755}\inst{\ref{inst:0072},\ref{inst:0294}}
\and         D.~Hatzidimitriou\orcit{0000-0002-5415-0464}\inst{\ref{inst:0295},\ref{inst:0087}}
\and         M.~Hauser\inst{\ref{inst:0020},\ref{inst:0298}}
\and         M.~Haywood\orcit{0000-0003-0434-0400}\inst{\ref{inst:0007}}
\and         A.~Helmer\inst{\ref{inst:0181}}
\and         A.~Helmi\orcit{0000-0003-3937-7641}\inst{\ref{inst:0172}}
\and       M.H.~Sarmiento\orcit{0000-0003-4252-5115}\inst{\ref{inst:0040}}
\and       S.L.~Hidalgo\orcit{0000-0002-0002-9298}\inst{\ref{inst:0303},\ref{inst:0304}}
\and         N.~H\l{}adczuk\orcit{0000-0001-9163-4209}\inst{\ref{inst:0039},\ref{inst:0306}}
\and         D.~Hobbs\orcit{0000-0002-2696-1366}\inst{\ref{inst:0011}}
\and         G.~Holland\inst{\ref{inst:0037}}
\and       H.E.~Huckle\inst{\ref{inst:0015}}
\and         K.~Jardine\inst{\ref{inst:0310}}
\and         G.~Jasniewicz\inst{\ref{inst:0311}}
\and         A.~Jean-Antoine Piccolo\orcit{0000-0001-8622-212X}\inst{\ref{inst:0047}}
\and     \'{O}.~Jim\'{e}nez-Arranz\orcit{0000-0001-7434-5165}\inst{\ref{inst:0019}}
\and         J.~Juaristi Campillo\inst{\ref{inst:0034}}
\and         F.~Julbe\inst{\ref{inst:0019}}
\and         L.~Karbevska\inst{\ref{inst:0072},\ref{inst:0317}}
\and         P.~Kervella\orcit{0000-0003-0626-1749}\inst{\ref{inst:0318}}
\and         S.~Khanna\orcit{0000-0002-2604-4277}\inst{\ref{inst:0172},\ref{inst:0005}}
\and       A.J.~Korn\orcit{0000-0002-3881-6756}\inst{\ref{inst:0026}}
\and      \'{A}~K\'{o}sp\'{a}l\orcit{'{u}t 15-17, 1121 B}\inst{\ref{inst:0153},\ref{inst:0020},\ref{inst:0154}}
\and         Z.~Kostrzewa-Rutkowska\inst{\ref{inst:0028},\ref{inst:0326}}
\and         K.~Kruszy\'{n}ska\orcit{0000-0002-2729-5369}\inst{\ref{inst:0327}}
\and         M.~Kun\orcit{0000-0002-7538-5166}\inst{\ref{inst:0153}}
\and         P.~Laizeau\inst{\ref{inst:0329}}
\and         S.~Lambert\orcit{0000-0001-6759-5502}\inst{\ref{inst:0084}}
\and       A.F.~Lanza\orcit{0000-0001-5928-7251}\inst{\ref{inst:0128}}
\and         Y.~Lasne\inst{\ref{inst:0181}}
\and      J.-F.~Le Campion\inst{\ref{inst:0016}}
\and         Y.~Lebreton\orcit{0000-0002-4834-2144}\inst{\ref{inst:0318},\ref{inst:0335}}
\and         T.~Lebzelter\orcit{0000-0002-0702-7551}\inst{\ref{inst:0165}}
\and         S.~Leccia\orcit{0000-0001-5685-6930}\inst{\ref{inst:0337}}
\and         N.~Leclerc\inst{\ref{inst:0007}}
\and         I.~Lecoeur-Taibi\orcit{0000-0003-0029-8575}\inst{\ref{inst:0072}}
\and         S.~Liao\orcit{0000-0002-9346-0211}\inst{\ref{inst:0340},\ref{inst:0005},\ref{inst:0342}}
\and       E.L.~Licata\orcit{0000-0002-5203-0135}\inst{\ref{inst:0005}}
\and     H.E.P.~Lindstr{\o}m\inst{\ref{inst:0005},\ref{inst:0345},\ref{inst:0346}}
\and       T.A.~Lister\orcit{0000-0002-3818-7769}\inst{\ref{inst:0347}}
\and         E.~Livanou\orcit{0000-0003-0628-2347}\inst{\ref{inst:0295}}
\and         A.~Lobel\orcit{0000-0001-5030-019X}\inst{\ref{inst:0064}}
\and         A.~Lorca\inst{\ref{inst:0040}}
\and         C.~Loup\inst{\ref{inst:0114}}
\and         P.~Madrero Pardo\inst{\ref{inst:0019}}
\and         A.~Magdaleno Romeo\inst{\ref{inst:0205}}
\and         S.~Managau\inst{\ref{inst:0181}}
\and       R.G.~Mann\orcit{0000-0002-0194-325X}\inst{\ref{inst:0101}}
\and         M.~Manteiga\orcit{0000-0002-7711-5581}\inst{\ref{inst:0356}}
\and       J.M.~Marchant\orcit{0000-0002-3678-3145}\inst{\ref{inst:0357}}
\and         M.~Marconi\orcit{0000-0002-1330-2927}\inst{\ref{inst:0337}}
\and         J.~Marcos\inst{\ref{inst:0082}}
\and     M.M.S.~Marcos Santos\inst{\ref{inst:0097}}
\and         D.~Mar\'{i}n Pina\orcit{0000-0001-6482-1842}\inst{\ref{inst:0019}}
\and         S.~Marinoni\orcit{0000-0001-7990-6849}\inst{\ref{inst:0132},\ref{inst:0133}}
\and         F.~Marocco\orcit{0000-0001-7519-1700}\inst{\ref{inst:0364}}
\and       D.J.~Marshall\orcit{0000-0003-3956-3524}\inst{\ref{inst:0365}}
\and         L.~Martin Polo\inst{\ref{inst:0097}}
\and       J.M.~Mart\'{i}n-Fleitas\orcit{0000-0002-8594-569X}\inst{\ref{inst:0040}}
\and         G.~Marton\orcit{0000-0002-1326-1686}\inst{\ref{inst:0153}}
\and         N.~Mary\inst{\ref{inst:0181}}
\and         A.~Masip\orcit{0000-0003-1419-0020}\inst{\ref{inst:0019}}
\and         D.~Massari\orcit{0000-0001-8892-4301}\inst{\ref{inst:0018}}
\and         A.~Mastrobuono-Battisti\orcit{0000-0002-2386-9142}\inst{\ref{inst:0007}}
\and         T.~Mazeh\orcit{0000-0002-3569-3391}\inst{\ref{inst:0258}}
\and         S.~Messina\orcit{0000-0002-2851-2468}\inst{\ref{inst:0128}}
\and         D.~Michalik\orcit{0000-0002-7618-6556}\inst{\ref{inst:0029}}
\and       N.R.~Millar\inst{\ref{inst:0037}}
\and         A.~Mints\orcit{0000-0002-8440-1455}\inst{\ref{inst:0123}}
\and         D.~Molina\orcit{0000-0003-4814-0275}\inst{\ref{inst:0019}}
\and         R.~Molinaro\orcit{0000-0003-3055-6002}\inst{\ref{inst:0337}}
\and         L.~Moln\'{a}r\orcit{0000-0002-8159-1599}\inst{\ref{inst:0153},\ref{inst:0381},\ref{inst:0154}}
\and         G.~Monari\orcit{0000-0002-6863-0661}\inst{\ref{inst:0114}}
\and         M.~Mongui\'{o}\orcit{0000-0002-4519-6700}\inst{\ref{inst:0019}}
\and         P.~Montegriffo\orcit{0000-0001-5013-5948}\inst{\ref{inst:0018}}
\and         A.~Montero\inst{\ref{inst:0040}}
\and         R.~Mor\orcit{0000-0002-8179-6527}\inst{\ref{inst:0019}}
\and         A.~Mora\inst{\ref{inst:0040}}
\and         R.~Morbidelli\orcit{0000-0001-7627-4946}\inst{\ref{inst:0005}}
\and         T.~Morel\orcit{0000-0002-8176-4816}\inst{\ref{inst:0099}}
\and         D.~Morris\inst{\ref{inst:0101}}
\and         T.~Muraveva\orcit{0000-0002-0969-1915}\inst{\ref{inst:0018}}
\and       C.P.~Murphy\inst{\ref{inst:0039}}
\and         I.~Musella\orcit{0000-0001-5909-6615}\inst{\ref{inst:0337}}
\and         Z.~Nagy\orcit{0000-0002-3632-1194}\inst{\ref{inst:0153}}
\and         L.~Noval\inst{\ref{inst:0181}}
\and         F.~Oca\~{n}a\inst{\ref{inst:0082},\ref{inst:0397}}
\and         A.~Ogden\inst{\ref{inst:0037}}
\and         C.~Ordenovic\inst{\ref{inst:0001}}
\and       J.O.~Osinde\inst{\ref{inst:0106}}
\and         C.~Pagani\orcit{0000-0001-5477-4720}\inst{\ref{inst:0179}}
\and         I.~Pagano\orcit{0000-0001-9573-4928}\inst{\ref{inst:0128}}
\and         L.~Palaversa\orcit{0000-0003-3710-0331}\inst{\ref{inst:0403},\ref{inst:0037}}
\and         L.~Pallas-Quintela\orcit{0000-0001-9296-3100}\inst{\ref{inst:0109}}
\and         A.~Panahi\orcit{0000-0001-5850-4373}\inst{\ref{inst:0258}}
\and         S.~Payne-Wardenaar\inst{\ref{inst:0034}}
\and         X.~Pe\~{n}alosa Esteller\inst{\ref{inst:0019}}
\and         A.~Penttil\"{ a}\orcit{0000-0001-7403-1721}\inst{\ref{inst:0135}}
\and         B.~Pichon\orcit{0000 0000 0062 1449}\inst{\ref{inst:0001}}
\and       A.M.~Piersimoni\orcit{0000-0002-8019-3708}\inst{\ref{inst:0236}}
\and      F.-X.~Pineau\orcit{0000-0002-2335-4499}\inst{\ref{inst:0114}}
\and         E.~Plachy\orcit{0000-0002-5481-3352}\inst{\ref{inst:0153},\ref{inst:0381},\ref{inst:0154}}
\and         G.~Plum\inst{\ref{inst:0007}}
\and         A.~Pr\v{s}a\orcit{0000-0002-1913-0281}\inst{\ref{inst:0416}}
\and         L.~Pulone\orcit{0000-0002-5285-998X}\inst{\ref{inst:0132}}
\and         E.~Racero\orcit{0000-0002-6101-9050}\inst{\ref{inst:0097},\ref{inst:0397}}
\and         S.~Ragaini\inst{\ref{inst:0018}}
\and         M.~Rainer\orcit{0000-0002-8786-2572}\inst{\ref{inst:0050},\ref{inst:0422}}
\and       C.M.~Raiteri\orcit{0000-0003-1784-2784}\inst{\ref{inst:0005}}
\and         P.~Ramos\orcit{0000-0002-5080-7027}\inst{\ref{inst:0019},\ref{inst:0114}}
\and         M.~Ramos-Lerate\inst{\ref{inst:0082}}
\and         S.~Regibo\inst{\ref{inst:0156}}
\and       P.J.~Richards\inst{\ref{inst:0428}}
\and         C.~Rios Diaz\inst{\ref{inst:0106}}
\and         V.~Ripepi\orcit{0000-0003-1801-426X}\inst{\ref{inst:0337}}
\and         A.~Riva\orcit{0000-0002-6928-8589}\inst{\ref{inst:0005}}
\and      H.-W.~Rix\orcit{0000-0003-4996-9069}\inst{\ref{inst:0020}}
\and         G.~Rixon\orcit{0000-0003-4399-6568}\inst{\ref{inst:0037}}
\and         N.~Robichon\orcit{0000-0003-4545-7517}\inst{\ref{inst:0007}}
\and       A.C.~Robin\orcit{0000-0001-8654-9499}\inst{\ref{inst:0141}}
\and         C.~Robin\inst{\ref{inst:0181}}
\and         M.~Roelens\orcit{0000-0003-0876-4673}\inst{\ref{inst:0038}}
\and     H.R.O.~Rogues\inst{\ref{inst:0246}}
\and         L.~Rohrbasser\inst{\ref{inst:0072}}
\and         M.~Romero-G\'{o}mez\orcit{0000-0003-3936-1025}\inst{\ref{inst:0019}}
\and         N.~Rowell\orcit{0000-0003-3809-1895}\inst{\ref{inst:0101}}
\and         F.~Royer\orcit{0000-0002-9374-8645}\inst{\ref{inst:0007}}
\and         D.~Ruz Mieres\orcit{0000-0002-9455-157X}\inst{\ref{inst:0037}}
\and       K.A.~Rybicki\orcit{0000-0002-9326-9329}\inst{\ref{inst:0327}}
\and         G.~Sadowski\orcit{0000-0002-3411-1003}\inst{\ref{inst:0048}}
\and         A.~S\'{a}ez N\'{u}\~{n}ez\inst{\ref{inst:0019}}
\and         A.~Sagrist\`{a} Sell\'{e}s\orcit{0000-0001-6191-2028}\inst{\ref{inst:0034}}
\and         J.~Sahlmann\orcit{0000-0001-9525-3673}\inst{\ref{inst:0106}}
\and         E.~Salguero\inst{\ref{inst:0107}}
\and         N.~Samaras\orcit{0000-0001-8375-6652}\inst{\ref{inst:0064},\ref{inst:0451}}
\and         V.~Sanchez Gimenez\orcit{0000-0003-1797-3557}\inst{\ref{inst:0019}}
\and         N.~Sanna\orcit{0000-0001-9275-9492}\inst{\ref{inst:0050}}
\and         R.~Santove\~{n}a\orcit{0000-0002-9257-2131}\inst{\ref{inst:0109}}
\and         M.~Sarasso\orcit{0000-0001-5121-0727}\inst{\ref{inst:0005}}
\and         E.~Sciacca\orcit{0000-0002-5574-2787}\inst{\ref{inst:0128}}
\and         M.~Segol\inst{\ref{inst:0246}}
\and       J.C.~Segovia\inst{\ref{inst:0097}}
\and         D.~S\'{e}gransan\orcit{0000-0003-2355-8034}\inst{\ref{inst:0038}}
\and         D.~Semeux\inst{\ref{inst:0255}}
\and         S.~Shahaf\orcit{0000-0001-9298-8068}\inst{\ref{inst:0462}}
\and       H.I.~Siddiqui\orcit{0000-0003-1853-6033}\inst{\ref{inst:0463}}
\and         A.~Siebert\orcit{0000-0001-8059-2840}\inst{\ref{inst:0114},\ref{inst:0262}}
\and         L.~Siltala\orcit{0000-0002-6938-794X}\inst{\ref{inst:0135}}
\and         A.~Silvelo\orcit{0000-0002-5126-6365}\inst{\ref{inst:0109}}
\and         E.~Slezak\inst{\ref{inst:0001}}
\and         I.~Slezak\inst{\ref{inst:0001}}
\and       R.L.~Smart\orcit{0000-0002-4424-4766}\inst{\ref{inst:0005}}
\and       O.N.~Snaith\inst{\ref{inst:0007}}
\and         E.~Solano\inst{\ref{inst:0472}}
\and         F.~Solitro\inst{\ref{inst:0068}}
\and         D.~Souami\orcit{0000-0003-4058-0815}\inst{\ref{inst:0318},\ref{inst:0475}}
\and         J.~Souchay\inst{\ref{inst:0084}}
\and         F.~Spoto\orcit{0000-0001-7319-5847}\inst{\ref{inst:0192}}
\and       I.A.~Steele\orcit{0000-0001-8397-5759}\inst{\ref{inst:0357}}
\and         H.~Steidelm\"{ u}ller\inst{\ref{inst:0042}}
\and       C.A.~Stephenson\inst{\ref{inst:0082},\ref{inst:0481}}
\and         M.~S\"{ u}veges\orcit{0000-0003-3017-5322}\inst{\ref{inst:0482}}
\and         J.~Surdej\orcit{0000-0002-7005-1976}\inst{\ref{inst:0099},\ref{inst:0484}}
\and         L.~Szabados\orcit{0000-0002-2046-4131}\inst{\ref{inst:0153}}
\and         E.~Szegedi-Elek\orcit{0000-0001-7807-6644}\inst{\ref{inst:0153}}
\and         F.~Taris\inst{\ref{inst:0084}}
\and       M.B.~Taylor\orcit{0000-0002-4209-1479}\inst{\ref{inst:0488}}
\and         R.~Teixeira\orcit{0000-0002-6806-6626}\inst{\ref{inst:0239}}
\and         L.~Tolomei\orcit{0000-0002-3541-3230}\inst{\ref{inst:0068}}
\and         N.~Tonello\orcit{0000-0003-0550-1667}\inst{\ref{inst:0280}}
\and         F.~Torra\orcit{0000-0002-8429-299X}\inst{\ref{inst:0061}}
\and         J.~Torra$^\dagger$\inst{\ref{inst:0019}}
\and         G.~Torralba Elipe\orcit{0000-0001-8738-194X}\inst{\ref{inst:0109}}
\and         M.~Trabucchi\orcit{0000-0002-1429-2388}\inst{\ref{inst:0495},\ref{inst:0038}}
\and       A.T.~Tsounis\inst{\ref{inst:0497}}
\and         C.~Turon\orcit{0000-0003-1236-5157}\inst{\ref{inst:0007}}
\and         A.~Ulla\orcit{0000-0001-6424-5005}\inst{\ref{inst:0499}}
\and         N.~Unger\orcit{0000-0003-3993-7127}\inst{\ref{inst:0038}}
\and       M.V.~Vaillant\inst{\ref{inst:0181}}
\and         E.~van Dillen\inst{\ref{inst:0246}}
\and         W.~van Reeven\inst{\ref{inst:0503}}
\and         O.~Vanel\orcit{0000-0002-7898-0454}\inst{\ref{inst:0007}}
\and         A.~Vecchiato\orcit{0000-0003-1399-5556}\inst{\ref{inst:0005}}
\and         Y.~Viala\inst{\ref{inst:0007}}
\and         D.~Vicente\orcit{0000-0002-1584-1182}\inst{\ref{inst:0280}}
\and         S.~Voutsinas\inst{\ref{inst:0101}}
\and         M.~Weiler\inst{\ref{inst:0019}}
\and         T.~Wevers\orcit{0000-0002-4043-9400}\inst{\ref{inst:0037},\ref{inst:0511}}
\and      \L{}.~Wyrzykowski\orcit{0000-0002-9658-6151}\inst{\ref{inst:0327}}
\and         A.~Yoldas\inst{\ref{inst:0037}}
\and         P.~Yvard\inst{\ref{inst:0246}}
\and         H.~Zhao\orcit{0000-0003-2645-6869}\inst{\ref{inst:0001}}
\and         J.~Zorec\inst{\ref{inst:0516}}
\and         S.~Zucker\orcit{0000-0003-3173-3138}\inst{\ref{inst:0191}}
\and         T.~Zwitter\orcit{0000-0002-2325-8763}\inst{\ref{inst:0518}}
}
\institute{
     Universit\'{e} C\^{o}te d'Azur, Observatoire de la C\^{o}te d'Azur, CNRS, Laboratoire Lagrange, Bd de l'Observatoire, CS 34229, 06304 Nice Cedex 4, France\relax                                                                                                                                                                                              \label{inst:0001}
\and INAF - Osservatorio Astrofisico di Torino, via Osservatorio 20, 10025 Pino Torinese (TO), Italy\relax                                                                                                                                                                                                                                                         \label{inst:0005}\vfill
\and INAF - Osservatorio astronomico di Padova, Vicolo Osservatorio 5, 35122 Padova, Italy\relax                                                                                                                                                                                                                                                                   \label{inst:0006}\vfill
\and GEPI, Observatoire de Paris, Universit\'{e} PSL, CNRS, 5 Place Jules Janssen, 92190 Meudon, France\relax                                                                                                                                                                                                                                                      \label{inst:0007}\vfill
\and Lund Observatory, Department of Astronomy and Theoretical Physics, Lund University, Box 43, 22100 Lund, Sweden\relax                                                                                                                                                                                                                                          \label{inst:0011}\vfill
\and University of Turin, Department of Physics, Via Pietro Giuria 1, 10125 Torino, Italy\relax                                                                                                                                                                                                                                                                    \label{inst:0014}\vfill
\and Mullard Space Science Laboratory, University College London, Holmbury St Mary, Dorking, Surrey RH5 6NT, United Kingdom\relax                                                                                                                                                                                                                                  \label{inst:0015}\vfill
\and Laboratoire d'astrophysique de Bordeaux, Univ. Bordeaux, CNRS, B18N, all{\'e}e Geoffroy Saint-Hilaire, 33615 Pessac, France\relax                                                                                                                                                                                                                             \label{inst:0016}\vfill
\and INAF - Osservatorio di Astrofisica e Scienza dello Spazio di Bologna, via Piero Gobetti 93/3, 40129 Bologna, Italy\relax                                                                                                                                                                                                                                      \label{inst:0018}\vfill
\and Institut de Ci\`{e}ncies del Cosmos (ICCUB), Universitat  de  Barcelona  (IEEC-UB), Mart\'{i} i  Franqu\`{e}s  1, 08028 Barcelona, Spain\relax                                                                                                                                                                                                                \label{inst:0019}\vfill
\and Max Planck Institute for Astronomy, K\"{ o}nigstuhl 17, 69117 Heidelberg, Germany\relax                                                                                                                                                                                                                                                                       \label{inst:0020}\vfill
\and Observational Astrophysics, Division of Astronomy and Space Physics, Department of Physics and Astronomy, Uppsala University, Box 516, 751 20 Uppsala, Sweden\relax                                                                                                                                                                                           \label{inst:0026}\vfill
\and Leiden Observatory, Leiden University, Niels Bohrweg 2, 2333 CA Leiden, The Netherlands\relax                                                                                                                                                                                                                                                                 \label{inst:0028}\vfill
\and European Space Agency (ESA), European Space Research and Technology Centre (ESTEC), Keplerlaan 1, 2201AZ, Noordwijk, The Netherlands\relax                                                                                                                                                                                                                    \label{inst:0029}\vfill
\and Univ. Grenoble Alpes, CNRS, IPAG, 38000 Grenoble, France\relax                                                                                                                                                                                                                                                                                                \label{inst:0032}\vfill
\and Astronomisches Rechen-Institut, Zentrum f\"{ u}r Astronomie der Universit\"{ a}t Heidelberg, M\"{ o}nchhofstr. 12-14, 69120 Heidelberg, Germany\relax                                                                                                                                                                                                         \label{inst:0034}\vfill
\and Institute of Astronomy, University of Cambridge, Madingley Road, Cambridge CB3 0HA, United Kingdom\relax                                                                                                                                                                                                                                                      \label{inst:0037}\vfill
\and Department of Astronomy, University of Geneva, Chemin Pegasi 51, 1290 Versoix, Switzerland\relax                                                                                                                                                                                                                                                              \label{inst:0038}\vfill
\and European Space Agency (ESA), European Space Astronomy Centre (ESAC), Camino bajo del Castillo, s/n, Urbanizacion Villafranca del Castillo, Villanueva de la Ca\~{n}ada, 28692 Madrid, Spain\relax                                                                                                                                                             \label{inst:0039}\vfill
\and Aurora Technology for European Space Agency (ESA), Camino bajo del Castillo, s/n, Urbanizacion Villafranca del Castillo, Villanueva de la Ca\~{n}ada, 28692 Madrid, Spain\relax                                                                                                                                                                               \label{inst:0040}\vfill
\and Lohrmann Observatory, Technische Universit\"{ a}t Dresden, Mommsenstra{\ss}e 13, 01062 Dresden, Germany\relax                                                                                                                                                                                                                                                 \label{inst:0042}\vfill
\and CNES Centre Spatial de Toulouse, 18 avenue Edouard Belin, 31401 Toulouse Cedex 9, France\relax                                                                                                                                                                                                                                                                \label{inst:0047}\vfill
\and Institut d'Astronomie et d'Astrophysique, Universit\'{e} Libre de Bruxelles CP 226, Boulevard du Triomphe, 1050 Brussels, Belgium\relax                                                                                                                                                                                                                       \label{inst:0048}\vfill
\and F.R.S.-FNRS, Rue d'Egmont 5, 1000 Brussels, Belgium\relax                                                                                                                                                                                                                                                                                                     \label{inst:0049}\vfill
\and INAF - Osservatorio Astrofisico di Arcetri, Largo Enrico Fermi 5, 50125 Firenze, Italy\relax                                                                                                                                                                                                                                                                  \label{inst:0050}\vfill
\and European Space Agency (ESA, retired)\relax                                                                                                                                                                                                                                                                                                                    \label{inst:0057}\vfill
\and DAPCOM for Institut de Ci\`{e}ncies del Cosmos (ICCUB), Universitat  de  Barcelona  (IEEC-UB), Mart\'{i} i  Franqu\`{e}s  1, 08028 Barcelona, Spain\relax                                                                                                                                                                                                     \label{inst:0061}\vfill
\and Royal Observatory of Belgium, Ringlaan 3, 1180 Brussels, Belgium\relax                                                                                                                                                                                                                                                                                        \label{inst:0064}\vfill
\and ALTEC S.p.a, Corso Marche, 79,10146 Torino, Italy\relax                                                                                                                                                                                                                                                                                                       \label{inst:0068}\vfill
\and S\`{a}rl, Geneva, Switzerland\relax                                                                                                                                                                                                                                                                                                                           \label{inst:0071}\vfill
\and Department of Astronomy, University of Geneva, Chemin d'Ecogia 16, 1290 Versoix, Switzerland\relax                                                                                                                                                                                                                                                            \label{inst:0072}\vfill
\and Gaia DPAC Project Office, ESAC, Camino bajo del Castillo, s/n, Urbanizacion Villafranca del Castillo, Villanueva de la Ca\~{n}ada, 28692 Madrid, Spain\relax                                                                                                                                                                                                  \label{inst:0079}\vfill
\and Telespazio UK S.L. for European Space Agency (ESA), Camino bajo del Castillo, s/n, Urbanizacion Villafranca del Castillo, Villanueva de la Ca\~{n}ada, 28692 Madrid, Spain\relax                                                                                                                                                                              \label{inst:0082}\vfill
\and SYRTE, Observatoire de Paris, Universit\'{e} PSL, CNRS,  Sorbonne Universit\'{e}, LNE, 61 avenue de l'Observatoire 75014 Paris, France\relax                                                                                                                                                                                                                  \label{inst:0084}\vfill
\and National Observatory of Athens, I. Metaxa and Vas. Pavlou, Palaia Penteli, 15236 Athens, Greece\relax                                                                                                                                                                                                                                                         \label{inst:0087}\vfill
\and IMCCE, Observatoire de Paris, Universit\'{e} PSL, CNRS, Sorbonne Universit{\'e}, Univ. Lille, 77 av. Denfert-Rochereau, 75014 Paris, France\relax                                                                                                                                                                                                             \label{inst:0089}\vfill
\and Serco Gesti\'{o}n de Negocios for European Space Agency (ESA), Camino bajo del Castillo, s/n, Urbanizacion Villafranca del Castillo, Villanueva de la Ca\~{n}ada, 28692 Madrid, Spain\relax                                                                                                                                                                   \label{inst:0097}\vfill
\and Institut d'Astrophysique et de G\'{e}ophysique, Universit\'{e} de Li\`{e}ge, 19c, All\'{e}e du 6 Ao\^{u}t, B-4000 Li\`{e}ge, Belgium\relax                                                                                                                                                                                                                    \label{inst:0099}\vfill
\and CRAAG - Centre de Recherche en Astronomie, Astrophysique et G\'{e}ophysique, Route de l'Observatoire Bp 63 Bouzareah 16340 Algiers, Algeria\relax                                                                                                                                                                                                             \label{inst:0100}\vfill
\and Institute for Astronomy, University of Edinburgh, Royal Observatory, Blackford Hill, Edinburgh EH9 3HJ, United Kingdom\relax                                                                                                                                                                                                                                  \label{inst:0101}\vfill
\and RHEA for European Space Agency (ESA), Camino bajo del Castillo, s/n, Urbanizacion Villafranca del Castillo, Villanueva de la Ca\~{n}ada, 28692 Madrid, Spain\relax                                                                                                                                                                                            \label{inst:0106}\vfill
\and ATG Europe for European Space Agency (ESA), Camino bajo del Castillo, s/n, Urbanizacion Villafranca del Castillo, Villanueva de la Ca\~{n}ada, 28692 Madrid, Spain\relax                                                                                                                                                                                      \label{inst:0107}\vfill
\and CIGUS CITIC - Department of Computer Science and Information Technologies, University of A Coru\~{n}a, Campus de Elvi\~{n}a s/n, A Coru\~{n}a, 15071, Spain\relax                                                                                                                                                                                             \label{inst:0109}\vfill
\and Universit\'{e} de Strasbourg, CNRS, Observatoire astronomique de Strasbourg, UMR 7550,  11 rue de l'Universit\'{e}, 67000 Strasbourg, France\relax                                                                                                                                                                                                            \label{inst:0114}\vfill
\and Kavli Institute for Cosmology Cambridge, Institute of Astronomy, Madingley Road, Cambridge, CB3 0HA\relax                                                                                                                                                                                                                                                     \label{inst:0117}\vfill
\and Leibniz Institute for Astrophysics Potsdam (AIP), An der Sternwarte 16, 14482 Potsdam, Germany\relax                                                                                                                                                                                                                                                          \label{inst:0123}\vfill
\and CENTRA, Faculdade de Ci\^{e}ncias, Universidade de Lisboa, Edif. C8, Campo Grande, 1749-016 Lisboa, Portugal\relax                                                                                                                                                                                                                                            \label{inst:0126}\vfill
\and Department of Informatics, Donald Bren School of Information and Computer Sciences, University of California, Irvine, 5226 Donald Bren Hall, 92697-3440 CA Irvine, United States\relax                                                                                                                                                                        \label{inst:0127}\vfill
\and INAF - Osservatorio Astrofisico di Catania, via S. Sofia 78, 95123 Catania, Italy\relax                                                                                                                                                                                                                                                                       \label{inst:0128}\vfill
\and Dipartimento di Fisica e Astronomia ""Ettore Majorana"", Universit\`{a} di Catania, Via S. Sofia 64, 95123 Catania, Italy\relax                                                                                                                                                                                                                               \label{inst:0129}\vfill
\and INAF - Osservatorio Astronomico di Roma, Via Frascati 33, 00078 Monte Porzio Catone (Roma), Italy\relax                                                                                                                                                                                                                                                       \label{inst:0132}\vfill
\and Space Science Data Center - ASI, Via del Politecnico SNC, 00133 Roma, Italy\relax                                                                                                                                                                                                                                                                             \label{inst:0133}\vfill
\and Department of Physics, University of Helsinki, P.O. Box 64, 00014 Helsinki, Finland\relax                                                                                                                                                                                                                                                                     \label{inst:0135}\vfill
\and Finnish Geospatial Research Institute FGI, Geodeetinrinne 2, 02430 Masala, Finland\relax                                                                                                                                                                                                                                                                      \label{inst:0136}\vfill
\and Institut UTINAM CNRS UMR6213, Universit\'{e} Bourgogne Franche-Comt\'{e}, OSU THETA Franche-Comt\'{e} Bourgogne, Observatoire de Besan\c{c}on, BP1615, 25010 Besan\c{c}on Cedex, France\relax                                                                                                                                                                 \label{inst:0141}\vfill
\and HE Space Operations BV for European Space Agency (ESA), Keplerlaan 1, 2201AZ, Noordwijk, The Netherlands\relax                                                                                                                                                                                                                                                \label{inst:0144}\vfill
\and Dpto. de Inteligencia Artificial, UNED, c/ Juan del Rosal 16, 28040 Madrid, Spain\relax                                                                                                                                                                                                                                                                       \label{inst:0146}\vfill
\and Konkoly Observatory, Research Centre for Astronomy and Earth Sciences, E\"{ o}tv\"{ o}s Lor{\'a}nd Research Network (ELKH), MTA Centre of Excellence, Konkoly Thege Mikl\'{o}s \'{u}t 15-17, 1121 Budapest, Hungary\relax                                                                                                                                     \label{inst:0153}\vfill
\and ELTE E\"{ o}tv\"{ o}s Lor\'{a}nd University, Institute of Physics, 1117, P\'{a}zm\'{a}ny P\'{e}ter s\'{e}t\'{a}ny 1A, Budapest, Hungary\relax                                                                                                                                                                                                                 \label{inst:0154}\vfill
\and Instituut voor Sterrenkunde, KU Leuven, Celestijnenlaan 200D, 3001 Leuven, Belgium\relax                                                                                                                                                                                                                                                                      \label{inst:0156}\vfill
\and Department of Astrophysics/IMAPP, Radboud University, P.O.Box 9010, 6500 GL Nijmegen, The Netherlands\relax                                                                                                                                                                                                                                                   \label{inst:0157}\vfill
\and University of Vienna, Department of Astrophysics, T\"{ u}rkenschanzstra{\ss}e 17, A1180 Vienna, Austria\relax                                                                                                                                                                                                                                                 \label{inst:0165}\vfill
\and Institute of Physics, Laboratory of Astrophysics, Ecole Polytechnique F\'ed\'erale de Lausanne (EPFL), Observatoire de Sauverny, 1290 Versoix, Switzerland\relax                                                                                                                                                                                              \label{inst:0167}\vfill
\and Kapteyn Astronomical Institute, University of Groningen, Landleven 12, 9747 AD Groningen, The Netherlands\relax                                                                                                                                                                                                                                               \label{inst:0172}\vfill
\and School of Physics and Astronomy / Space Park Leicester, University of Leicester, University Road, Leicester LE1 7RH, United Kingdom\relax                                                                                                                                                                                                                     \label{inst:0179}\vfill
\and Thales Services for CNES Centre Spatial de Toulouse, 18 avenue Edouard Belin, 31401 Toulouse Cedex 9, France\relax                                                                                                                                                                                                                                            \label{inst:0181}\vfill
\and Depto. Estad\'istica e Investigaci\'on Operativa. Universidad de C\'adiz, Avda. Rep\'ublica Saharaui s/n, 11510 Puerto Real, C\'adiz, Spain\relax                                                                                                                                                                                                             \label{inst:0185}\vfill
\and Center for Research and Exploration in Space Science and Technology, University of Maryland Baltimore County, 1000 Hilltop Circle, Baltimore MD, USA\relax                                                                                                                                                                                                    \label{inst:0187}\vfill
\and GSFC - Goddard Space Flight Center, Code 698, 8800 Greenbelt Rd, 20771 MD Greenbelt, United States\relax                                                                                                                                                                                                                                                      \label{inst:0188}\vfill
\and EURIX S.r.l., Corso Vittorio Emanuele II 61, 10128, Torino, Italy\relax                                                                                                                                                                                                                                                                                       \label{inst:0190}\vfill
\and Porter School of the Environment and Earth Sciences, Tel Aviv University, Tel Aviv 6997801, Israel\relax                                                                                                                                                                                                                                                      \label{inst:0191}\vfill
\and Harvard-Smithsonian Center for Astrophysics, 60 Garden St., MS 15, Cambridge, MA 02138, USA\relax                                                                                                                                                                                                                                                             \label{inst:0192}\vfill
\and HE Space Operations BV for European Space Agency (ESA), Camino bajo del Castillo, s/n, Urbanizacion Villafranca del Castillo, Villanueva de la Ca\~{n}ada, 28692 Madrid, Spain\relax                                                                                                                                                                          \label{inst:0194}\vfill
\and Instituto de Astrof\'{i}sica e Ci\^{e}ncias do Espa\c{c}o, Universidade do Porto, CAUP, Rua das Estrelas, PT4150-762 Porto, Portugal\relax                                                                                                                                                                                                                    \label{inst:0195}\vfill
\and LFCA/DAS,Universidad de Chile,CNRS,Casilla 36-D, Santiago, Chile\relax                                                                                                                                                                                                                                                                                        \label{inst:0197}\vfill
\and SISSA - Scuola Internazionale Superiore di Studi Avanzati, via Bonomea 265, 34136 Trieste, Italy\relax                                                                                                                                                                                                                                                        \label{inst:0200}\vfill
\and Telespazio for CNES Centre Spatial de Toulouse, 18 avenue Edouard Belin, 31401 Toulouse Cedex 9, France\relax                                                                                                                                                                                                                                                 \label{inst:0205}\vfill
\and University of Turin, Department of Computer Sciences, Corso Svizzera 185, 10149 Torino, Italy\relax                                                                                                                                                                                                                                                           \label{inst:0209}\vfill
\and Dpto. de Matem\'{a}tica Aplicada y Ciencias de la Computaci\'{o}n, Univ. de Cantabria, ETS Ingenieros de Caminos, Canales y Puertos, Avda. de los Castros s/n, 39005 Santander, Spain\relax                                                                                                                                                                   \label{inst:0210}\vfill
\and Centro de Astronom\'{i}a - CITEVA, Universidad de Antofagasta, Avenida Angamos 601, Antofagasta 1270300, Chile\relax                                                                                                                                                                                                                                          \label{inst:0218}\vfill
\and DLR Gesellschaft f\"{ u}r Raumfahrtanwendungen (GfR) mbH M\"{ u}nchener Stra{\ss}e 20 , 82234 We{\ss}ling\relax                                                                                                                                                                                                                                               \label{inst:0223}\vfill
\and Centre for Astrophysics Research, University of Hertfordshire, College Lane, AL10 9AB, Hatfield, United Kingdom\relax                                                                                                                                                                                                                                         \label{inst:0225}\vfill
\and University of Turin, Mathematical Department ""G.Peano"", Via Carlo Alberto 10, 10123 Torino, Italy\relax                                                                                                                                                                                                                                                     \label{inst:0231}\vfill
\and INAF - Osservatorio Astronomico d'Abruzzo, Via Mentore Maggini, 64100 Teramo, Italy\relax                                                                                                                                                                                                                                                                     \label{inst:0236}\vfill
\and Instituto de Astronomia, Geof\`{i}sica e Ci\^{e}ncias Atmosf\'{e}ricas, Universidade de S\~{a}o Paulo, Rua do Mat\~{a}o, 1226, Cidade Universitaria, 05508-900 S\~{a}o Paulo, SP, Brazil\relax                                                                                                                                                                \label{inst:0239}\vfill
\and APAVE SUDEUROPE SAS for CNES Centre Spatial de Toulouse, 18 avenue Edouard Belin, 31401 Toulouse Cedex 9, France\relax                                                                                                                                                                                                                                        \label{inst:0246}\vfill
\and M\'{e}socentre de calcul de Franche-Comt\'{e}, Universit\'{e} de Franche-Comt\'{e}, 16 route de Gray, 25030 Besan\c{c}on Cedex, France\relax                                                                                                                                                                                                                  \label{inst:0249}\vfill
\and Theoretical Astrophysics, Division of Astronomy and Space Physics, Department of Physics and Astronomy, Uppsala University, Box 516, 751 20 Uppsala, Sweden\relax                                                                                                                                                                                             \label{inst:0253}\vfill
\and ATOS for CNES Centre Spatial de Toulouse, 18 avenue Edouard Belin, 31401 Toulouse Cedex 9, France\relax                                                                                                                                                                                                                                                       \label{inst:0255}\vfill
\and School of Physics and Astronomy, Tel Aviv University, Tel Aviv 6997801, Israel\relax                                                                                                                                                                                                                                                                          \label{inst:0258}\vfill
\and Astrophysics Research Centre, School of Mathematics and Physics, Queen's University Belfast, Belfast BT7 1NN, UK\relax                                                                                                                                                                                                                                        \label{inst:0260}\vfill
\and Centre de Donn\'{e}es Astronomique de Strasbourg, Strasbourg, France\relax                                                                                                                                                                                                                                                                                    \label{inst:0262}\vfill
\and Institute for Computational Cosmology, Department of Physics, Durham University, Durham DH1 3LE, UK\relax                                                                                                                                                                                                                                                     \label{inst:0266}\vfill
\and European Southern Observatory, Karl-Schwarzschild-Str. 2, 85748 Garching, Germany\relax                                                                                                                                                                                                                                                                       \label{inst:0267}\vfill
\and Max-Planck-Institut f\"{ u}r Astrophysik, Karl-Schwarzschild-Stra{\ss}e 1, 85748 Garching, Germany\relax                                                                                                                                                                                                                                                      \label{inst:0268}\vfill
\and Data Science and Big Data Lab, Pablo de Olavide University, 41013, Seville, Spain\relax                                                                                                                                                                                                                                                                       \label{inst:0272}\vfill
\and Barcelona Supercomputing Center (BSC), Pla\c{c}a Eusebi G\"{ u}ell 1-3, 08034-Barcelona, Spain\relax                                                                                                                                                                                                                                                          \label{inst:0280}\vfill
\and ETSE Telecomunicaci\'{o}n, Universidade de Vigo, Campus Lagoas-Marcosende, 36310 Vigo, Galicia, Spain\relax                                                                                                                                                                                                                                                   \label{inst:0285}\vfill
\and Asteroid Engineering Laboratory, Space Systems, Lule\aa{} University of Technology, Box 848, S-981 28 Kiruna, Sweden\relax                                                                                                                                                                                                                                    \label{inst:0289}\vfill
\and Vera C Rubin Observatory,  950 N. Cherry Avenue, Tucson, AZ 85719, USA\relax                                                                                                                                                                                                                                                                                  \label{inst:0294}\vfill
\and Department of Astrophysics, Astronomy and Mechanics, National and Kapodistrian University of Athens, Panepistimiopolis, Zografos, 15783 Athens, Greece\relax                                                                                                                                                                                                  \label{inst:0295}\vfill
\and TRUMPF Photonic Components GmbH, Lise-Meitner-Stra{\ss}e 13,  89081 Ulm, Germany\relax                                                                                                                                                                                                                                                                        \label{inst:0298}\vfill
\and IAC - Instituto de Astrofisica de Canarias, Via L\'{a}ctea s/n, 38200 La Laguna S.C., Tenerife, Spain\relax                                                                                                                                                                                                                                                   \label{inst:0303}\vfill
\and Department of Astrophysics, University of La Laguna, Via L\'{a}ctea s/n, 38200 La Laguna S.C., Tenerife, Spain\relax                                                                                                                                                                                                                                          \label{inst:0304}\vfill
\and Faculty of Aerospace Engineering, Delft University of Technology, Kluyverweg 1, 2629 HS Delft, The Netherlands\relax                                                                                                                                                                                                                                          \label{inst:0306}\vfill
\and Radagast Solutions\relax                                                                                                                                                                                                                                                                                                                                      \label{inst:0310}\vfill
\and Laboratoire Univers et Particules de Montpellier, CNRS Universit\'{e} Montpellier, Place Eug\`{e}ne Bataillon, CC72, 34095 Montpellier Cedex 05, France\relax                                                                                                                                                                                                 \label{inst:0311}\vfill
\and Universit\'{e} de Caen Normandie, C\^{o}te de Nacre Boulevard Mar\'{e}chal Juin, 14032 Caen, France\relax                                                                                                                                                                                                                                                     \label{inst:0317}\vfill
\and LESIA, Observatoire de Paris, Universit\'{e} PSL, CNRS, Sorbonne Universit\'{e}, Universit\'{e} de Paris, 5 Place Jules Janssen, 92190 Meudon, France\relax                                                                                                                                                                                                   \label{inst:0318}\vfill
\and SRON Netherlands Institute for Space Research, Niels Bohrweg 4, 2333 CA Leiden, The Netherlands\relax                                                                                                                                                                                                                                                         \label{inst:0326}\vfill
\and Astronomical Observatory, University of Warsaw,  Al. Ujazdowskie 4, 00-478 Warszawa, Poland\relax                                                                                                                                                                                                                                                             \label{inst:0327}\vfill
\and Scalian for CNES Centre Spatial de Toulouse, 18 avenue Edouard Belin, 31401 Toulouse Cedex 9, France\relax                                                                                                                                                                                                                                                    \label{inst:0329}\vfill
\and Universit\'{e} Rennes, CNRS, IPR (Institut de Physique de Rennes) - UMR 6251, 35000 Rennes, France\relax                                                                                                                                                                                                                                                      \label{inst:0335}\vfill
\and INAF - Osservatorio Astronomico di Capodimonte, Via Moiariello 16, 80131, Napoli, Italy\relax                                                                                                                                                                                                                                                                 \label{inst:0337}\vfill
\and Shanghai Astronomical Observatory, Chinese Academy of Sciences, 80 Nandan Road, Shanghai 200030, People's Republic of China\relax                                                                                                                                                                                                                             \label{inst:0340}\vfill
\and University of Chinese Academy of Sciences, No.19(A) Yuquan Road, Shijingshan District, Beijing 100049, People's Republic of China\relax                                                                                                                                                                                                                       \label{inst:0342}\vfill
\and Niels Bohr Institute, University of Copenhagen, Juliane Maries Vej 30, 2100 Copenhagen {\O}, Denmark\relax                                                                                                                                                                                                                                                    \label{inst:0345}\vfill
\and DXC Technology, Retortvej 8, 2500 Valby, Denmark\relax                                                                                                                                                                                                                                                                                                        \label{inst:0346}\vfill
\and Las Cumbres Observatory, 6740 Cortona Drive Suite 102, Goleta, CA 93117, USA\relax                                                                                                                                                                                                                                                                            \label{inst:0347}\vfill
\and CIGUS CITIC, Department of Nautical Sciences and Marine Engineering, University of A Coru\~{n}a, Paseo de Ronda 51, 15071, A Coru\~{n}a, Spain\relax                                                                                                                                                                                                          \label{inst:0356}\vfill
\and Astrophysics Research Institute, Liverpool John Moores University, 146 Brownlow Hill, Liverpool L3 5RF, United Kingdom\relax                                                                                                                                                                                                                                  \label{inst:0357}\vfill
\and IPAC, Mail Code 100-22, California Institute of Technology, 1200 E. California Blvd., Pasadena, CA 91125, USA\relax                                                                                                                                                                                                                                           \label{inst:0364}\vfill
\and IRAP, Universit\'{e} de Toulouse, CNRS, UPS, CNES, 9 Av. colonel Roche, BP 44346, 31028 Toulouse Cedex 4, France\relax                                                                                                                                                                                                                                        \label{inst:0365}\vfill
\and MTA CSFK Lend\"{ u}let Near-Field Cosmology Research Group, Konkoly Observatory, MTA Research Centre for Astronomy and Earth Sciences, Konkoly Thege Mikl\'{o}s \'{u}t 15-17, 1121 Budapest, Hungary\relax                                                                                                                                                    \label{inst:0381}\vfill
\and Departmento de F\'{i}sica de la Tierra y Astrof\'{i}sica, Universidad Complutense de Madrid, 28040 Madrid, Spain\relax                                                                                                                                                                                                                                        \label{inst:0397}\vfill
\and Ru{\dj}er Bo\v{s}kovi\'{c} Institute, Bijeni\v{c}ka cesta 54, 10000 Zagreb, Croatia\relax                                                                                                                                                                                                                                                                     \label{inst:0403}\vfill
\and Villanova University, Department of Astrophysics and Planetary Science, 800 E Lancaster Avenue, Villanova PA 19085, USA\relax                                                                                                                                                                                                                                 \label{inst:0416}\vfill
\and INAF - Osservatorio Astronomico di Brera, via E. Bianchi, 46, 23807 Merate (LC), Italy\relax                                                                                                                                                                                                                                                                  \label{inst:0422}\vfill
\and STFC, Rutherford Appleton Laboratory, Harwell, Didcot, OX11 0QX, United Kingdom\relax                                                                                                                                                                                                                                                                         \label{inst:0428}\vfill
\and Charles University, Faculty of Mathematics and Physics, Astronomical Institute of Charles University, V Holesovickach 2, 18000 Prague, Czech Republic\relax                                                                                                                                                                                                   \label{inst:0451}\vfill
\and Department of Particle Physics and Astrophysics, Weizmann Institute of Science, Rehovot 7610001, Israel\relax                                                                                                                                                                                                                                                 \label{inst:0462}\vfill
\and Department of Astrophysical Sciences, 4 Ivy Lane, Princeton University, Princeton NJ 08544, USA\relax                                                                                                                                                                                                                                                         \label{inst:0463}\vfill
\and Departamento de Astrof\'{i}sica, Centro de Astrobiolog\'{i}a (CSIC-INTA), ESA-ESAC. Camino Bajo del Castillo s/n. 28692 Villanueva de la Ca\~{n}ada, Madrid, Spain\relax                                                                                                                                                                                      \label{inst:0472}\vfill
\and naXys, University of Namur, Rempart de la Vierge, 5000 Namur, Belgium\relax                                                                                                                                                                                                                                                                                   \label{inst:0475}\vfill
\and CGI Deutschland B.V. \& Co. KG, Mornewegstr. 30, 64293 Darmstadt, Germany\relax                                                                                                                                                                                                                                                                               \label{inst:0481}\vfill
\and Institute of Global Health, University of Geneva\relax                                                                                                                                                                                                                                                                                                        \label{inst:0482}\vfill
\and Astronomical Observatory Institute, Faculty of Physics, Adam Mickiewicz University, Pozna\'{n}, Poland\relax                                                                                                                                                                                                                                                  \label{inst:0484}\vfill
\and H H Wills Physics Laboratory, University of Bristol, Tyndall Avenue, Bristol BS8 1TL, United Kingdom\relax                                                                                                                                                                                                                                                    \label{inst:0488}\vfill
\and Department of Physics and Astronomy G. Galilei, University of Padova, Vicolo dell'Osservatorio 3, 35122, Padova, Italy\relax                                                                                                                                                                                                                                  \label{inst:0495}\vfill
\and CERN, Geneva, Switzerland\relax                                                                                                                                                                                                                                                                                                                               \label{inst:0497}\vfill
\and Applied Physics Department, Universidade de Vigo, 36310 Vigo, Spain\relax                                                                                                                                                                                                                                                                                     \label{inst:0499}\vfill
\and Association of Universities for Research in Astronomy, 1331 Pennsylvania Ave. NW, Washington, DC 20004, USA\relax                                                                                                                                                                                                                                             \label{inst:0503}\vfill
\and European Southern Observatory, Alonso de C\'ordova 3107, Casilla 19, Santiago, Chile\relax                                                                                                                                                                                                                                                                    \label{inst:0511}\vfill
\and Sorbonne Universit\'{e}, CNRS, UMR7095, Institut d'Astrophysique de Paris, 98bis bd. Arago, 75014 Paris, France\relax                                                                                                                                                                                                                                         \label{inst:0516}\vfill
\and Faculty of Mathematics and Physics, University of Ljubljana, Jadranska ulica 19, 1000 Ljubljana, Slovenia\relax                                                                                                                                                                                                                                               \label{inst:0518}\vfill
}